\documentclass[structabstract]{aa}  
\usepackage{graphicx}

\usepackage{txfonts}
\usepackage{natbib} % \citep{} etc.
\usepackage{longtable}
\usepackage{hyperref} % for \url{}
\usepackage{amstext} % for \text{} in equations
\bibpunct{(}{)}{,}{a}{}{,} %%% !!! This is to remove coma between Author and year !!!
\makeindex

\begin{document}

   \title{A VLBA survey of the core shift effect in AGN jets}
   \subtitle{I. Evidence of dominating synchrotron opacity}

   \author{
           K.~V.~Sokolovsky\inst{1,2}
           \and
           Y.~Y.~Kovalev\inst{2,1}
           \and
           A.~B.~Pushkarev\inst{1,3,4}
           \and
           A.~P.~Lobanov\inst{1}
          }

   \institute{
           Max-Planck-Institut f\"ur Radioastronomie, Auf dem H\"ugel 69, D-53121 Bonn, Germany
           \and
           Astro Space Center of Lebedev Physical Institute, Profsoyuznaya Str. 84/32, 117997 Moscow, Russia
           \and
           Pulkovo Astronomical Observatory, Pulkovskoe Chaussee 65/1, 196140 St. Petersburg, Russia
           \and
           Crimean Astrophysical Observatory, 98409 Nauchny, Crimea, Ukraine
           }

   \offprints{K.~V.~Sokolovsky\\ \email{ksokolov@mpifr-bonn.mpg.de}}

   \date{Received November 4, 2010; accepted XXXX XX, 2011}

% \abstract{}{}{}{}{} 
% 5 {} token are mandatory
 
  \abstract
   {
The effect of a frequency dependent shift of the VLBI core position (known
as the ``core shift'') was predicted more than three decades ago and 
has since been observed in a few sources, but often within a narrow frequency range.
This effect has important astrophysical and astrometric applications.
}
   {
To achieve a broader understanding of the core shift effect 
and the physics behind it, we conducted a dedicated survey 
with NRAO's Very Long Baseline Array (VLBA).
}
   {
We used the VLBA to image 20 pre-selected sources simultaneously at 
nine frequencies in the $1.4$--$15.4$~GHz range. The core position at each
frequency was measured by referencing it to a bright, optically
thin feature in the jet. 
}
   {
A significant core shift has been successfully measured in each of the twenty sources observed. 
The median value of the core shift is found to be $1.21$~mas if measured between $1.4$
and $15.4$~GHz, and $0.24$~mas between $5.0$ and $15.4$~GHz.
The core position, $r_c$, as a function of frequency, $\nu$,
is found to be consistent with an $r_c \propto \nu^{-1}$ law. 
This behavior is predicted by the Blandford \& K\"onigl model of 
a purely synchrotron self-absorbed conical jet in equipartition. 
No systematic deviation from unity of the power law index in
the $r_c(\nu)$ relation has been convincingly detected. 
}
   {
We conclude that neither free-free absorption nor 
gradients in pressure and/or density in the jet itself and in the ambient medium surrounding the jet 
play a significant role in the sources observed within
the $1.4$--$15.4$~GHz frequency range.
These results support the interpretation of the parsec-scale
core as a continuous Blandford-K\"onigl type jet with smooth gradients of
physical properties along it.
}


\maketitle
%
%________________________________________________________________

\section{Introduction}
\label{sec:intro}

In VLBI images of relativistic AGN jets, the ``core'' is the most
compact feature near the apparent base of the jet. It is identified with the
surface at which the optical depth $\tau_\nu \approx 1$ (photosphere) in 
a continuous jet flow with gradually changing physical properties \citep{1979ApJ...232...34B}. 
The position of this surface is a function of the observation frequency, $\nu$,
and the exact form of this function depends on the absorption mechanism.
The above interpretation of the core is supported by observations of 
the ``core shift'' effect serendipitously discovered by
\cite{1984ApJ...276...56M} and later investigated by, among others, 
\cite{1995ApJ...443...35Z}, \cite{1998A&A...330...79L},
\cite{2008A&A...483..759K}, and \cite{2009MNRAS.400...26O}.
Additional support comes from observations of the related increase in 
the apparent core size toward lower frequencies
\citep{1994ApJ...432..103U,2008arXiv0811.2926Y}.

An alternative interpretation of the VLBI core as a standing shock in the jet
is discussed by \cite{2006AIPC..856....1M,2008ASPC..386..437M}. This interpretation may be relevant for
high-frequency observations of some sources. The standing shock model does
not predict a shift in the apparent core position with frequency unless
one resorts to an assumption that there is a series of multiple standing
shocks, and different shocks are observed as the core at different frequencies.
In some objects the core could be a region where the jet bends such that it
is more aligned with the line of sight leading to an increased Doppler
beaming factor relative to jet regions farther upstream \citep{2006AIPC..856....1M}. 
It has even been discussed sometimes, that the radio core may not be part of the jet at all.

The most likely absorption mechanism acting in the compact jet is synchrotron
self-absorption \citep{1981ApJ...243..700K}. If the jet is freely expanding
and there is an equipartition between the particle and magnetic field energy
densities, the position $r_c(\nu)$ of the $\tau_\nu = 1$ surface will be $r_c(\nu) \propto \nu^{-1}$. 
If the above assumptions do not hold (e.g., synchrotron self-absorption is not the dominating absorption 
mechanism), we can expect a different dependency of $r_c(\nu)$.

\begin{table*}[t!]
\renewcommand{\thefootnote}{\alph{footnote}}
\begin{center}
\caption{The large core shift source sample observed with the VLBA}
\label{tab:obslog}
\begin{tabular}{clcclcc}
\hline
\hline
IAU name     &Alias     &  R.A. (J2000)          & Dec. (J2000)          & ~~z     & Optical class &  VLBA epoch   \\ 
\hline 
\object{0148$+$274} &                  &   $01$:$51$:$27.146174$ & $+27$:$44$:$41.79363$ & $1.26 $ & QSO & 2007-03-01 \\ %1996MNRAS.282.1274H - redshift reference S2 0148+27 
\object{0342$+$147} &                  &   $03$:$45$:$06.416545$ & $+14$:$53$:$49.55818$ & $1.556$ & QSO & 2007-06-01 \\ %2005ApJ...626...95S  TEX 0342+147
\object{0425$+$048} &\object{OF\,42}   &   $04$:$27$:$47.570531$ & $+04$:$57$:$08.32555$ & $0.517$\footnotemark[1] & AGN & 2007-04-30 \\ %2003ARep...47..458A - not in VV catalog !!! 	OF +042
\object{0507$+$179} &                  &   $05$:$10$:$02.369133$ & $+18$:$00$:$41.58160$ & $0.416$ & AGN & 2007-05-03 \\ %1998AJ....115.1253P  PKS 0507+17
\object{0610$+$260} &\object{3C\,154}  &   $06$:$13$:$50.139161$ & $+26$:$04$:$36.71971$ & $0.580$ & QSO & 2007-03-01 \\ %1994ApJS...93....1A  3C 154.0
\object{0839$+$187} &                  &   $08$:$42$:$05.094175$ & $+18$:$35$:$40.99050$ & $1.272$ & QSO & 2007-06-01 \\ %1994ApJS...90....1E  DW 0839+18 
\object{0952$+$179} &                  &   $09$:$54$:$56.823616$ & $+17$:$43$:$31.22204$ & $1.478$ & QSO & 2007-04-30 \\ %1992ApJS...79....1T  PKS 0952+179
\object{1004$+$141} &                  &   $10$:$07$:$41.498089$ & $+13$:$56$:$29.60070$ & $2.707$ & QSO & 2007-05-03 \\ %1976ApJS...31..143W  PKS 1004+141
\object{1011$+$250} &                  &   $10$:$13$:$53.428771$ & $+24$:$49$:$16.44062$ & $1.636$ & QSO & 2007-03-01 \\ %1992ApJS...79....1T  B2 1011+25
\object{1049$+$215} &                  &   $10$:$51$:$48.789077$ & $+21$:$19$:$52.31374$ & $1.300$ & QSO & 2007-06-01 \\ %1976ApJS...31..143W  PKS 1049+21
\object{1219$+$285} &\object{W~Comae}  &   $12$:$21$:$31.690524$ & $+28$:$13$:$58.50011$ & $0.161$\footnotemark[2] & BL~Lac & 2007-04-30 \\ %W Com 2008A&A...477..513F !!!!zphot!!!! The commonly used redshift of z = 0.102 for 1219+285 is almost certainly wrong...
\object{1406$-$076} &                  &   $14$:$08$:$56.481198$ & $-07$:$52$:$26.66661$ & $1.493$ & QSO & 2007-05-03 \\ %1979ApJ...232..400P PKS 1406-076
\object{1458$+$718} &\object{3C\,309.1}&   $14$:$59$:$07.583927$ & $+71$:$40$:$19.86646$ & $0.904$ & QSO & 2007-03-01 \\ %1996ApJS..102....1B  3C 309.1
\object{1642$+$690} &                  &   $16$:$42$:$07.848505$ & $+68$:$56$:$39.75639$ & $0.751$ & QSO & 2007-04-30 \\ %1996ApJS..107..541L  4C 69.21
\object{1655$+$077} &                  &   $16$:$58$:$09.011464$ & $+07$:$41$:$27.54034$ & $0.621$ & QSO & 2007-06-01 \\ %1986MNRAS.218..331W  PKS 1655+077
\object{1803$+$784} &                  &   $18$:$00$:$45.683905$ & $+78$:$28$:$04.01839$ & $0.680$ & QSO & 2007-05-03 \\ %2001AJ....122..565R  S5 1803+78
\object{1830$+$285} &                  &   $18$:$32$:$50.185622$ & $+28$:$33$:$35.95514$ & $0.594$ & QSO & 2007-03-01 \\ %1974ApJ...190..271W  B2 1830+28A
\object{1845$+$797} &\object{3C\,390.3}&   $18$:$42$:$08.989895$ & $+79$:$46$:$17.12825$ & $0.056$ & AGN & 2007-06-01 \\ %2006A&A...454..459K  3C 390.3
\object{2201$+$315} &                  &   $22$:$03$:$14.975788$ & $+31$:$45$:$38.26990$ & $0.298$ & QSO & 2007-04-30 \\ %1992A&A...262..401B  B2 2201+31A
\object{2320$+$506} &                  &   $23$:$22$:$25.982173$ & $+50$:$57$:$51.96364$ & $1.279$ & QSO & 2007-05-03 \\ %2005ApJ...626...95S  TEX 2320+506
\hline
\end{tabular}
\end{center}

{\bf Column designation:}
Col.~1~-- IAU source name (B1950),
Col.~2~-- commonly used source name,
Cols.~3~and~4~-- VLBI position, for details see
\url{http://astrogeo.org/vlbi/solutions/rfc_2010c/} and
\cite{2002ApJS..141...13B,2003AJ....126.2562F,2005AJ....129.1163P,2006AJ....131.1872P,2007AJ....133.1236K,2008AJ....136..580P,RDV2009},
Cols.~5~and~6~-- redshift and optical class from \cite{2010A&A...518A..10V},
Col.~7~-- epoch of multifrequency VLBA observations reported in this paper.\\
\footnotemark[1]~Spectroscopic redshift obtained by \cite{2003ARep...47..458A}.\\
\footnotemark[2]~Photometric redshift, see \cite{2008A&A...477..513F}.

\renewcommand{\thefootnote}{\arabic{footnote}}
\end{table*}

% Why is it important to study the stupid core-shift?
Investigation of the core shift effect is important for gaining deeper
understanding of the structure and physical conditions in ultracompact AGN jets.
It may also provide information about the pressure and density gradients in ambient medium surrounding
VLBI-scale jets -- the broad-line region (BLR) and the inner part of the narrow-line region 
(NLR) (e.g.,~\citealt{2007Ap&SS.311..263L}).
The impact of the effect on interpretations of
current and future radio Very Long Baseline Interferometry (VLBI)
astrometric measurements has been discussed by
\cite{Rioja_etal05} and \cite{2009A&A...505L...1P}.
The core shift is expected to introduce a systematic offsets of approximately $0.1$~mas
between the optical (6000~\AA) and radio ($8$~GHz) positions of reference extragalactic sources
\citep{2008A&A...483..759K}, which needs to be taken into account for accurate 
radio--optical reference frame alignment in the era of modern space-based
astrometric missions \citep{2007HiA....14..481L} such as GAIA \citep{2001A&A...369..339P}
and for accurate spacecraft navigation with VLBI (e.g.,
\citealt{1994tdar.nasa...46H,2004ivsg.conf..258S,2009epsc.conf..199P}).

In this paper, we present observational results from a dedicated nine-frequency 
($1.4$--$15.4$~GHz) VLBA survey of the core shift effect in 20 compact 
extragalactic radio sources. 
The selection of targets, the VLBA observation setup 
and our data reduction strategy are described in 
Section~\ref{sec:obs_and_sample}. 
Section~\ref{sec:core_shift_measurement} presents the core shift 
measurement technique. 
In Section~\ref{sec:core_shift_discussion}, we discuss the results 
and their implication for the dominating absorption mechanism in 
the core region. Section~\ref{summary} presents a brief summary of this work.
Throughout this paper, we adopt a $\Lambda$-CDM cosmology, with the
following values for the cosmological parameters: $H_0 = 0.71$, 
$\Omega_\mathrm{m} = 0.27$, and $\Omega_{\Lambda} = 0.73$
\citep{WMAP5_cosmology}.

%__________________________________________________________________

\section{VLBA observations}
\label{sec:obs_and_sample}

\subsection{Sample selection}
\label{sec:sample_selection}

A blind survey for sources with measurable core shifts would be an extraordinarily 
time consuming project because it requires performing simultaneous
multifrequency, high-resolution, high dynamic range VLBI observations.
Therefore, we turned to an existing large-scale geodetic VLBI database 
to find promising candidates for the survey.
\cite{2008A&A...483..759K} have imaged and analyzed 277 sources from the Research 
and Development VLBA program (RDV, see \citealt*{1997ApJS..111...95F} and \citealt*{RDV2009})
observations made in 2002--2003. RDV experiments involve simultaneous dual 
frequency $2.3$/$8.6$~GHz ($S$/$X$ band) observations with the VLBA and up to 
ten other globally distributed antennas. The data processing technique and 
imaging results are described by \cite{PK07}. 
The core shift was measured by \cite{2008A&A...483..759K}
by referencing the core position to optically thin jet features,
the positions of which are not expected to change with frequency.

During the preparation of the \cite{2008A&A...483..759K} work, 18
sources with large core shifts between $2.3$ and $8.6$~GHz were selected for
detailed multifrequency study with the VLBA, which we report here.
These sources were selected in such a way that they not only show 
large core shifts in this frequency range, but are also particularly suitable 
for measuring it by referencing the core position to bright jet features 
-- the technique used in this paper.
This procedure is needed because the absolute position information is
lost during the phase self-calibration process necessary for VLBI
imaging; therefore, it is not known a~priori how images at different
frequencies should be aligned.
Two more sources, 2201$+$315 and 3C\,309.1 (1458$+$718), were added to the list 
in order to continue our previous core shift studies
\citep{1998A&A...330...79L,2002astro.ph.11200R,BH103}.

%__________________________________________________________________

\subsection{VLBA observation setup}     
\label{sec:obs}

\begin{table}[tb!]
\begin{center}
\caption{IF central frequencies}
\label{tab:IFs}
\begin{tabular}{ccr}
\hline
\hline
IEEE Band & IF & Frequency \\ %(MHz) \\
          &    &   (MHz) \\
\hline
 $L$        &  1 &  $1404.49$ \\
 $L$        &  2 &  $1412.49$ \\
 $L$        &  3 &  $1658.49$ \\
 $L$        &  4 &  $1666.49$ \\
            &    &            \\
 $S$        &  1 &  $2275.49$ \\
 $S$        &  2 &  $2283.49$ \\
 $S$        &  3 &  $2383.49$ \\
 $S$        &  4 &  $2391.49$ \\
            &    &            \\
 $C$        &  1 &  $4604.49$ \\
 $C$        &  2 &  $4612.49$ \\
 $C$        &  3 &  $4999.49$ \\
 $C$        &  4 &  $5007.49$ \\
            &    &            \\
 $X$        &  1 &  $8104.49$ \\
 $X$        &  2 &  $8112.49$ \\
 $X$        &  3 &  $8425.49$ \\
 $X$        &  4 &  $8433.49$ \\
            &    &            \\
 $K_u$      &  1 &  $15353.49$  \\
 $K_u$      &  2 &  $15361.49$  \\
 $K_u$      &  3 &  $15369.49$  \\
 $K_u$      &  4 &  $15377.49$  \\
\hline
\end{tabular}
\end{center}

{\bf Column designation:}
Col.~1~-- Radio band name according to the IEEE radar band nomenclature,
Col.~2~-- number of the frequency channel (IF),
Col.~3~-- central frequency of the frequency channel (IF).
\end{table}

Twenty sources selected as promising candidates for a detailed core shift
investigation were observed with the VLBA \citep{1994IAUS..158..117N} during four 24\,h sessions in
March -- June 2007. The list of the observed sources is presented in Table~\ref{tab:obslog}. 
Each source was observed simultaneously using $L$, $S$, $C$, $X$, and $K_u$ band
receivers (according to the Institute of Electrical and Electronics
Engineers or IEEE nomenclature, see Table~\ref{tab:IFs}) in the $1.4$--$15.4$~GHz range.
In each band four $8$~MHz-wide frequency channels (IFs, Table~\ref{tab:IFs})
were recorded in both right and left circular polarizations with 2-bit sampling
and a total aggregate bit rate of 256\,Mbits/sec. 
The correlation of the data was performed at the VLBA Array Operation Center in Socorro, NM,
USA, with an averaging time of 2~seconds.
Data in $L$, $S$, $C$, and $X$ bands were divided into two sub-bands (two IFs in each
sub-band) centered at $1.4$, $1.7$, $2.3$, $2.4$, $4.6$, $5.0$, $8.1$, and
$8.4$~GHz, respectively. The $K_u$ band was not divided into sub-bands in order to achieve 
similar image sensitivity to other frequencies in a comparable
integration time. All four IFs in the $K_u$ band were stacked together around $15.4$~GHz.
The central frequencies of the sub-bands were chosen
in such a way that at least one sub-band in each band is centered at a frequency for
which antenna gain curve measurements are available. A special procedure
was applied to ensure accurate amplitude calibration of all
sub-bands (Section~\ref{sec:calibration_imaging}).

%__________________________________________________________________

\subsection{VLBA data calibration and imaging}
\label{sec:calibration_imaging}

The initial calibration was conducted in {\it AIPS}
\citep{aips} following the standard VLBA calibration
procedure involving {\it (i)}a priori amplitude calibration with measured antenna gain
curves and system temperatures, {\it (ii)}phase calibration using the phase-cal signal 
injected during observations, and {\it (iii)}fringe fitting. The fringe fitting was 
performed by the task {\it FRING}. A separate solution for the group delay and phase rate 
was made for each frequency channel (IF). As the final step in calibration, bandpass 
corrections were applied utilizing the task {\it BPASS}.

%\onltab{3}{
\begin{table}[tb!]
\begin{center}
\caption{Amplitude corrections for the BK134 VLBA experiment}
\label{tab:amp_corrs}
\begin{tabular}{cccccl}
\hline
\hline
Ant. & Band & IF & Epoch & Polarization & Corr. \\
\hline
BR   &  $L$  &  1   & All  &  LCP  & $1.17$                \\
BR   &  $L$  &  2   & All  &  LCP  & $1.13$                \\
LA   &  $L$  &  1   & All  &  LCP  & $0.89$                \\
LA   &  $L$  &  2   & All  &  LCP  & $0.90$                \\
HN   &  $L$  &  1-4 & 2007-06-01 &  LCP, RCP  & $0.75$     \\
OV   &  $L$  &  1   & All  &  RCP    & $1.17$              \\
OV   &  $L$  &  2   & All  &  RCP    & $1.15$              \\
     &       &      &      &       &                       \\
BR   &  $S$  &  3   & All  &  LCP  & $1.13$                \\
LA   &  $S$  &  1   & All  &  LCP  & $1.12$                \\
LA   &  $S$  &  2   & All  &  LCP  & $1.09$                \\
LA   &  $S$  &  3   & All  &  LCP  & $1.19$                \\
LA   &  $S$  &  4   & All  &  LCP  & $1.23$                \\
HN   &  $S$  &  3   & All  &  LCP  & $0.90$                \\
HN   &  $S$  &  4   & All  &  LCP  & $0.89$                \\
KP   &  $S$  &  1   & All  &  LCP  & $1.09$                \\
KP   &  $S$  &  2   & All  &  LCP  & $1.13$                \\
NL   &  $S$  &  3   & All  &  LCP  & $1.09$                \\
MK   &  $S$  &  3   & All  &  LCP  & $1.13$                \\
MK   &  $S$  &  4   & All  &  LCP  & $1.14$                \\
LA   &  $S$  &  1   & All  &  RCP  & $0.85$                \\
LA   &  $S$  &  2   & All  &  RCP  & $0.85$                \\
LA   &  $S$  &  3   & All  &  RCP  & $0.76$                \\
LA   &  $S$  &  4   & All  &  RCP  & $0.79$                \\
OV   &  $S$  &  3   & All  &  RCP  & $0.86$                \\
OV   &  $S$  &  4   & All  &  RCP  & $0.89$                \\
     &       &      &      &       &                       \\
KP   &  $C$  &  1   & All  &  LCP  & $1.10$                \\
KP   &  $C$  &  2   & All  &  LCP  & $1.11$                \\
BR   &  $C$  &  1   & All  &  LCP  & $0.91$                \\
KP   &  $C$  &  1   & All  &  RCP  & $1.11$                \\
KP   &  $C$  &  2   & All  &  RCP  & $1.09$                \\
MK   &  $C$  &  1   & All  &  RCP  & $1.24$                \\
MK   &  $C$  &  2   & All  &  RCP  & $1.17$                \\
     &       &      &      &       &                       \\
OV   &  $X$  &  1   & All  &  LCP  & $1.21$                \\
OV   &  $X$  &  2   & All  &  LCP  & $1.20$                \\
KP   &  $X$  &  1   & All  &  RCP  & $0.91$                \\
KP   &  $X$  &  2   & All  &  RCP  & $0.93$                \\
     &       &      &      &       &                       \\
BR   & $K_u$ &  1   & All  &  LCP  & $0.94$                \\ 
BR   & $K_u$ &  4   & All  &  LCP  & $0.95$                \\ 
SC   & $K_u$ &  1   & All  &  LCP  & $1.05$                \\ 
KP   & $K_u$ &  2   & All  &  RCP  & $1.42$                \\
\hline
\end{tabular}
\end{center}

{\bf Column designation:}
Col.~1~-- Antenna name,
Col.~2~-- radio band name,
Col.~3~-- number of the frequency channel (IF),
Col.~4~-- epoch to which the amplitude correction was applied,
Col.~5~-- polarization (right or left circular),
Col.~6~-- amplitude correction coefficient.

\end{table}
%} % end of \onltab{3}{

The sources were imaged independently in each band, using the CLEAN
algorithm \citep{1974A&AS...15..417H} implemented in the {\it Difmap} software
\citep{difmap}.
Global amplitude corrections for left and right circular polarization for
each IF at each antenna were determined by comparing the total intensity
CLEAN model obtained with the initially calibrated data (again using {\it Difmap}).
The amplitude corrections obtained were averaged over all the sources
observed in the experiment with the exception of 0610$+$260 and 1830$+$285.
These two sources have steep radio spectra \citep{1999A&AS..139..545K,2002PASA...19...83K}
and were not used for the amplitude corrections computation in order to avoid 
introducing systematic errors into the corrections. 
The significant amplitude correction factors that were found are listed in
Table~\ref{tab:amp_corrs}.
These amplitude corrections were introduced into the dataset using the {\it AIPS}
task {\it CLCOR}.
After that $L$, $S$, $C$, and $X$ band data were split into two
sub-bands as described above, and final total intensity imaging for each 
sub-band was conducted independently using {\it Difmap}.
Naturally weighted CLEAN images of the observed sources between 1.4 and
$15.4$~GHz are presented in Fig.~\ref{images} available in the electronic
version of the paper.

The typical VLBA amplitude calibration accuracy in the $1$--$10$~GHz 
frequency range is estimated to be $\sim5$\,\% according to the VLBA Observational Status 
Summary\footnote{\url{http://www.vlba.nrao.edu/astro/obstatus/current/}}.
\cite{2005AJ....130.2473K} and \cite{2008AJ....136..580P}
have confirmed this estimate to be correct for $S$, $X$, and $K_u$ bands
by comparing VLBA integrated flux density with a single-dish flux density
measured from 1 to 22~GHz quasi-simultaneously at RATAN-600 
\citep[see description of the program
in][]{1999A&AS..139..545K,2002PASA...19...83K}
and the data from the University of Michigan
Radio Astronomy Observatory (UMRAO) monitoring program
\citep{1985ApJS...59..513A,1992ApJ...399...16A,2003ASPC..300..159A}.
This agrees with the VLBA $K_u$ band amplitude calibration accuracy
reported by \cite{2002ApJ...568...99H}.
We checked and confirmed the above estimates for this particular dataset 
by using our RATAN-600 data at 1-22~GHz and 
VLA measurements at 5 and 8~GHz from the VLA/VLBA Polarization Calibration
Program\footnote{\url{http://www.aoc.nrao.edu/~smyers/calibration/}}.
This check was particularly efficient with 
our slowly varying calibrator \object{0923+392} observed in each of
the four VLBA sessions. It has negligible kiloparsec-scale emission, 
which makes this comparison possible.

\onlfig{1}{
\clearpage
%\addtocounter{figure}{-1}
\begin{figure*}
 \centering

 \includegraphics[width=0.33\textwidth,angle=0,trim=0.5cm 3cm 1.6cm 5cm,clip]{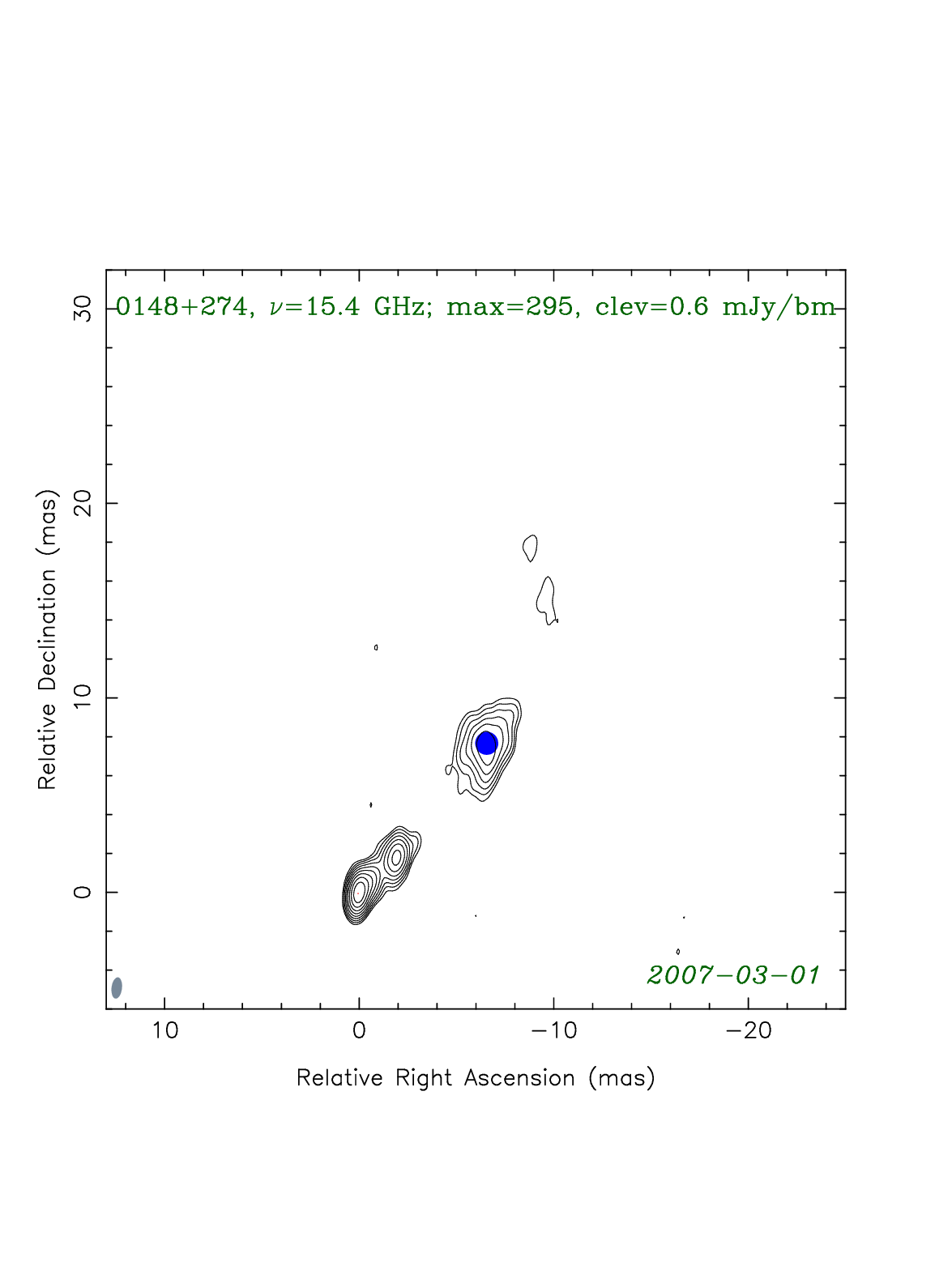}
 \includegraphics[width=0.33\textwidth,angle=0,trim=0.5cm 3cm 1.6cm 5cm,clip]{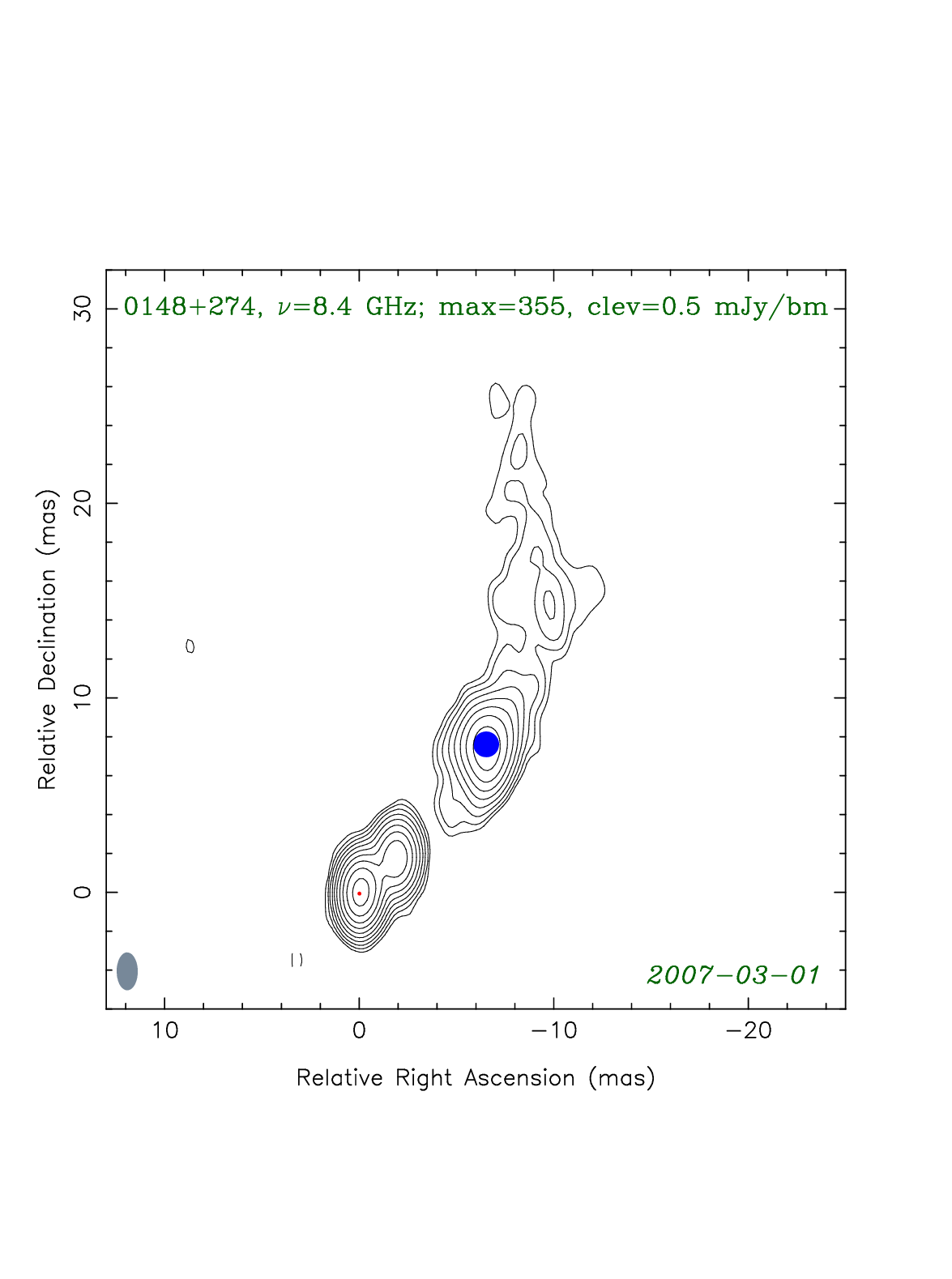}
 \includegraphics[width=0.33\textwidth,angle=0,trim=0.5cm 3cm 1.6cm 5cm,clip]{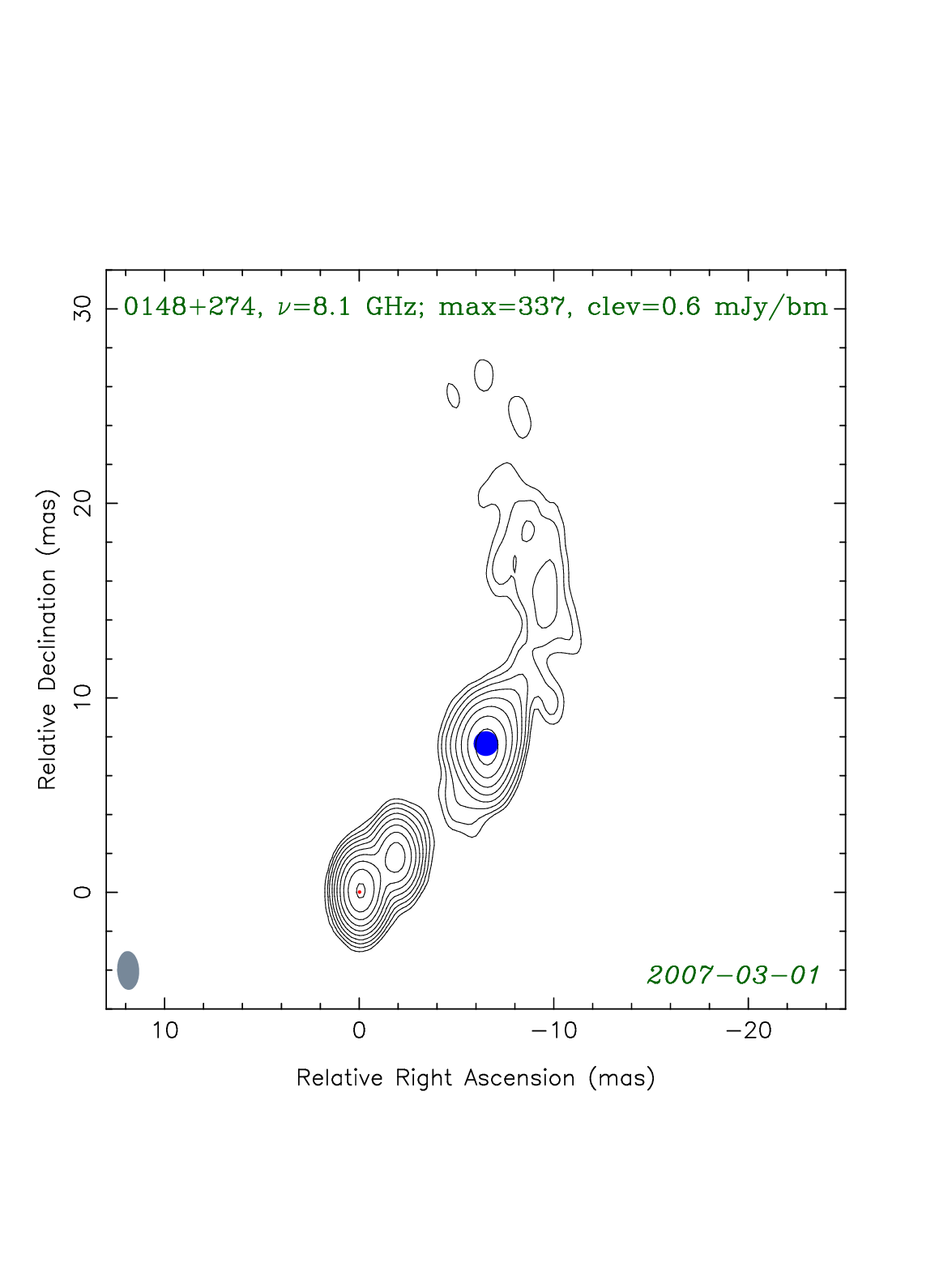}

 \includegraphics[width=0.33\textwidth,angle=0,trim=0.5cm 3cm 1.6cm 5cm,clip]{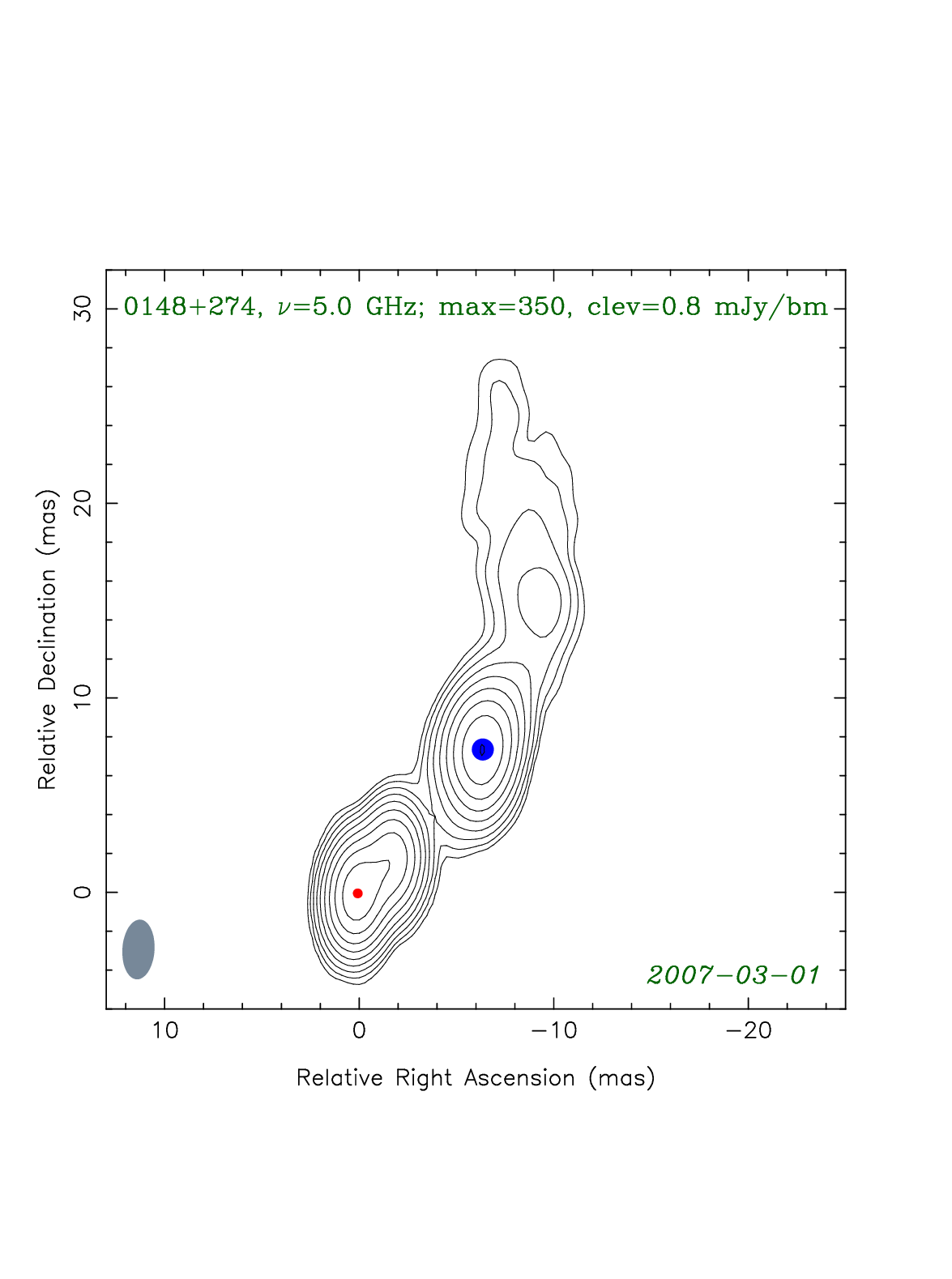}
 \includegraphics[width=0.33\textwidth,angle=0,trim=0.5cm 3cm 1.6cm 5cm,clip]{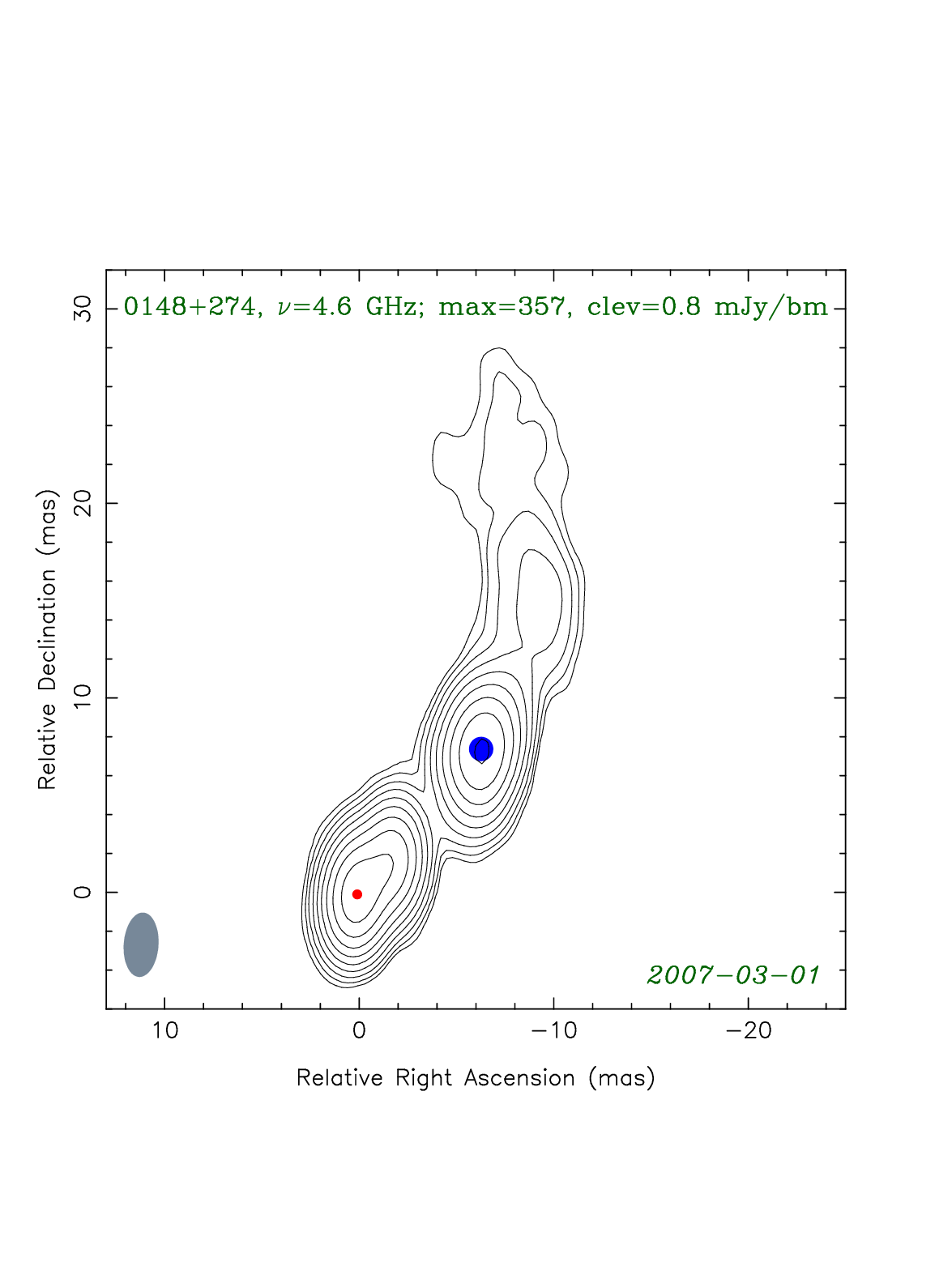}
 \includegraphics[width=0.33\textwidth,angle=0,trim=0.5cm 3cm 1.6cm 5cm,clip]{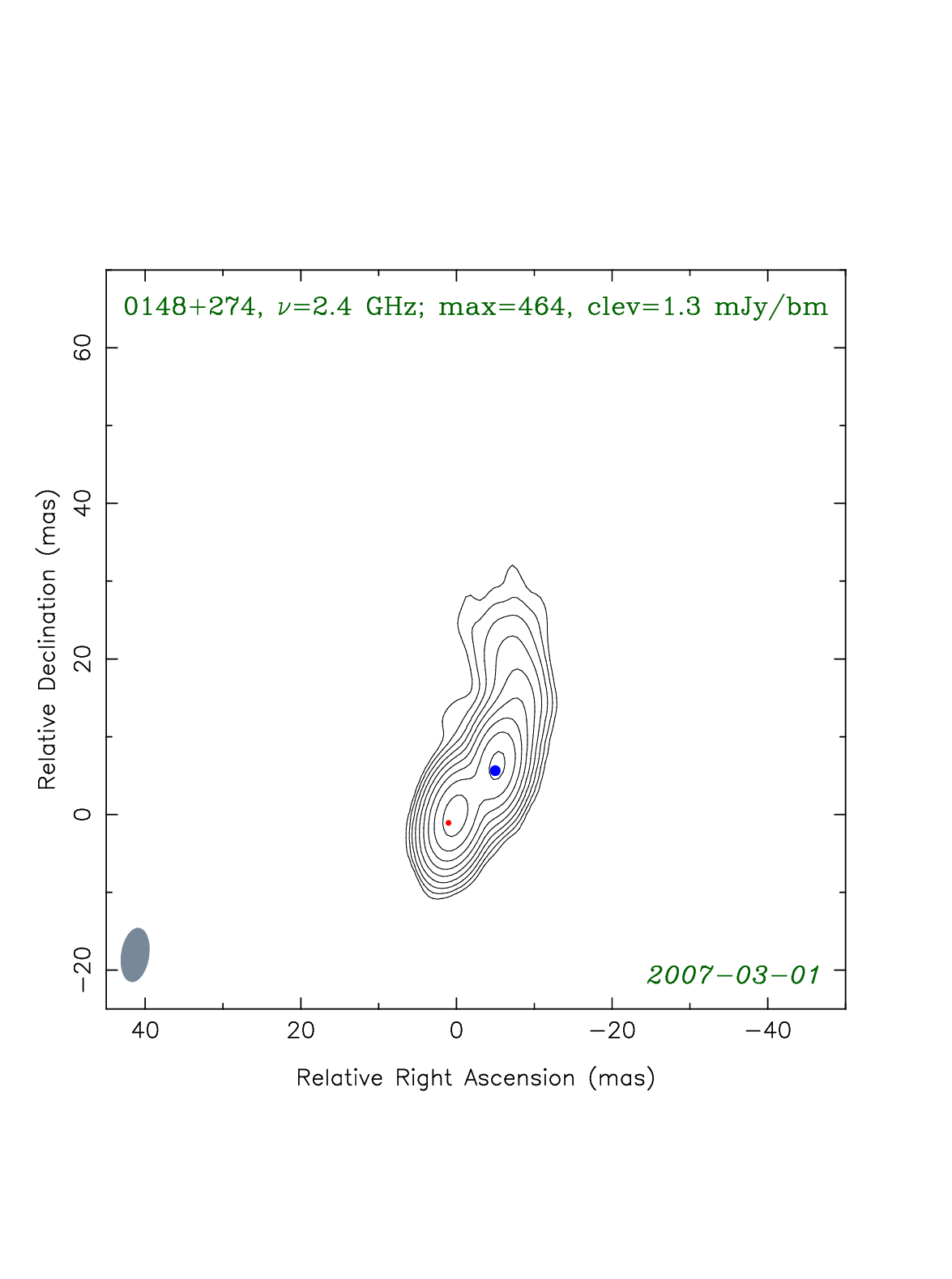}

 \includegraphics[width=0.33\textwidth,angle=0,trim=0.5cm 3cm 1.6cm 5cm,clip]{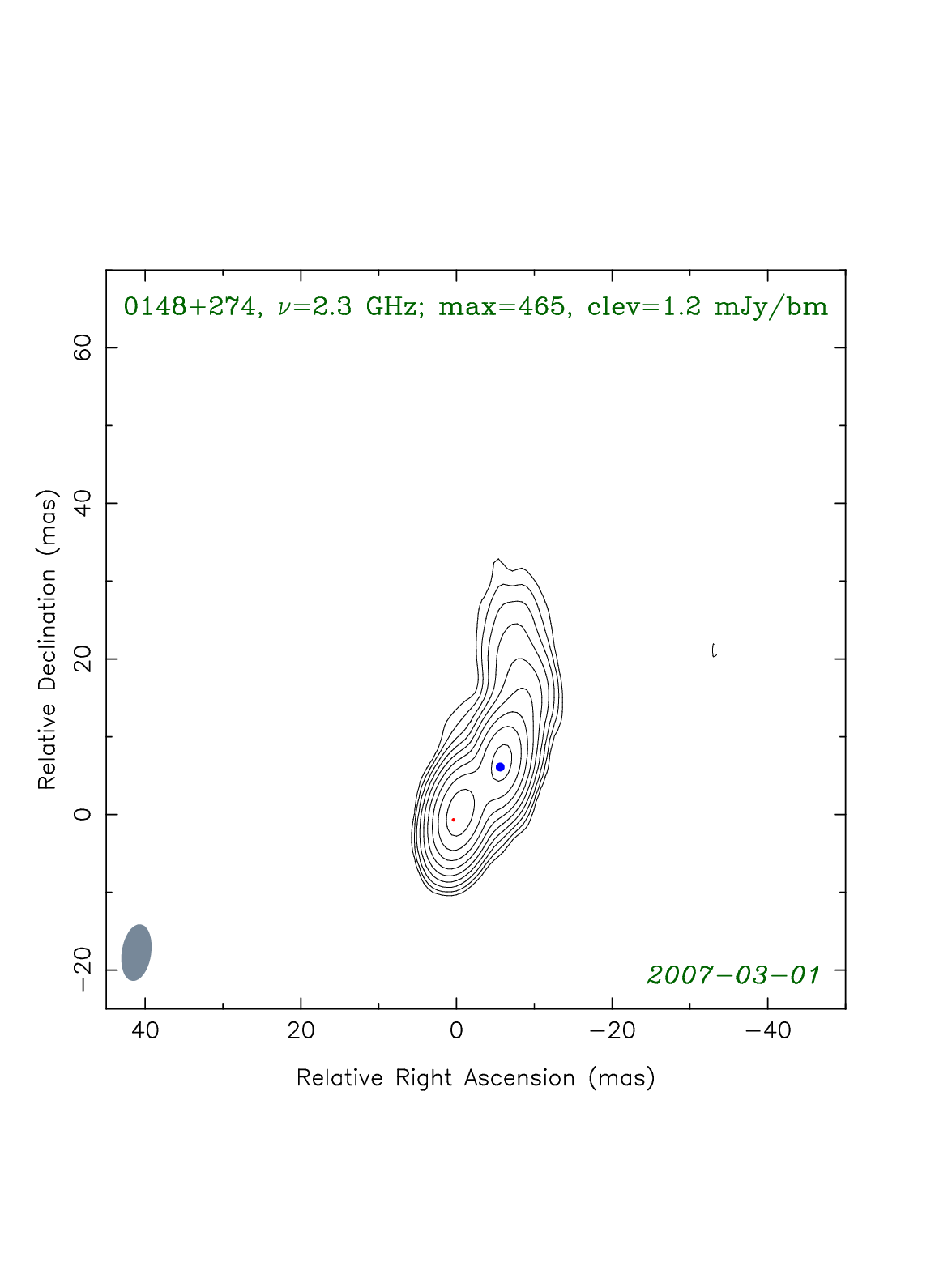}
 \includegraphics[width=0.33\textwidth,angle=0,trim=0.5cm 3cm 1.6cm 5cm,clip]{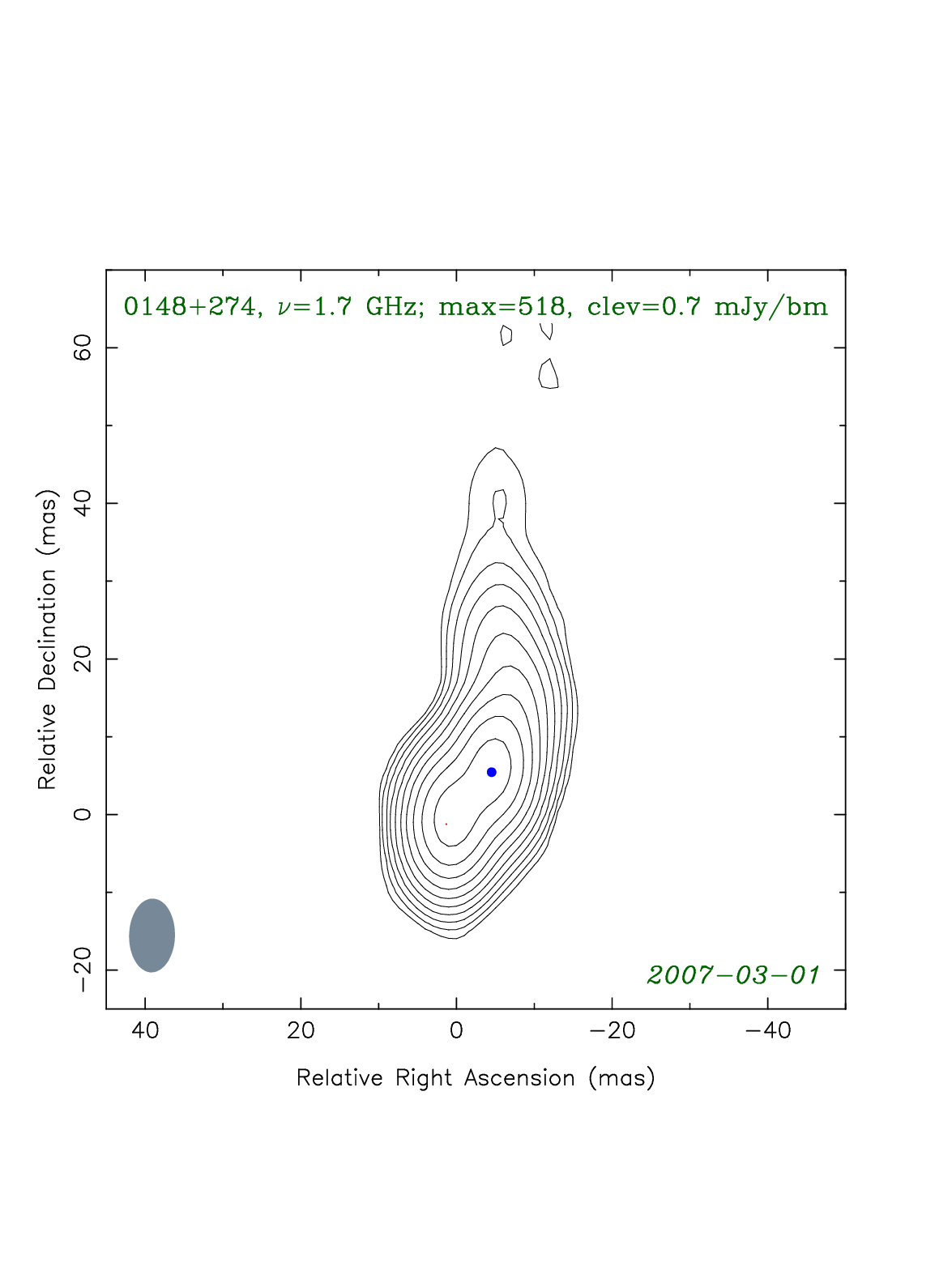}
 \includegraphics[width=0.33\textwidth,angle=0,trim=0.5cm 3cm 1.6cm 5cm,clip]{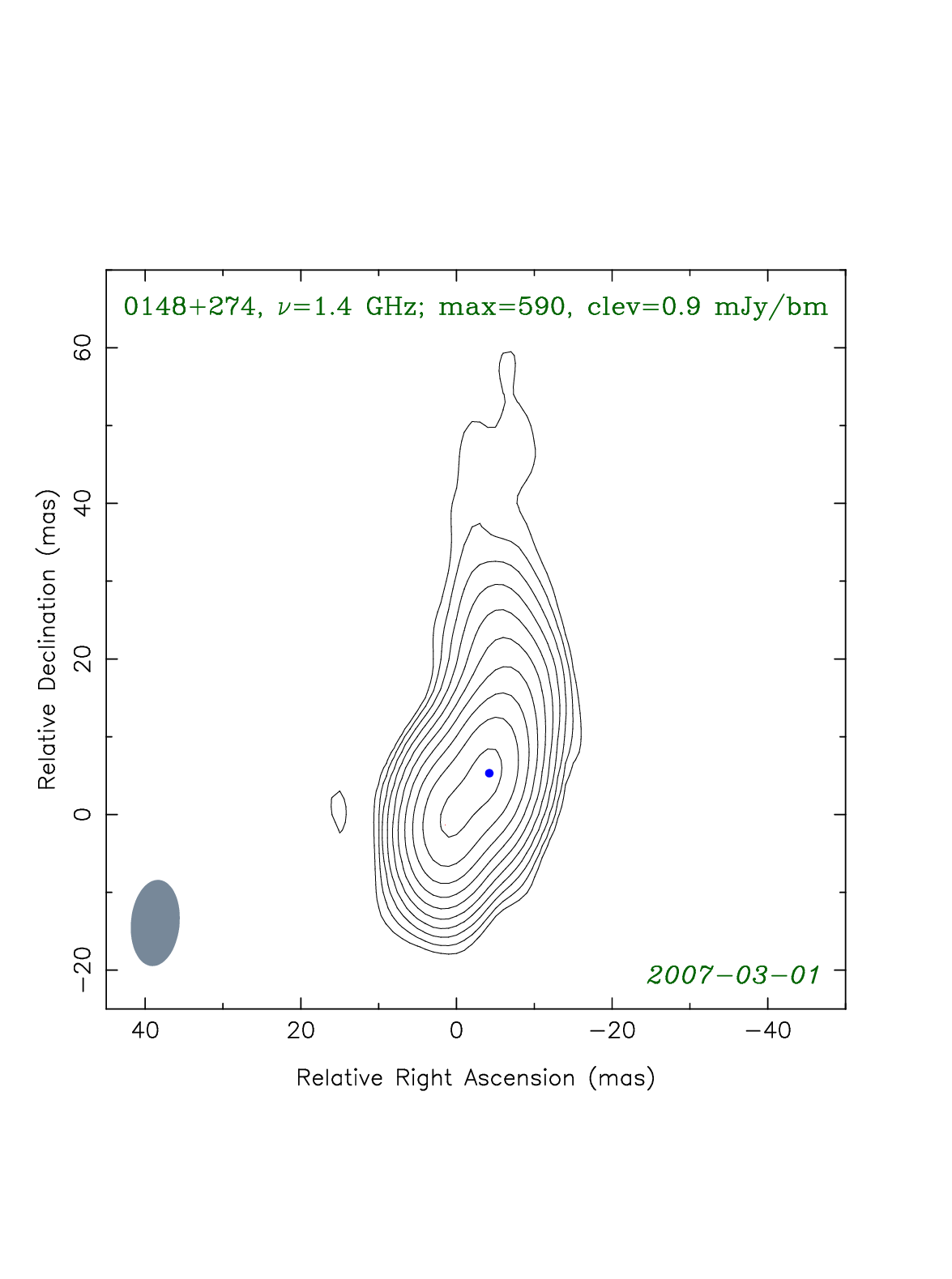}

\caption{\label{images}
Naturally weighted CLEAN images of the observed sources between $1.4$ and $15$~GHz.
The lowest contour
value `clev' is chosen at four times the rms noise,
the peak brightness is given by `max'. The contour levels
increase by factors of two. The dashed contours indicate negative brightness.
The beam's full width at half maximum (FWHM) is shown in the bottom left corner of the images in gray.
An epoch of observation is shown in the bottom right corner.
Red and blue spots indicate the positions and sizes (FWHM) of Gaussian model components for the core
and the jet features, respectively.
}
\end{figure*}

\clearpage
\addtocounter{figure}{-1}
\begin{figure*}
 \centering

 \includegraphics[width=0.33\textwidth,angle=0,trim=0.5cm 3cm 1.6cm 5cm,clip]{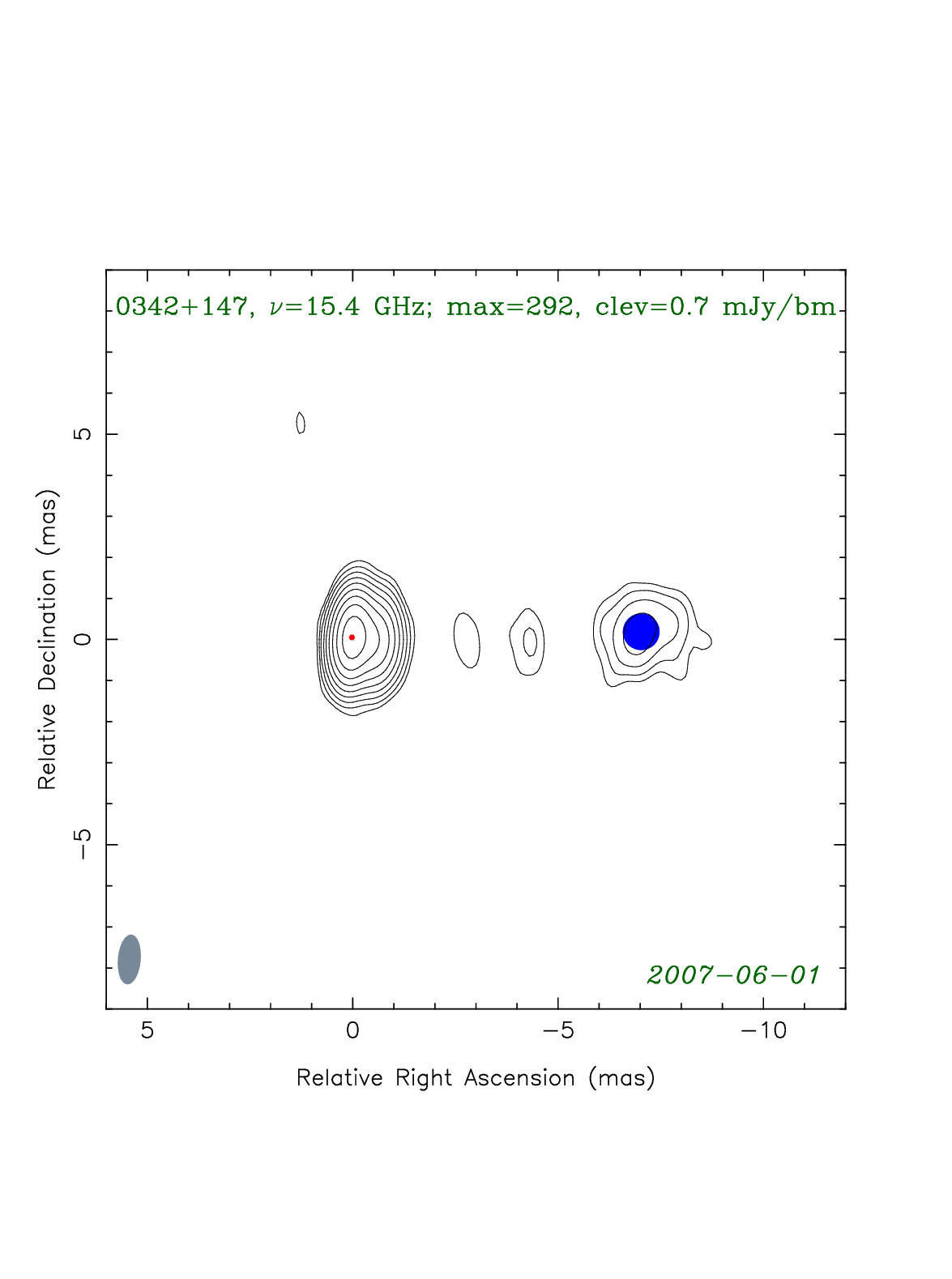}
 \includegraphics[width=0.33\textwidth,angle=0,trim=0.5cm 3cm 1.6cm 5cm,clip]{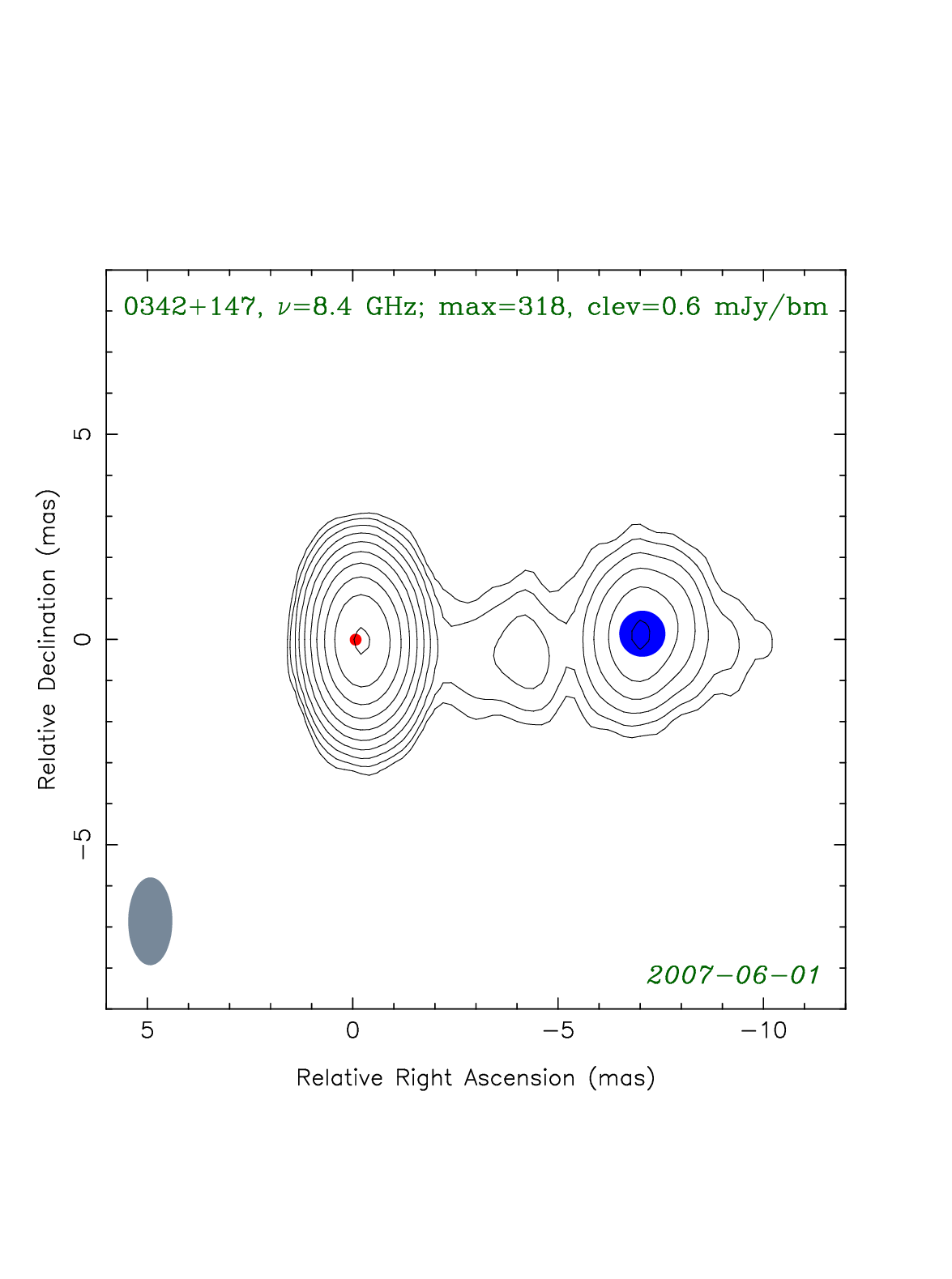}
 \includegraphics[width=0.33\textwidth,angle=0,trim=0.5cm 3cm 1.6cm 5cm,clip]{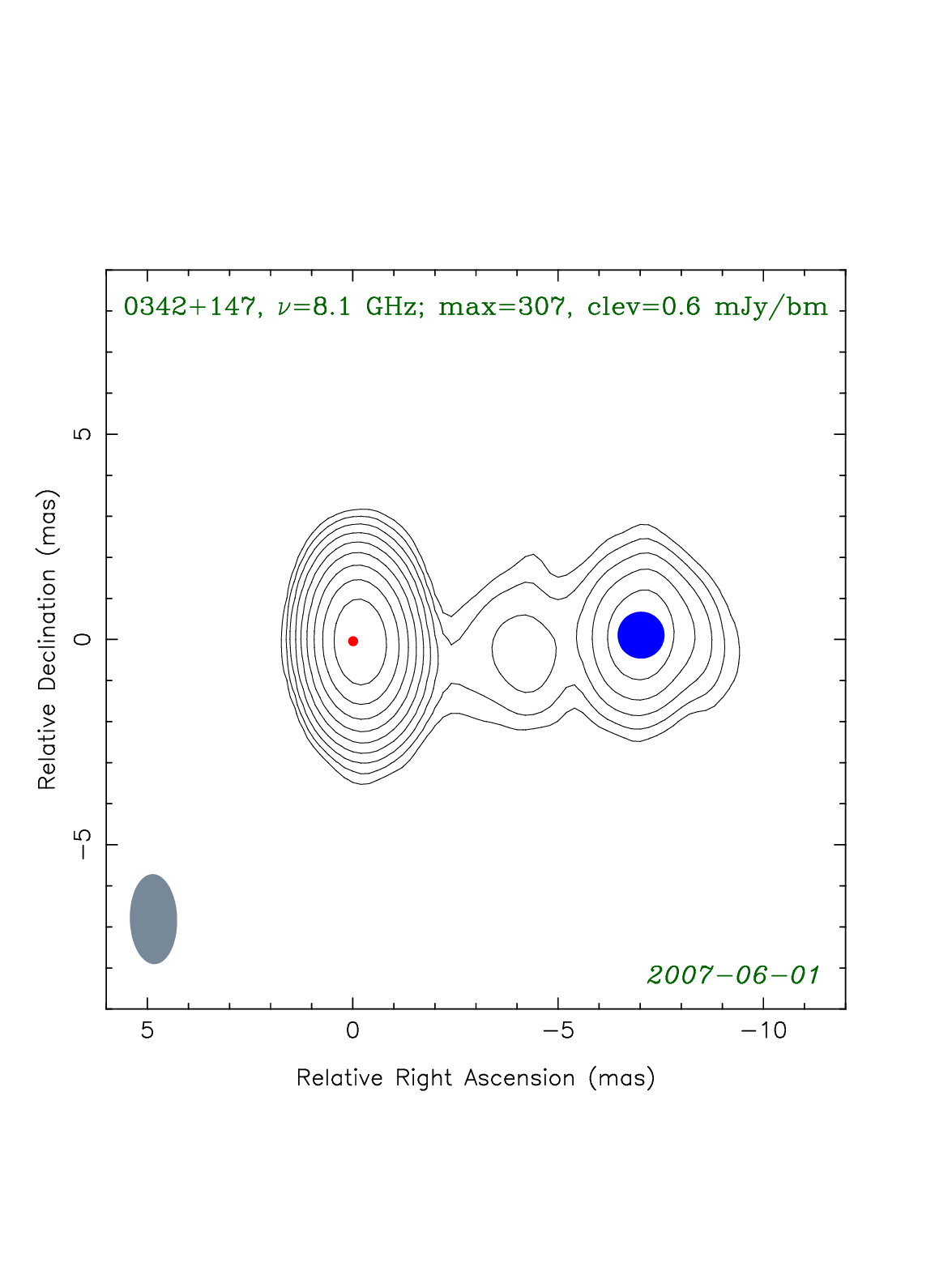}

 \includegraphics[width=0.33\textwidth,angle=0,trim=0.5cm 3cm 1.6cm 5cm,clip]{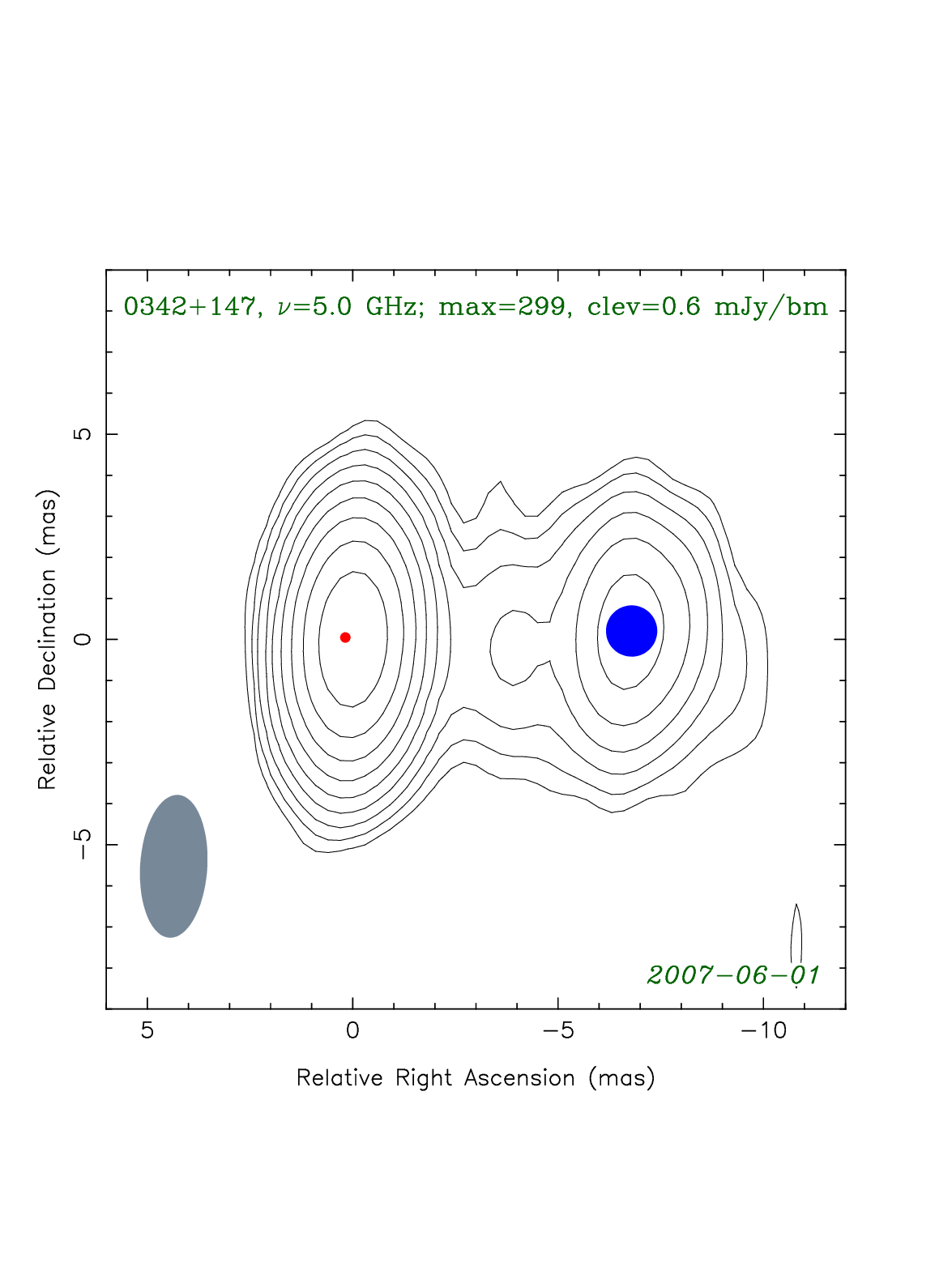}
 \includegraphics[width=0.33\textwidth,angle=0,trim=0.5cm 3cm 1.6cm 5cm,clip]{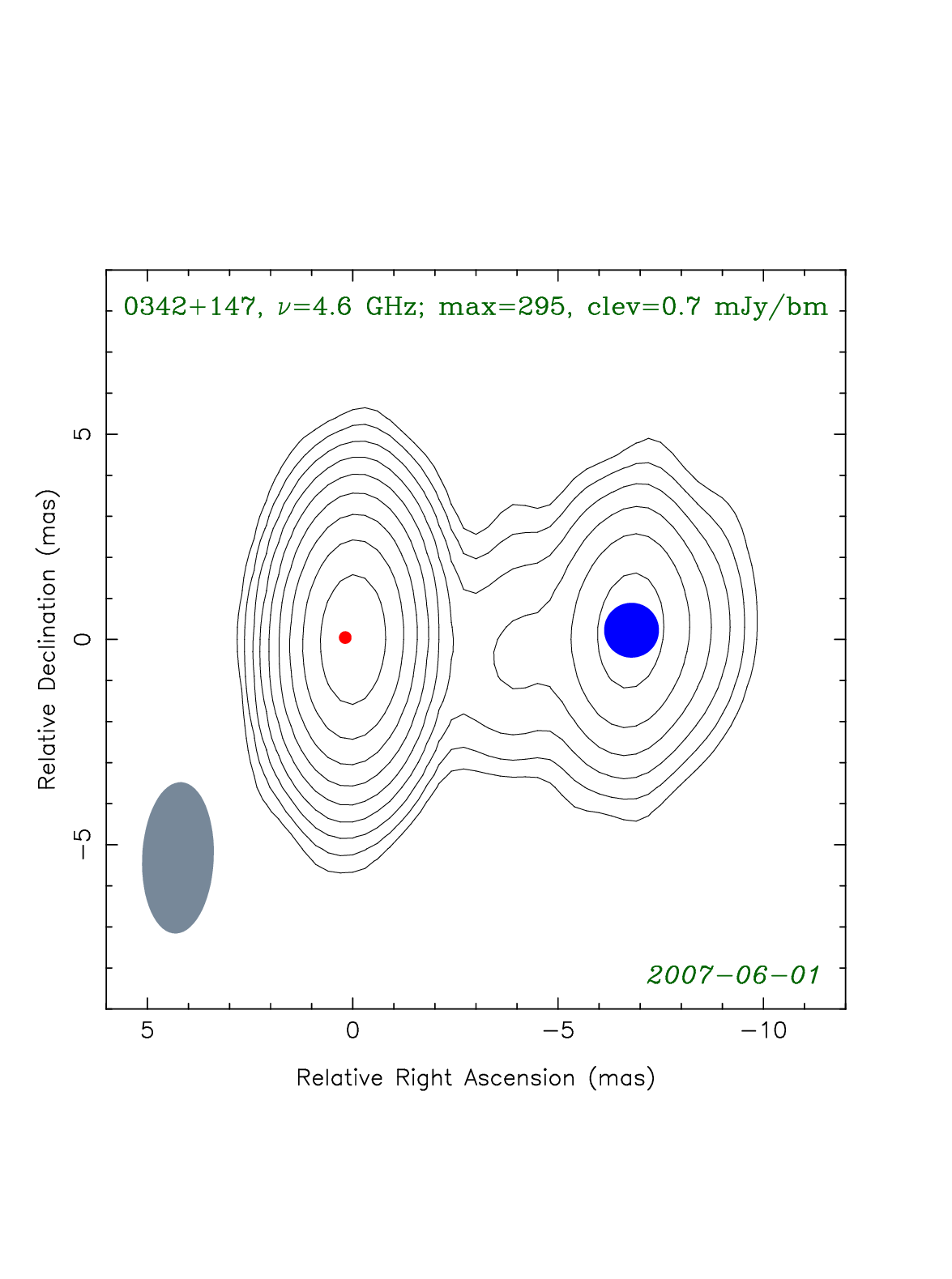}
 \includegraphics[width=0.33\textwidth,angle=0,trim=0.5cm 3cm 1.6cm 5cm,clip]{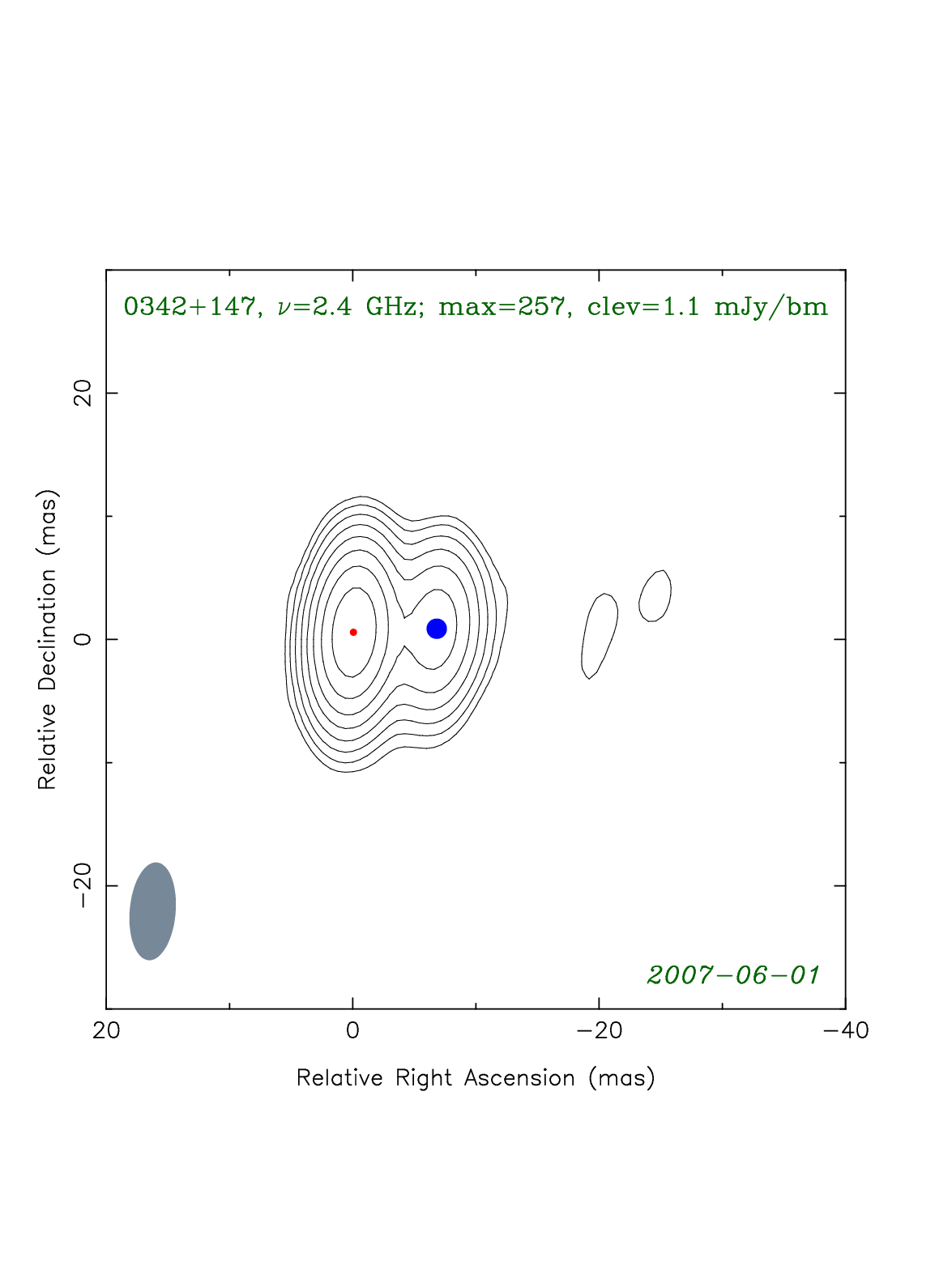}

 \includegraphics[width=0.33\textwidth,angle=0,trim=0.5cm 3cm 1.6cm 5cm,clip]{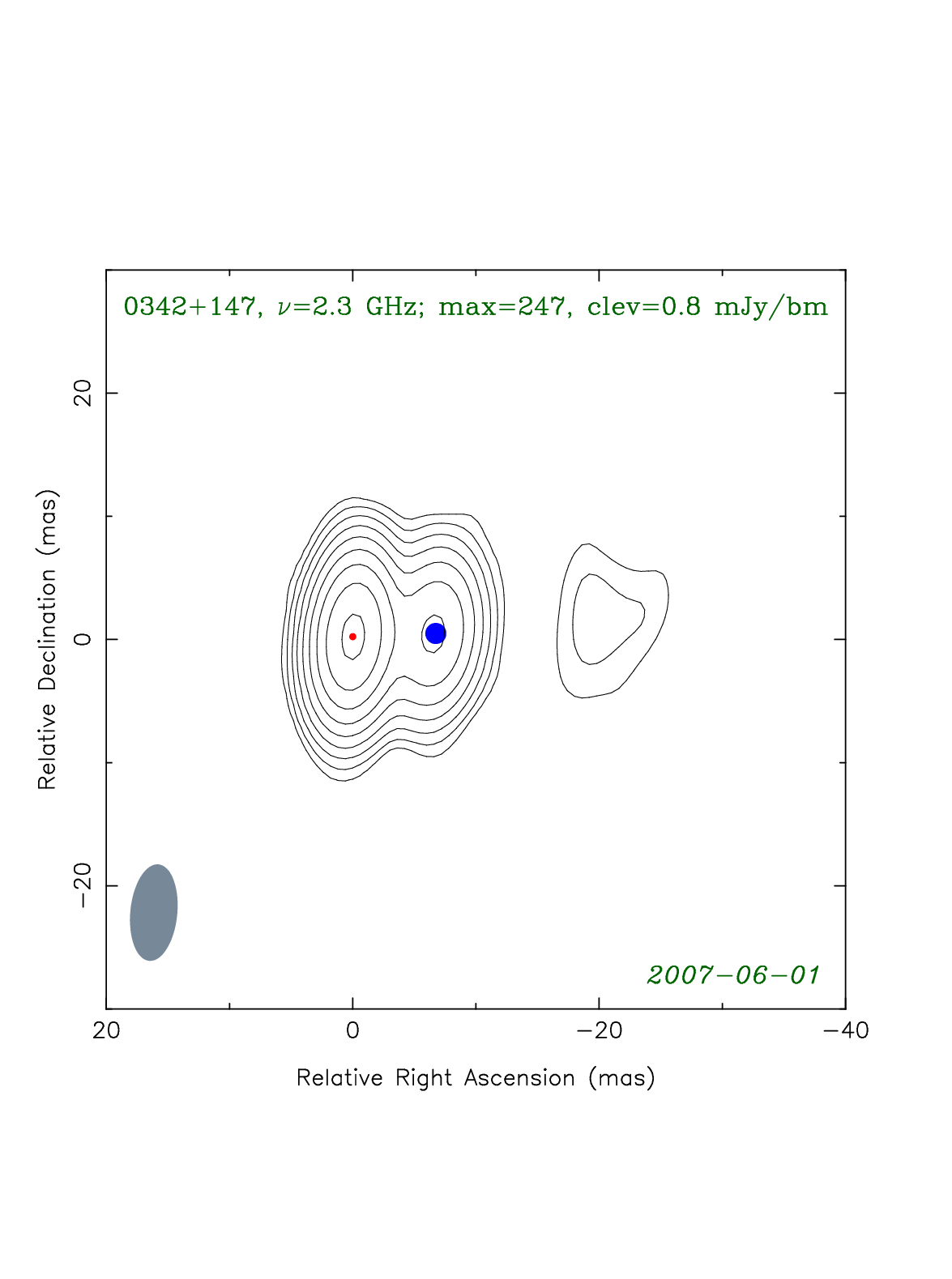}
 \includegraphics[width=0.33\textwidth,angle=0,trim=0.5cm 3cm 1.6cm 5cm,clip]{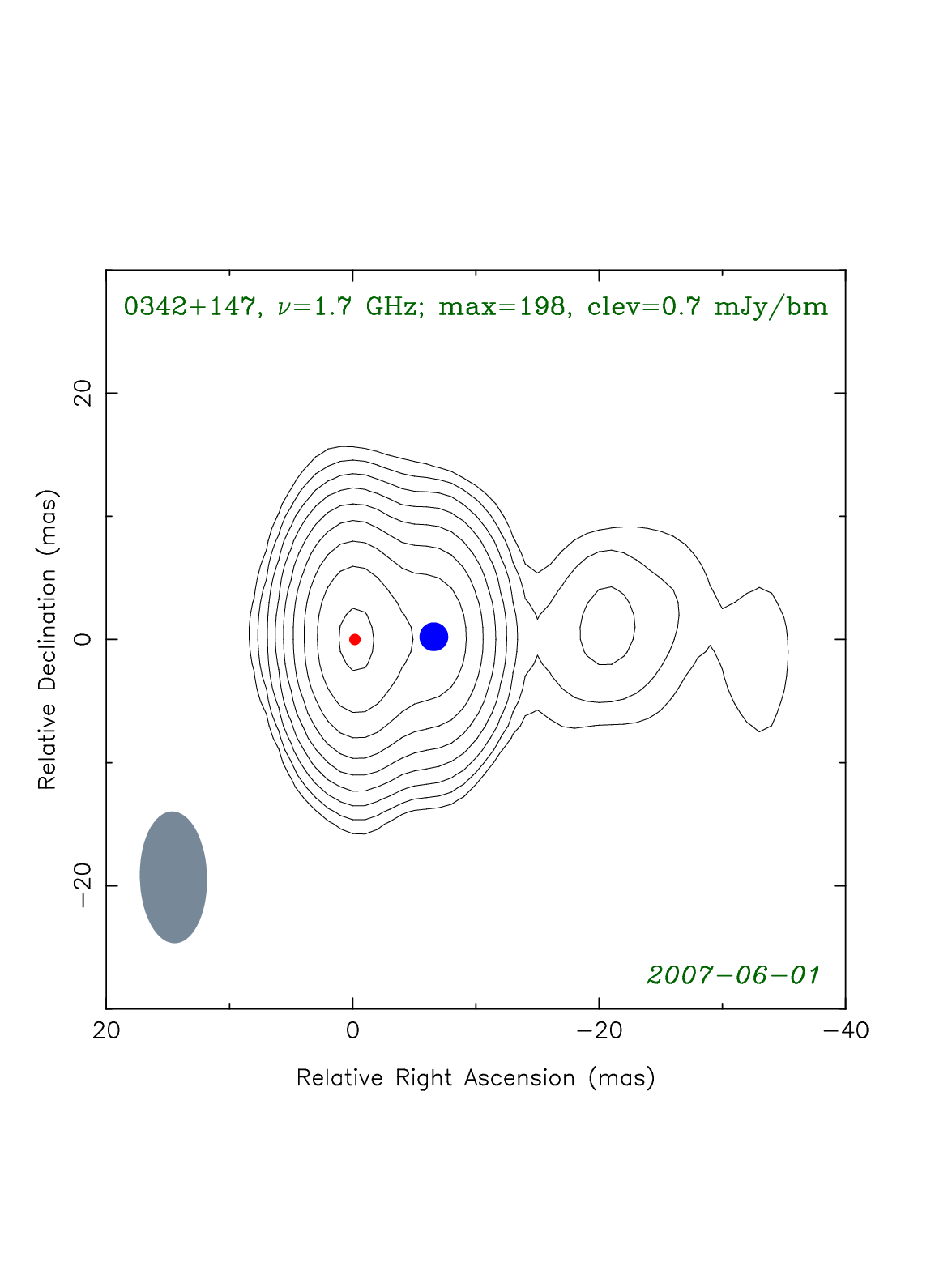}
 \includegraphics[width=0.33\textwidth,angle=0,trim=0.5cm 3cm 1.6cm 5cm,clip]{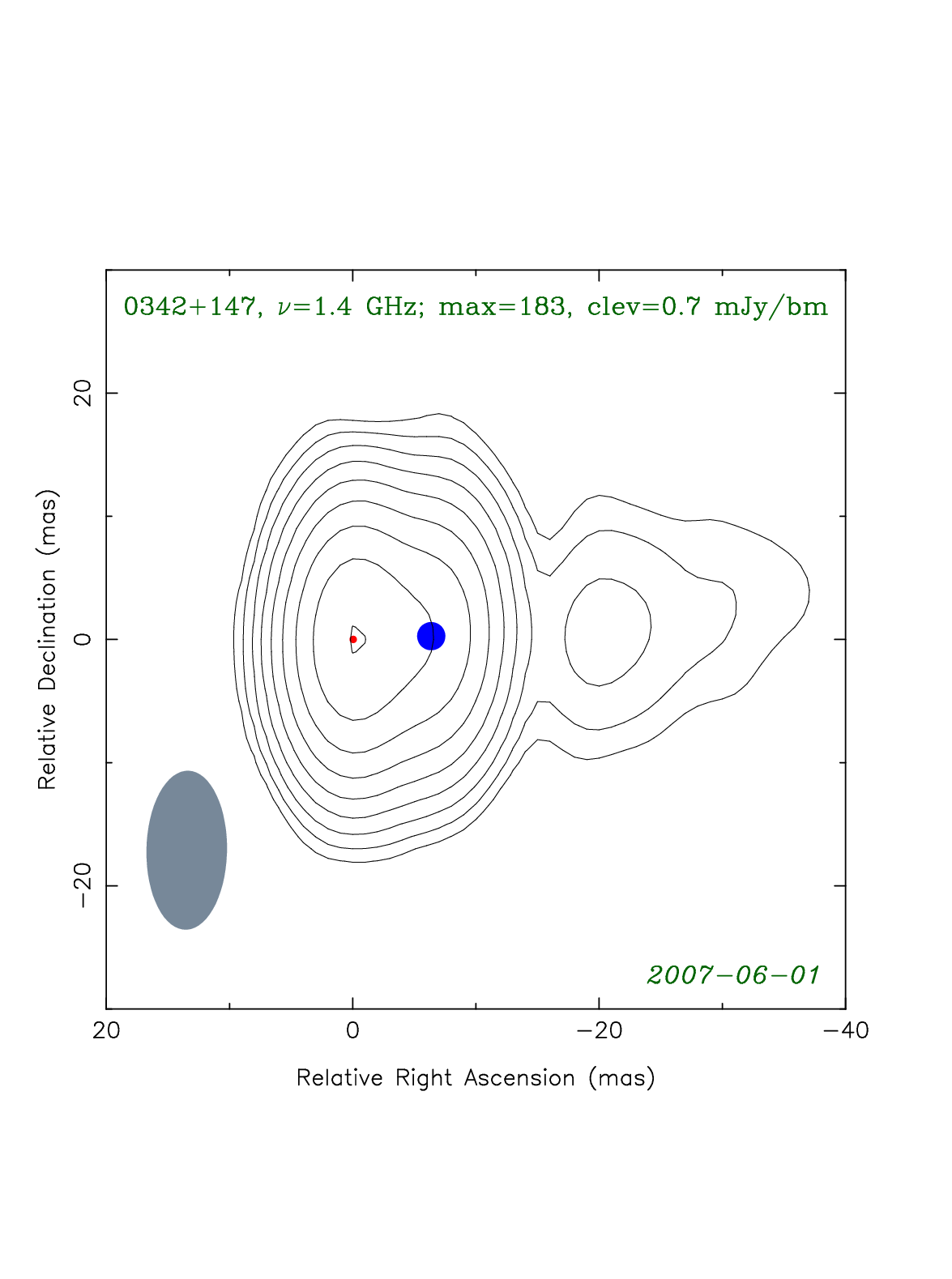}

 \caption{continued...}
\end{figure*}

\clearpage
\addtocounter{figure}{-1}
\begin{figure*}
 \centering

 \includegraphics[width=0.33\textwidth,angle=0,trim=0.5cm 3cm 1.6cm 5cm,clip]{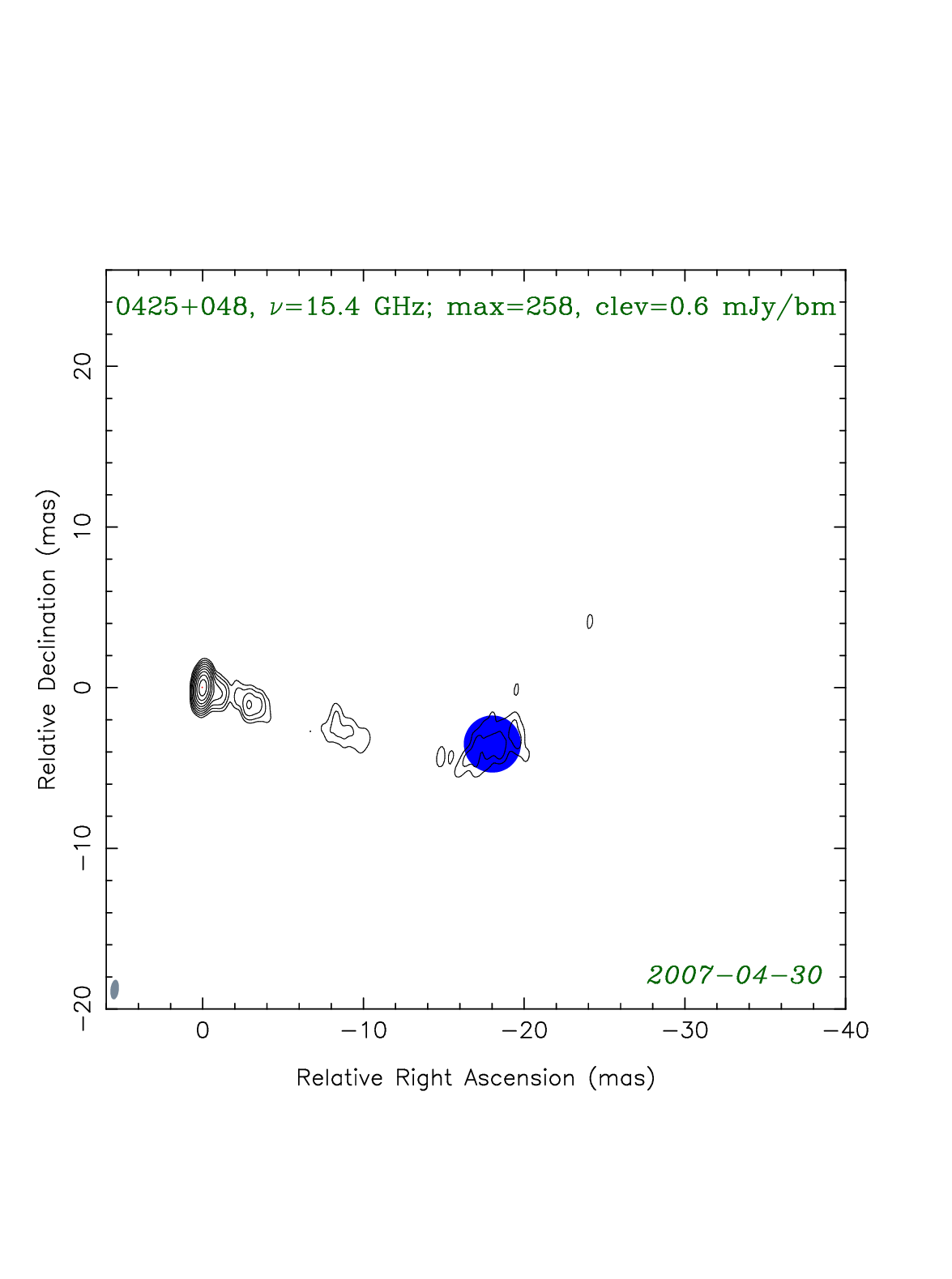}
 \includegraphics[width=0.33\textwidth,angle=0,trim=0.5cm 3cm 1.6cm 5cm,clip]{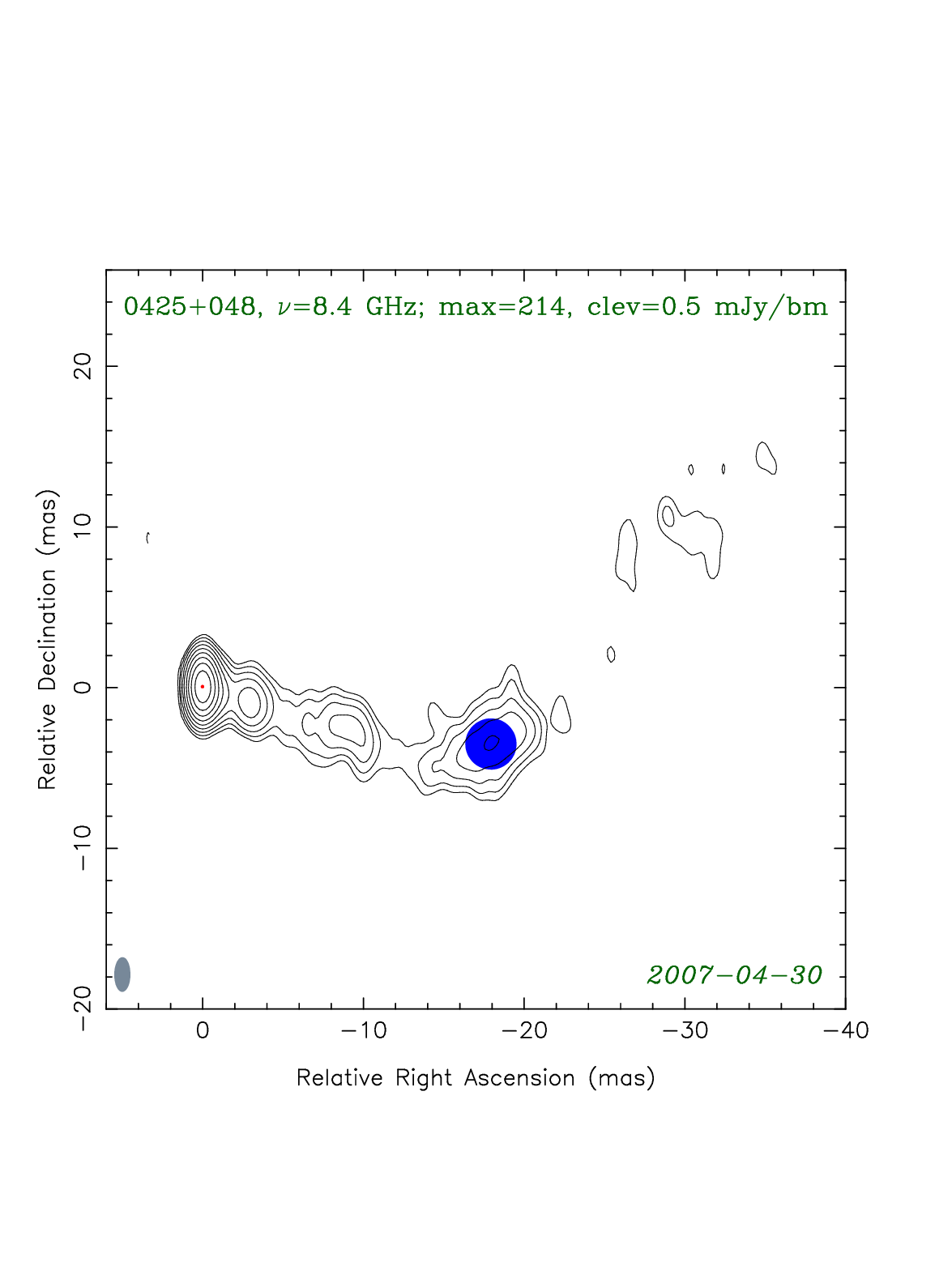}
 \includegraphics[width=0.33\textwidth,angle=0,trim=0.5cm 3cm 1.6cm 5cm,clip]{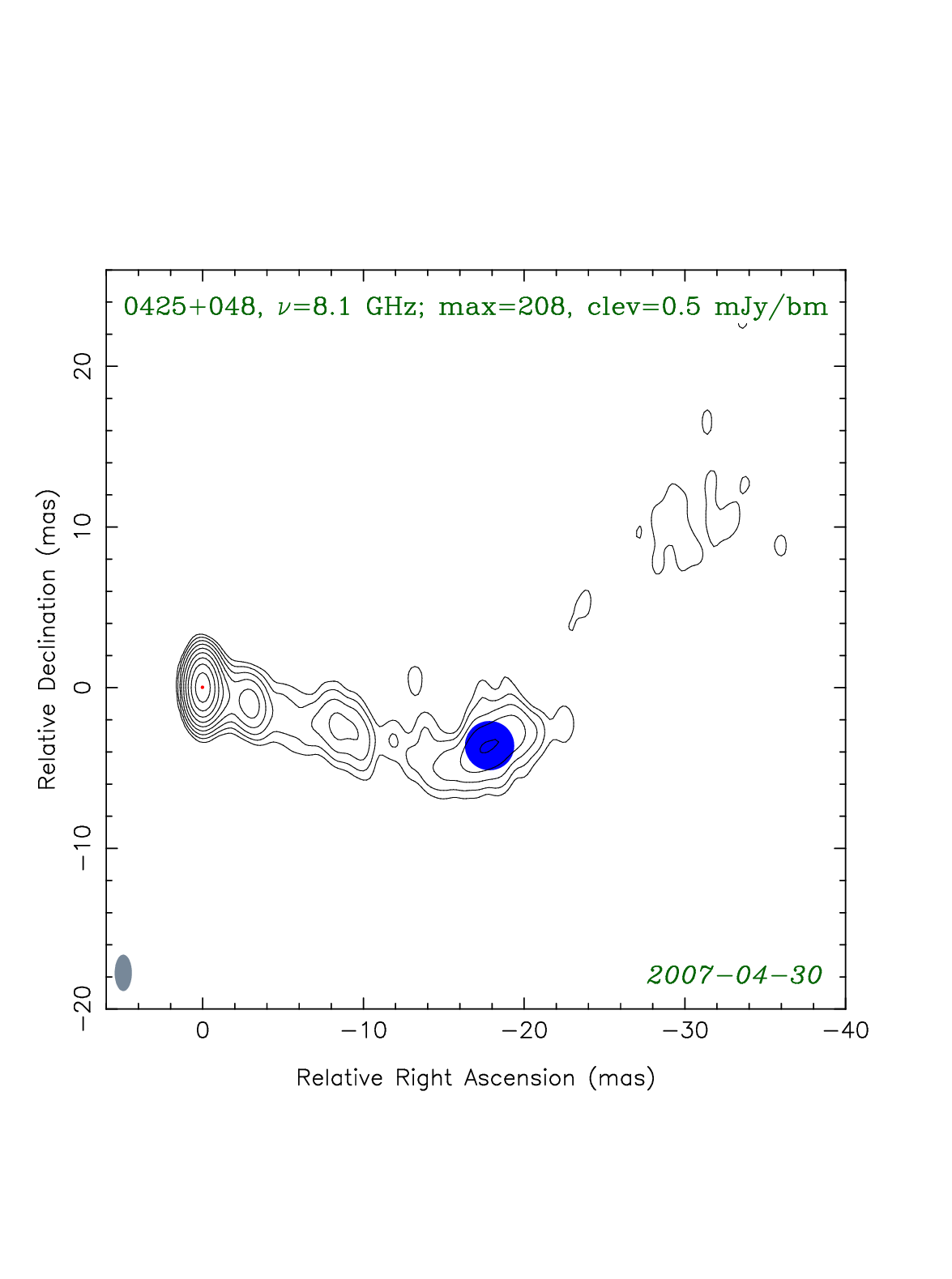}

 \includegraphics[width=0.33\textwidth,angle=0,trim=0.5cm 3cm 1.6cm 5cm,clip]{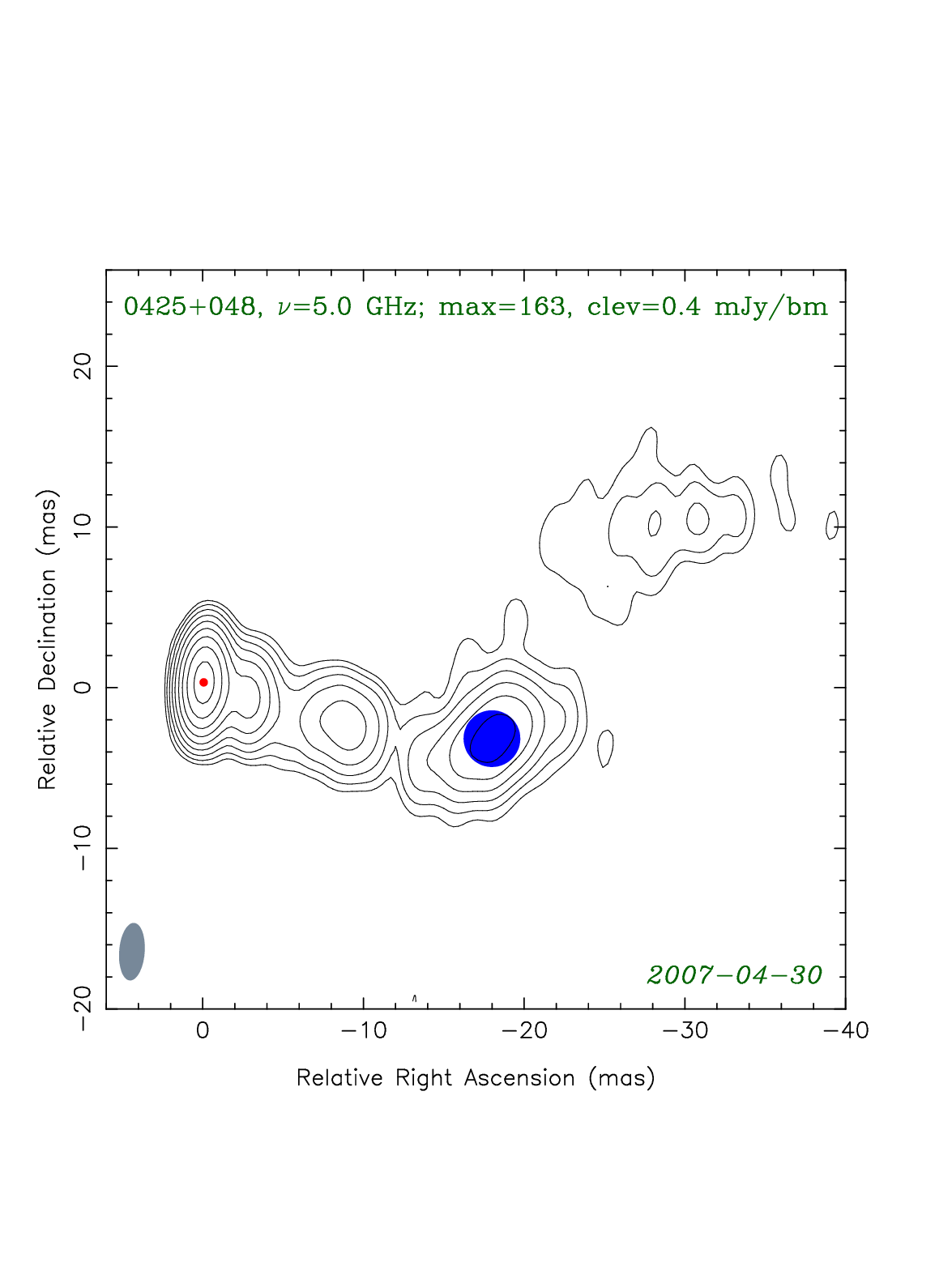}
 \includegraphics[width=0.33\textwidth,angle=0,trim=0.5cm 3cm 1.6cm 5cm,clip]{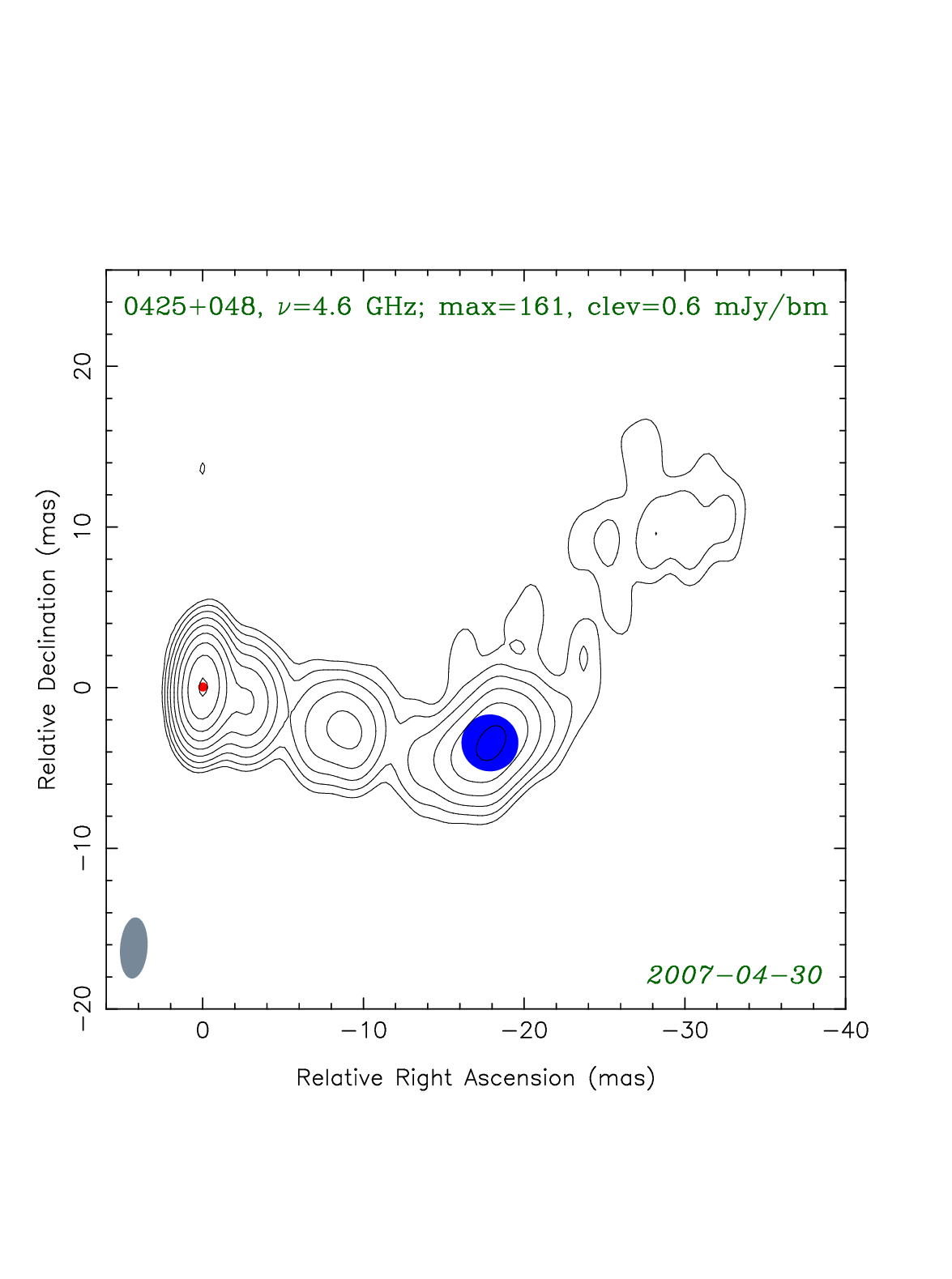}
 \includegraphics[width=0.33\textwidth,angle=0,trim=0.5cm 3cm 1.6cm 5cm,clip]{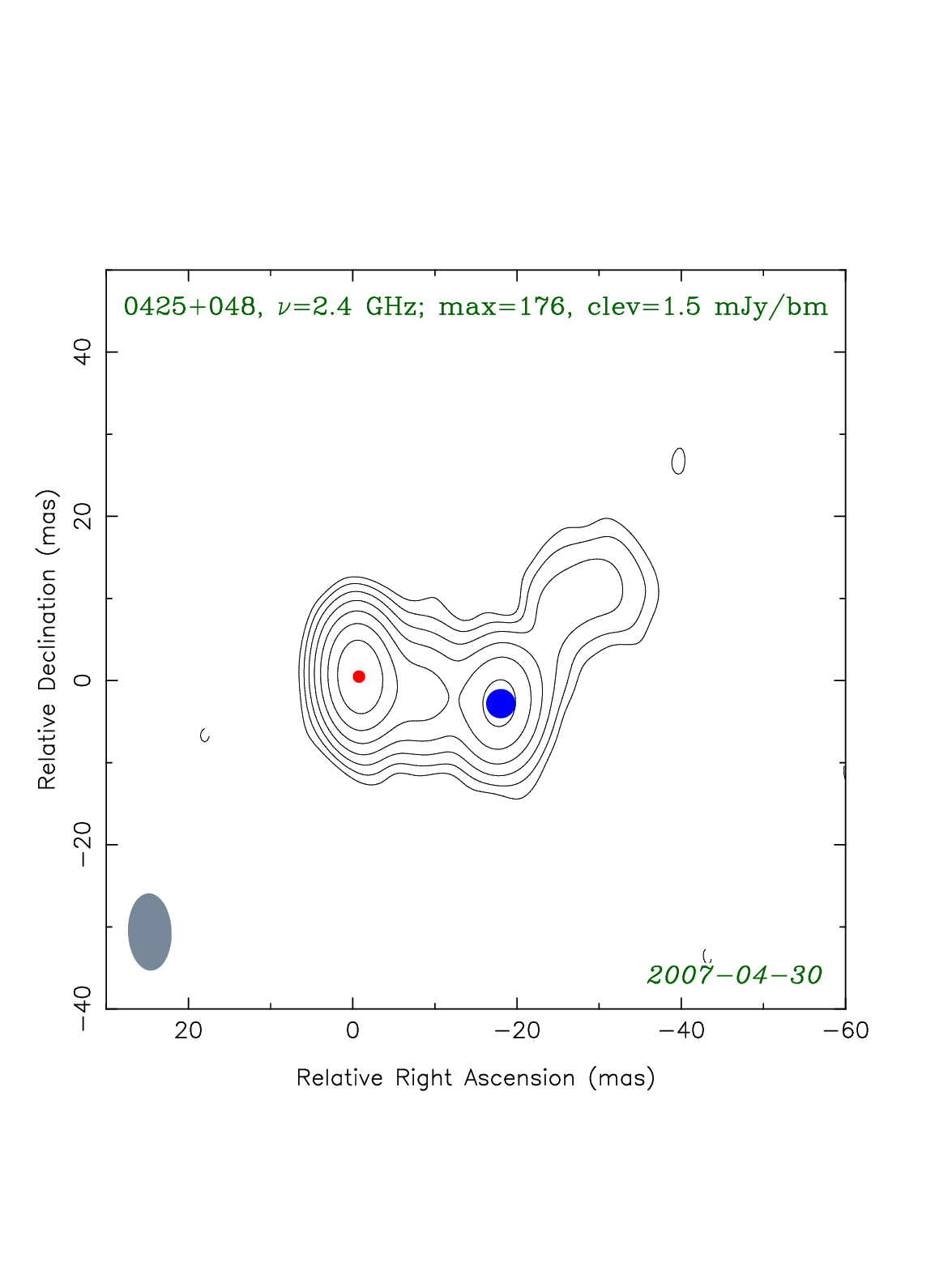}

 \includegraphics[width=0.33\textwidth,angle=0,trim=0.5cm 3cm 1.6cm 5cm,clip]{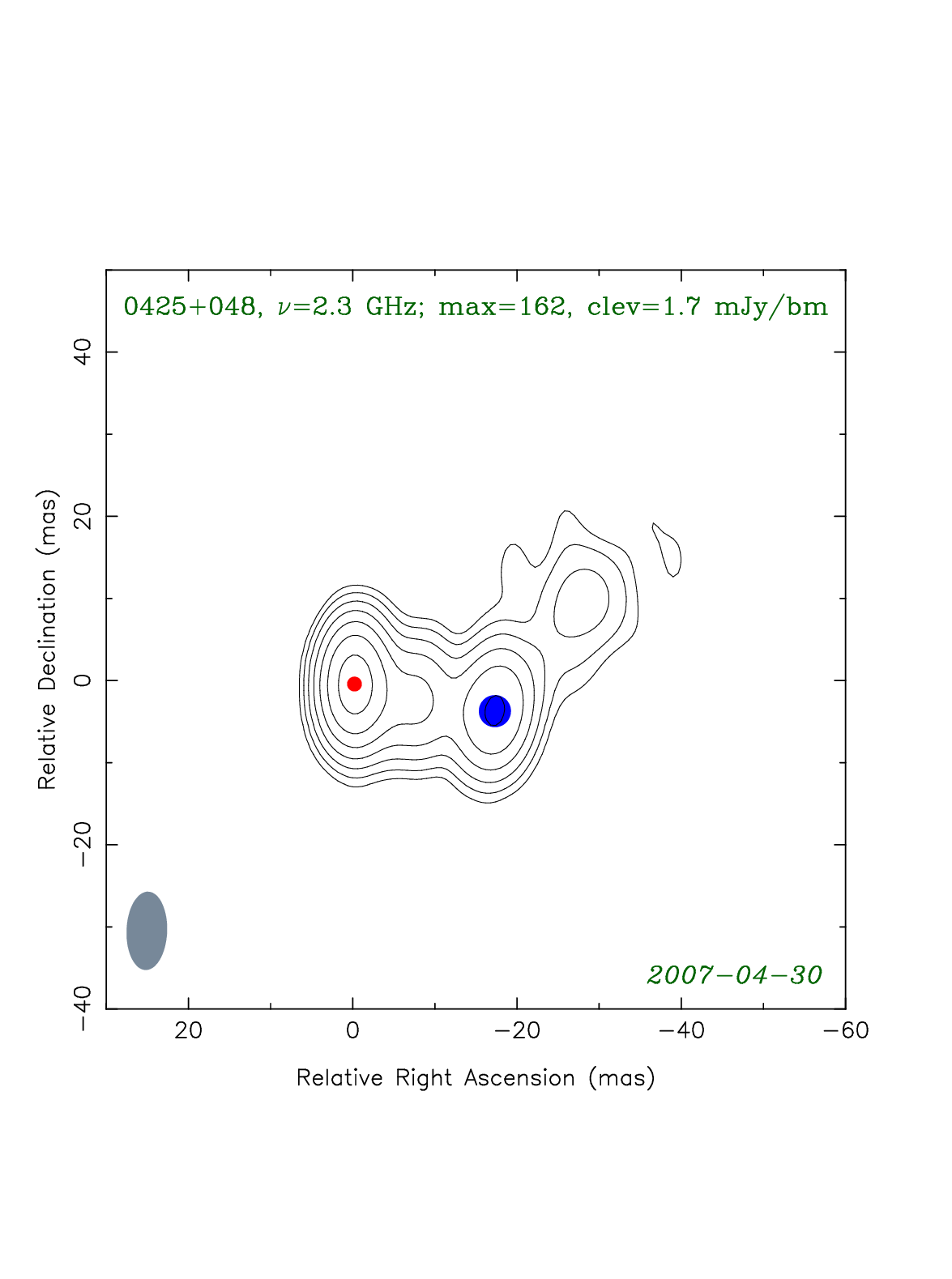}
 \includegraphics[width=0.33\textwidth,angle=0,trim=0.5cm 3cm 1.6cm 5cm,clip]{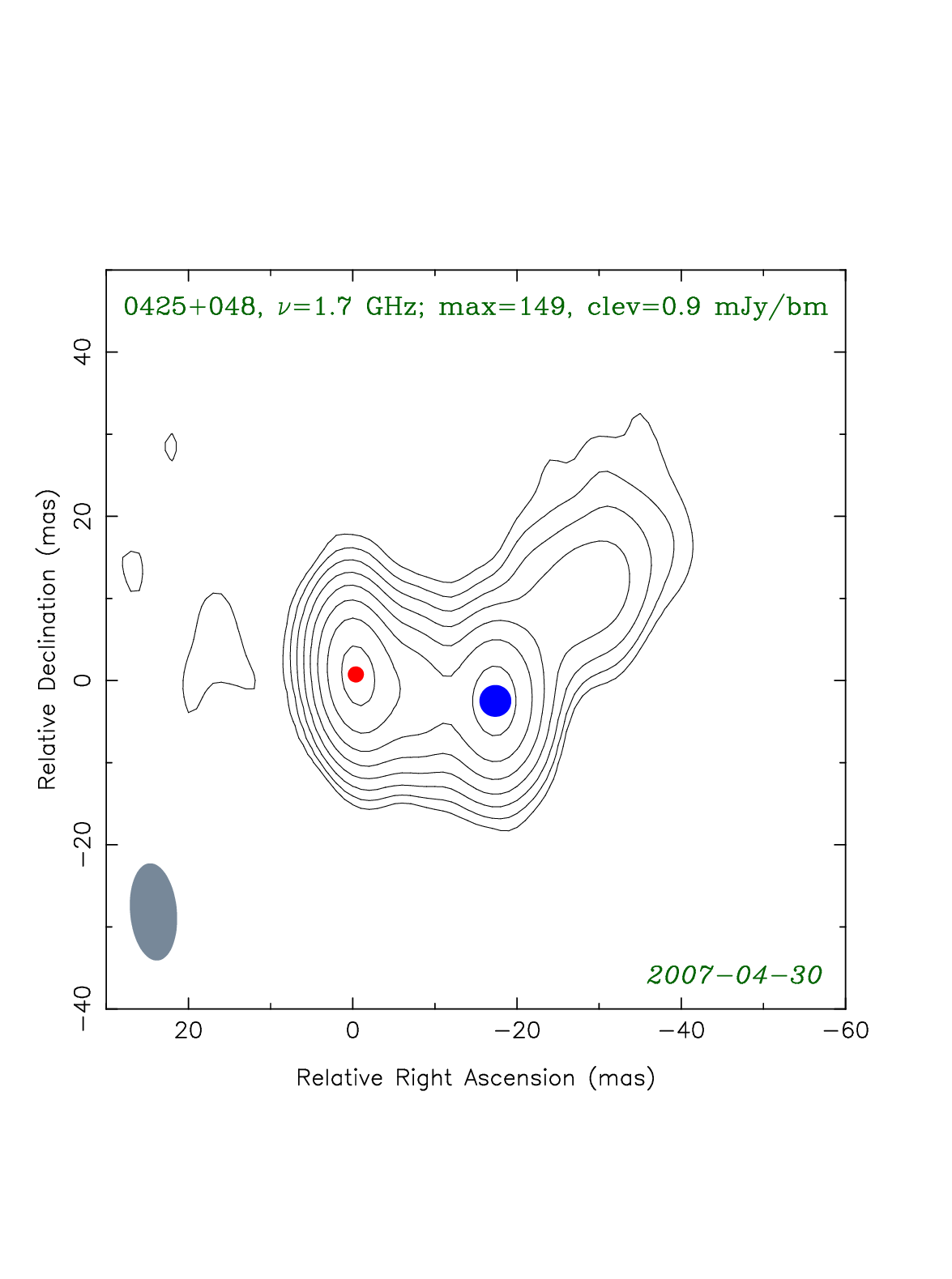}
 \includegraphics[width=0.33\textwidth,angle=0,trim=0.5cm 3cm 1.6cm 5cm,clip]{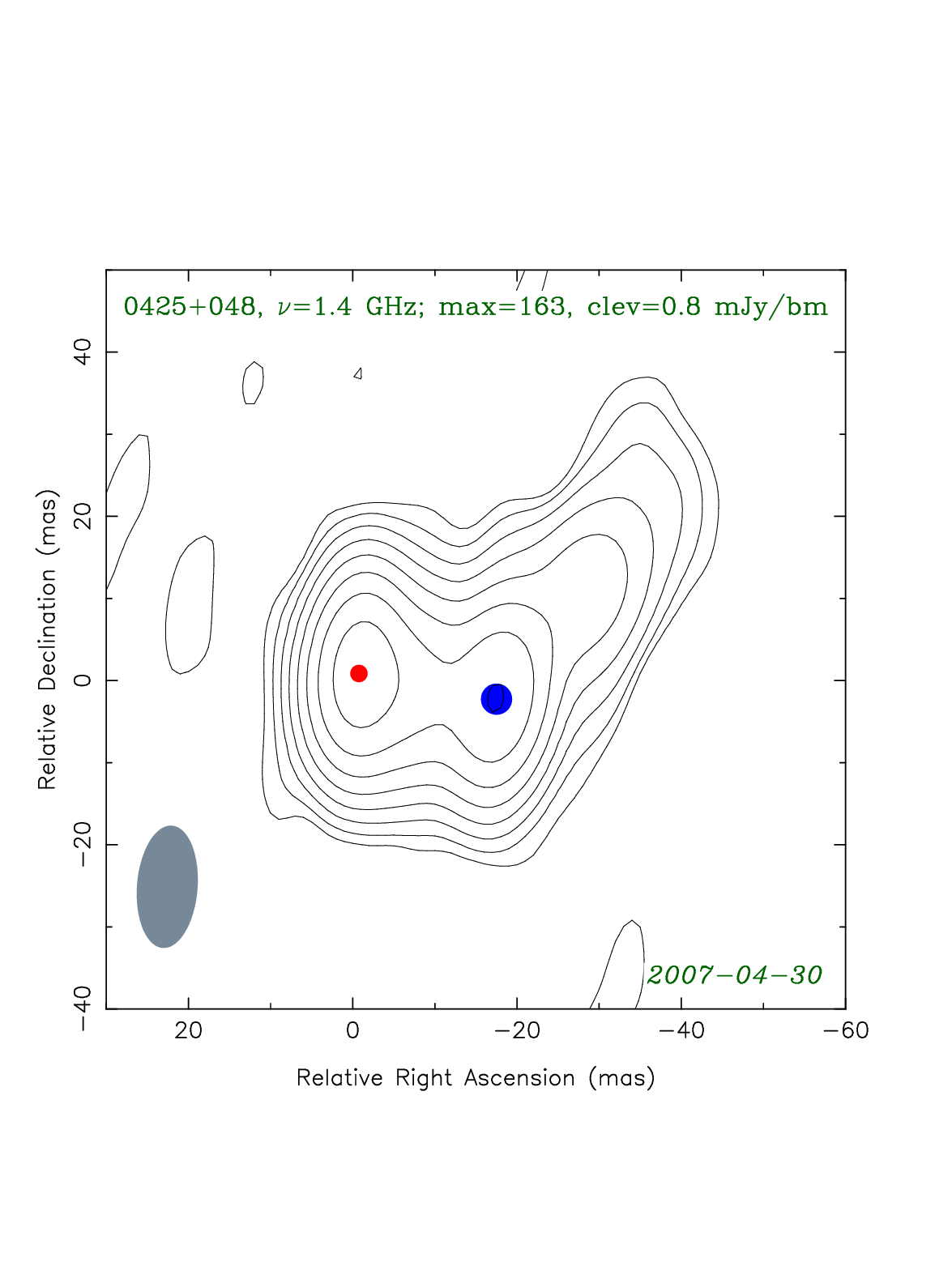}

 \caption{continued...}
\end{figure*}

\clearpage
\addtocounter{figure}{-1}
\begin{figure*}
 \centering

 \includegraphics[width=0.33\textwidth,angle=0,trim=0.5cm 3cm 1.6cm 5cm,clip]{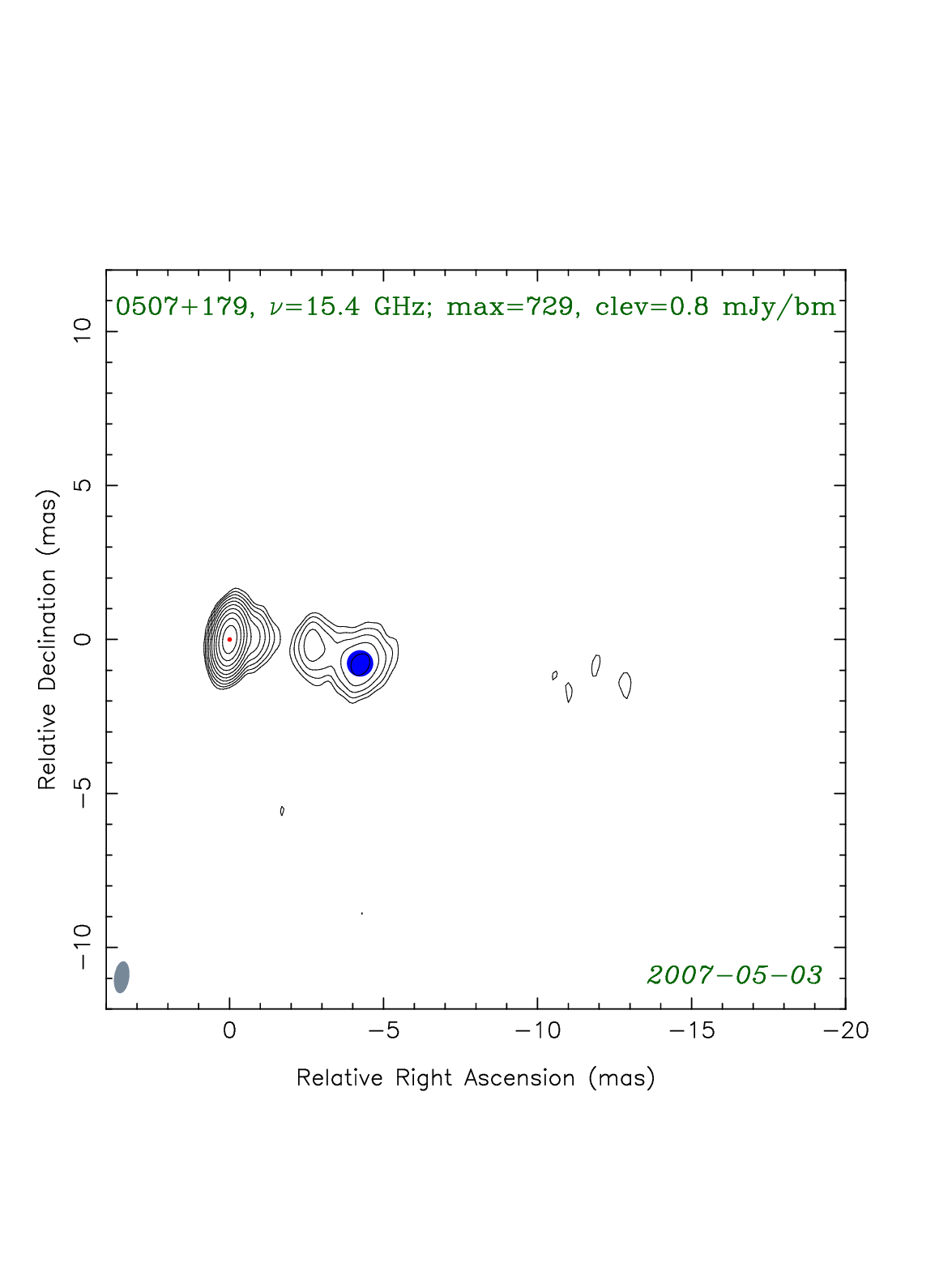}
 \includegraphics[width=0.33\textwidth,angle=0,trim=0.5cm 3cm 1.6cm 5cm,clip]{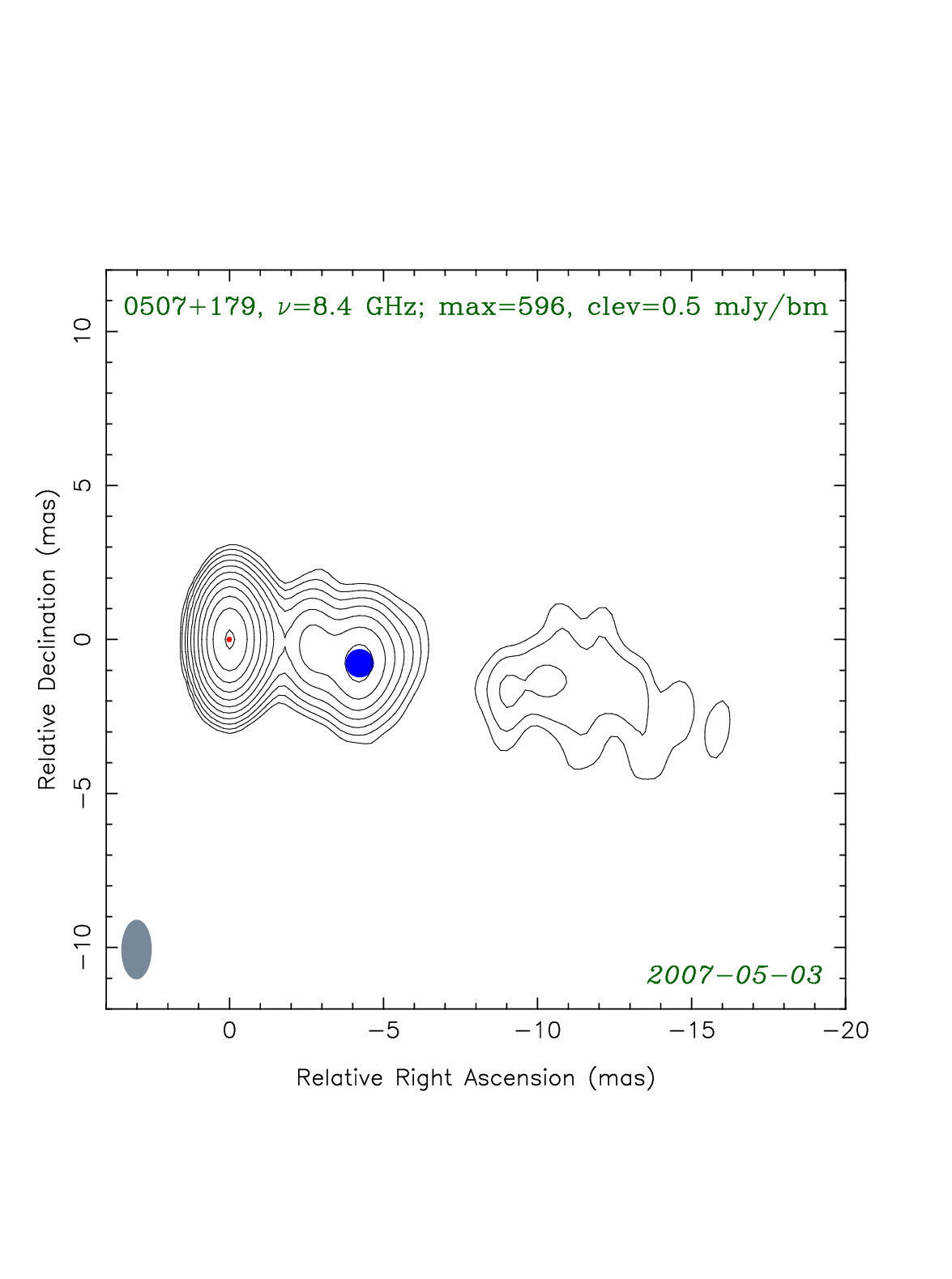}
 \includegraphics[width=0.33\textwidth,angle=0,trim=0.5cm 3cm 1.6cm 5cm,clip]{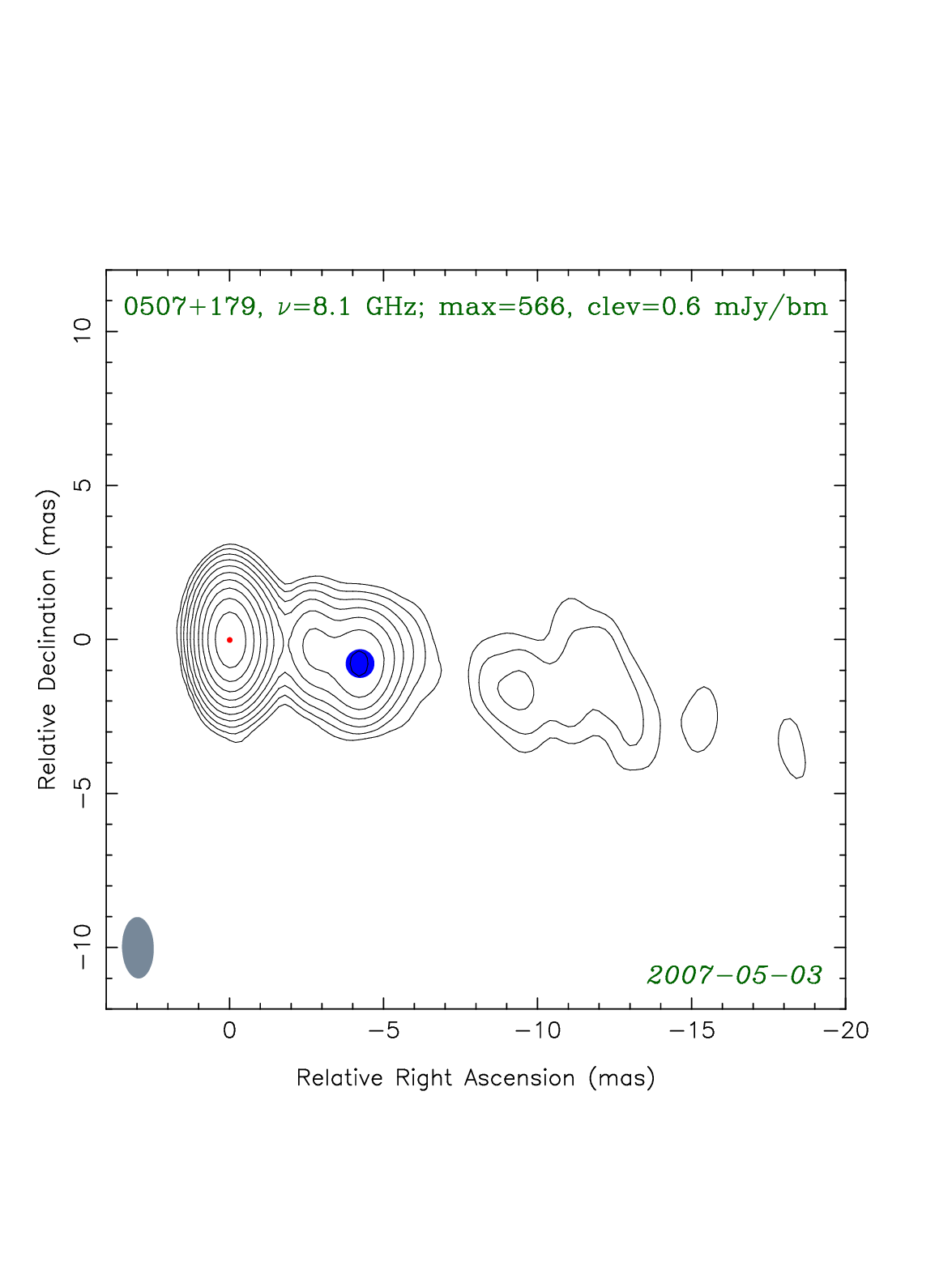}

 \includegraphics[width=0.33\textwidth,angle=0,trim=0.5cm 3cm 1.6cm 5cm,clip]{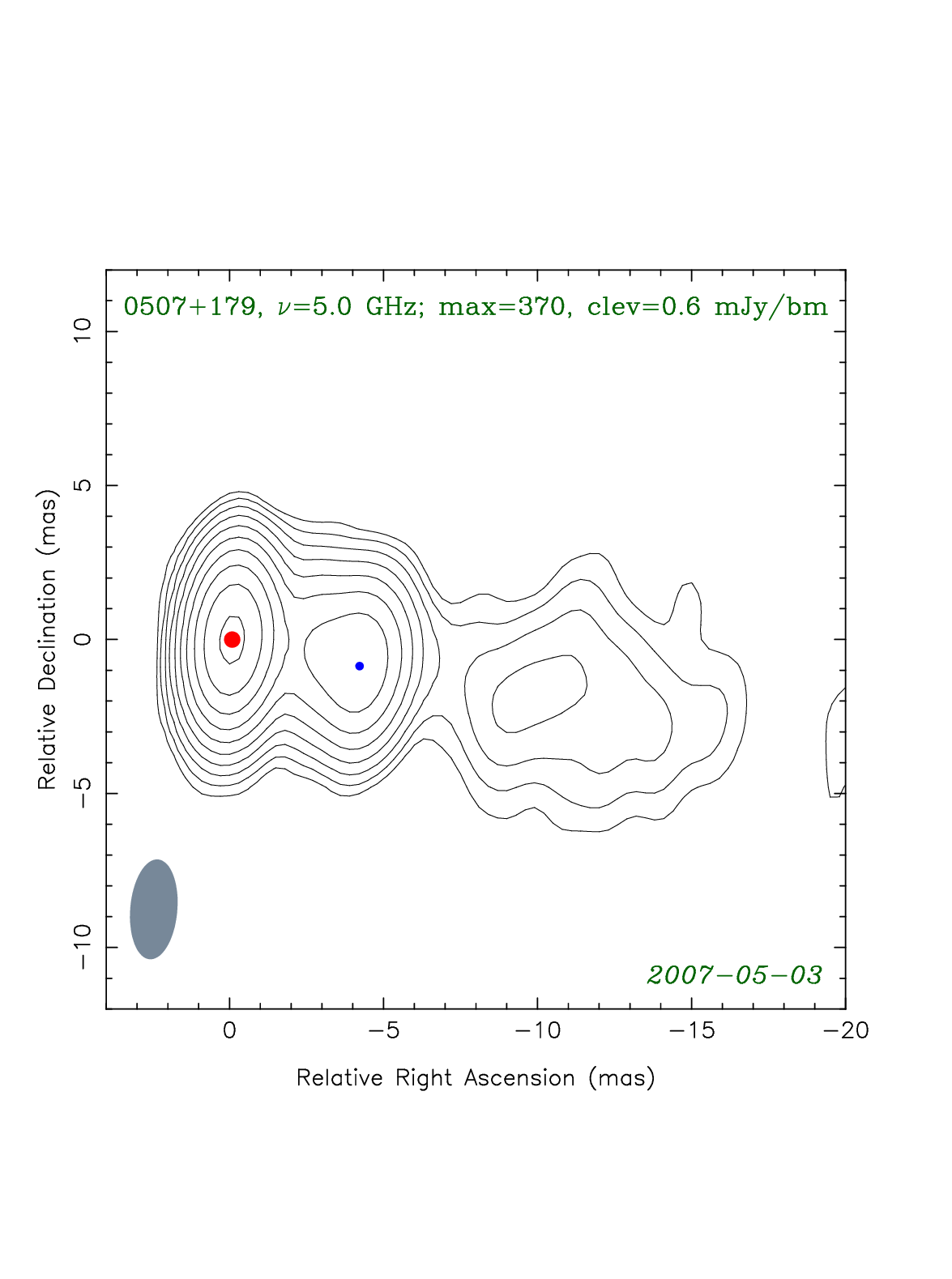}
 \includegraphics[width=0.33\textwidth,angle=0,trim=0.5cm 3cm 1.6cm 5cm,clip]{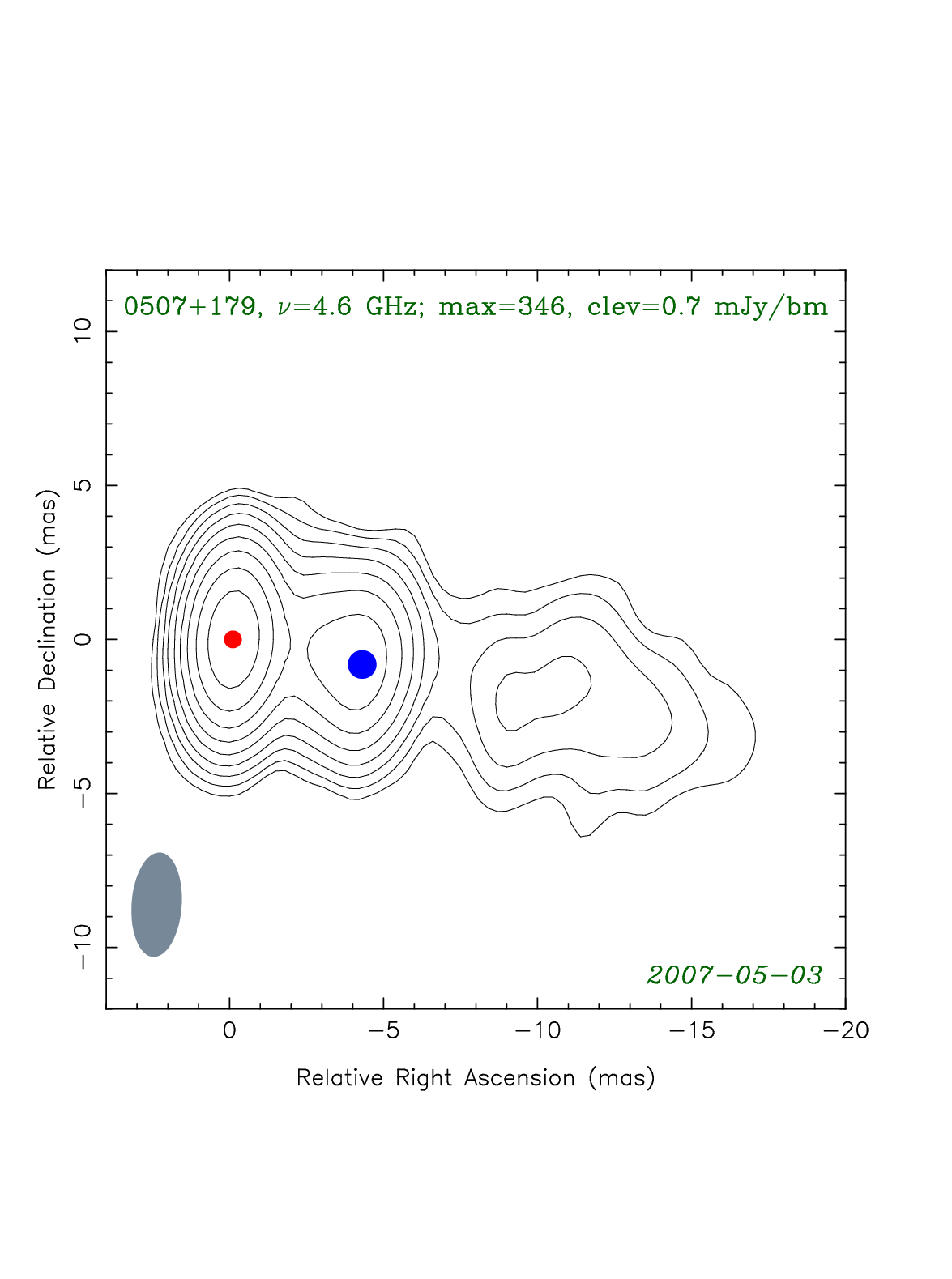}
 \includegraphics[width=0.33\textwidth,angle=0,trim=0.5cm 3cm 1.6cm 5cm,clip]{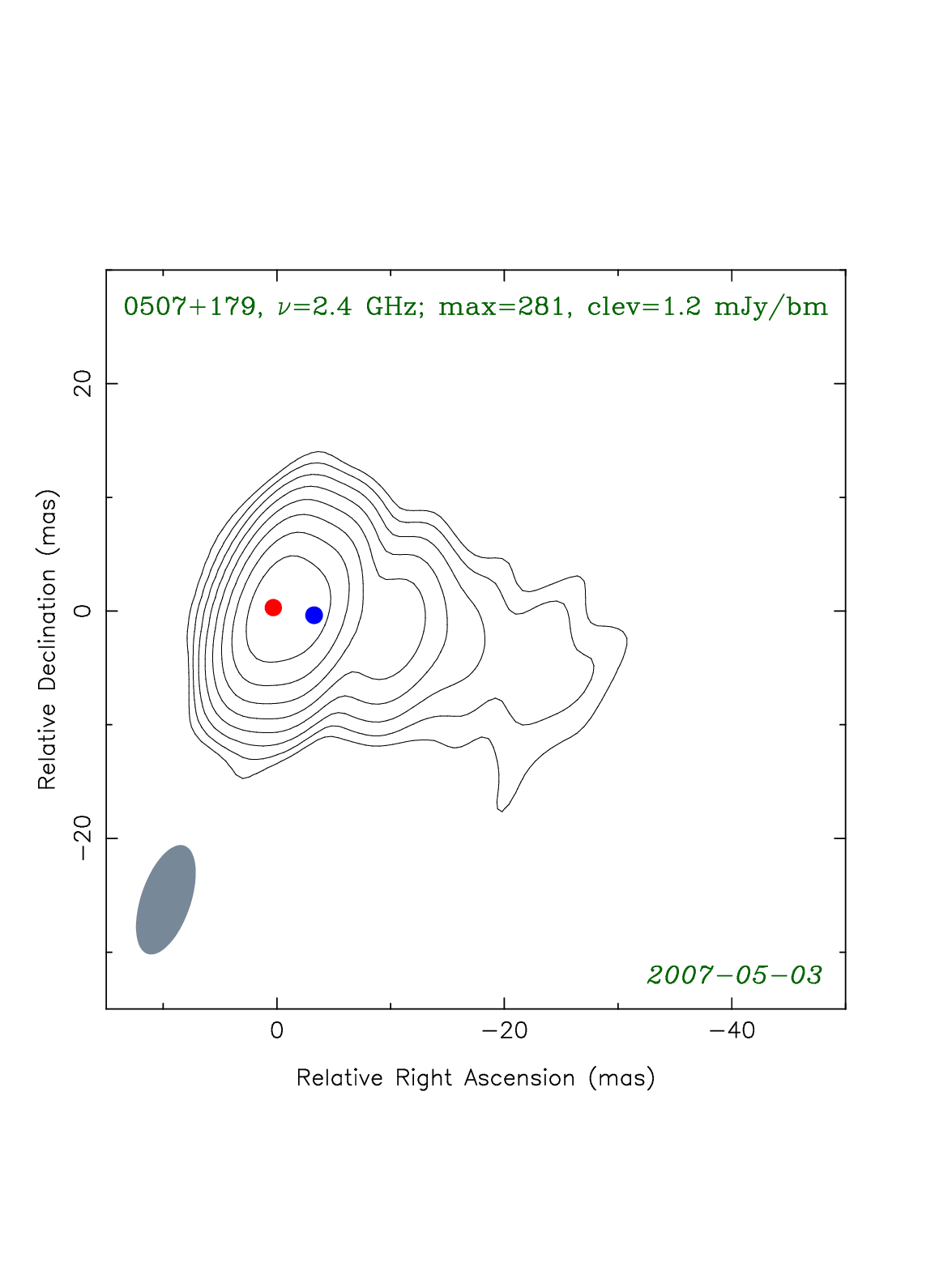}

 \includegraphics[width=0.33\textwidth,angle=0,trim=0.5cm 3cm 1.6cm 5cm,clip]{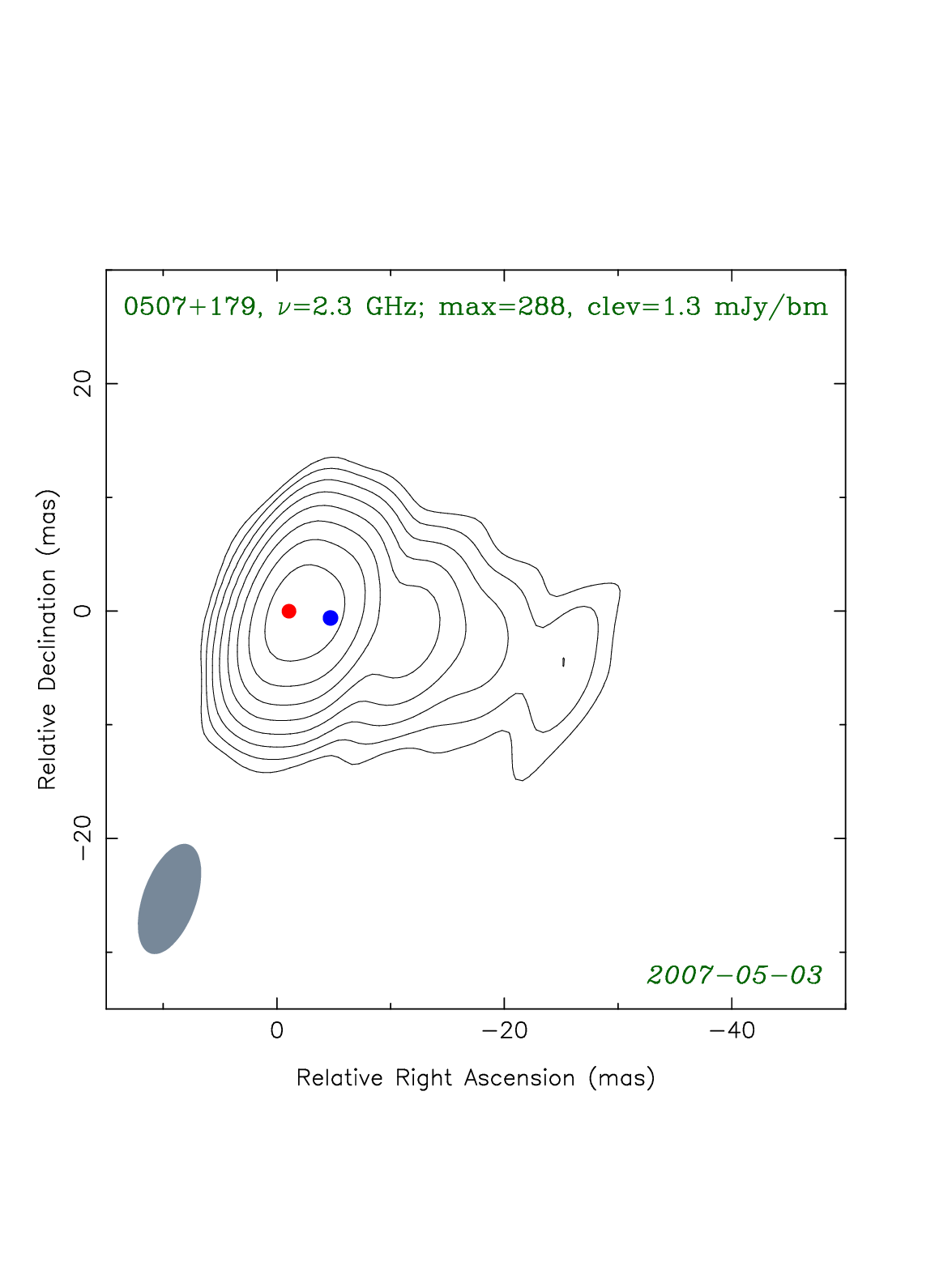}
 \includegraphics[width=0.33\textwidth,angle=0,trim=0.5cm 3cm 1.6cm 5cm,clip]{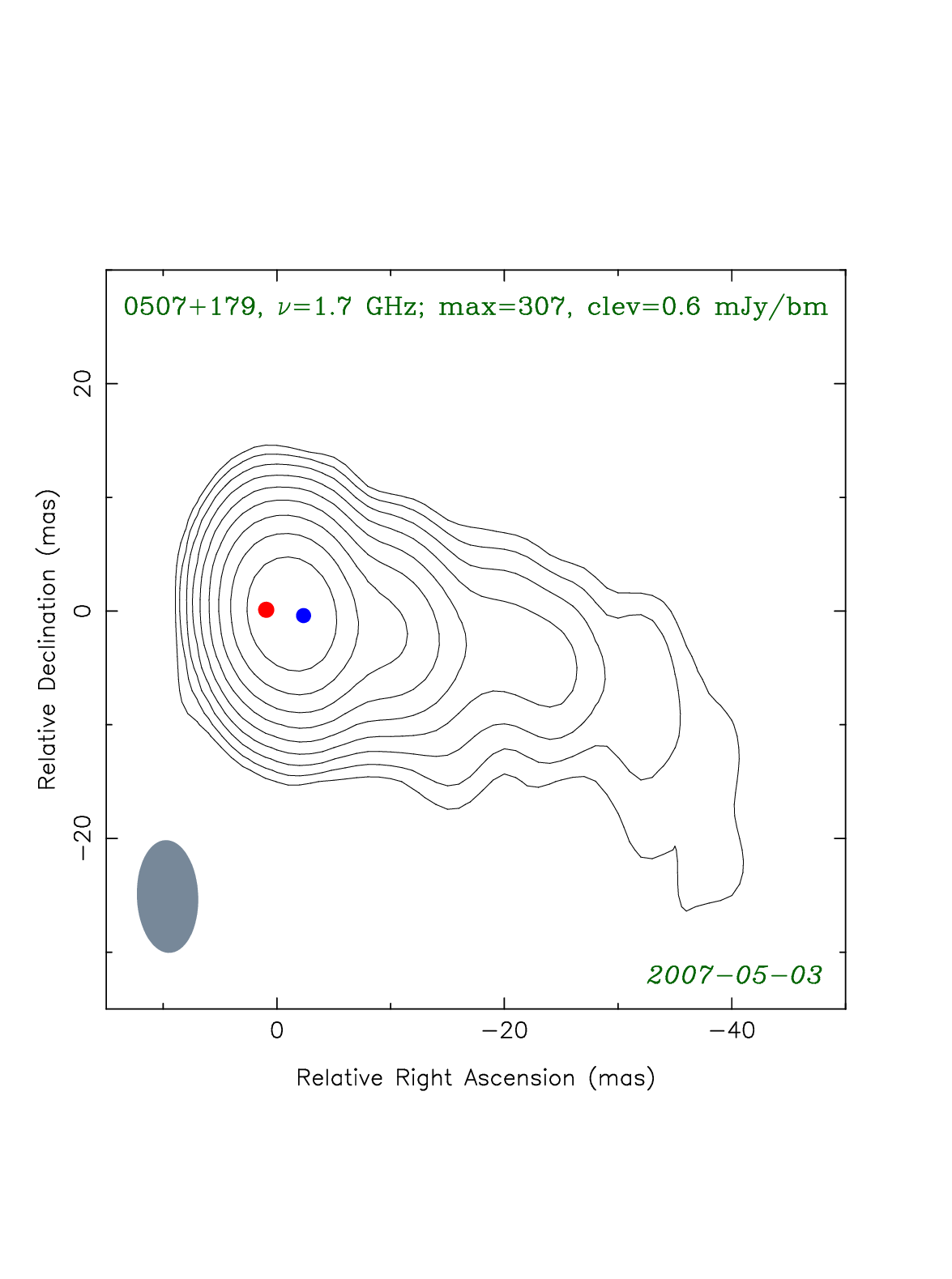}
 \includegraphics[width=0.33\textwidth,angle=0,trim=0.5cm 3cm 1.6cm 5cm,clip]{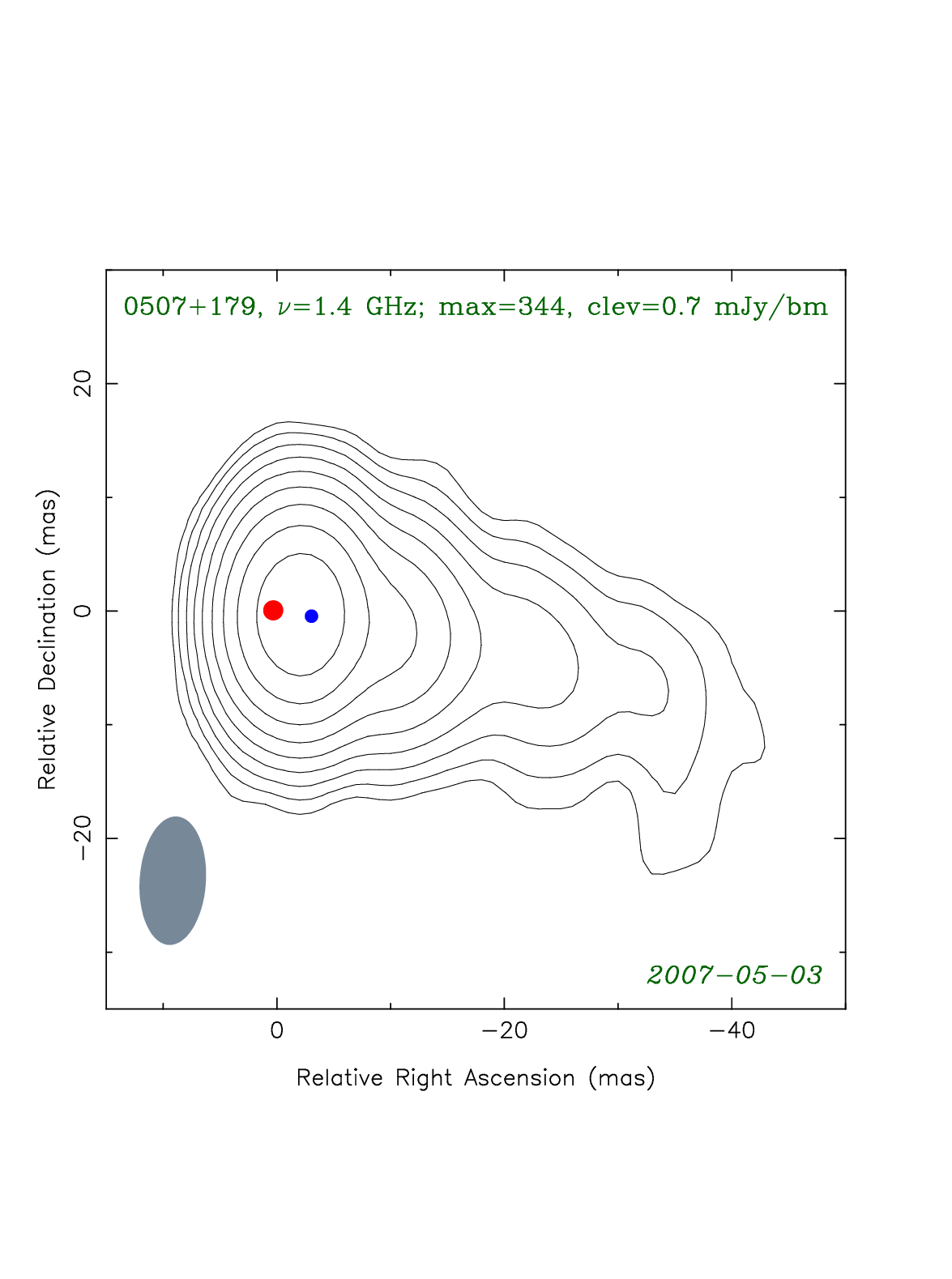}

 \caption{continued...}
\end{figure*}

\clearpage
\addtocounter{figure}{-1}
\begin{figure*}
 \centering

 \includegraphics[width=0.33\textwidth,angle=0,trim=0.5cm 3cm 1.6cm 5cm,clip]{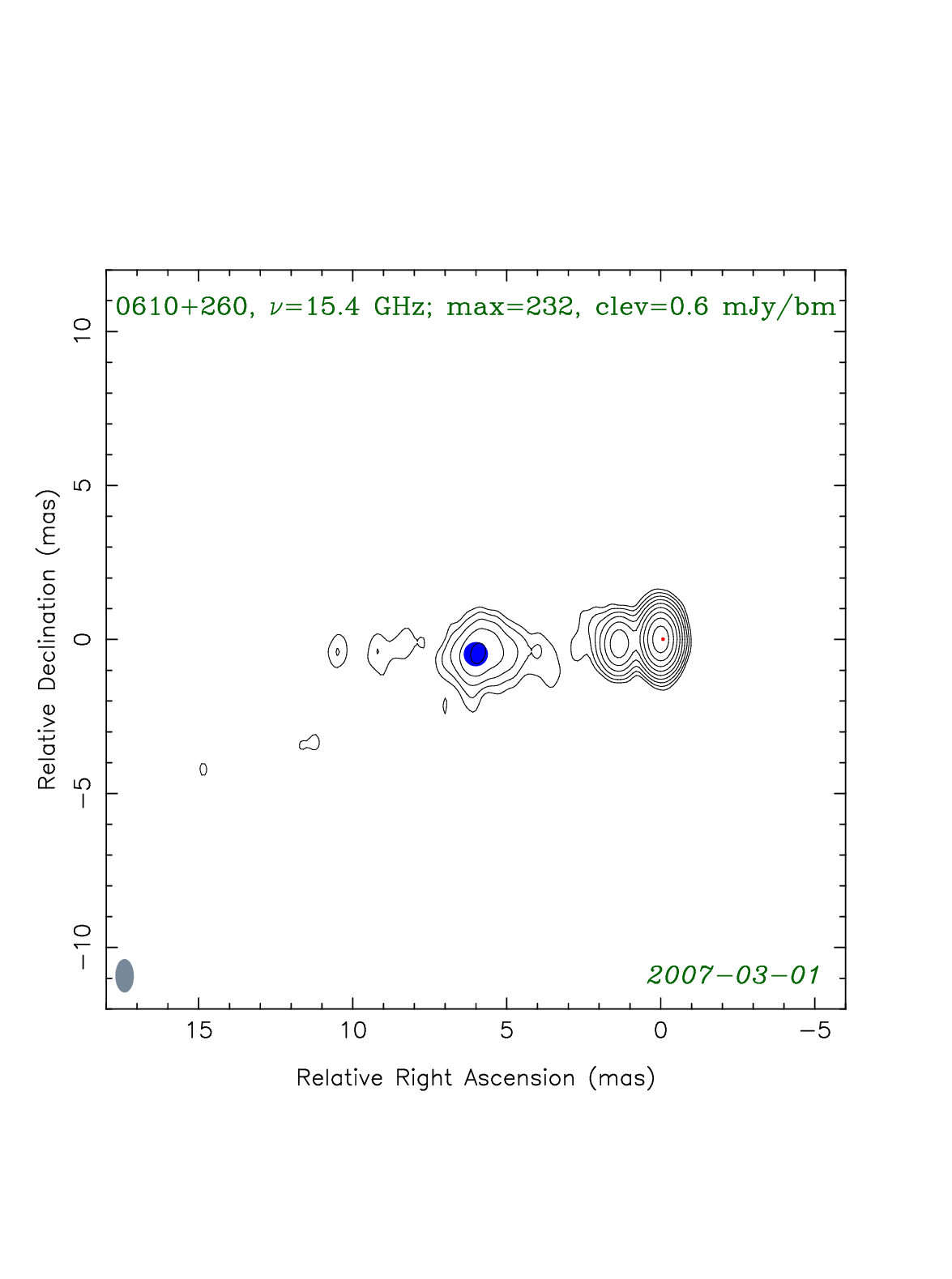}
 \includegraphics[width=0.33\textwidth,angle=0,trim=0.5cm 3cm 1.6cm 5cm,clip]{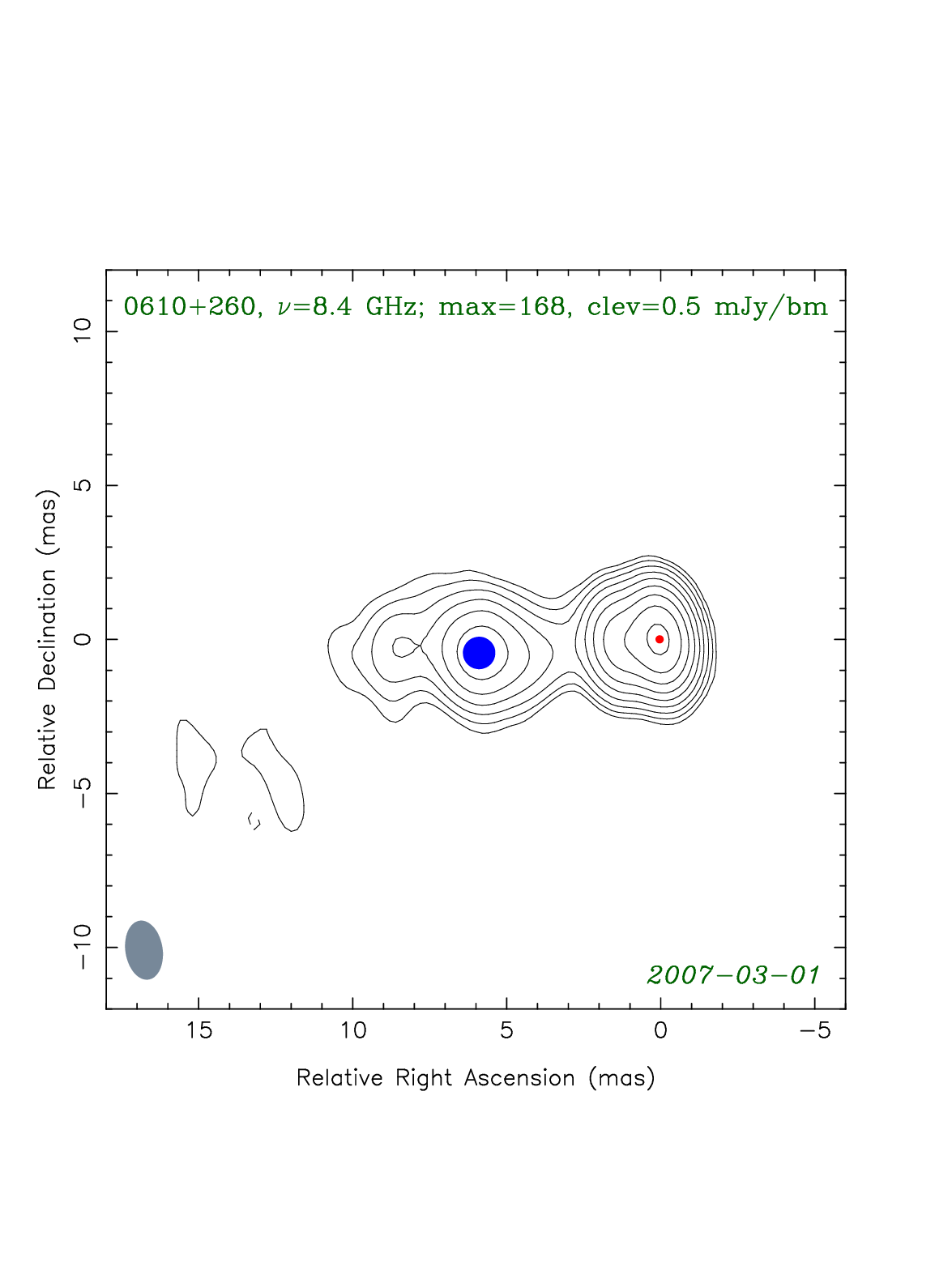}
 \includegraphics[width=0.33\textwidth,angle=0,trim=0.5cm 3cm 1.6cm 5cm,clip]{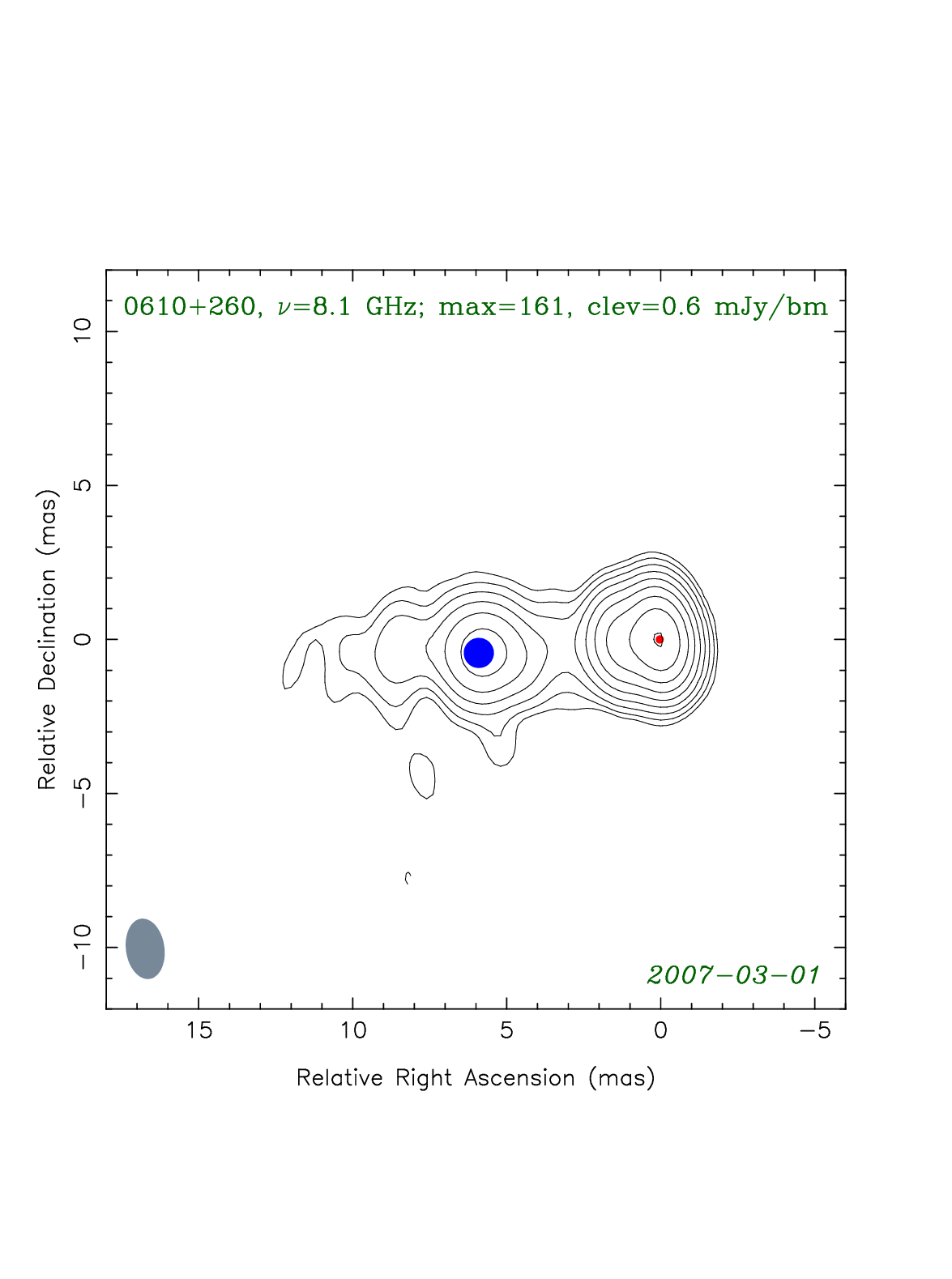}

 \includegraphics[width=0.33\textwidth,angle=0,trim=0.5cm 3cm 1.6cm 5cm,clip]{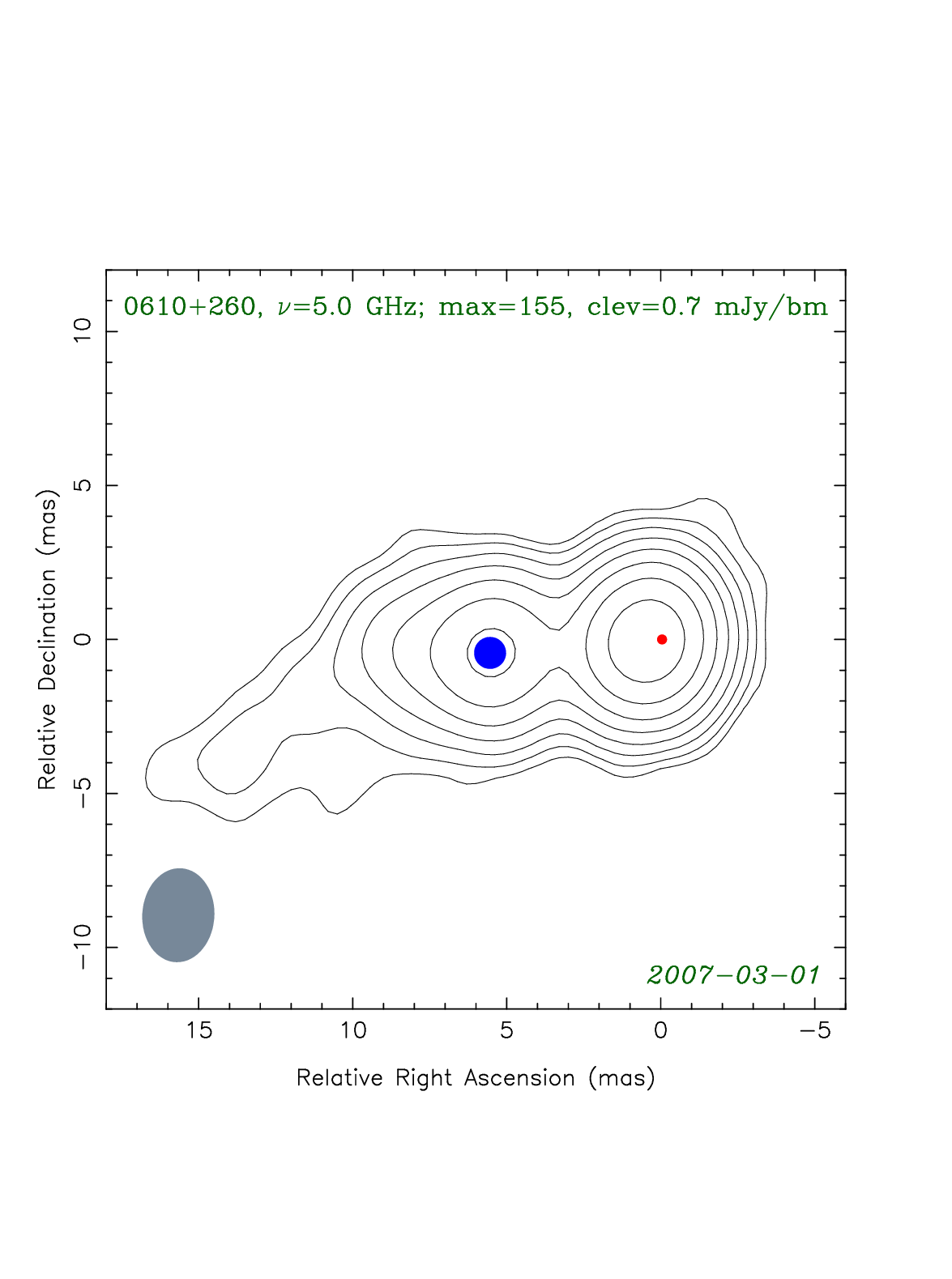}
 \includegraphics[width=0.33\textwidth,angle=0,trim=0.5cm 3cm 1.6cm 5cm,clip]{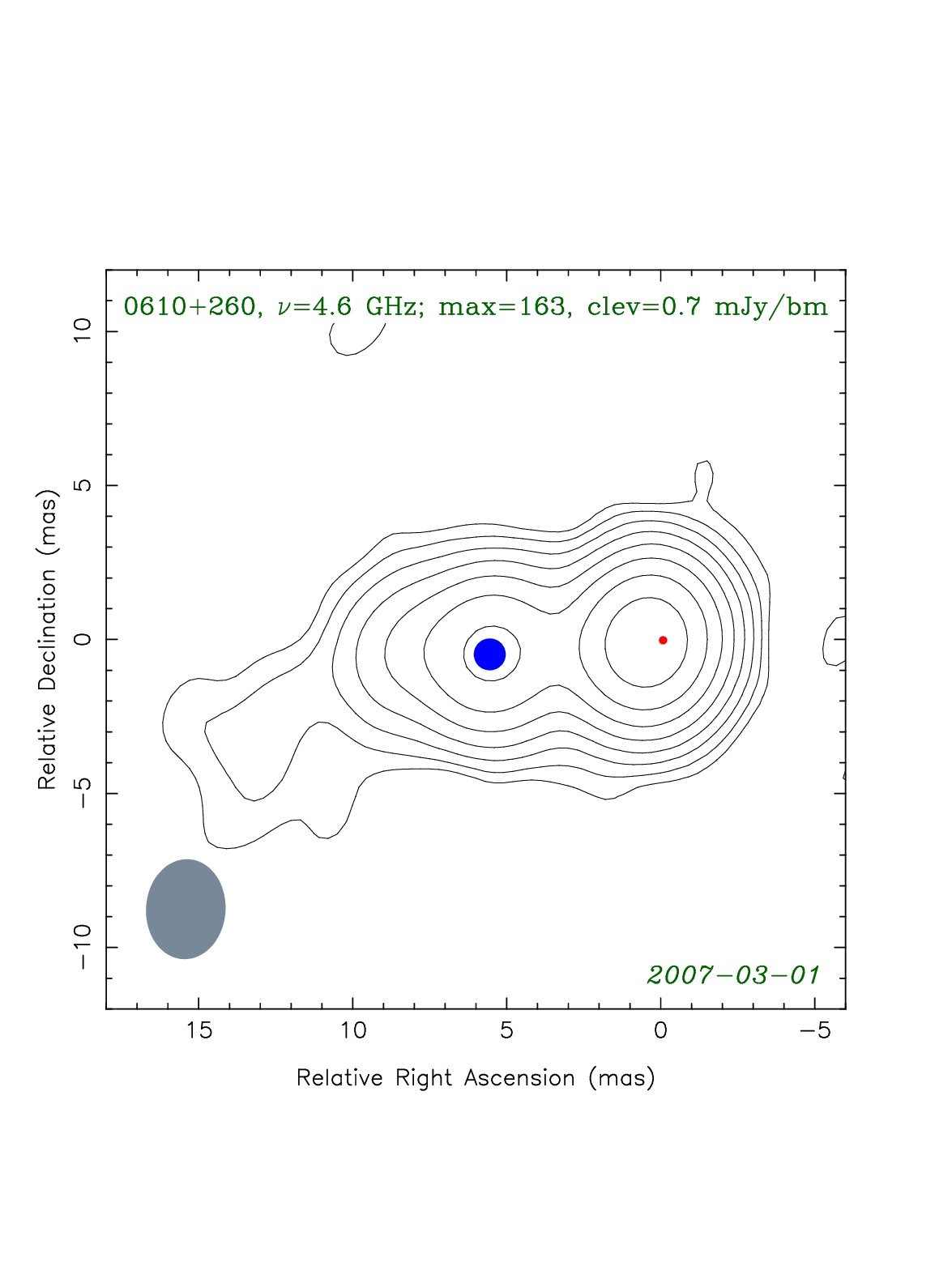}
 \includegraphics[width=0.33\textwidth,angle=0,trim=0.5cm 3cm 1.6cm 5cm,clip]{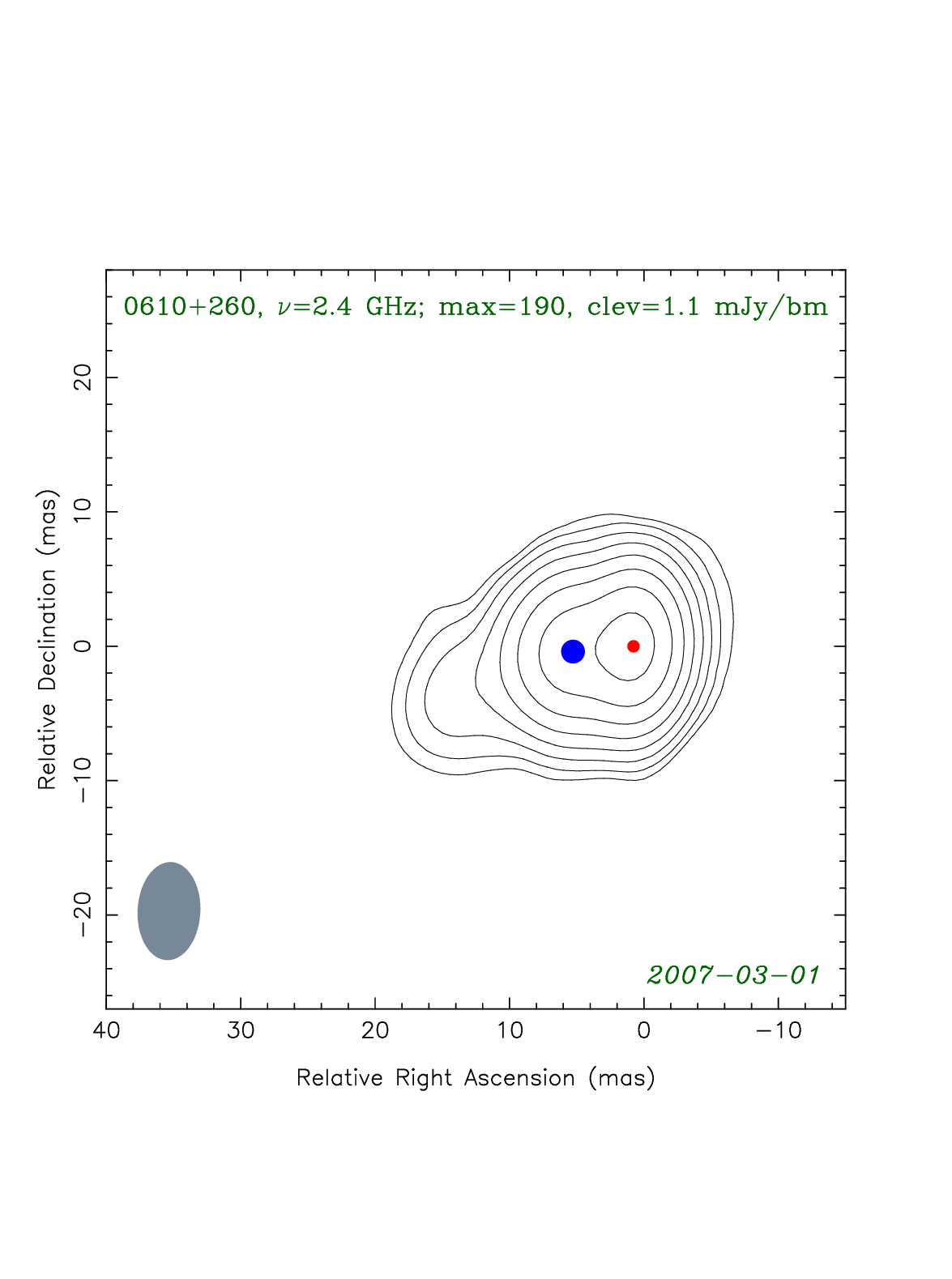}

 \includegraphics[width=0.33\textwidth,angle=0,trim=0.5cm 3cm 1.6cm 5cm,clip]{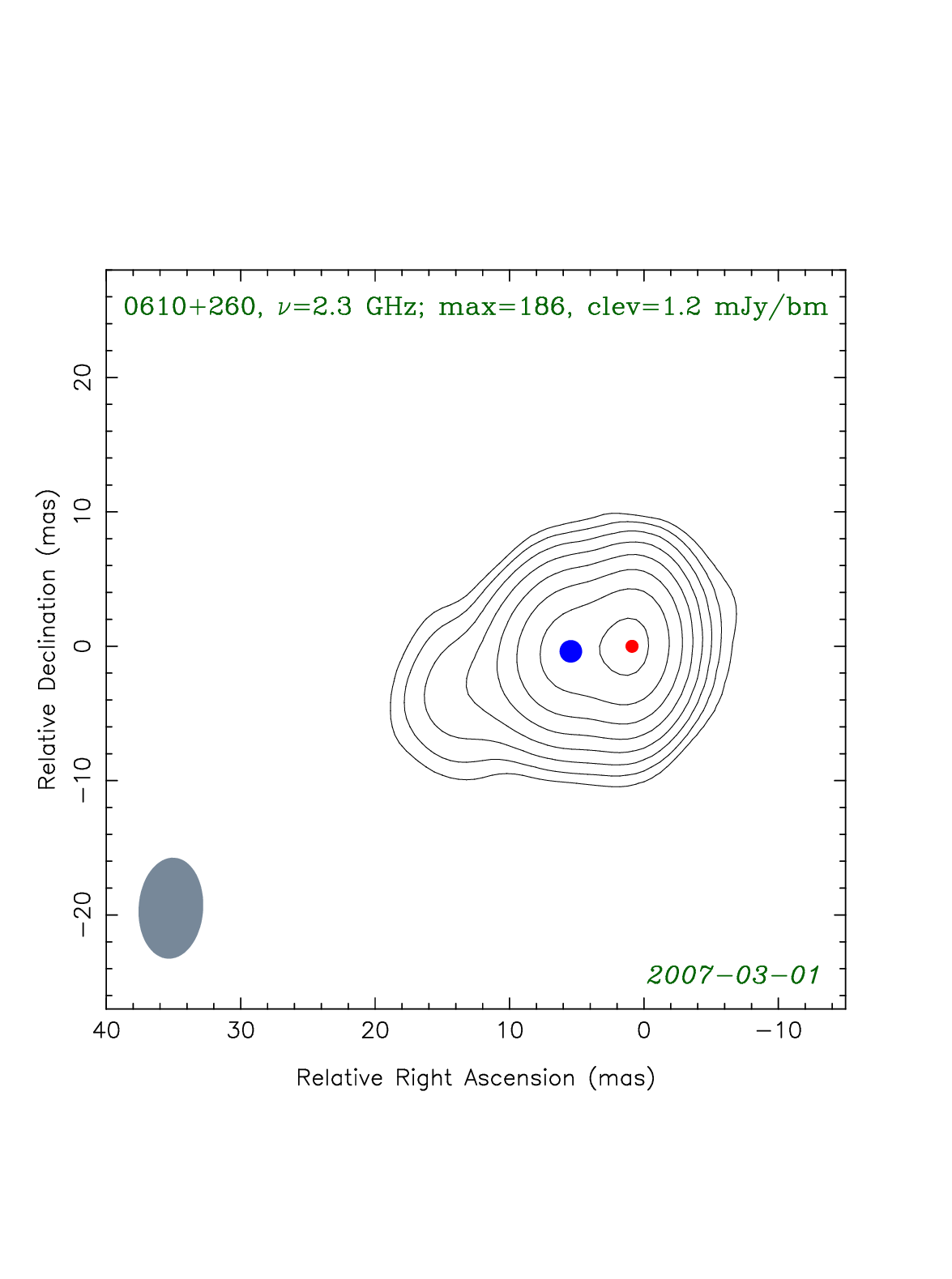}
 \includegraphics[width=0.33\textwidth,angle=0,trim=0.5cm 3cm 1.6cm 5cm,clip]{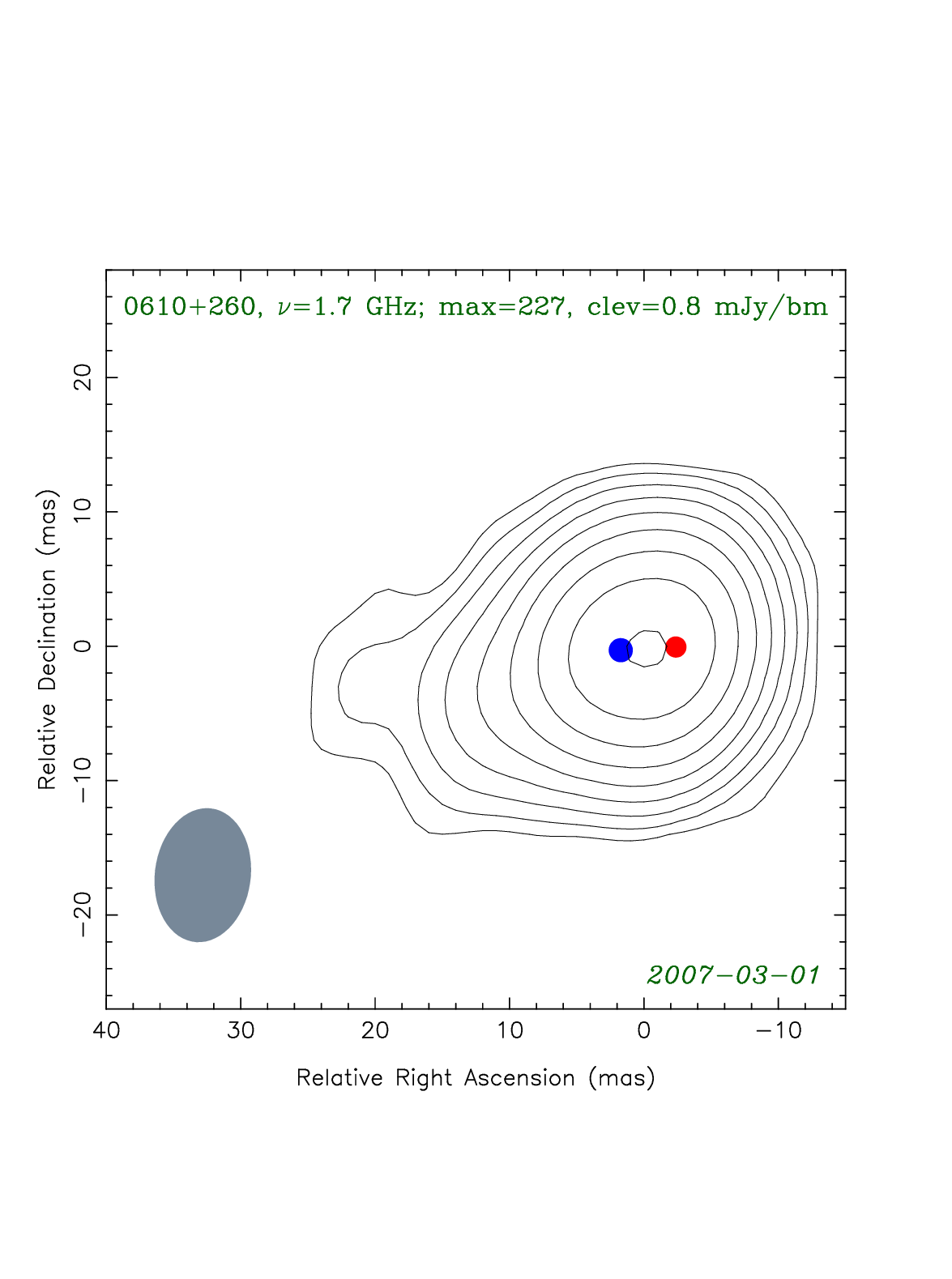}
 \includegraphics[width=0.33\textwidth,angle=0,trim=0.5cm 3cm 1.6cm 5cm,clip]{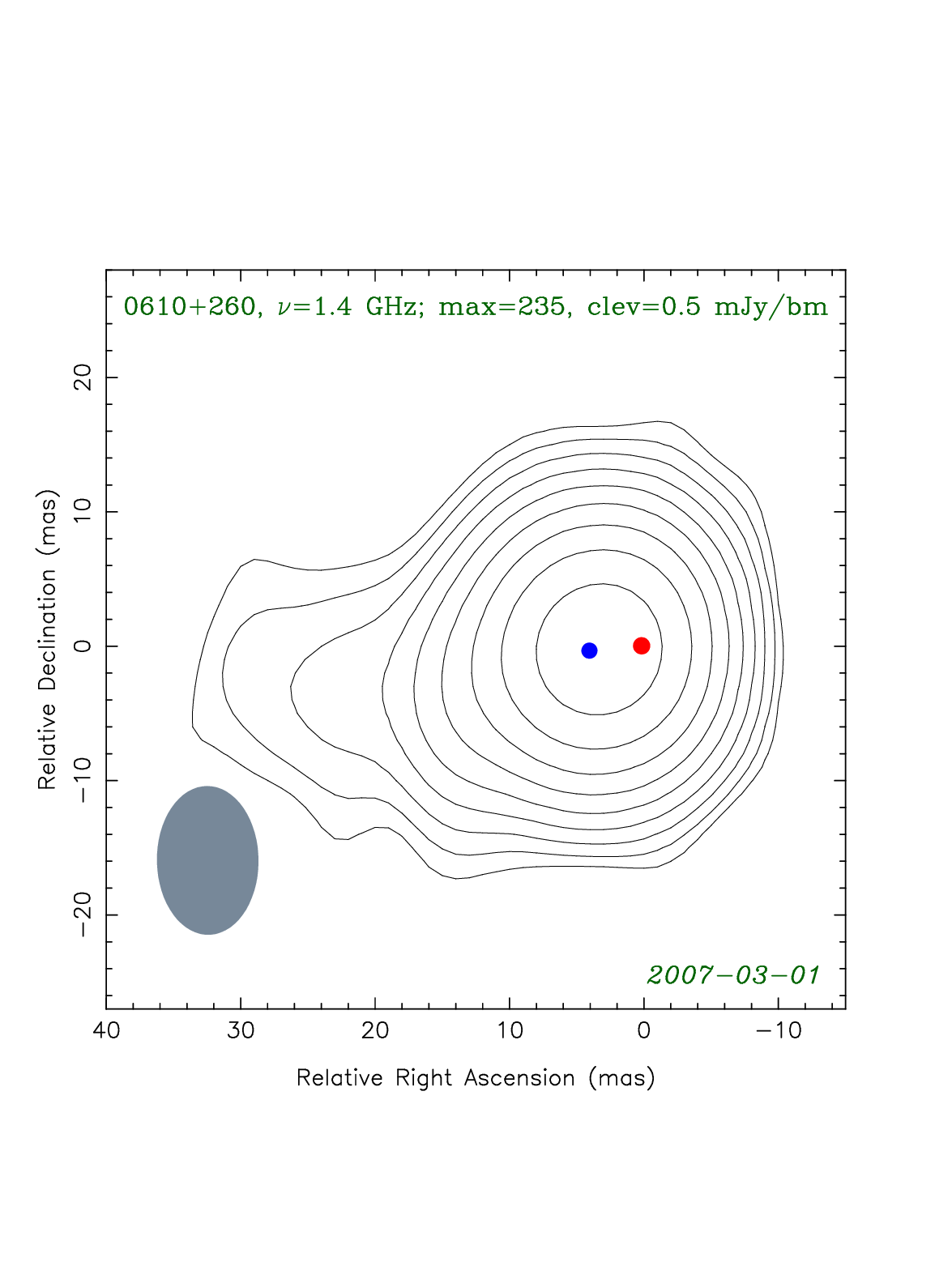}

 \caption{continued...}
\end{figure*}

\clearpage
\addtocounter{figure}{-1}
\begin{figure*}
 \centering

 \includegraphics[width=0.33\textwidth,angle=0,trim=0.5cm 3cm 1.6cm 5cm,clip]{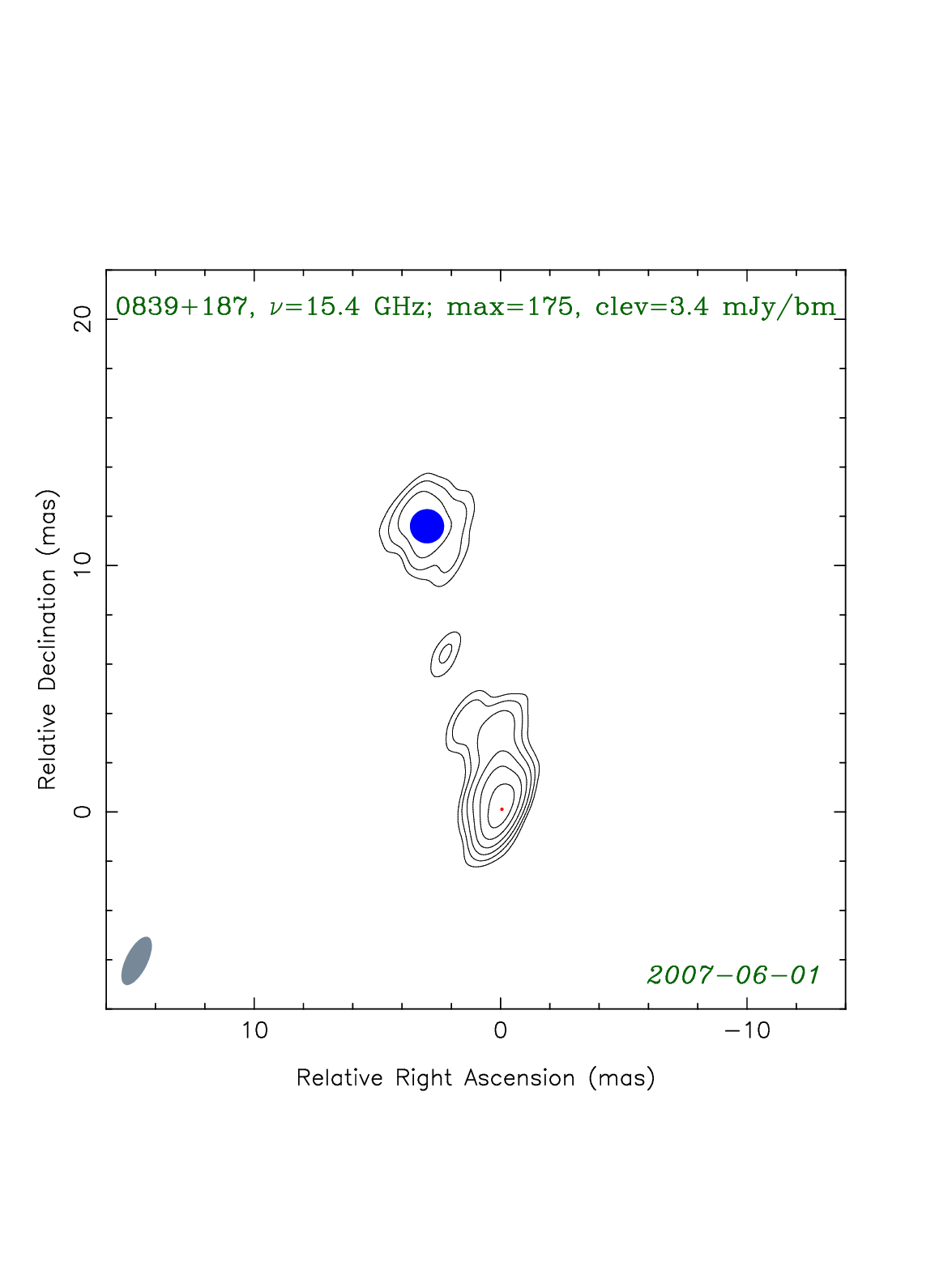}
 \includegraphics[width=0.33\textwidth,angle=0,trim=0.5cm 3cm 1.6cm 5cm,clip]{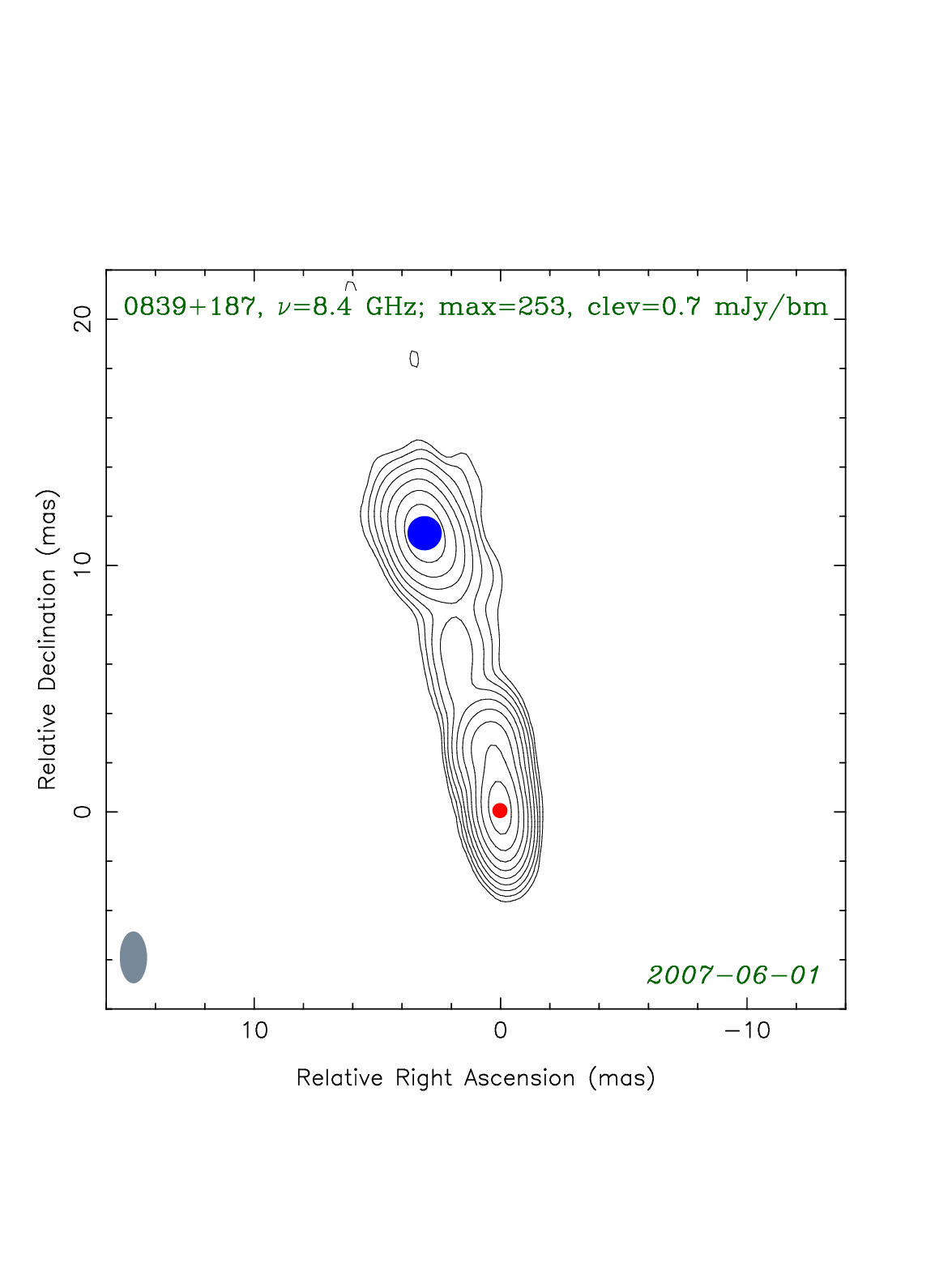}
 \includegraphics[width=0.33\textwidth,angle=0,trim=0.5cm 3cm 1.6cm 5cm,clip]{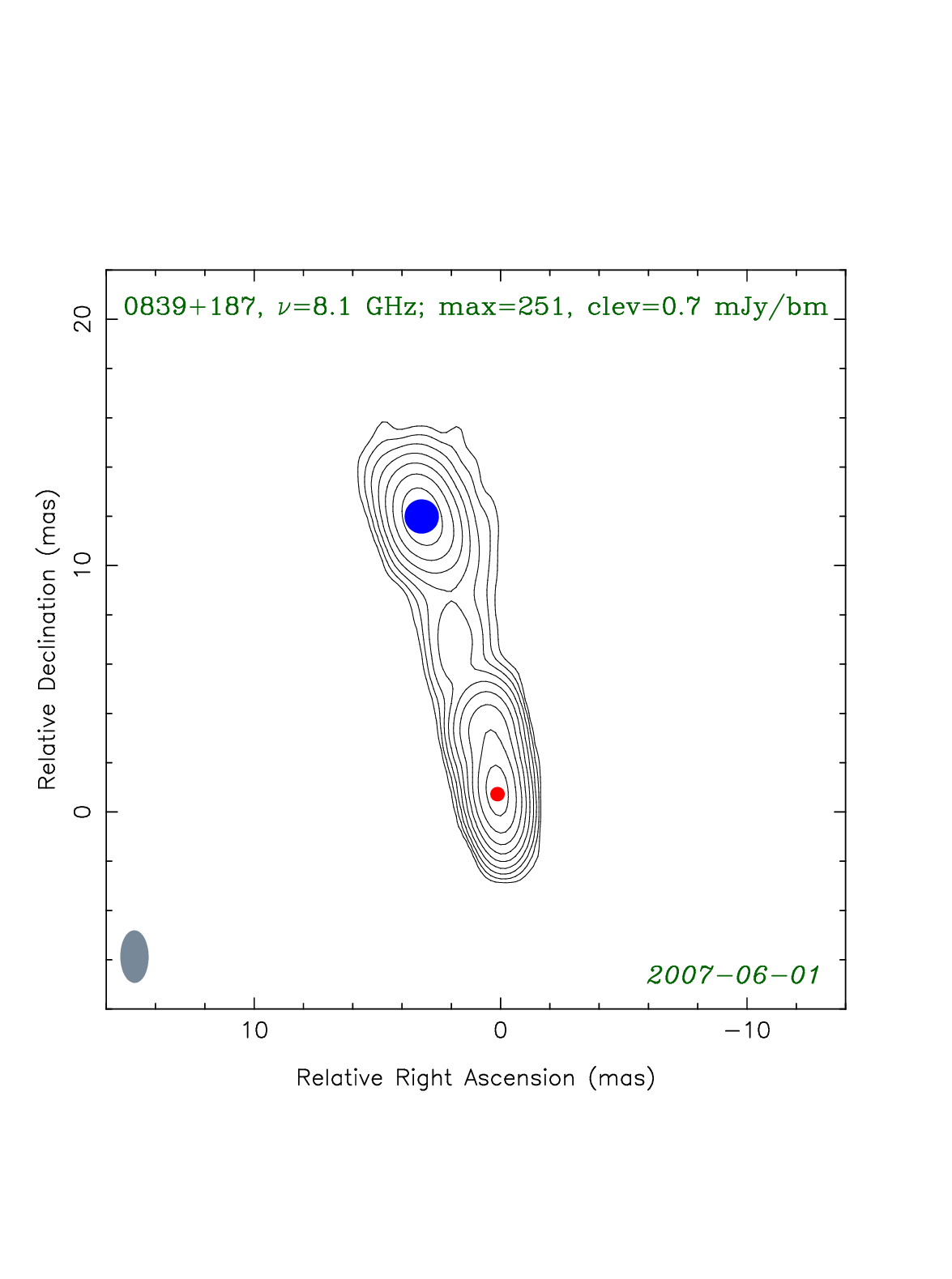}

 \includegraphics[width=0.33\textwidth,angle=0,trim=0.5cm 3cm 1.6cm 5cm,clip]{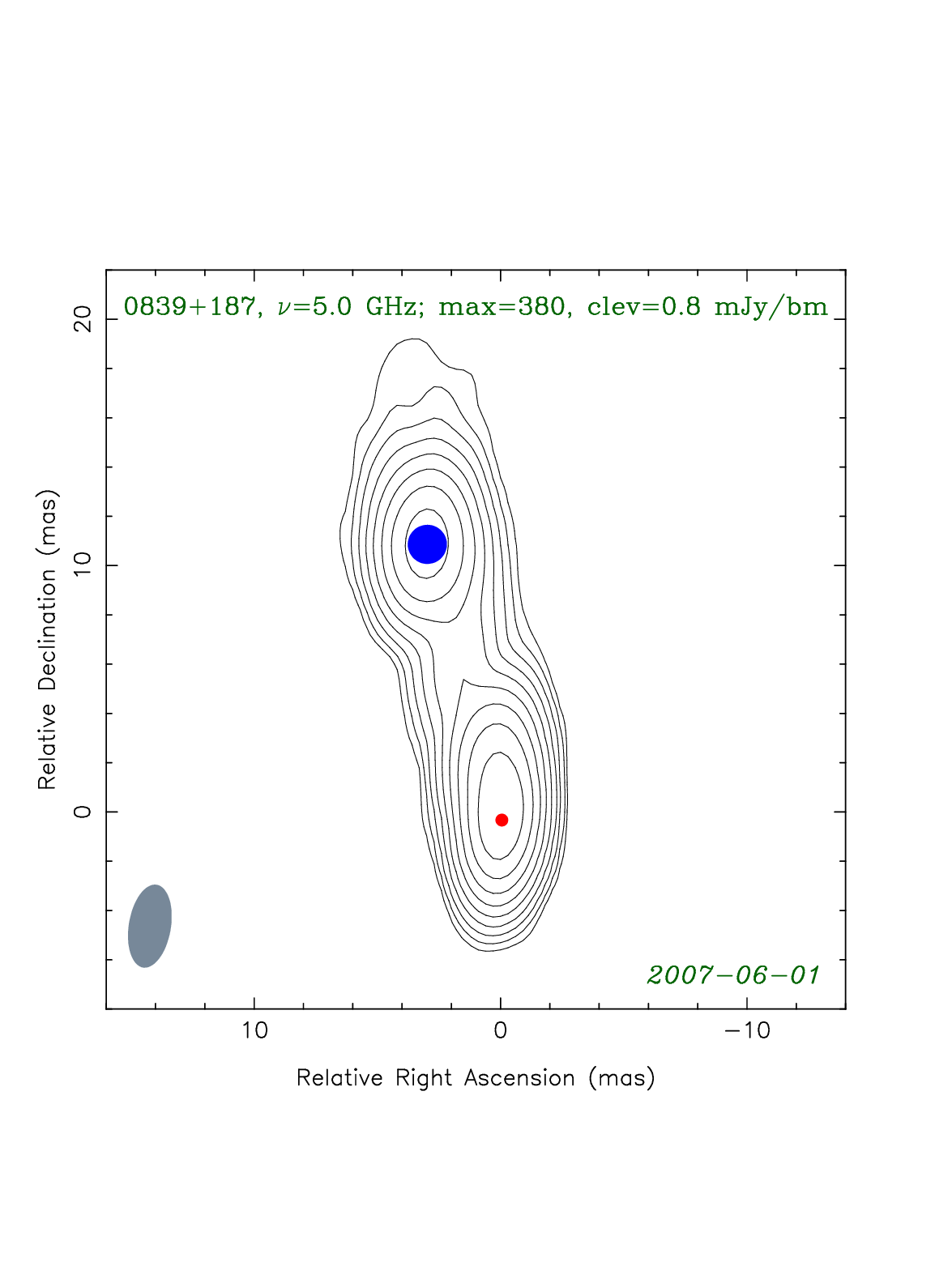}
 \includegraphics[width=0.33\textwidth,angle=0,trim=0.5cm 3cm 1.6cm 5cm,clip]{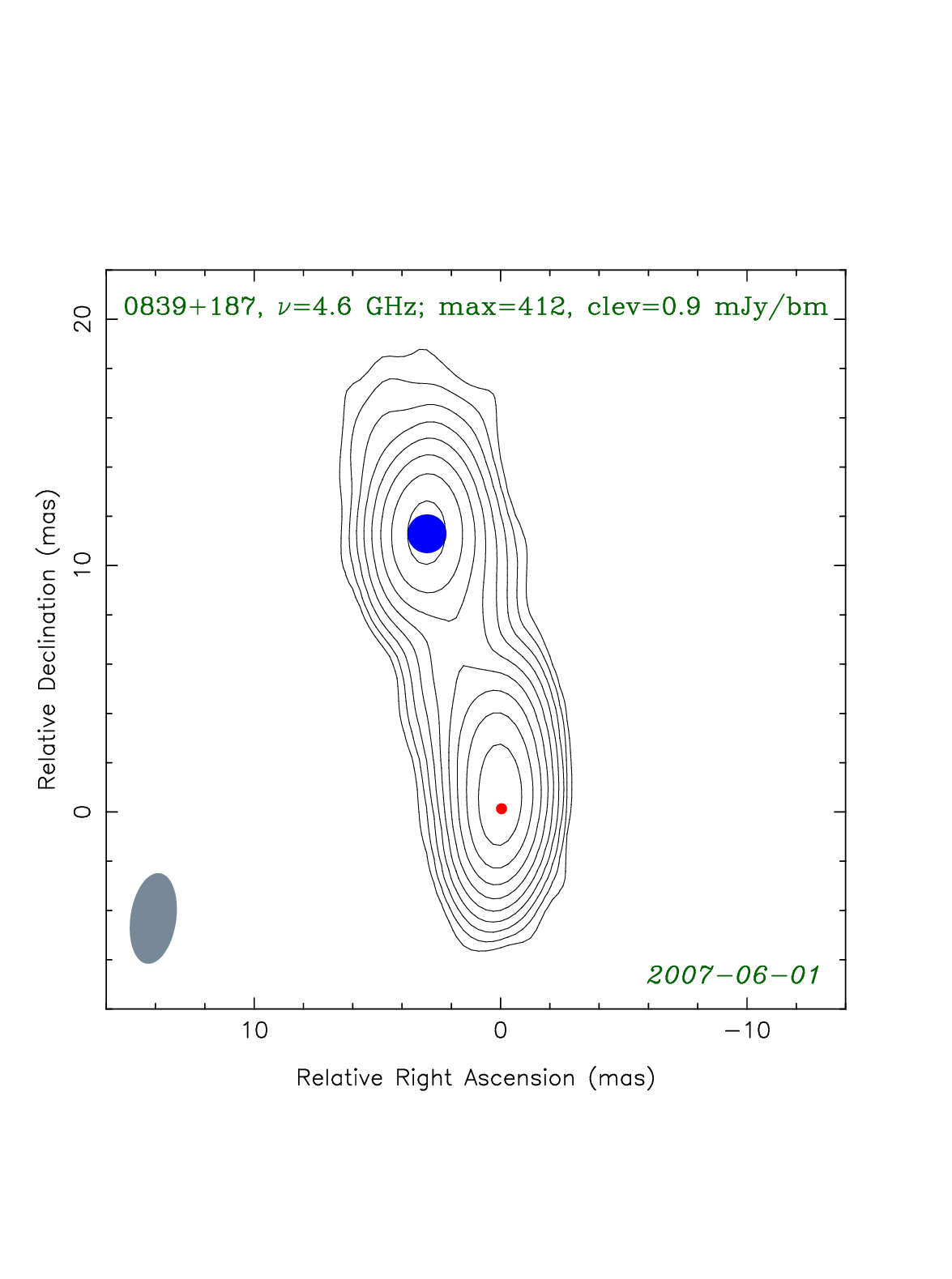}
 \includegraphics[width=0.33\textwidth,angle=0,trim=0.5cm 3cm 1.6cm 5cm,clip]{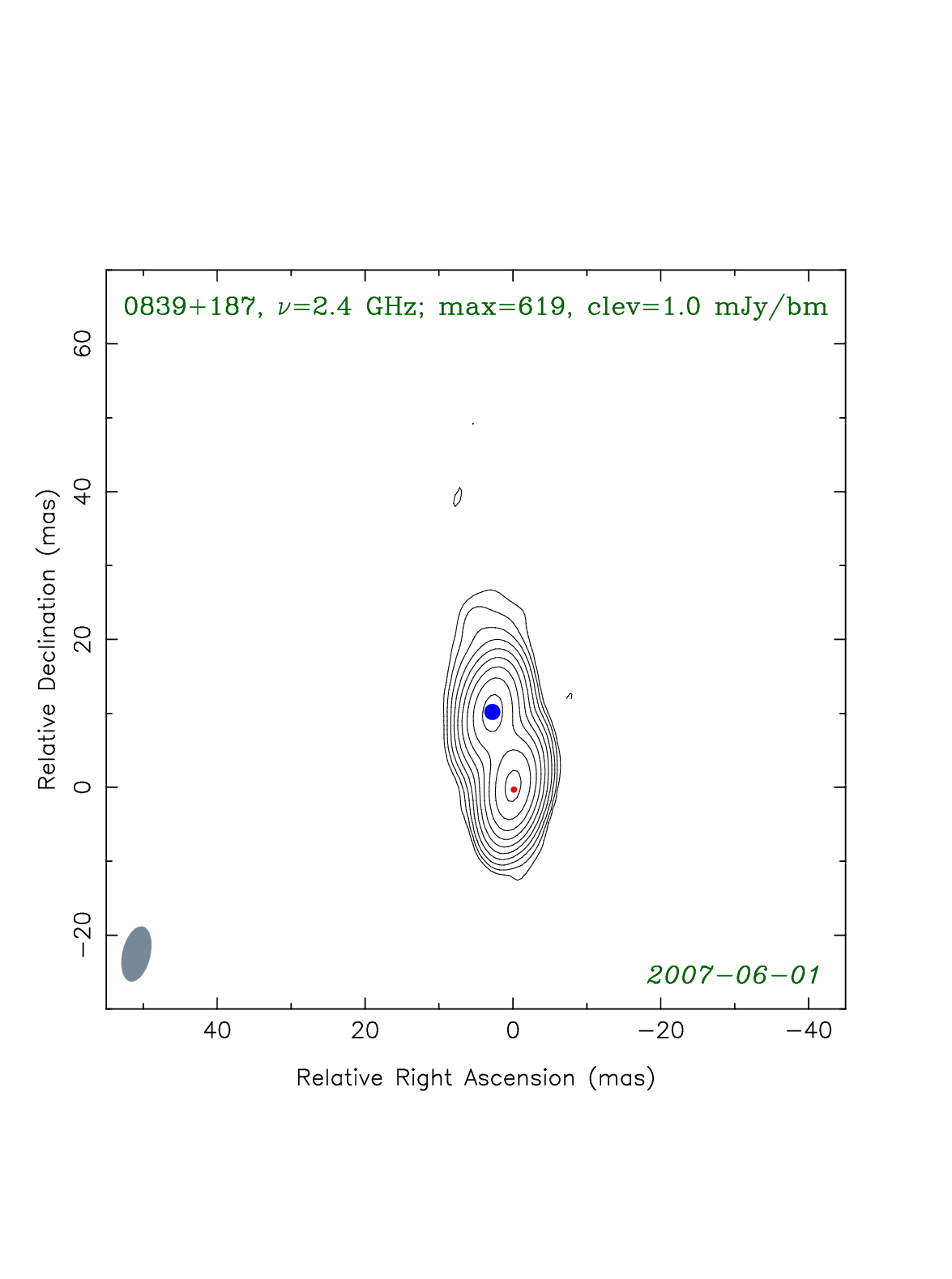}

 \includegraphics[width=0.33\textwidth,angle=0,trim=0.5cm 3cm 1.6cm 5cm,clip]{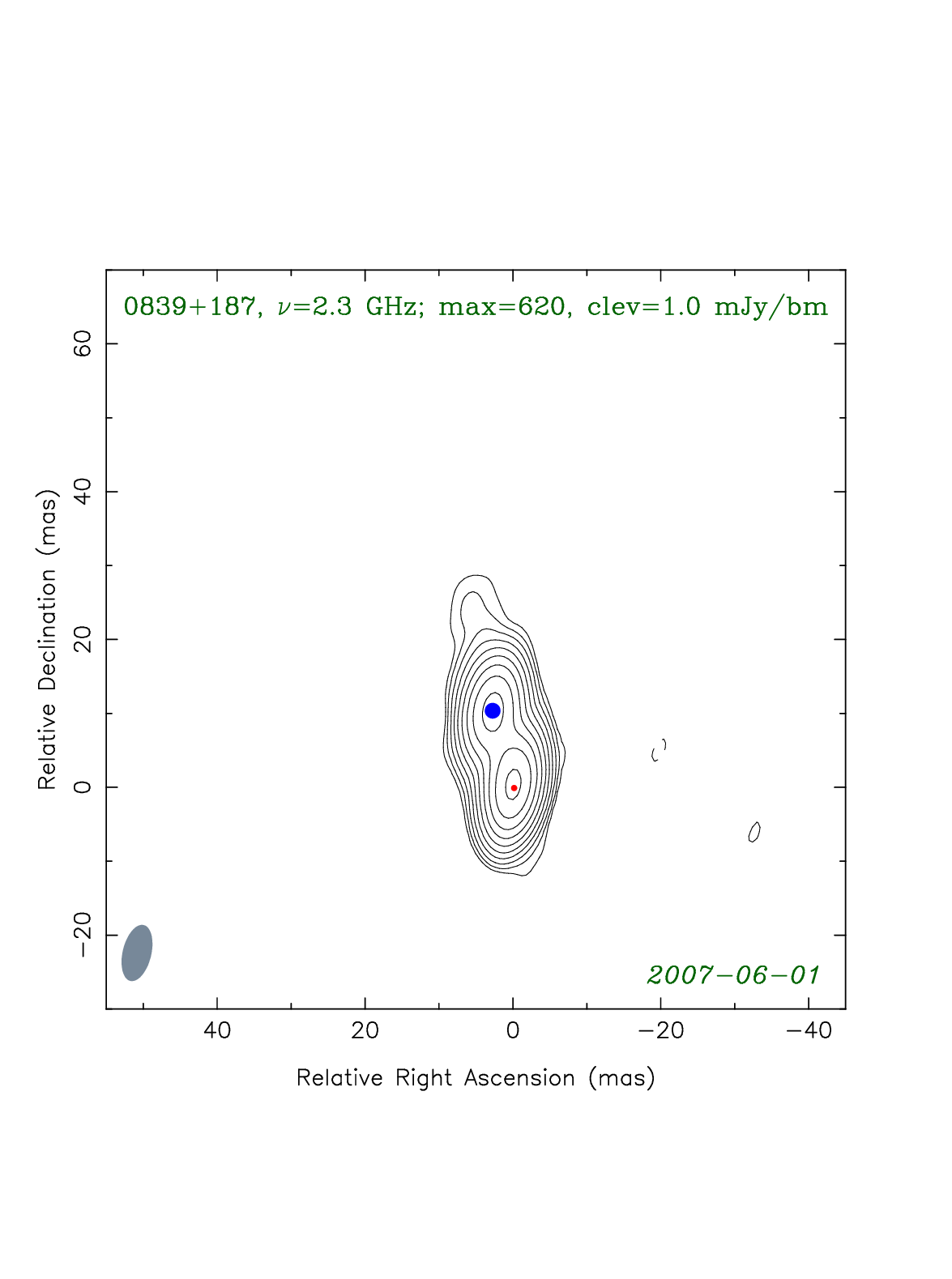}
 \includegraphics[width=0.33\textwidth,angle=0,trim=0.5cm 3cm 1.6cm 5cm,clip]{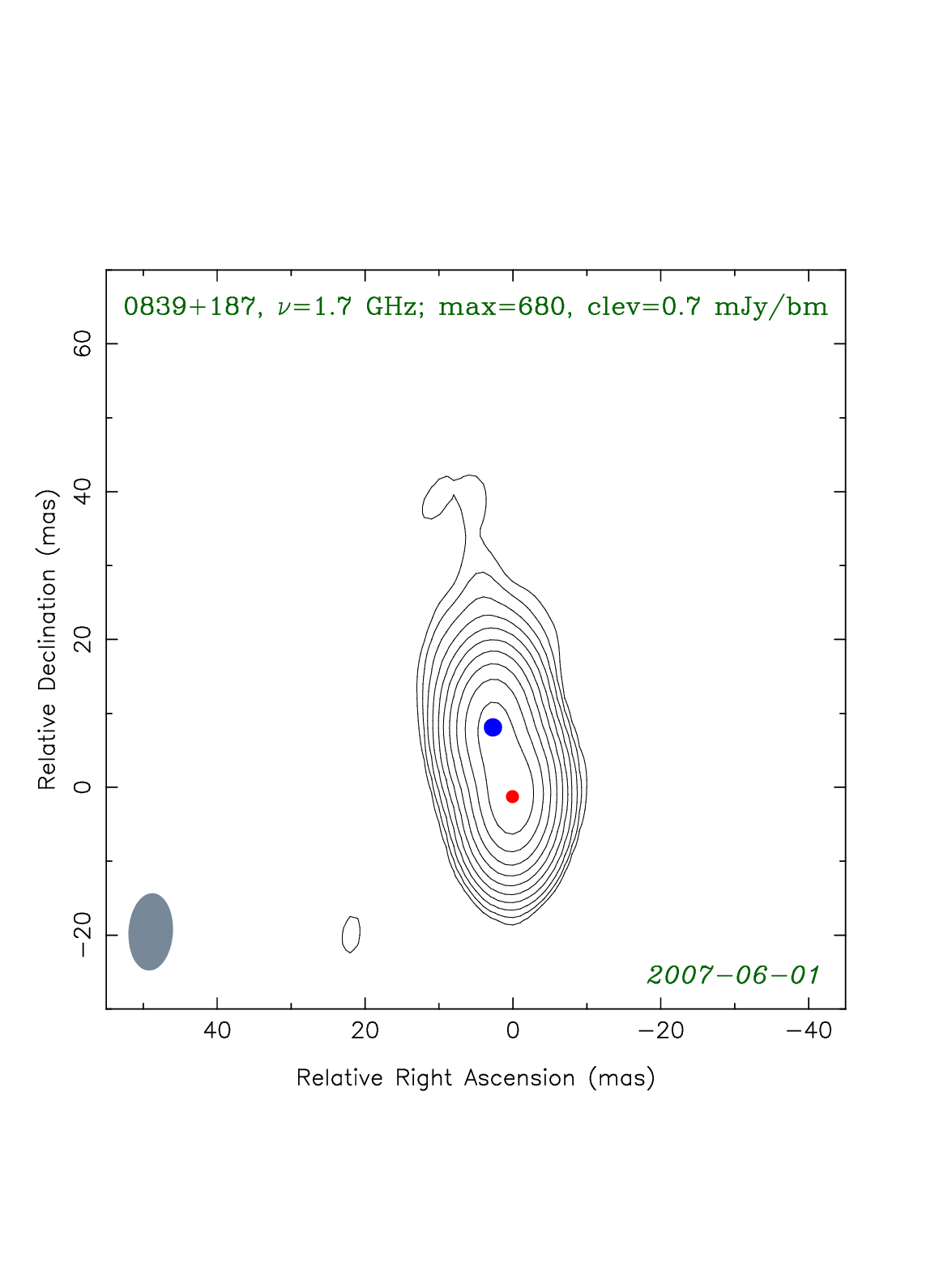}
 \includegraphics[width=0.33\textwidth,angle=0,trim=0.5cm 3cm 1.6cm 5cm,clip]{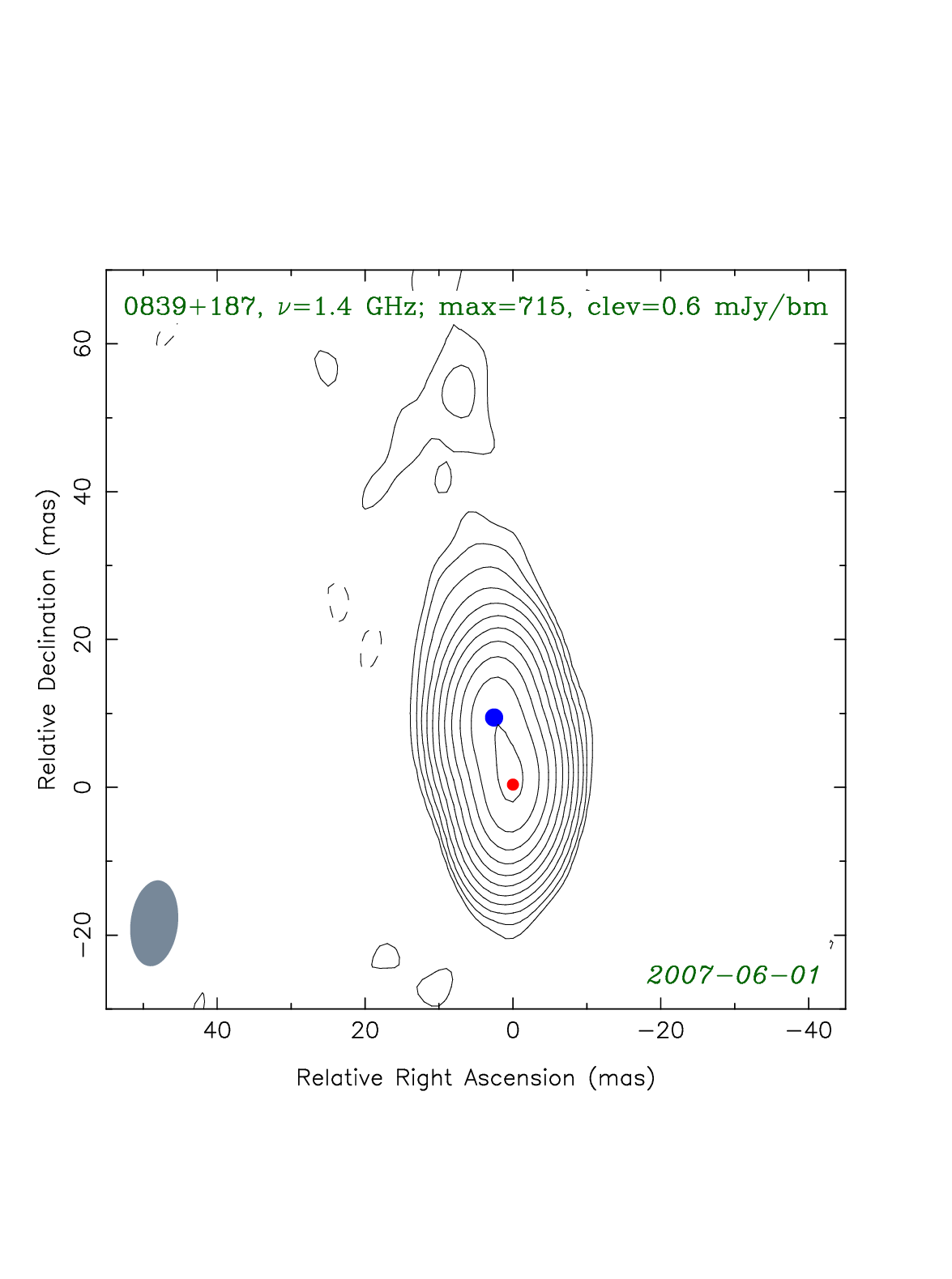}

 \caption{continued...}
\end{figure*}

\clearpage
\addtocounter{figure}{-1}
\begin{figure*}
 \centering

 \includegraphics[width=0.33\textwidth,angle=0,trim=0.5cm 3cm 1.6cm 5cm,clip]{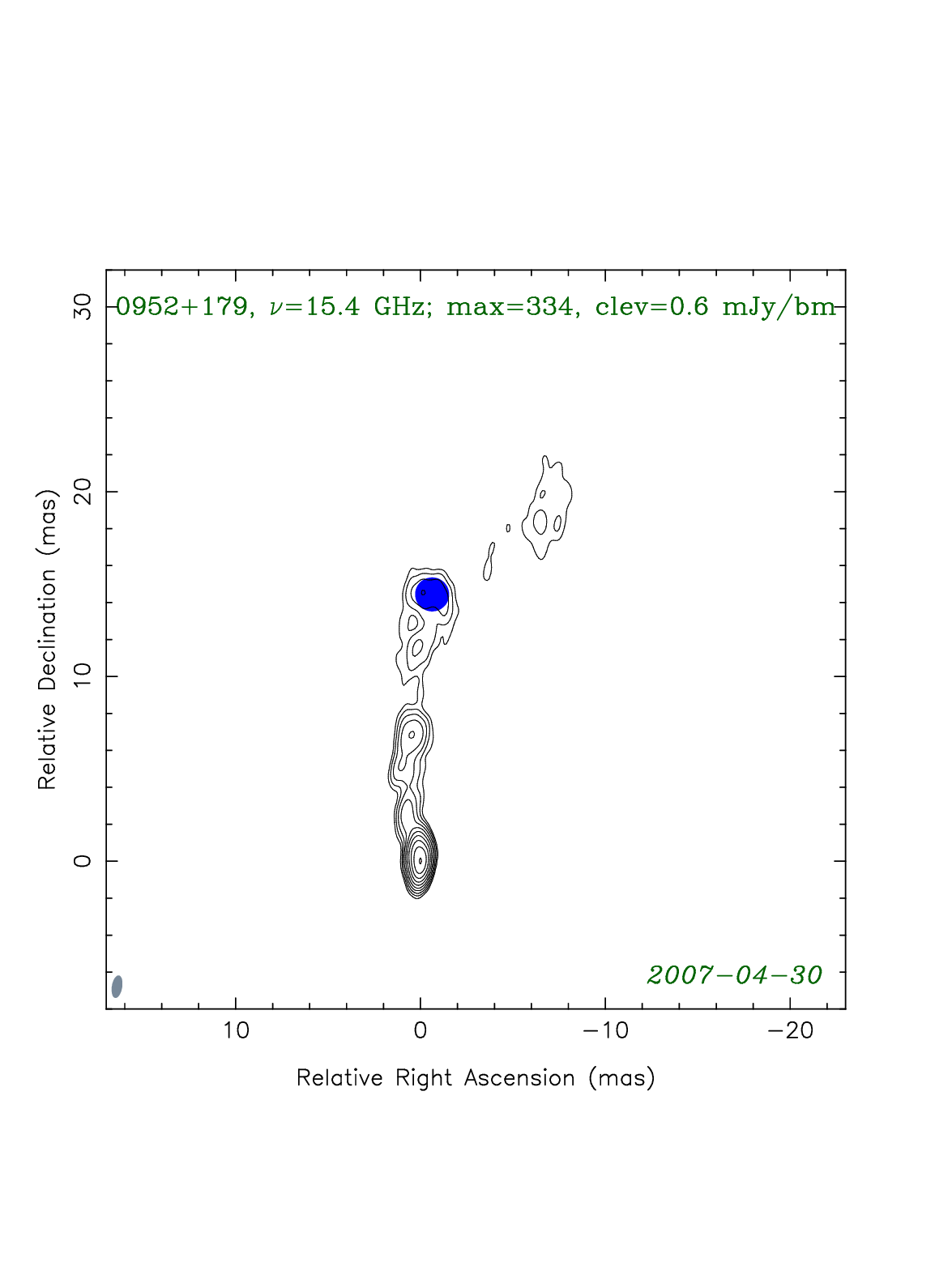}
 \includegraphics[width=0.33\textwidth,angle=0,trim=0.5cm 3cm 1.6cm 5cm,clip]{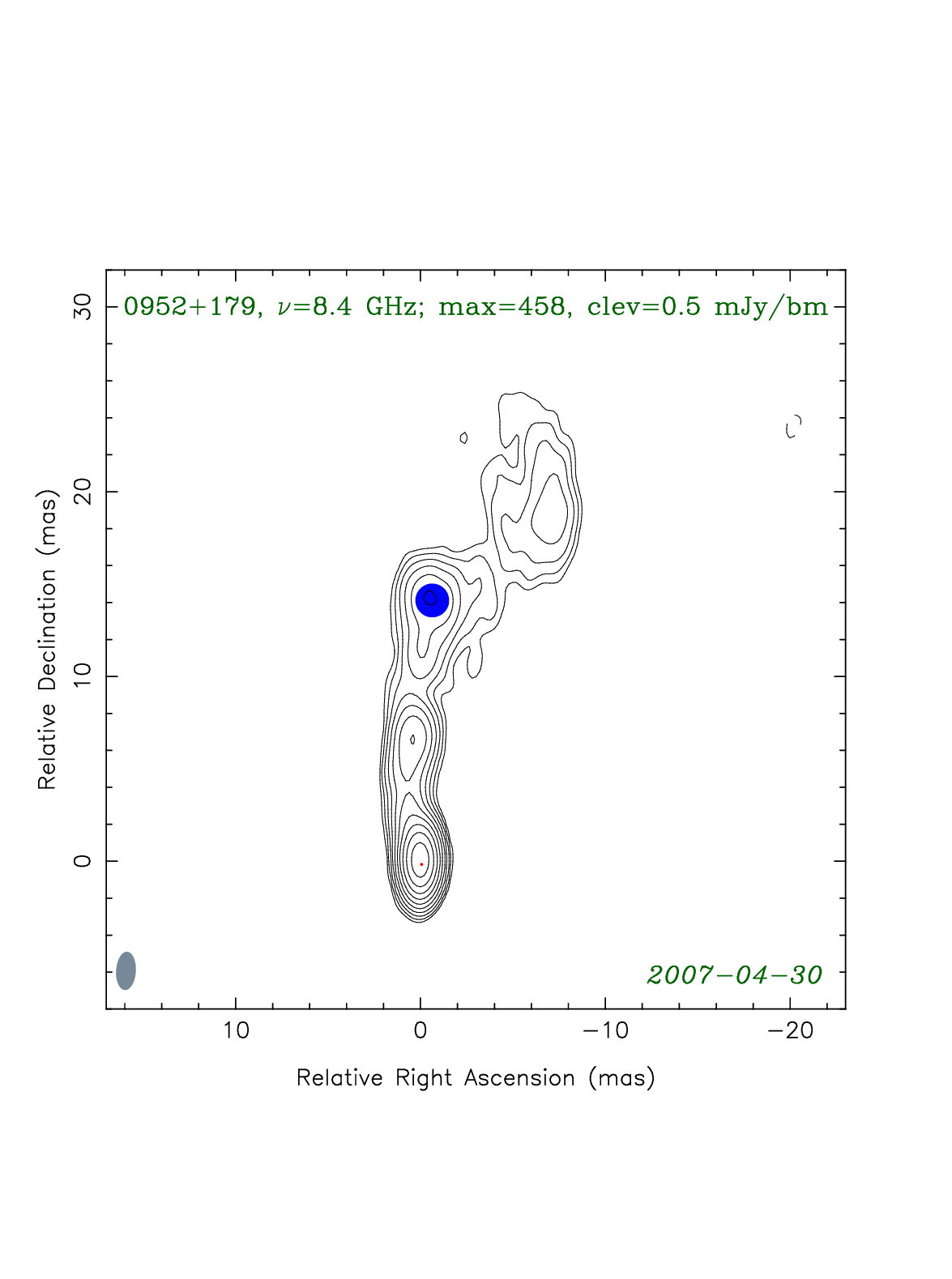}
 \includegraphics[width=0.33\textwidth,angle=0,trim=0.5cm 3cm 1.6cm 5cm,clip]{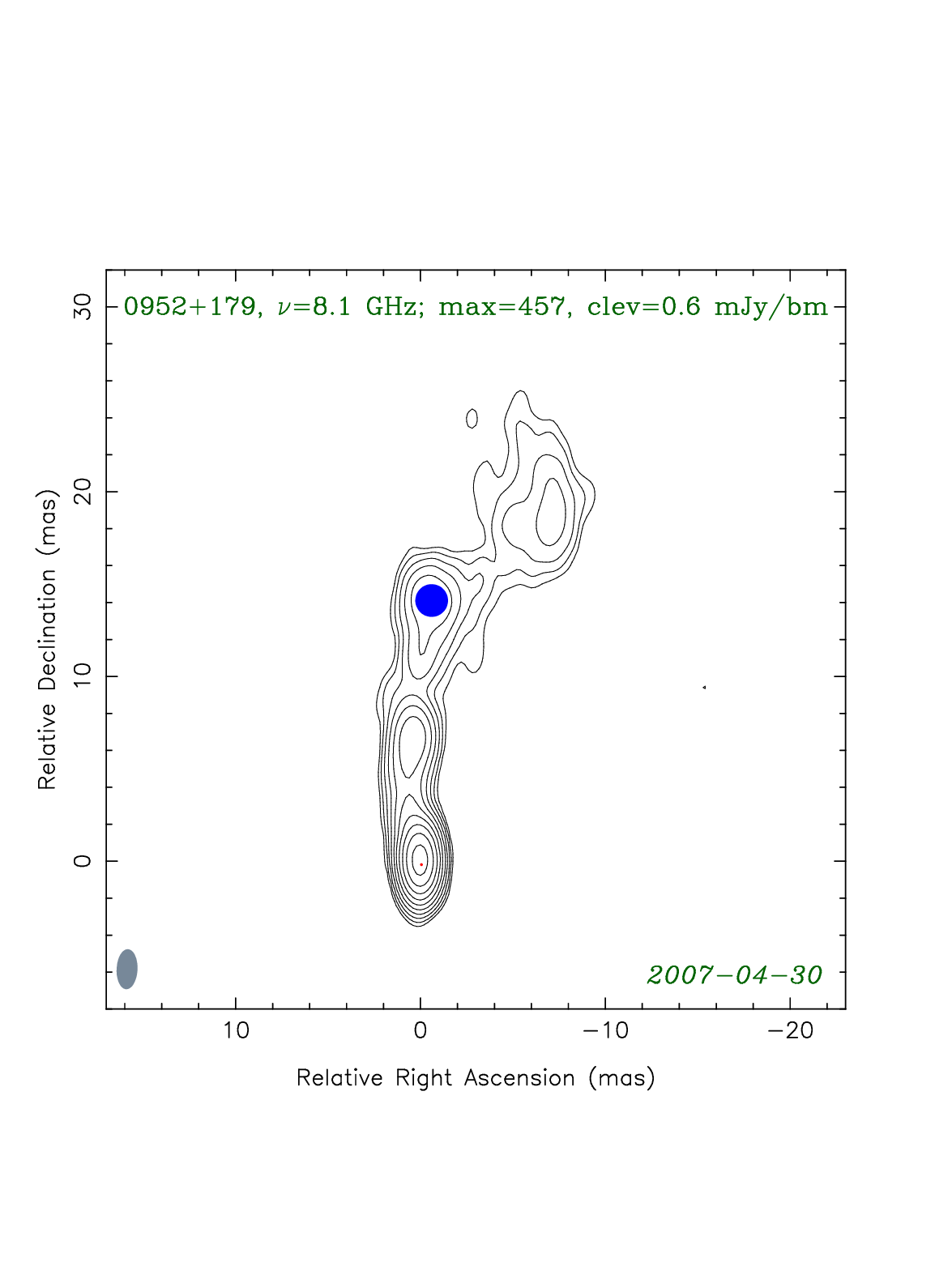}

 \includegraphics[width=0.33\textwidth,angle=0,trim=0.5cm 3cm 1.6cm 5cm,clip]{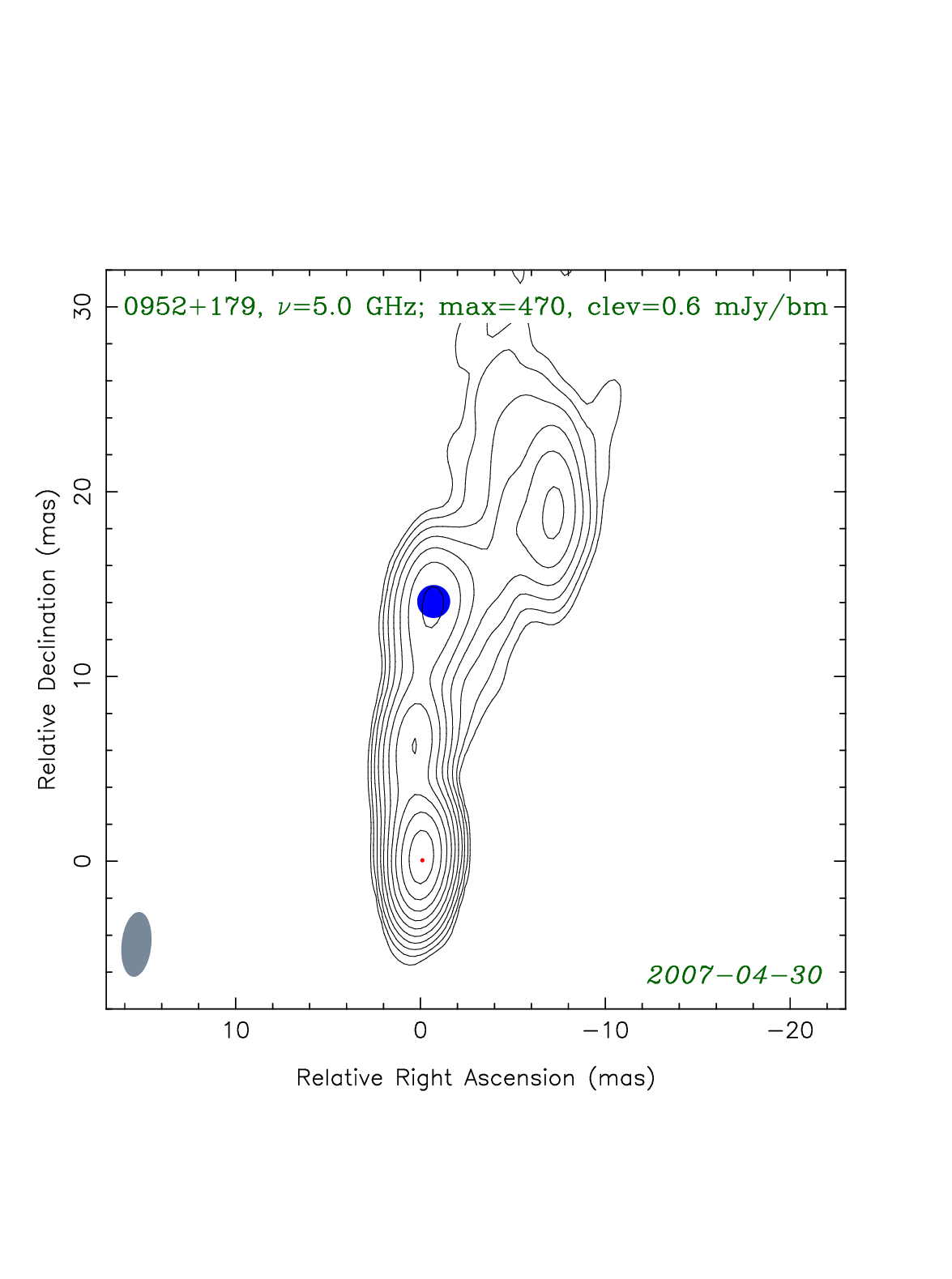}
 \includegraphics[width=0.33\textwidth,angle=0,trim=0.5cm 3cm 1.6cm 5cm,clip]{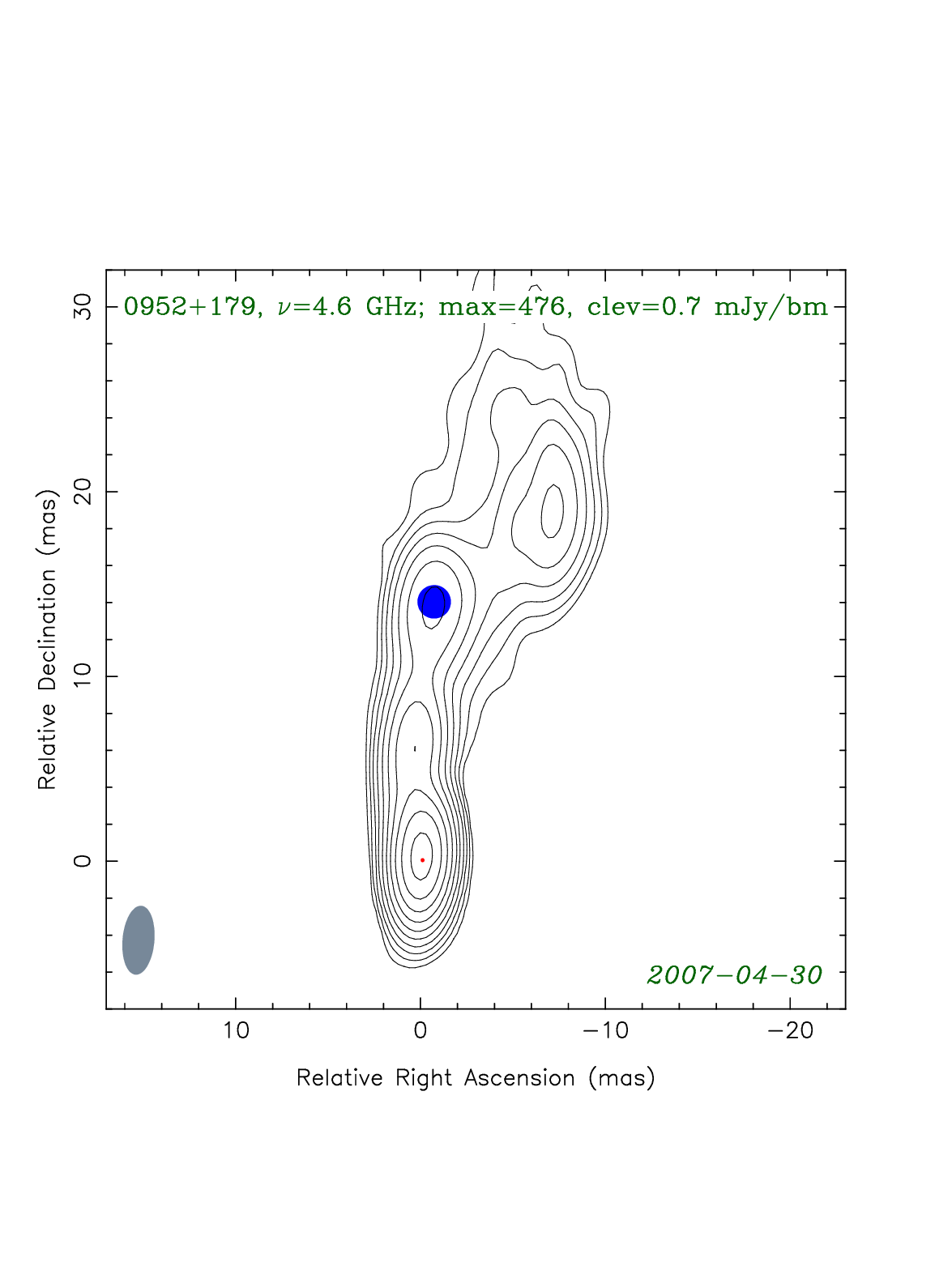}
 \includegraphics[width=0.33\textwidth,angle=0,trim=0.5cm 3cm 1.6cm 5cm,clip]{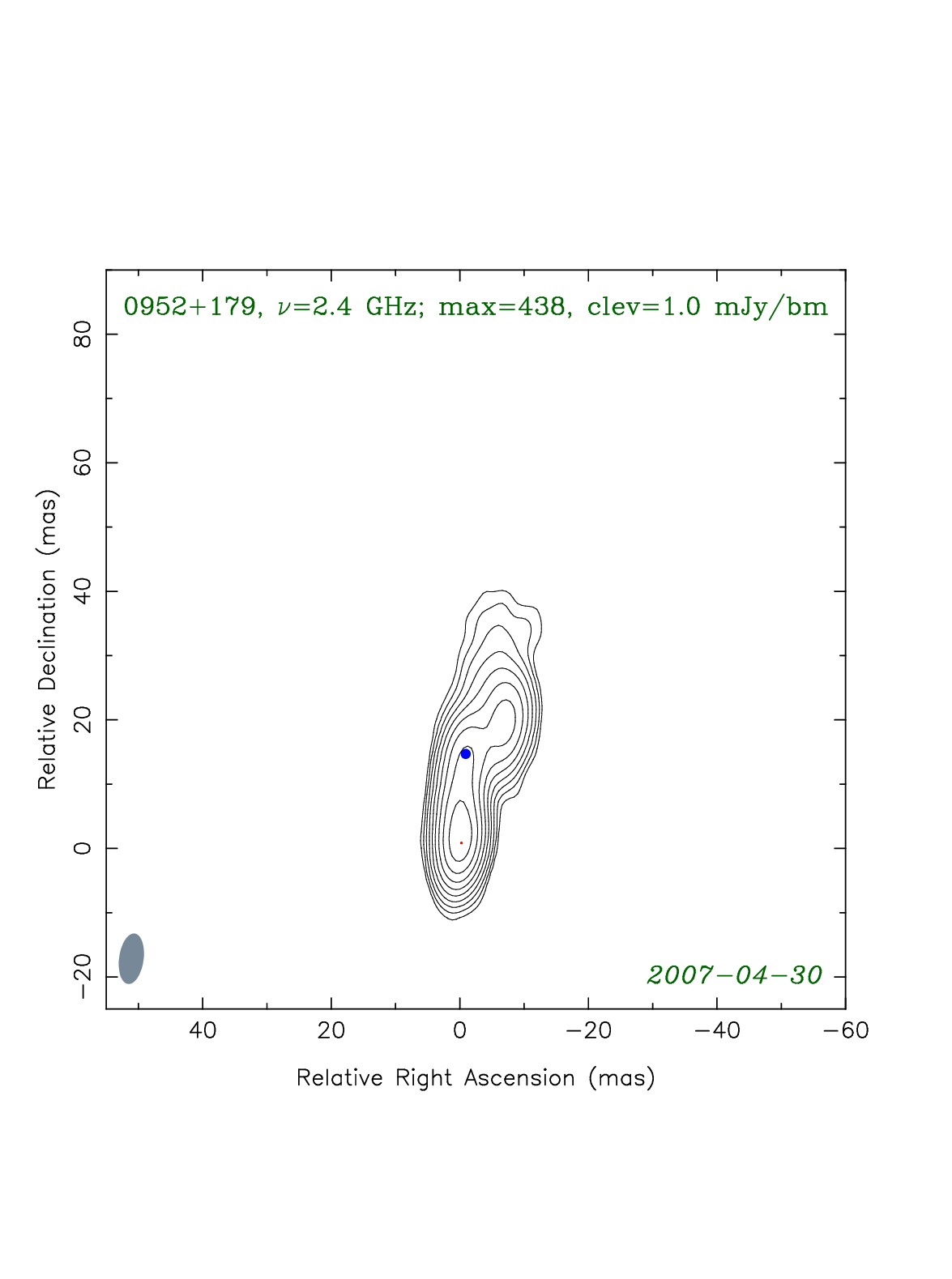}

 \includegraphics[width=0.33\textwidth,angle=0,trim=0.5cm 3cm 1.6cm 5cm,clip]{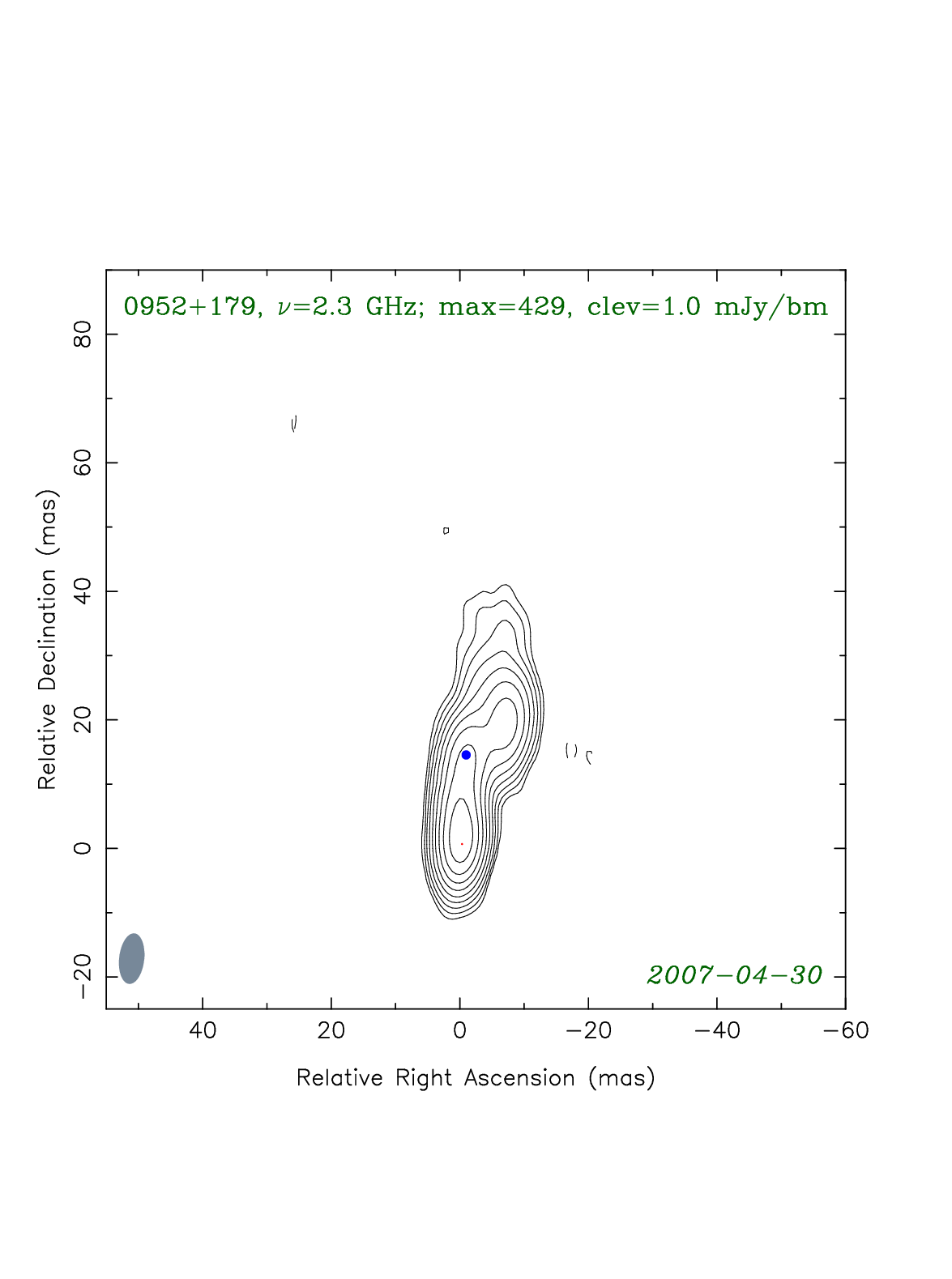}
 \includegraphics[width=0.33\textwidth,angle=0,trim=0.5cm 3cm 1.6cm 5cm,clip]{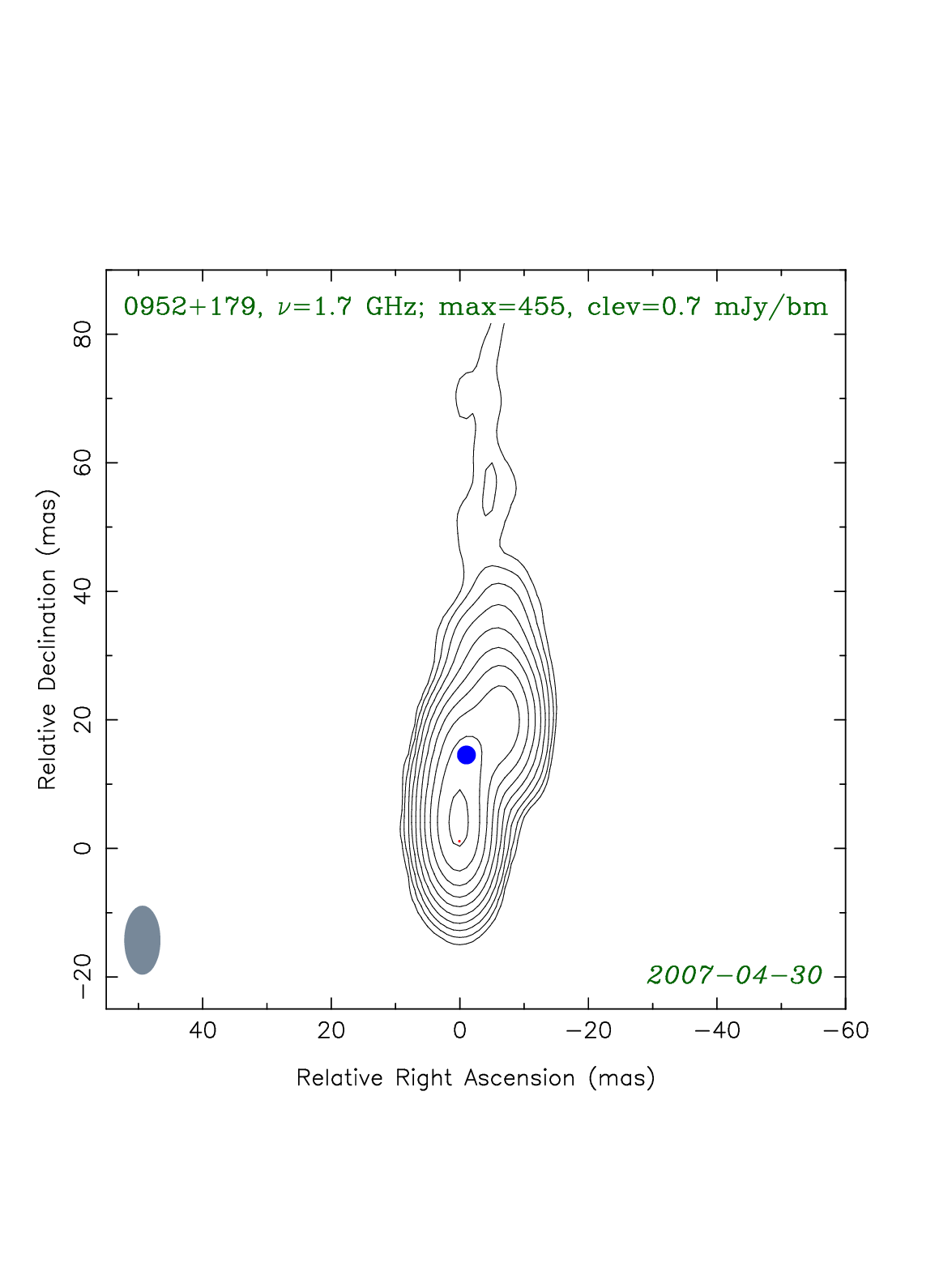}
 \includegraphics[width=0.33\textwidth,angle=0,trim=0.5cm 3cm 1.6cm 5cm,clip]{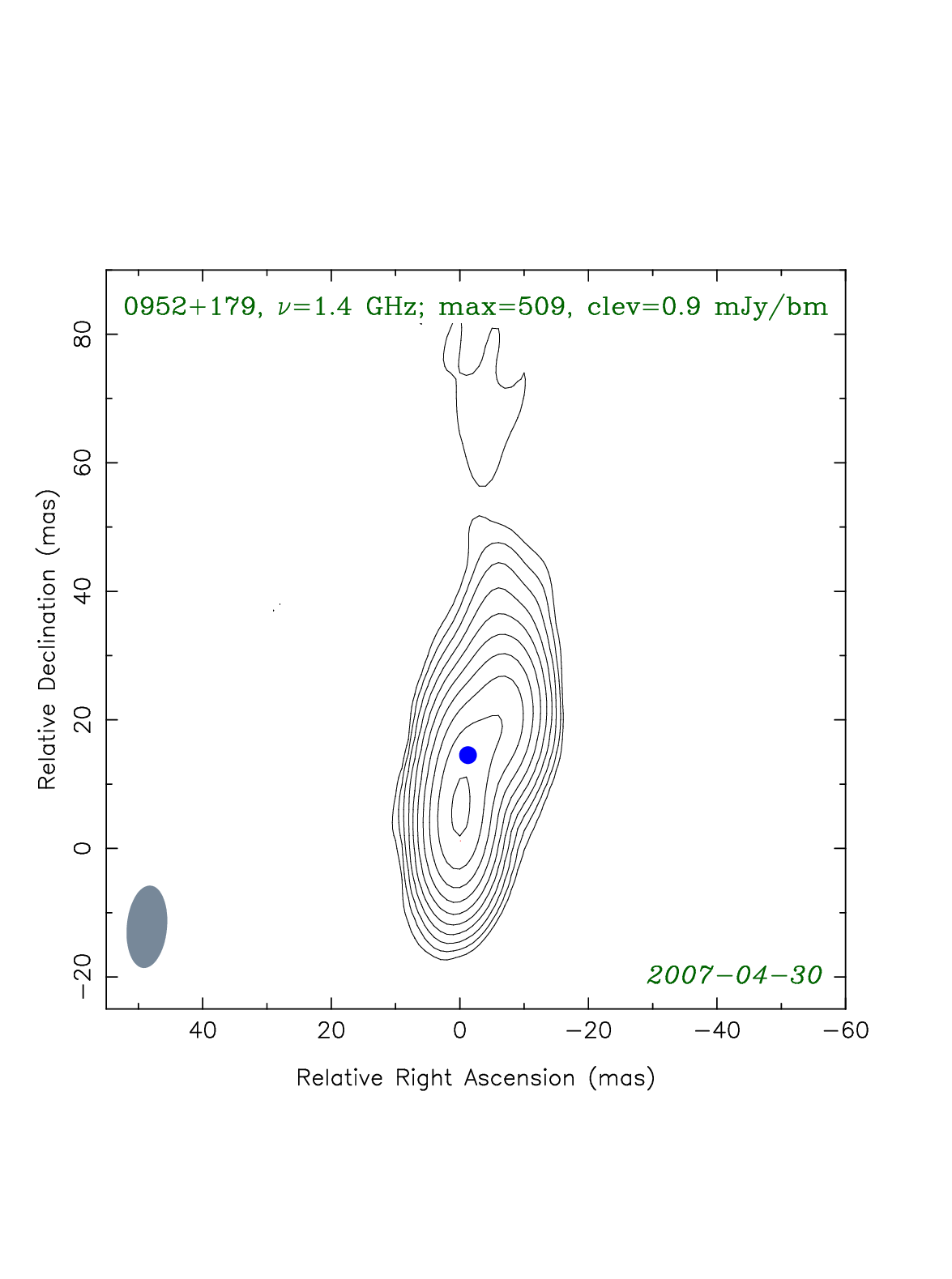}

 \caption{continued...}
\end{figure*}

\clearpage
\addtocounter{figure}{-1}
\begin{figure*}
 \centering

 \includegraphics[width=0.33\textwidth,angle=0,trim=0.5cm 3cm 1.6cm 5cm,clip]{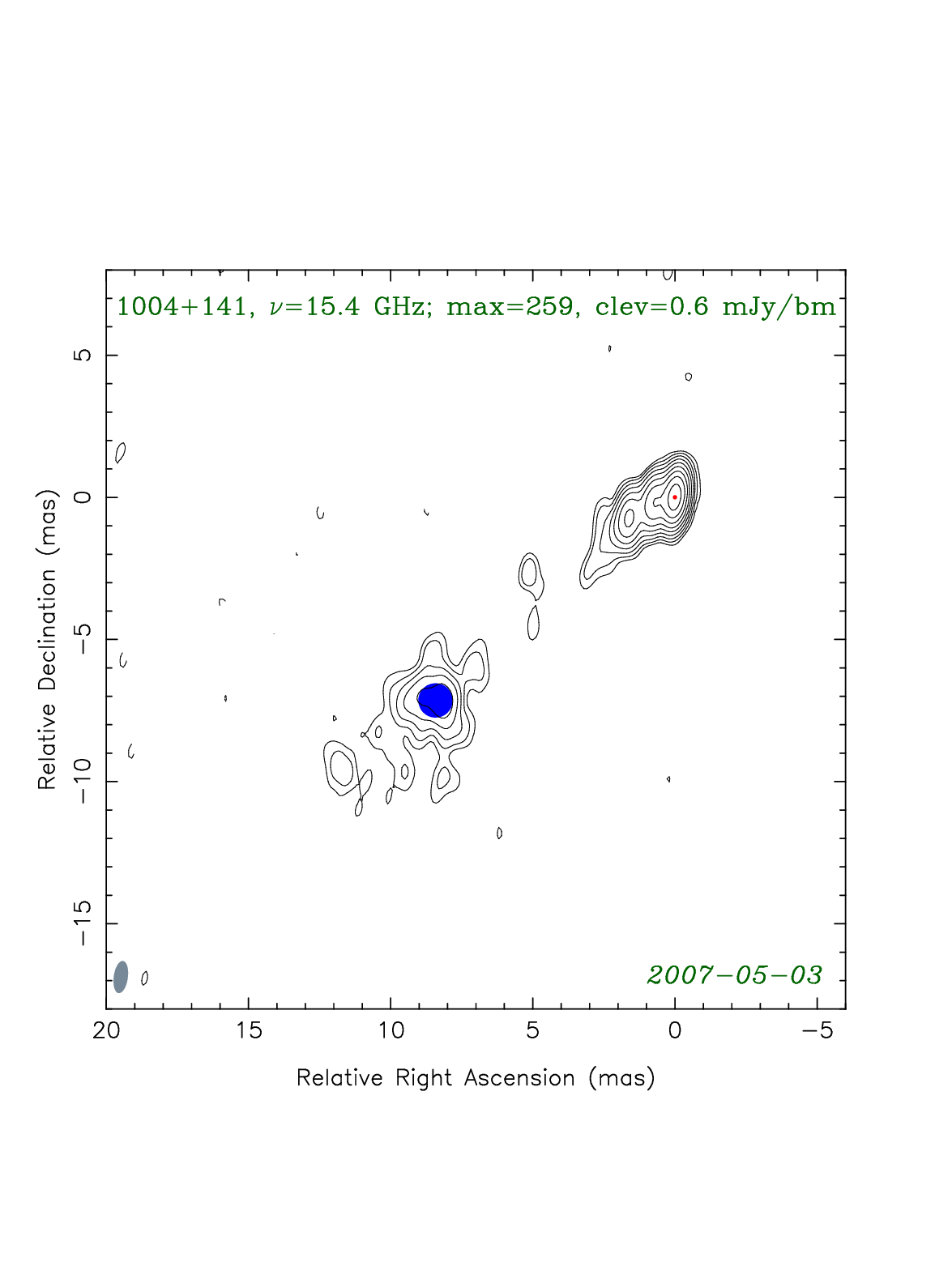}
 \includegraphics[width=0.33\textwidth,angle=0,trim=0.5cm 3cm 1.6cm 5cm,clip]{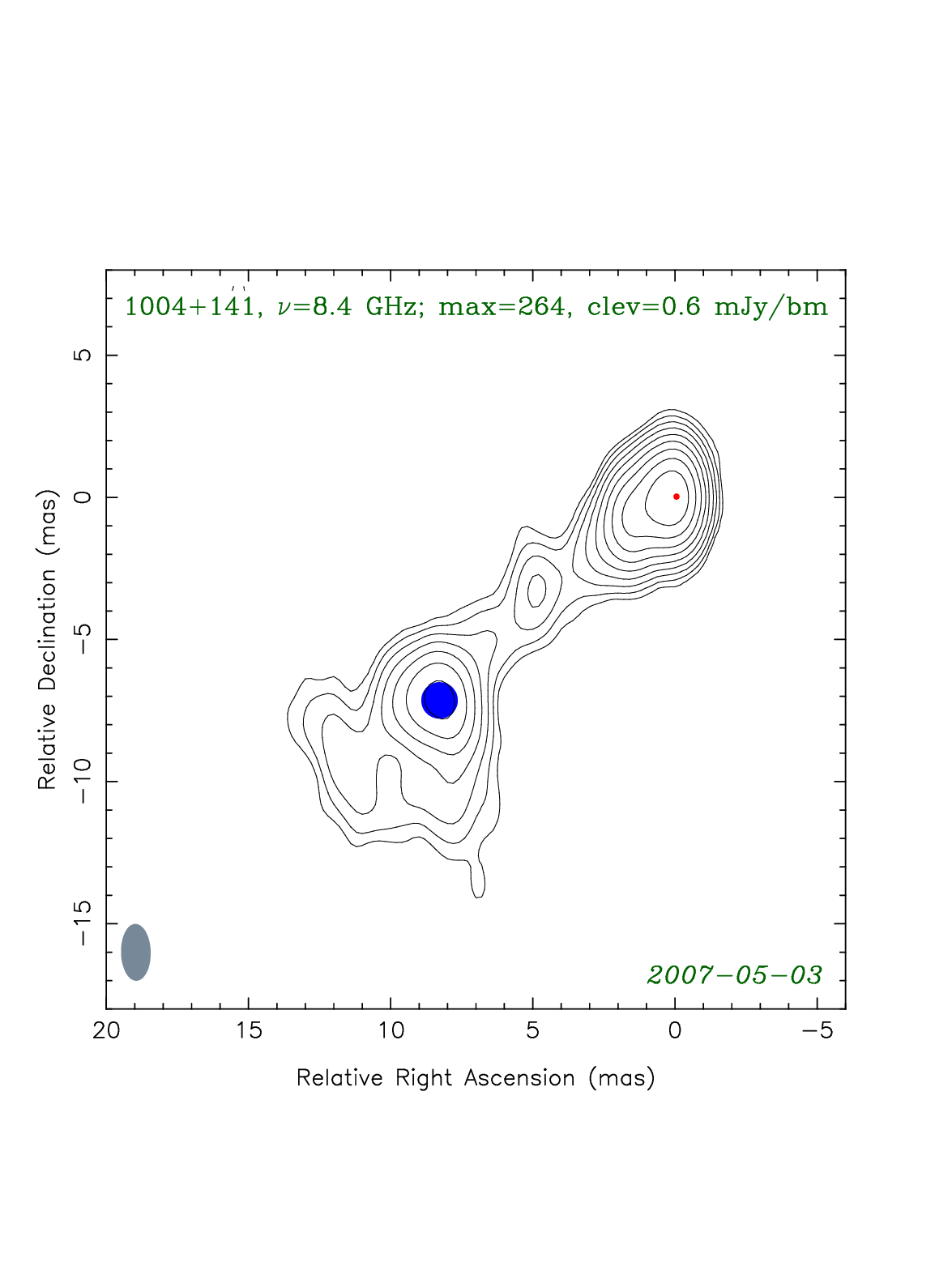}
 \includegraphics[width=0.33\textwidth,angle=0,trim=0.5cm 3cm 1.6cm 5cm,clip]{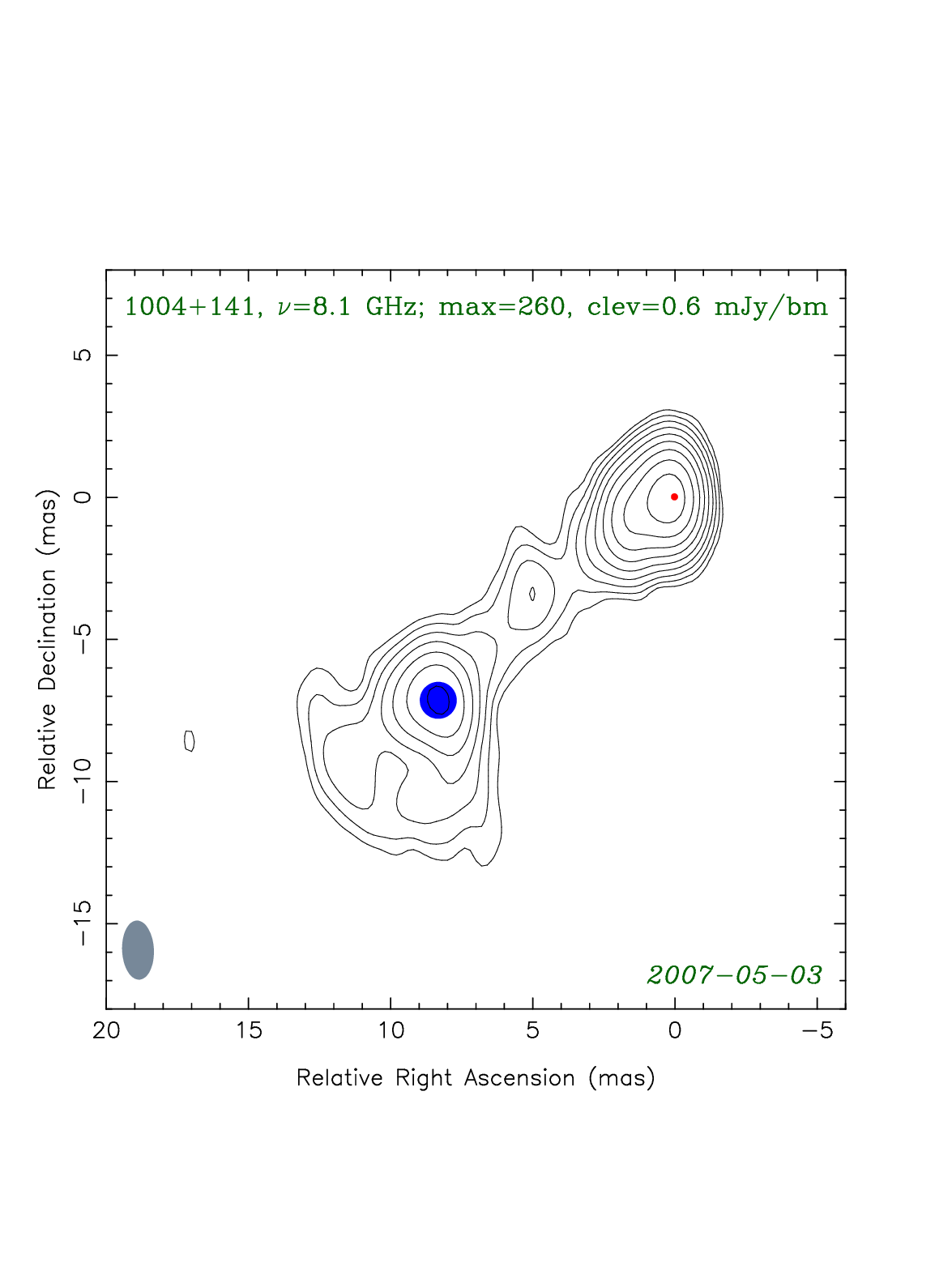}

 \includegraphics[width=0.33\textwidth,angle=0,trim=0.5cm 3cm 1.6cm 5cm,clip]{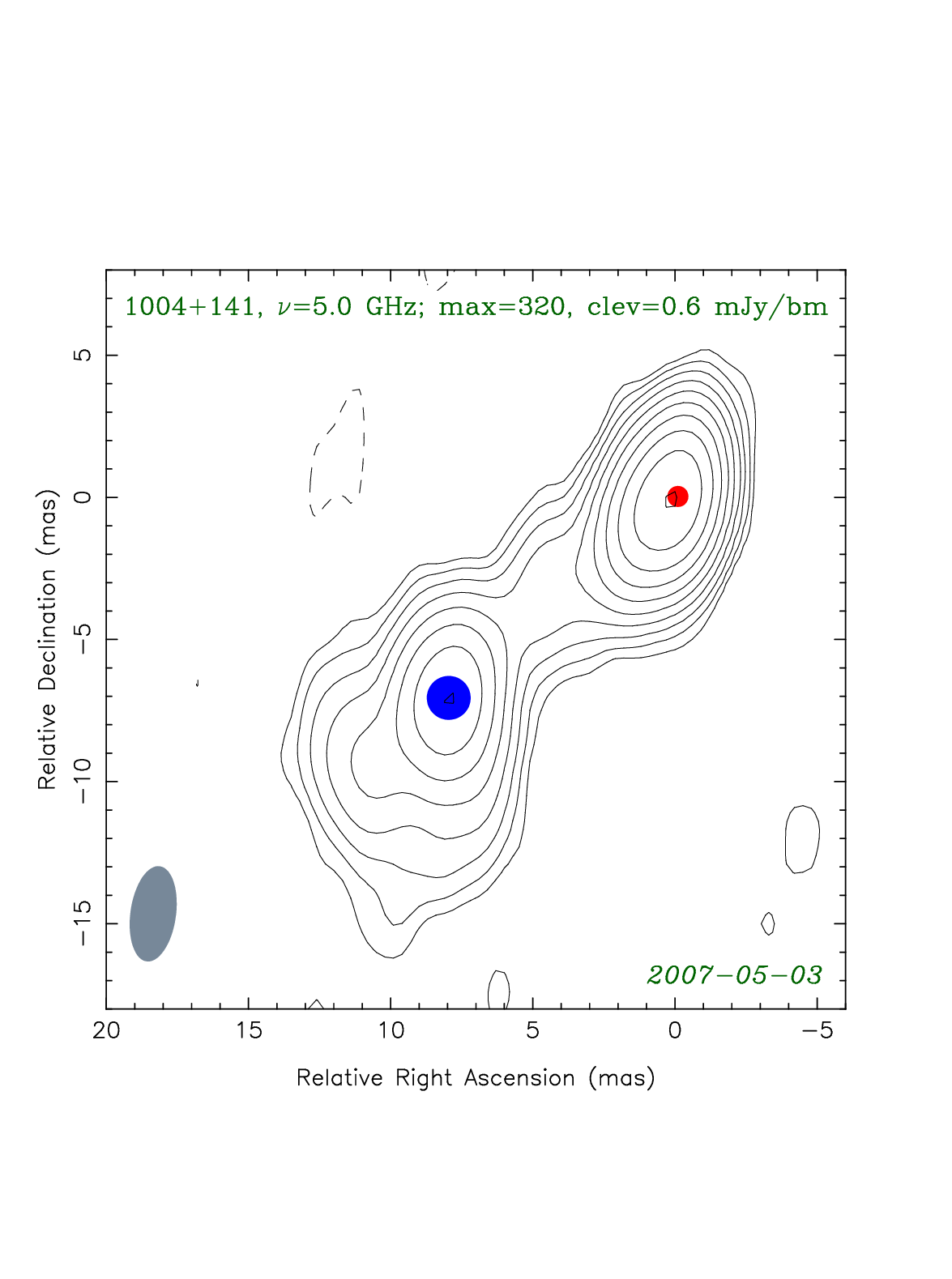}
 \includegraphics[width=0.33\textwidth,angle=0,trim=0.5cm 3cm 1.6cm 5cm,clip]{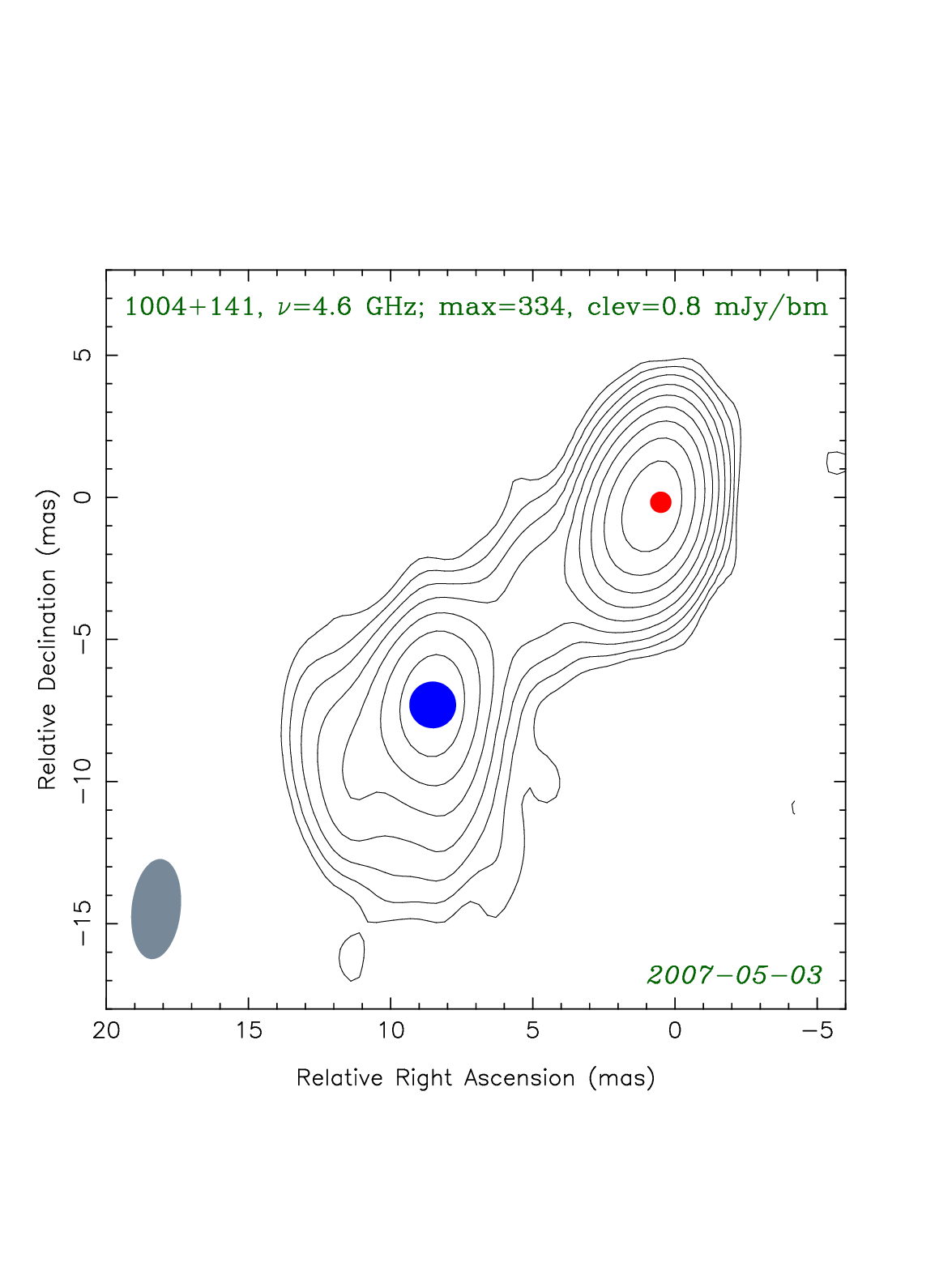}
 \includegraphics[width=0.33\textwidth,angle=0,trim=0.5cm 3cm 1.6cm 5cm,clip]{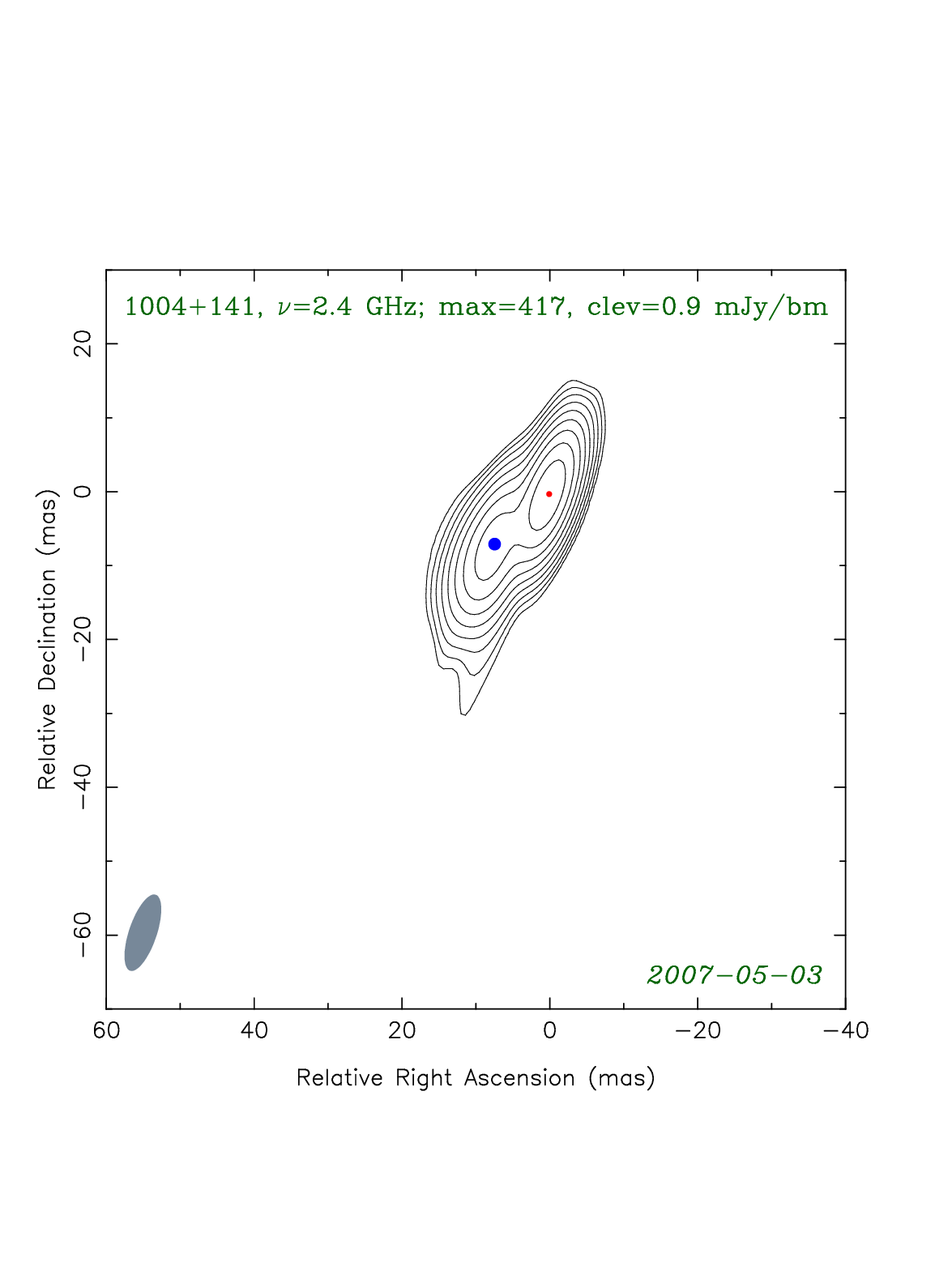}

 \includegraphics[width=0.33\textwidth,angle=0,trim=0.5cm 3cm 1.6cm 5cm,clip]{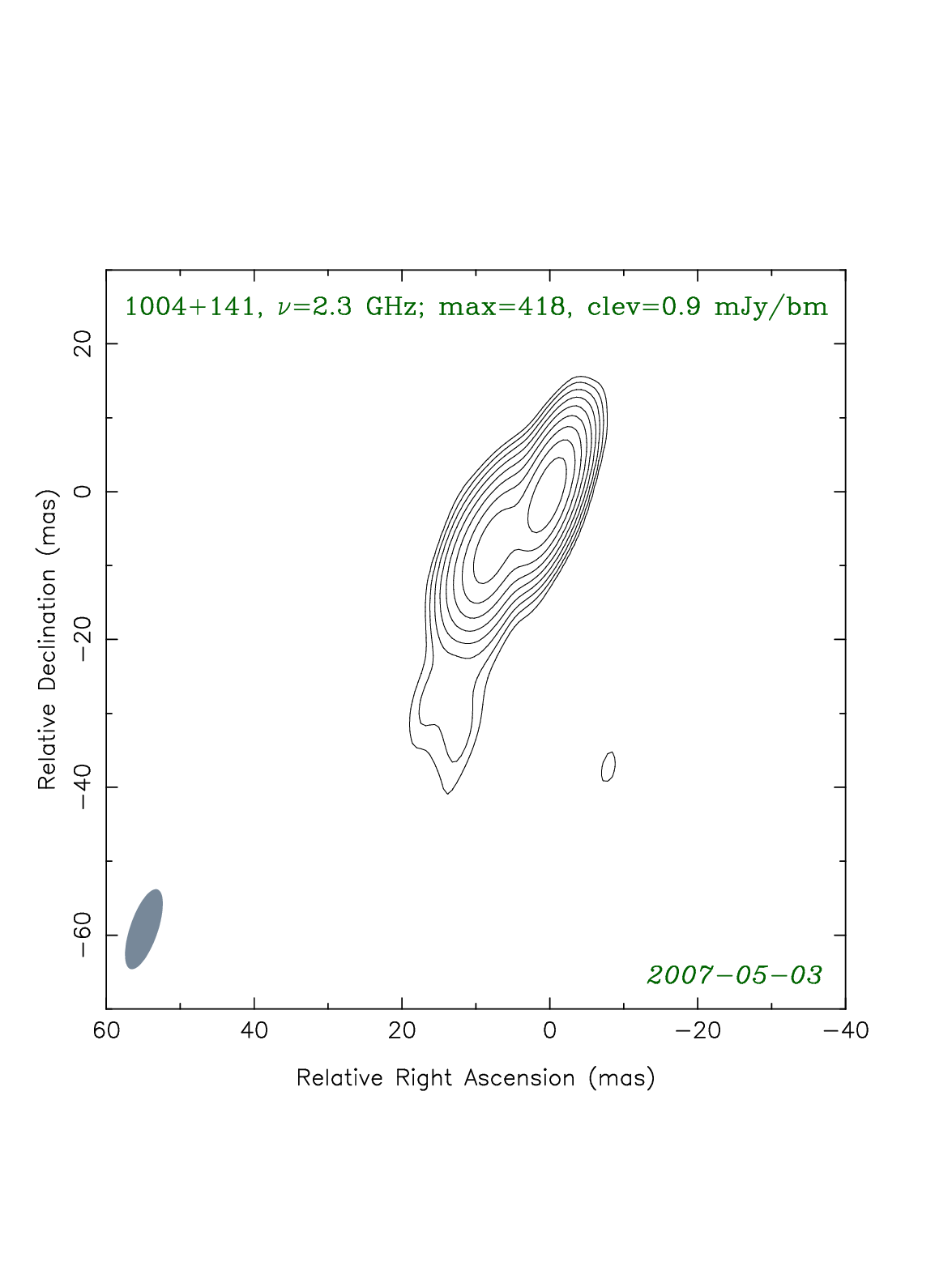}
 \includegraphics[width=0.33\textwidth,angle=0,trim=0.5cm 3cm 1.6cm 5cm,clip]{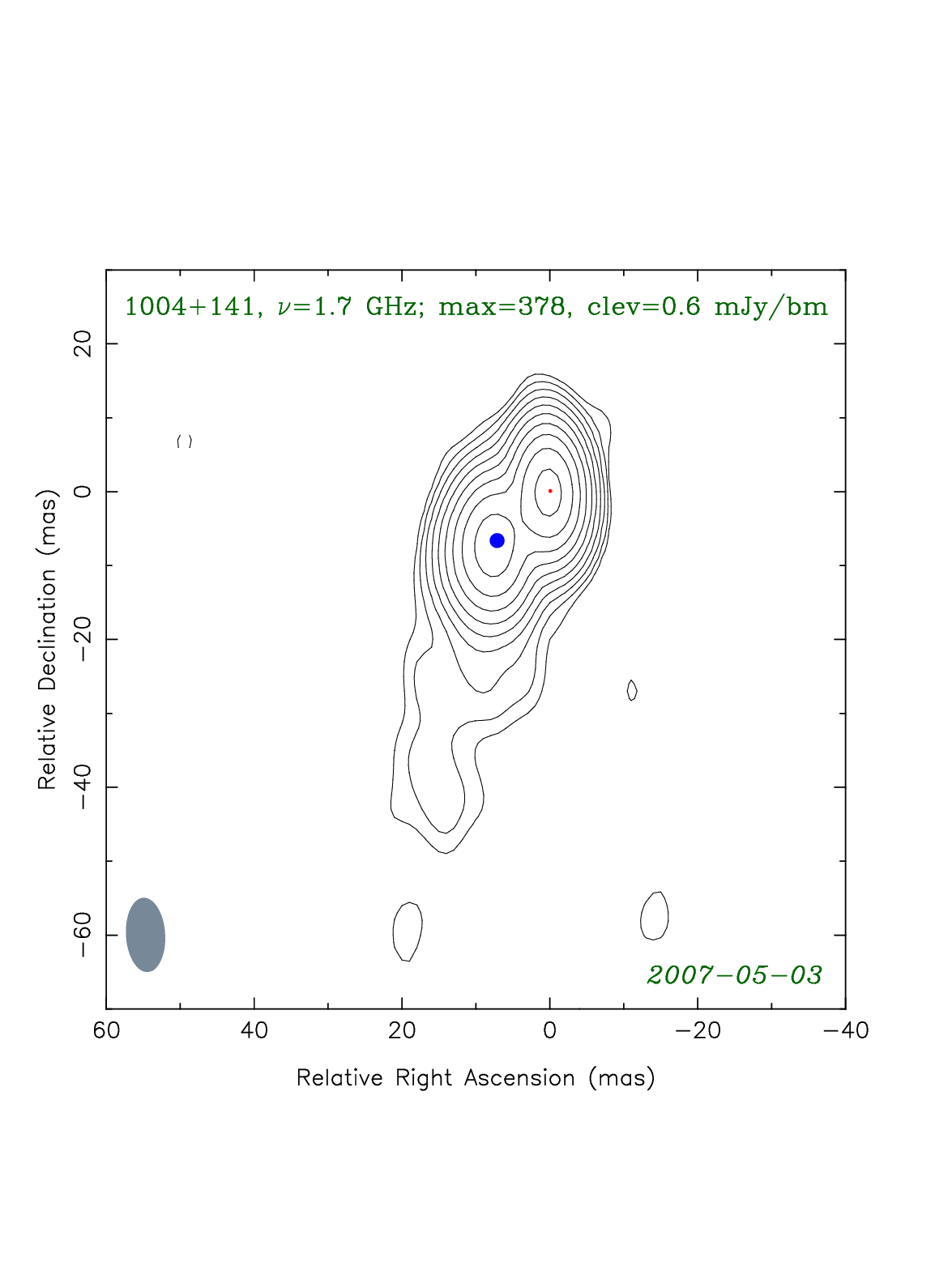}
 \includegraphics[width=0.33\textwidth,angle=0,trim=0.5cm 3cm 1.6cm 5cm,clip]{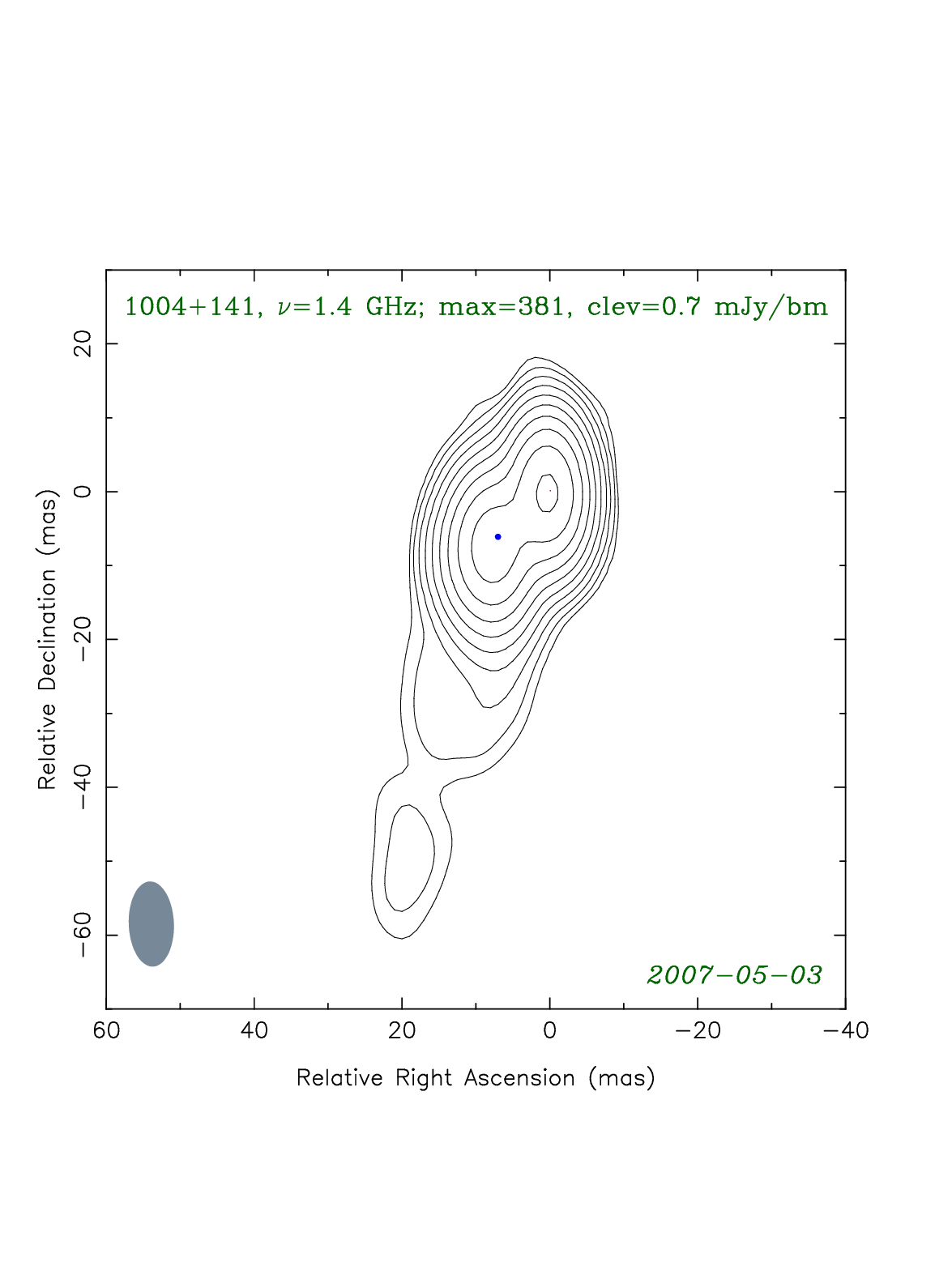}

 \caption{continued...}
\end{figure*}

\clearpage
\addtocounter{figure}{-1}
\begin{figure*}
 \centering

 \includegraphics[width=0.33\textwidth,angle=0,trim=0.5cm 3cm 1.6cm 5cm,clip]{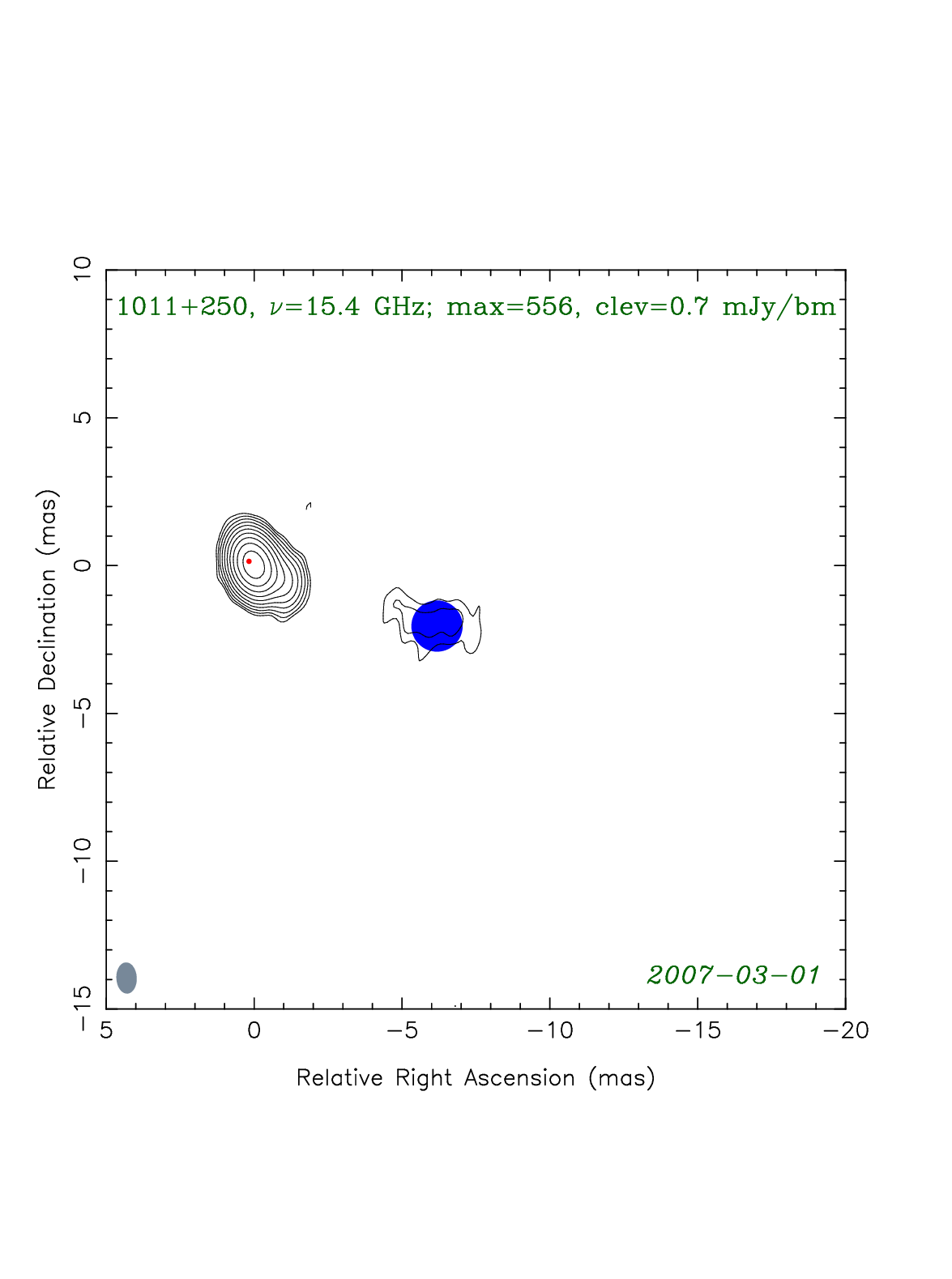}
 \includegraphics[width=0.33\textwidth,angle=0,trim=0.5cm 3cm 1.6cm 5cm,clip]{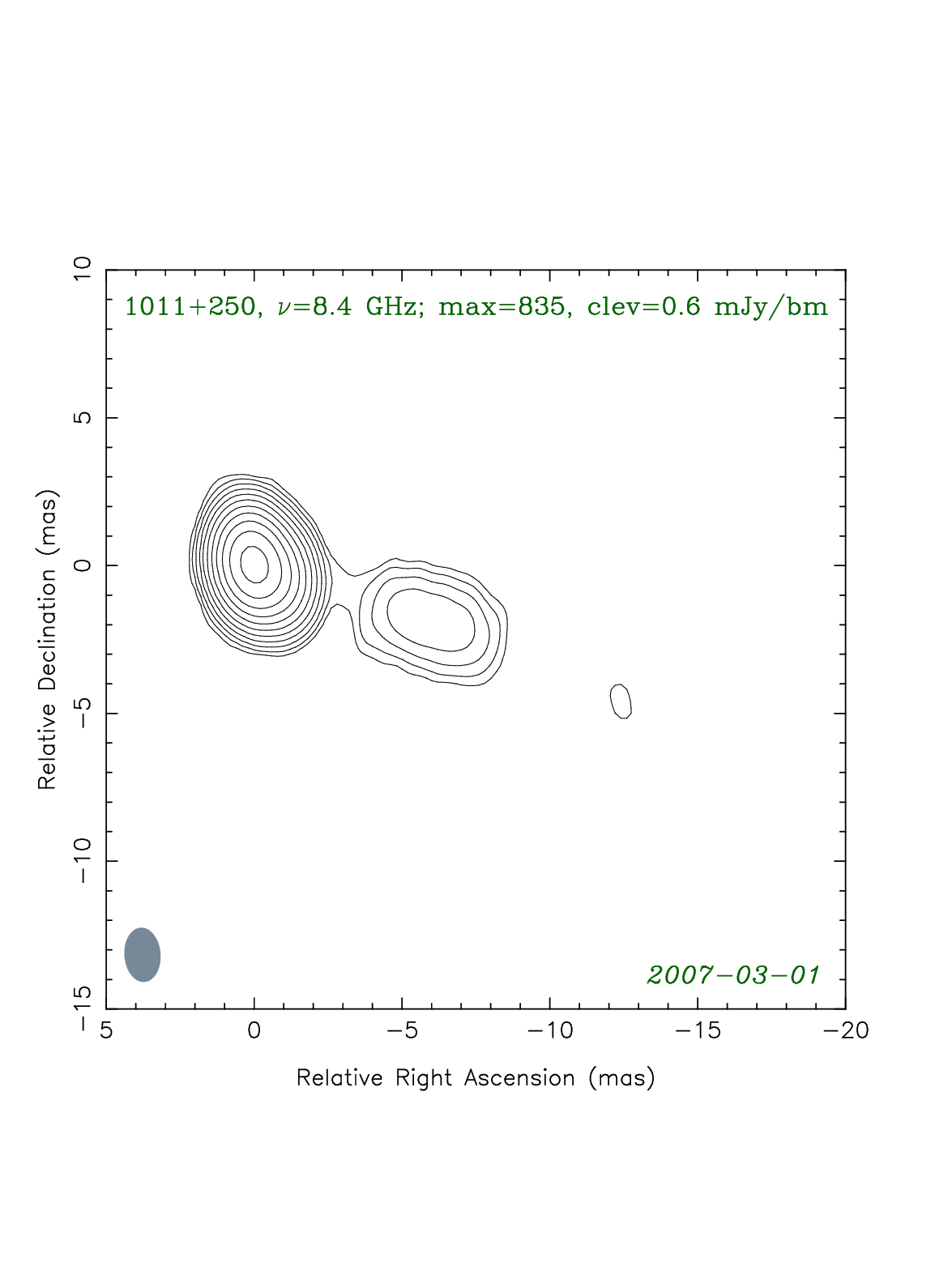}
 \includegraphics[width=0.33\textwidth,angle=0,trim=0.5cm 3cm 1.6cm 5cm,clip]{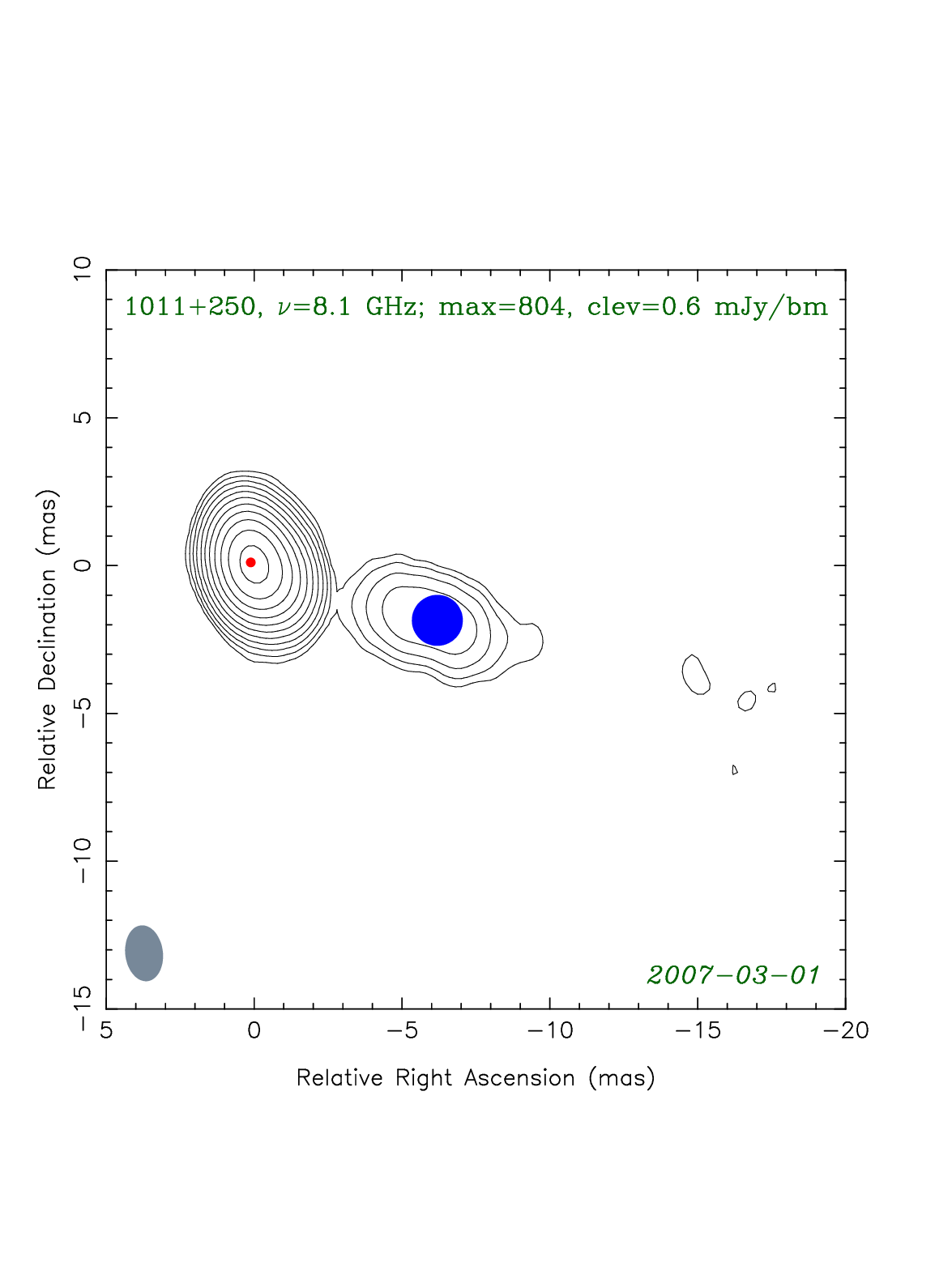}

 \includegraphics[width=0.33\textwidth,angle=0,trim=0.5cm 3cm 1.6cm 5cm,clip]{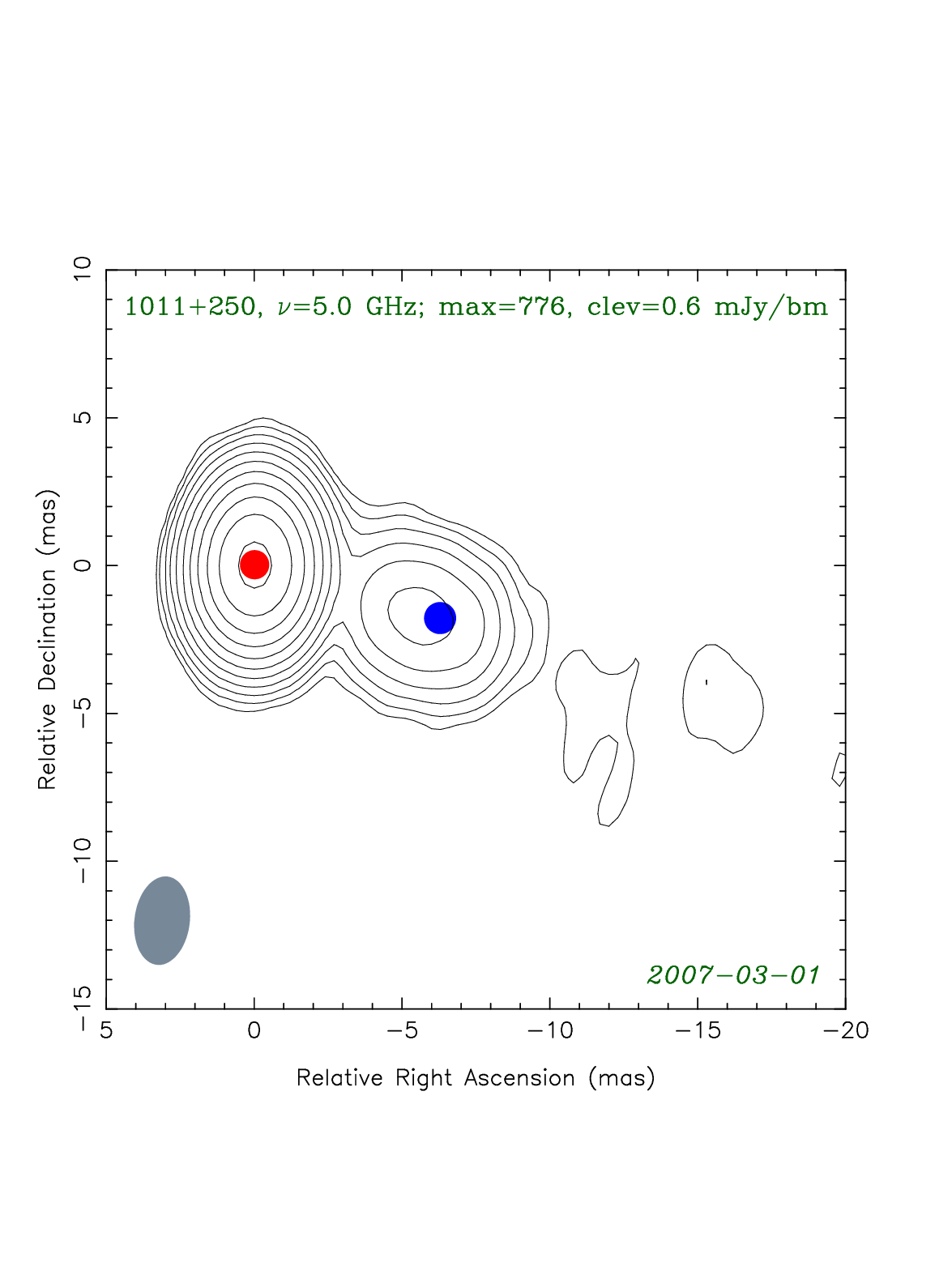}
 \includegraphics[width=0.33\textwidth,angle=0,trim=0.5cm 3cm 1.6cm 5cm,clip]{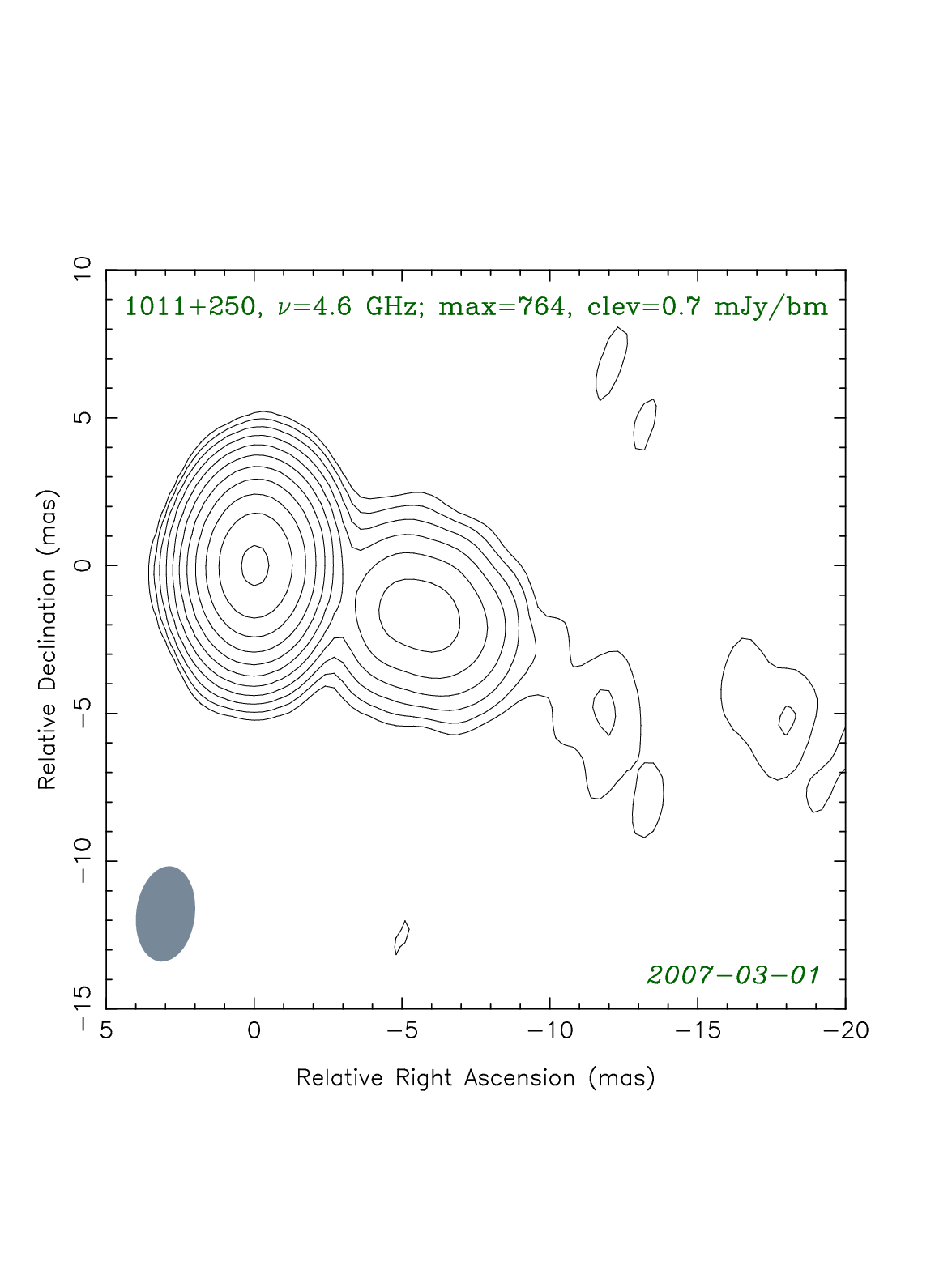}
 \includegraphics[width=0.33\textwidth,angle=0,trim=0.5cm 3cm 1.6cm 5cm,clip]{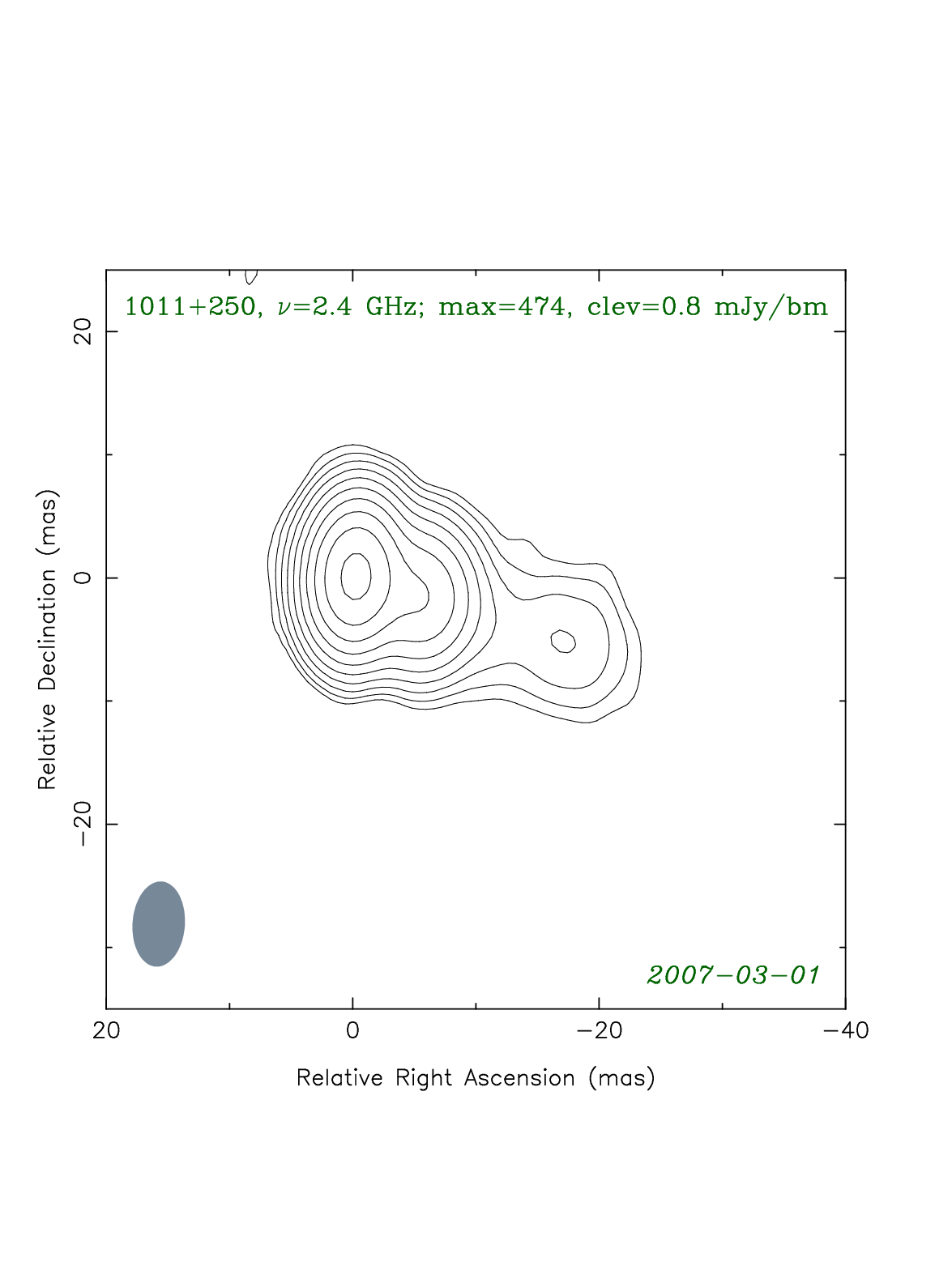}

 \includegraphics[width=0.33\textwidth,angle=0,trim=0.5cm 3cm 1.6cm 5cm,clip]{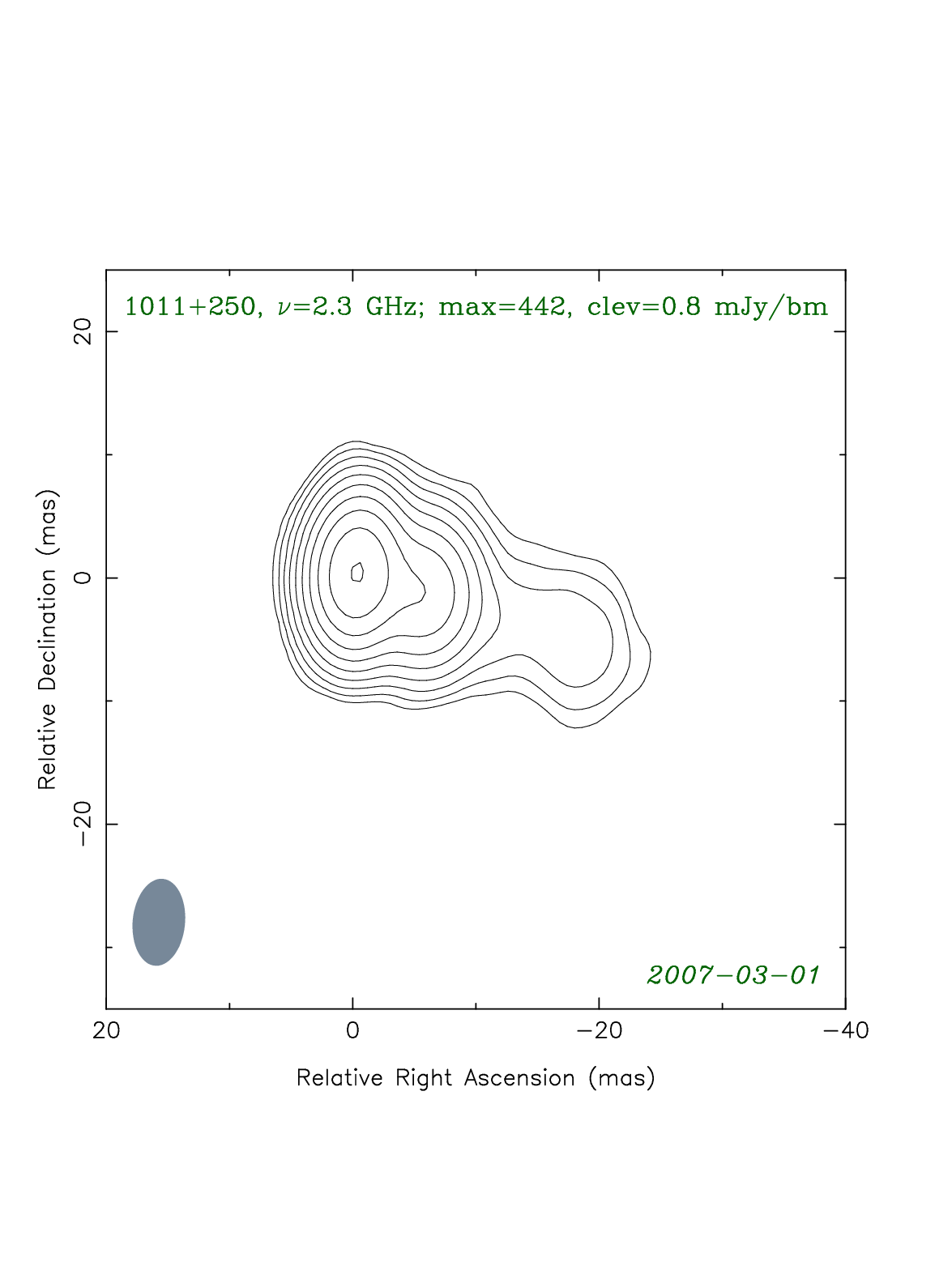}
 \includegraphics[width=0.33\textwidth,angle=0,trim=0.5cm 3cm 1.6cm 5cm,clip]{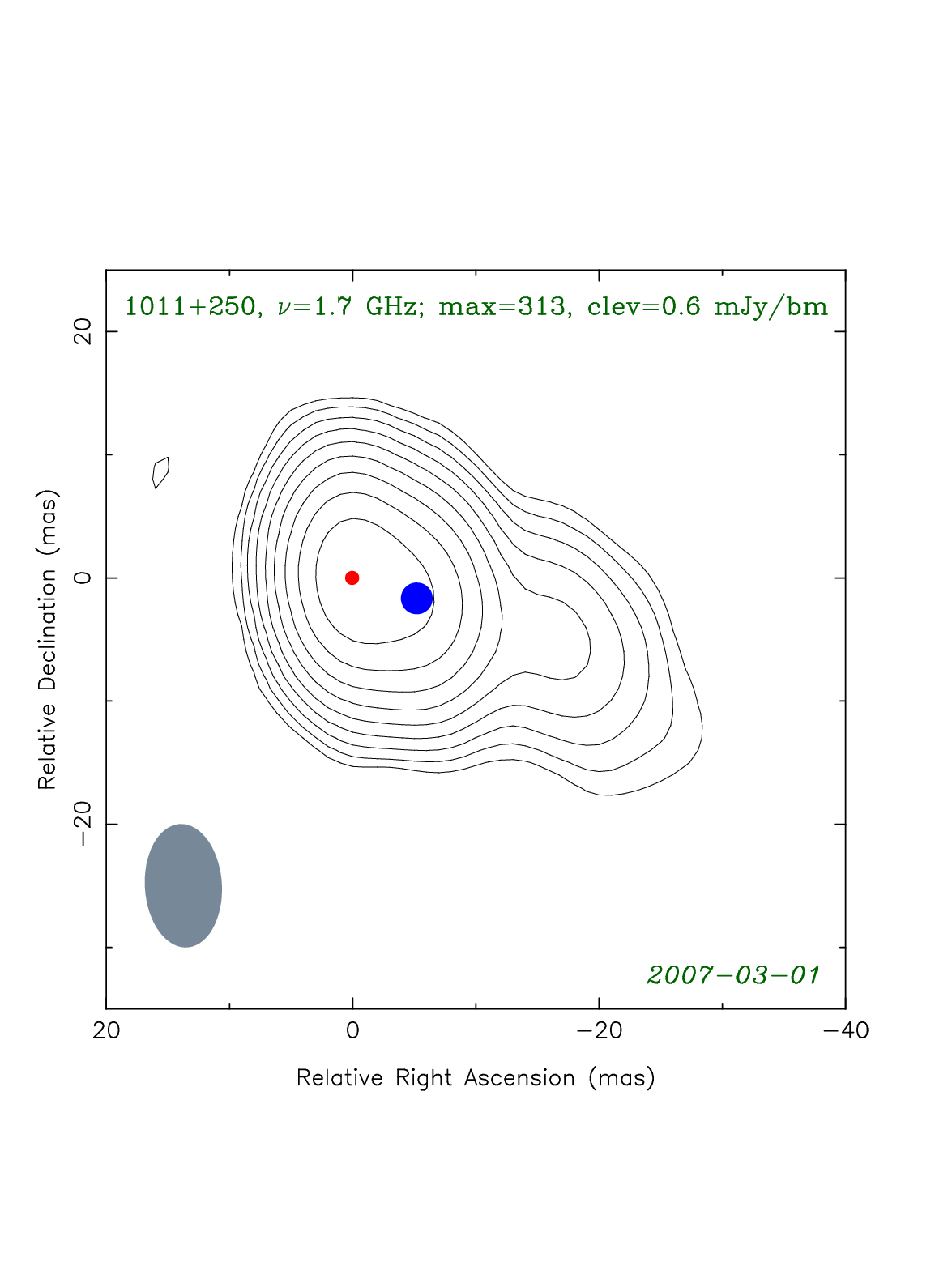}
 \includegraphics[width=0.33\textwidth,angle=0,trim=0.5cm 3cm 1.6cm 5cm,clip]{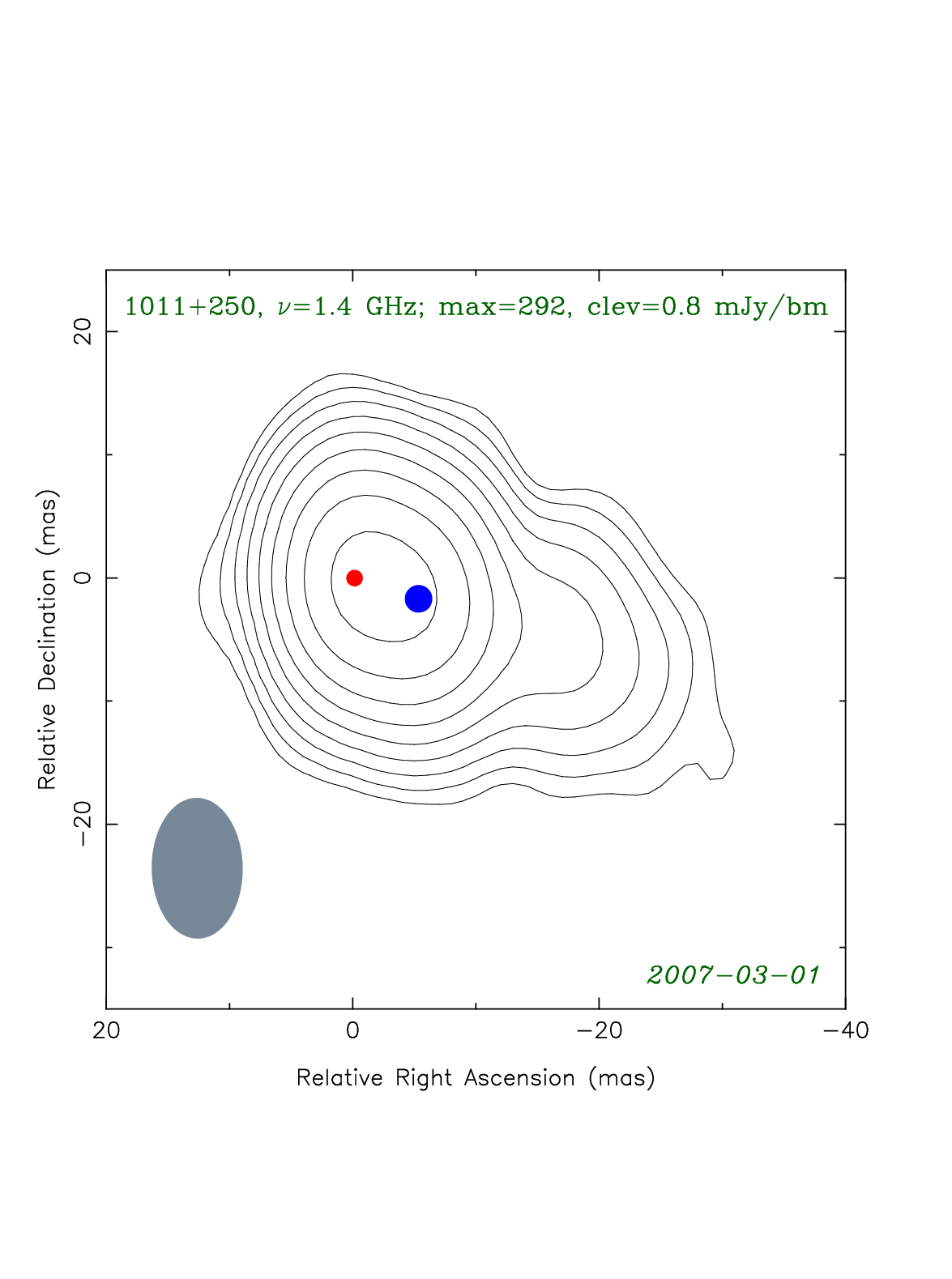}

 \caption{continued...}
\end{figure*}

\clearpage
\addtocounter{figure}{-1}
\begin{figure*}
 \centering

 \includegraphics[width=0.33\textwidth,angle=0,trim=0.5cm 3cm 1.6cm 5cm,clip]{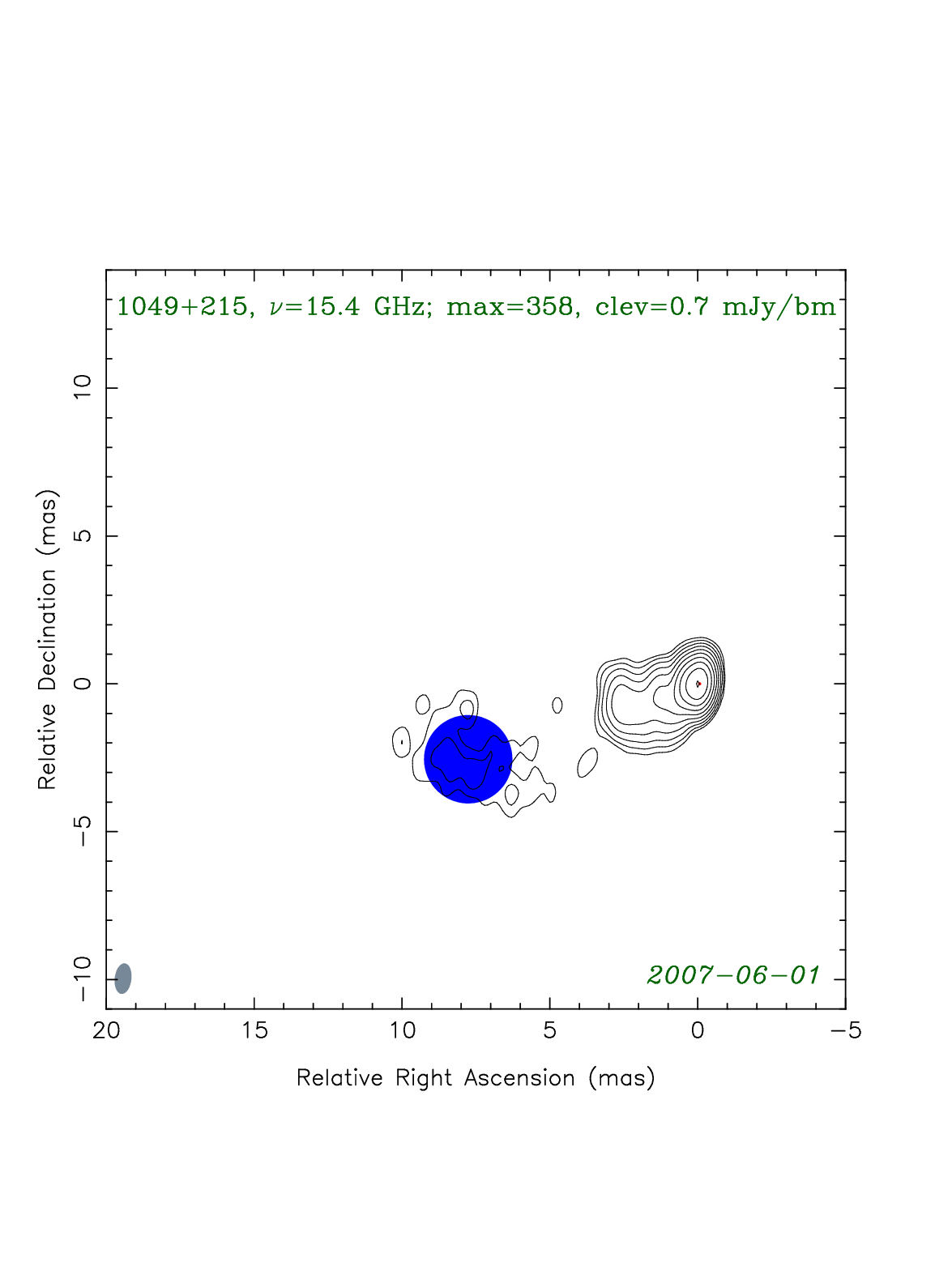}
 \includegraphics[width=0.33\textwidth,angle=0,trim=0.5cm 3cm 1.6cm 5cm,clip]{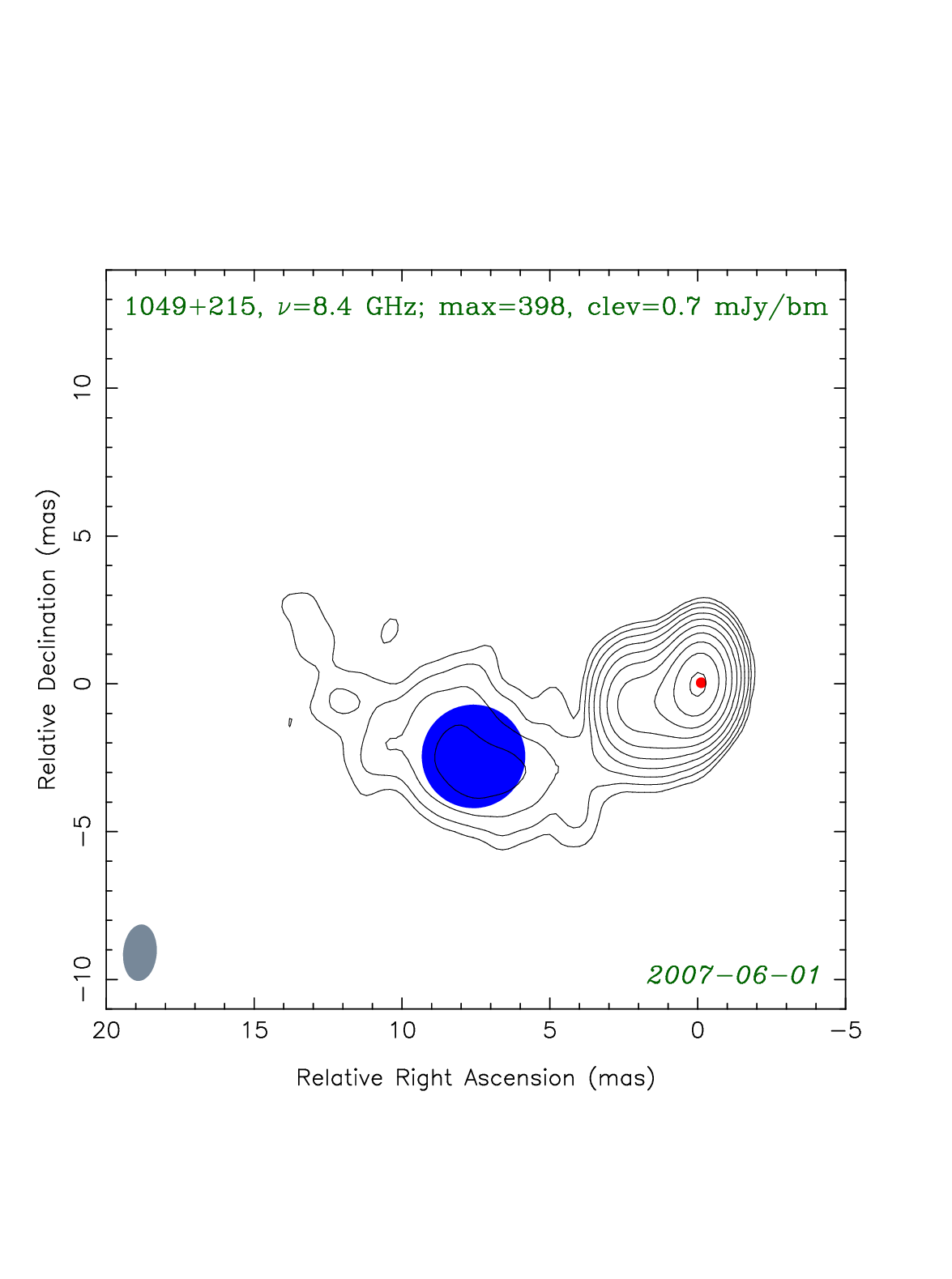}
 \includegraphics[width=0.33\textwidth,angle=0,trim=0.5cm 3cm 1.6cm 5cm,clip]{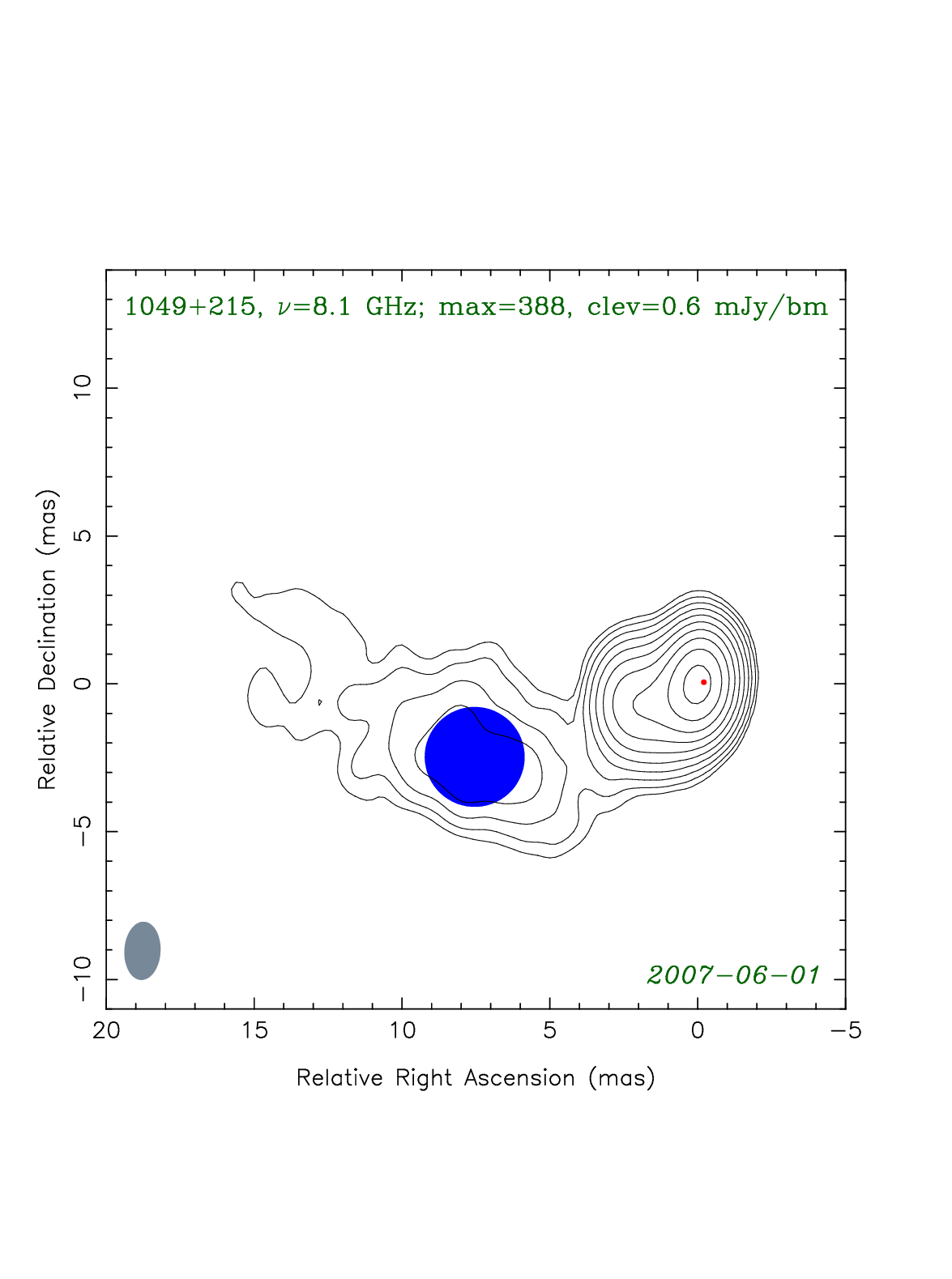}

 \includegraphics[width=0.33\textwidth,angle=0,trim=0.5cm 3cm 1.6cm 5cm,clip]{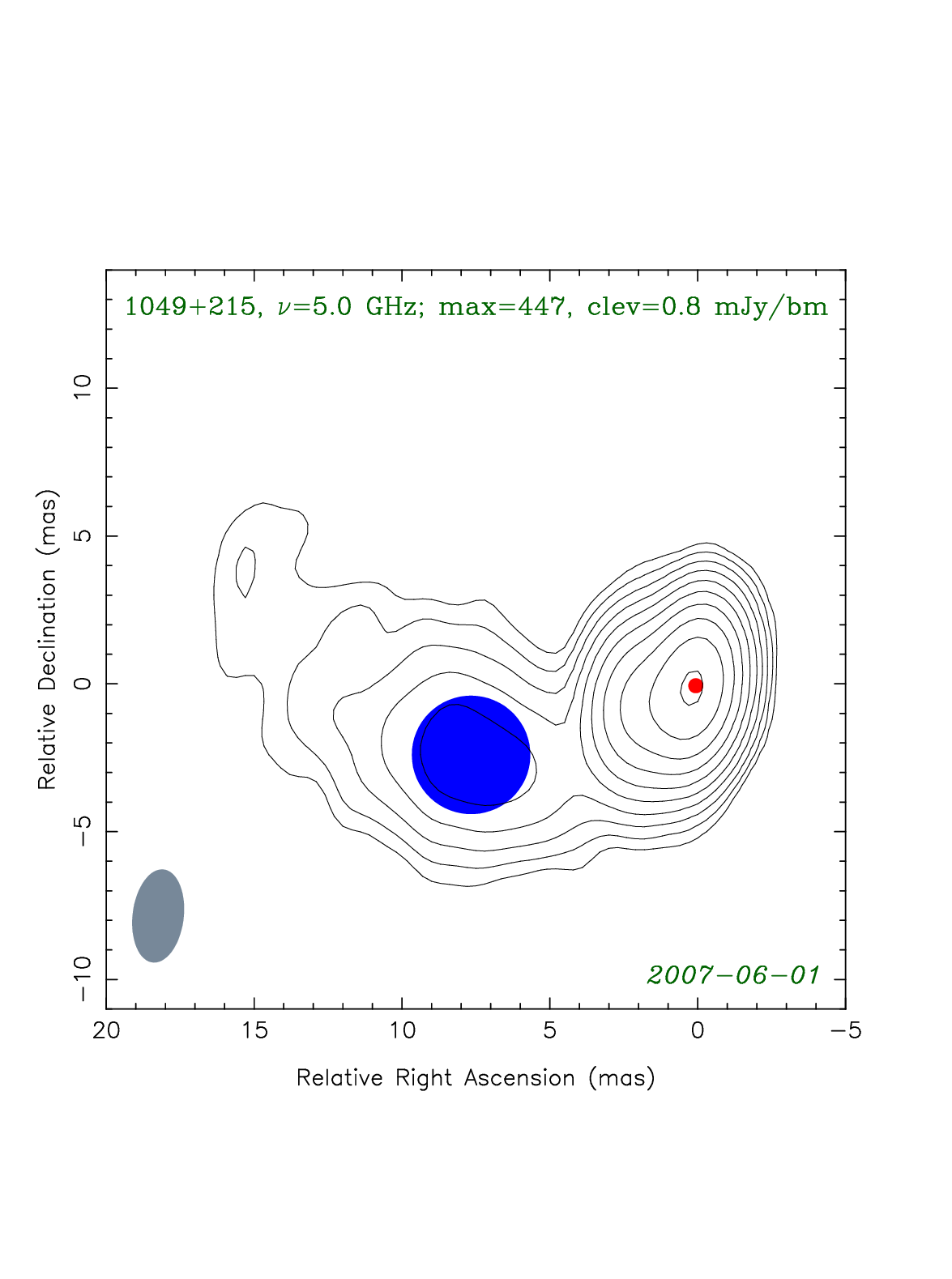}
 \includegraphics[width=0.33\textwidth,angle=0,trim=0.5cm 3cm 1.6cm 5cm,clip]{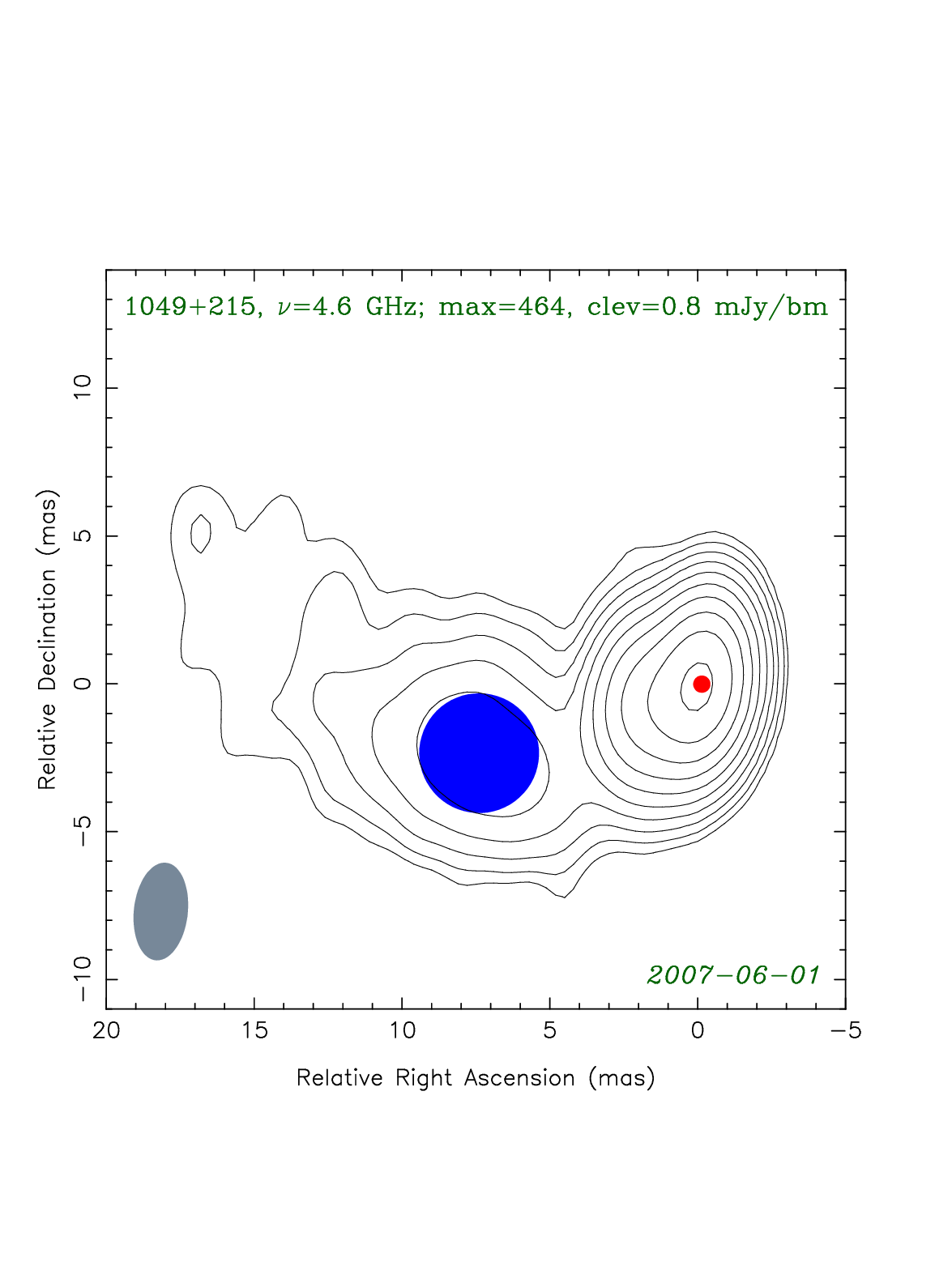}
 \includegraphics[width=0.33\textwidth,angle=0,trim=0.5cm 3cm 1.6cm 5cm,clip]{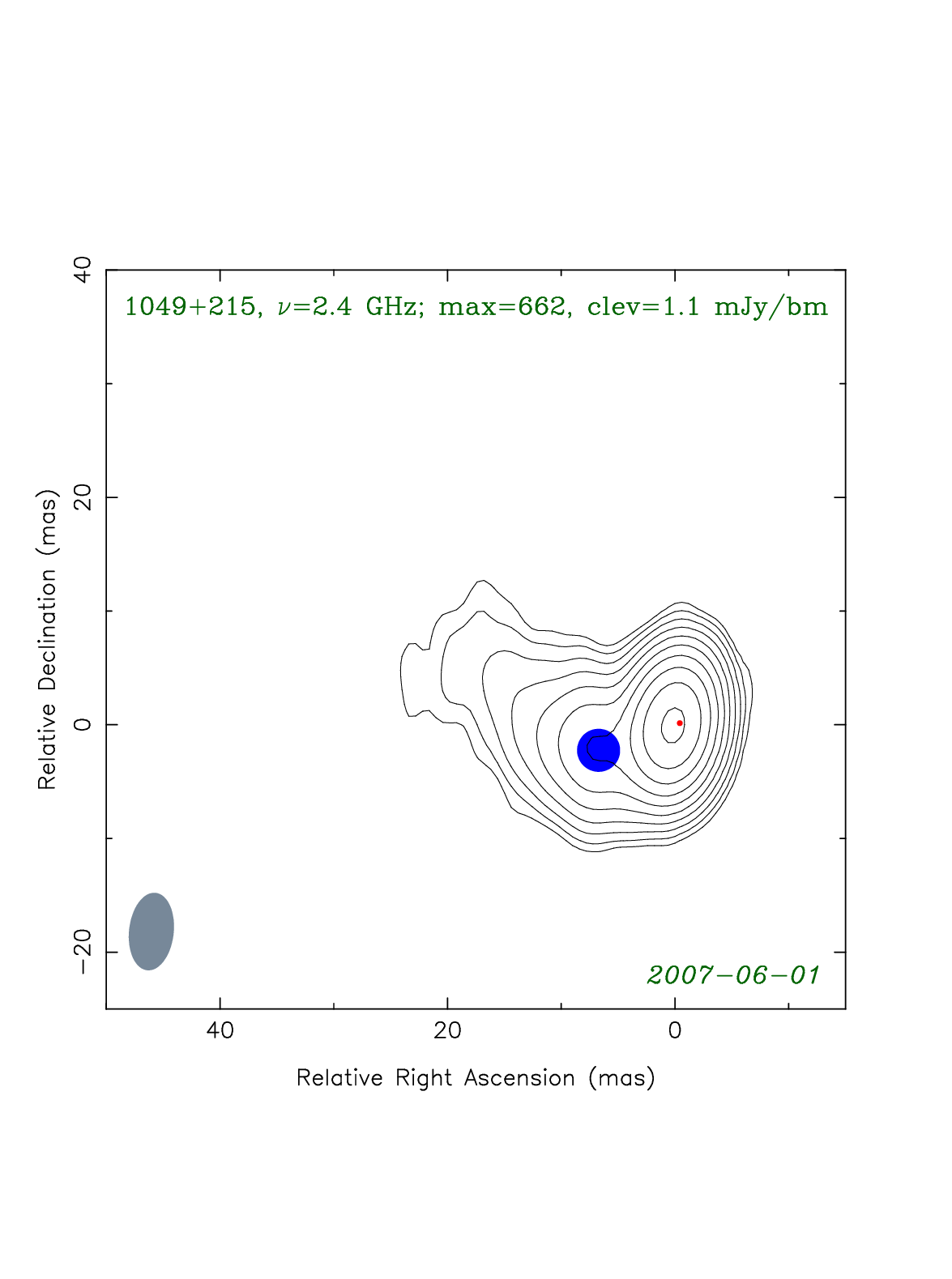}

 \includegraphics[width=0.33\textwidth,angle=0,trim=0.5cm 3cm 1.6cm 5cm,clip]{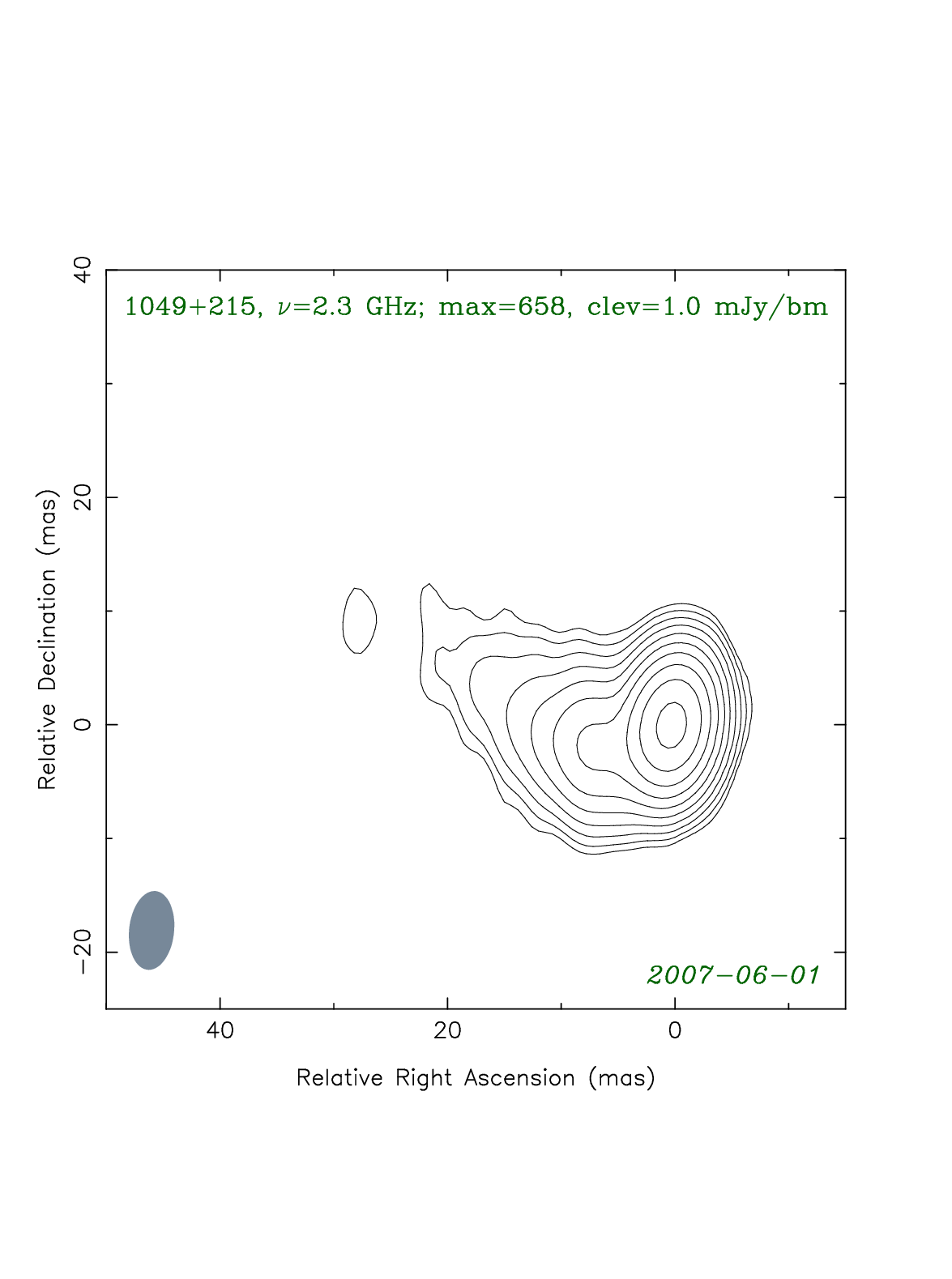}
 \includegraphics[width=0.33\textwidth,angle=0,trim=0.5cm 3cm 1.6cm 5cm,clip]{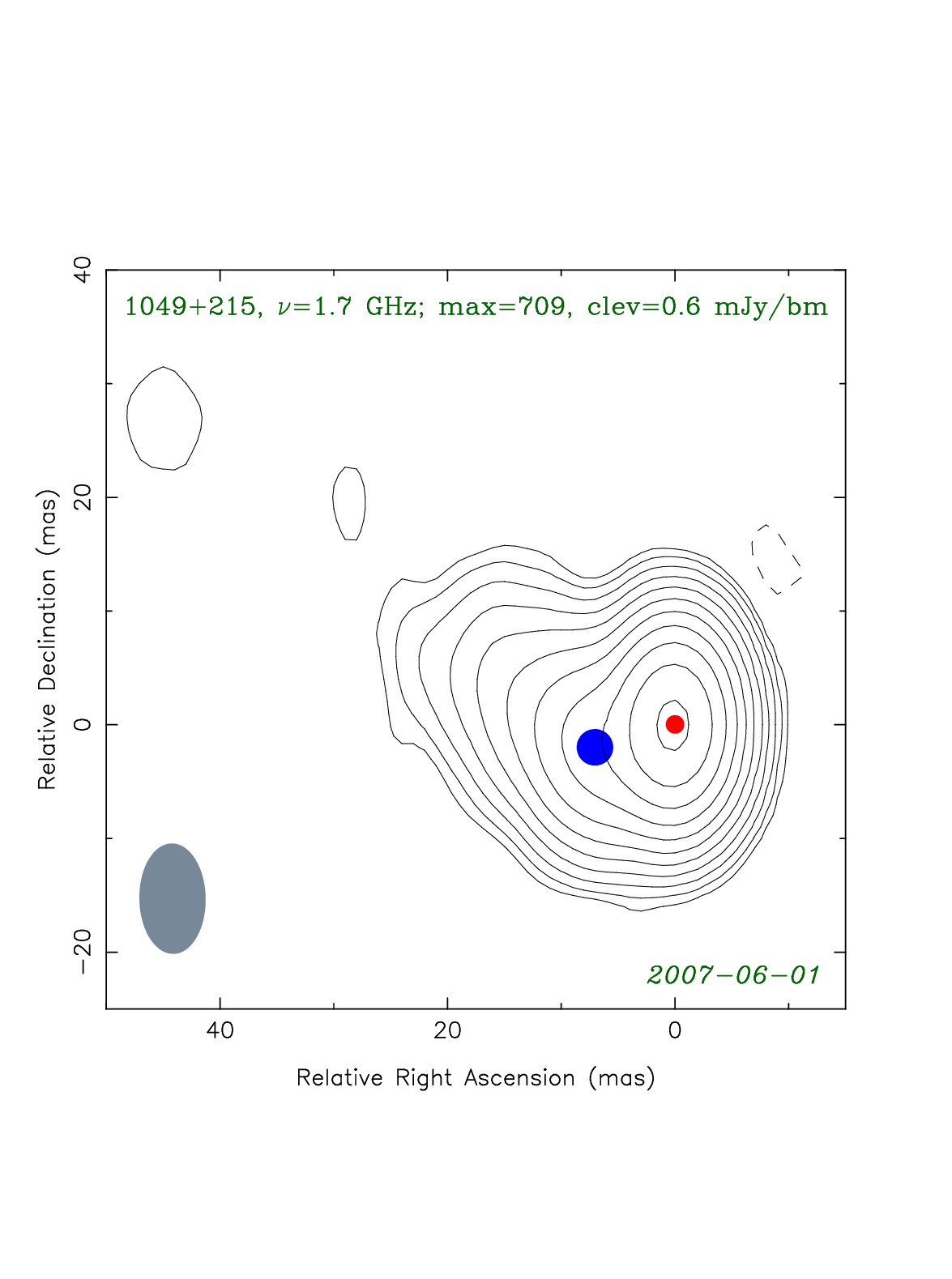}
 \includegraphics[width=0.33\textwidth,angle=0,trim=0.5cm 3cm 1.6cm 5cm,clip]{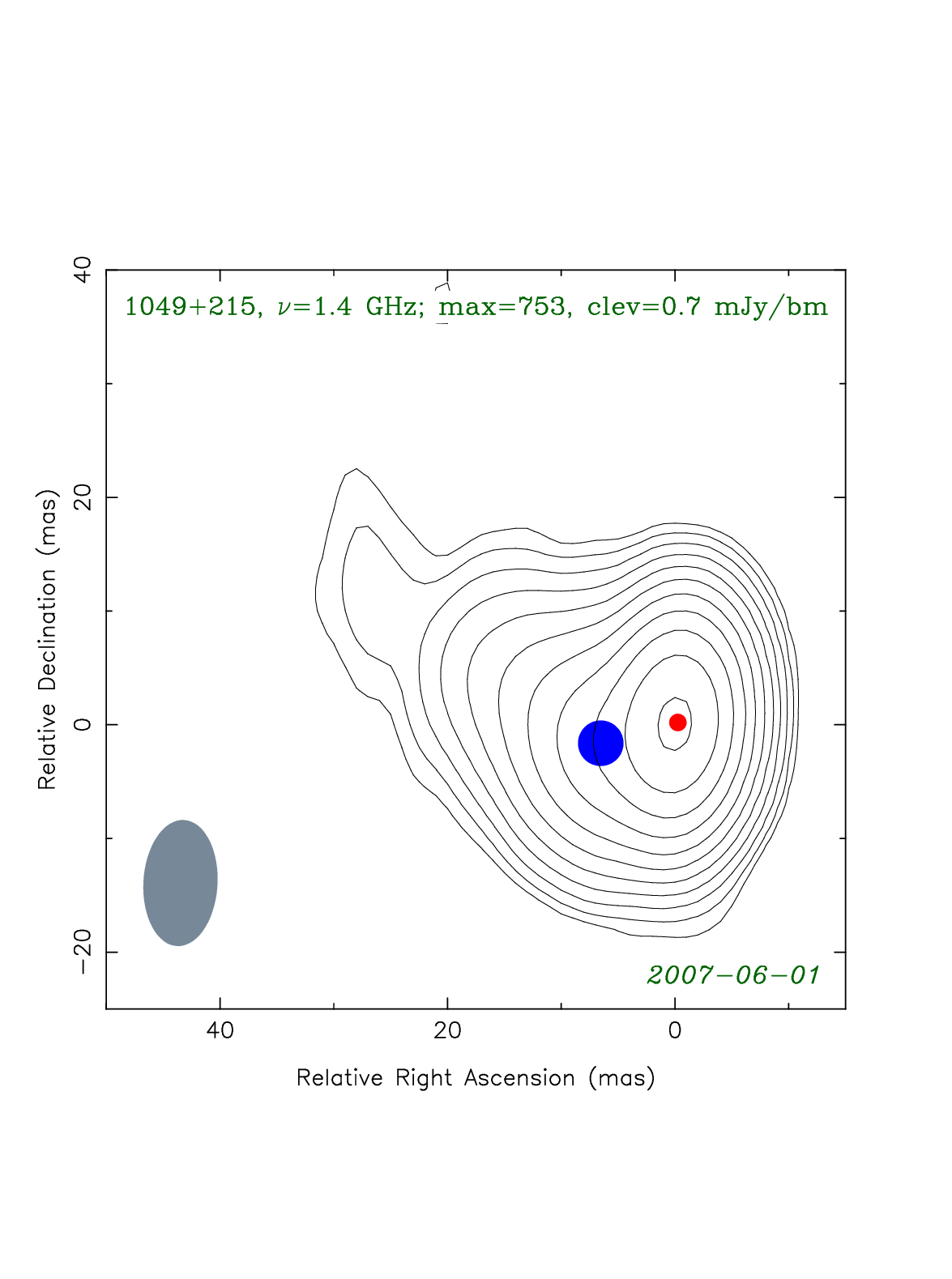}

 \caption{continued...}
\end{figure*}

\clearpage
\addtocounter{figure}{-1}
\begin{figure*}
 \centering

 \includegraphics[width=0.33\textwidth,angle=0,trim=0.5cm 3cm 1.6cm 5cm,clip]{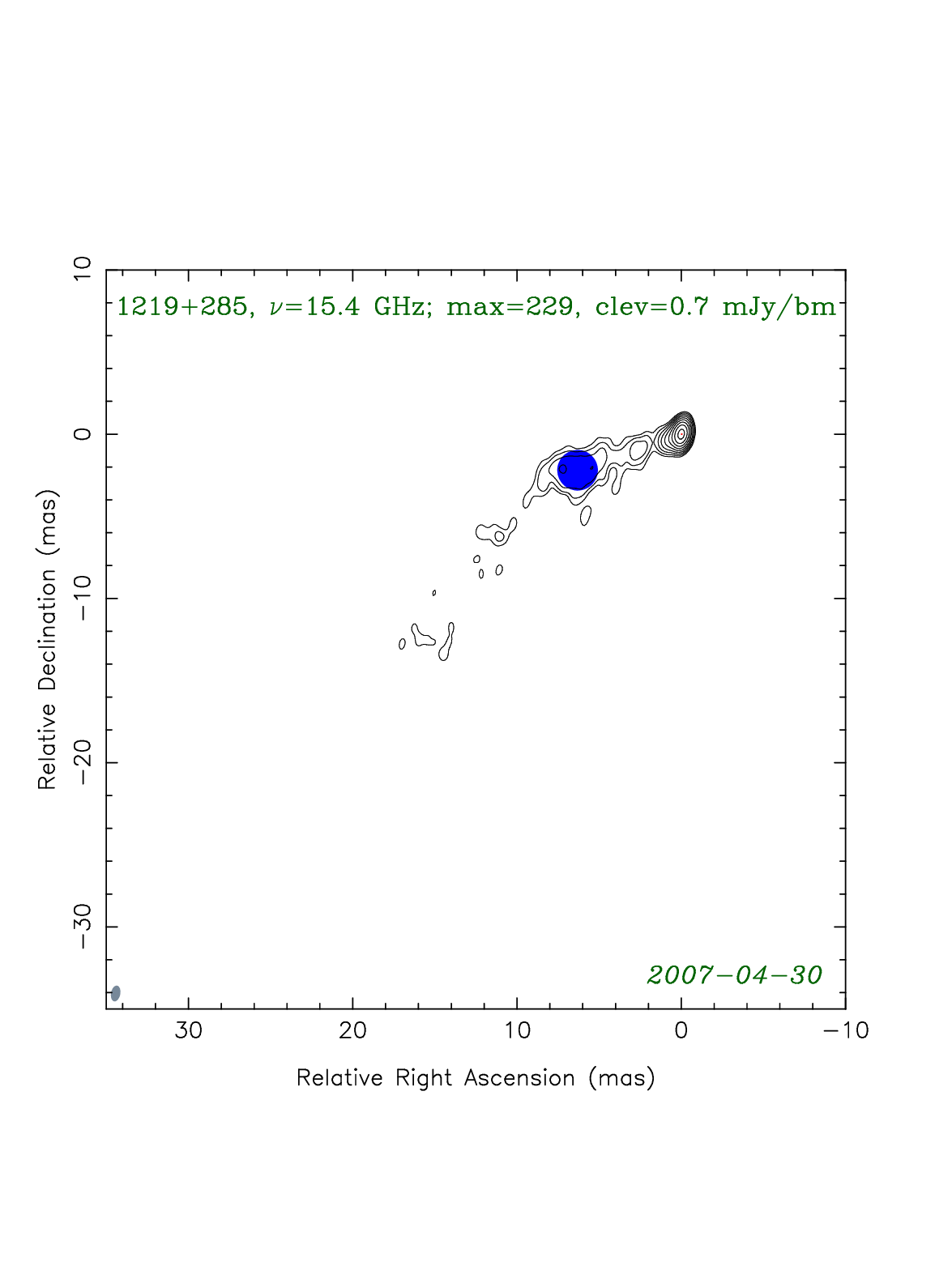}
 \includegraphics[width=0.33\textwidth,angle=0,trim=0.5cm 3cm 1.6cm 5cm,clip]{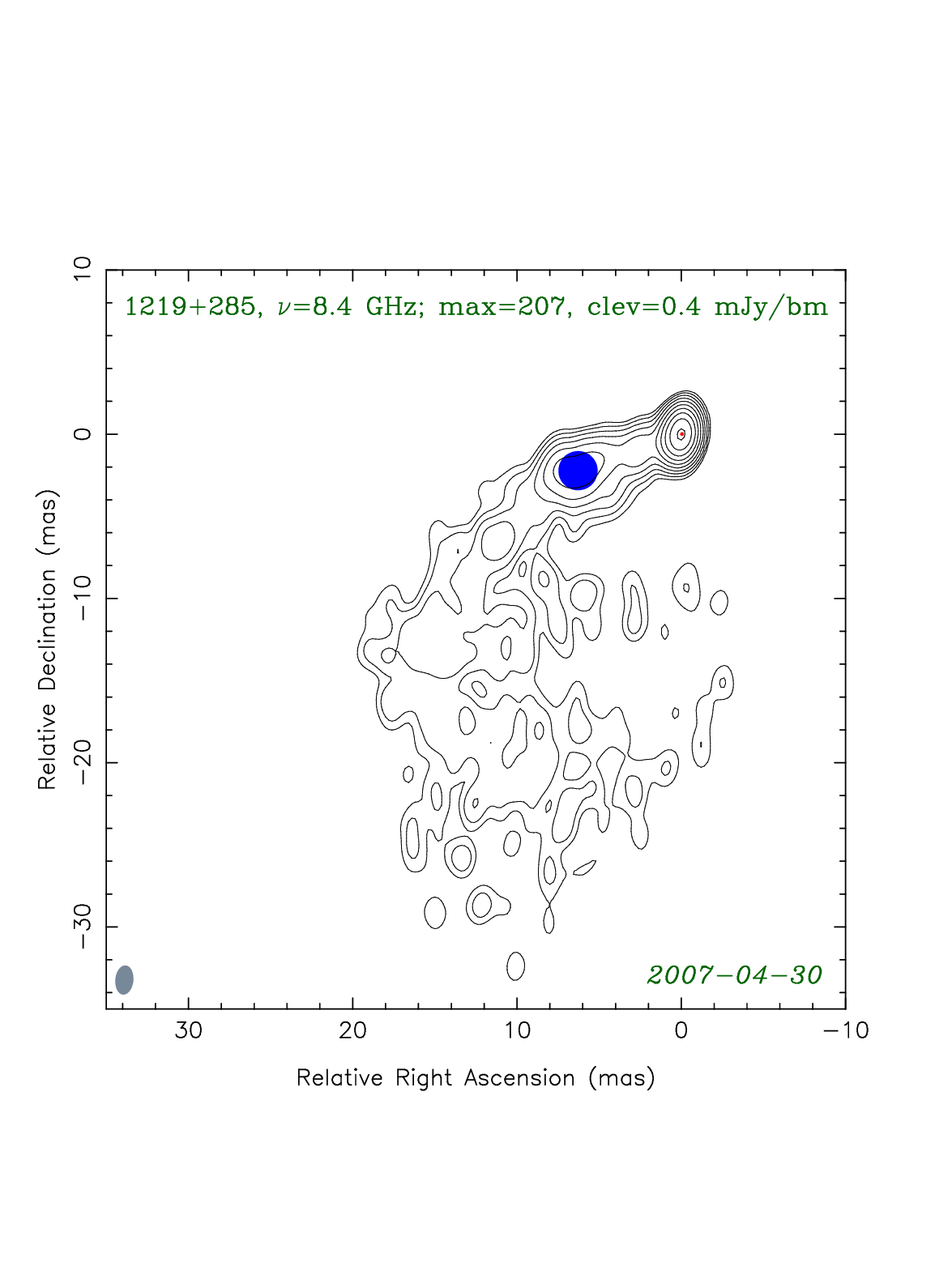}
 \includegraphics[width=0.33\textwidth,angle=0,trim=0.5cm 3cm 1.6cm 5cm,clip]{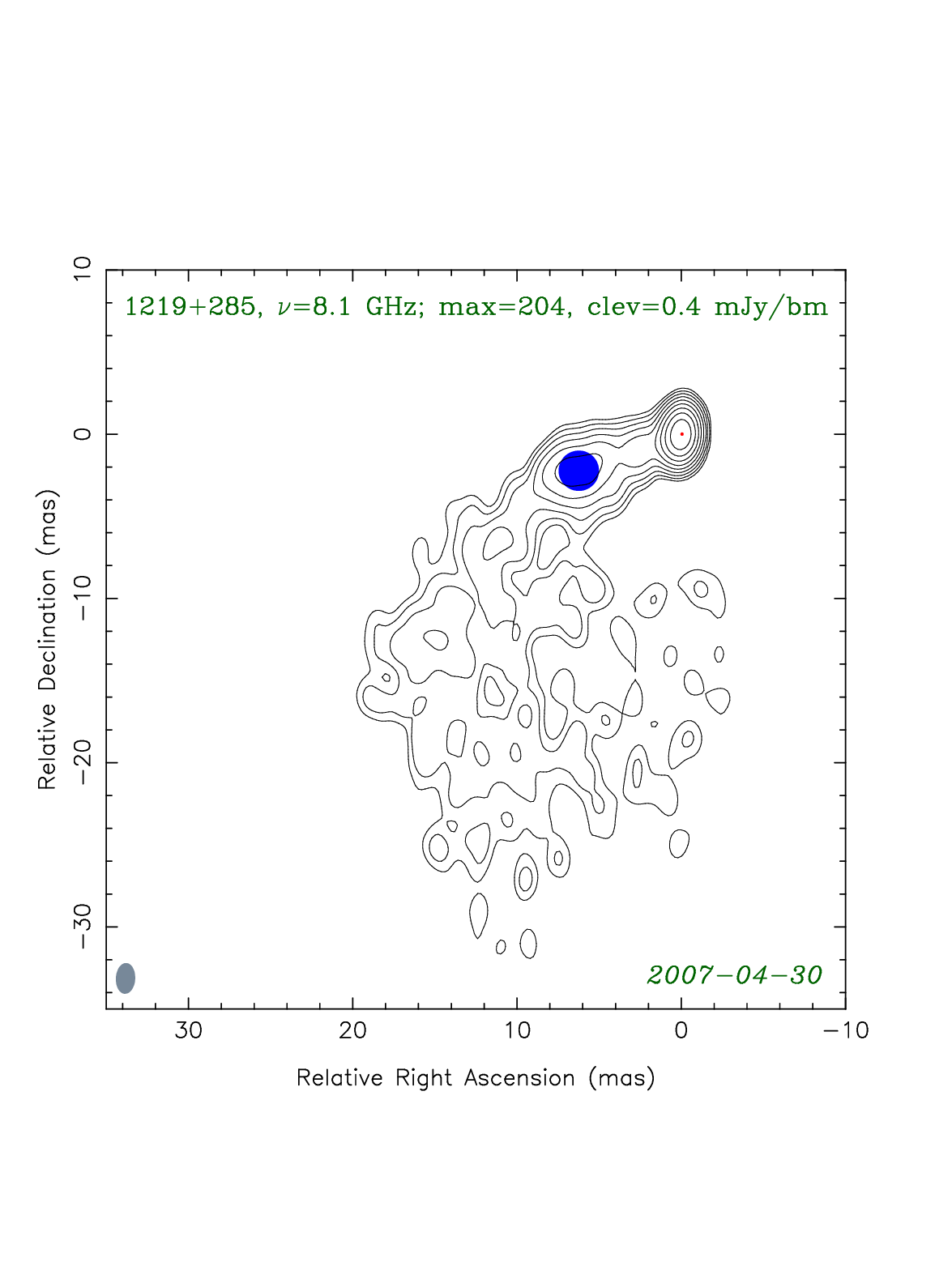}

 \includegraphics[width=0.33\textwidth,angle=0,trim=0.5cm 3cm 1.6cm 5cm,clip]{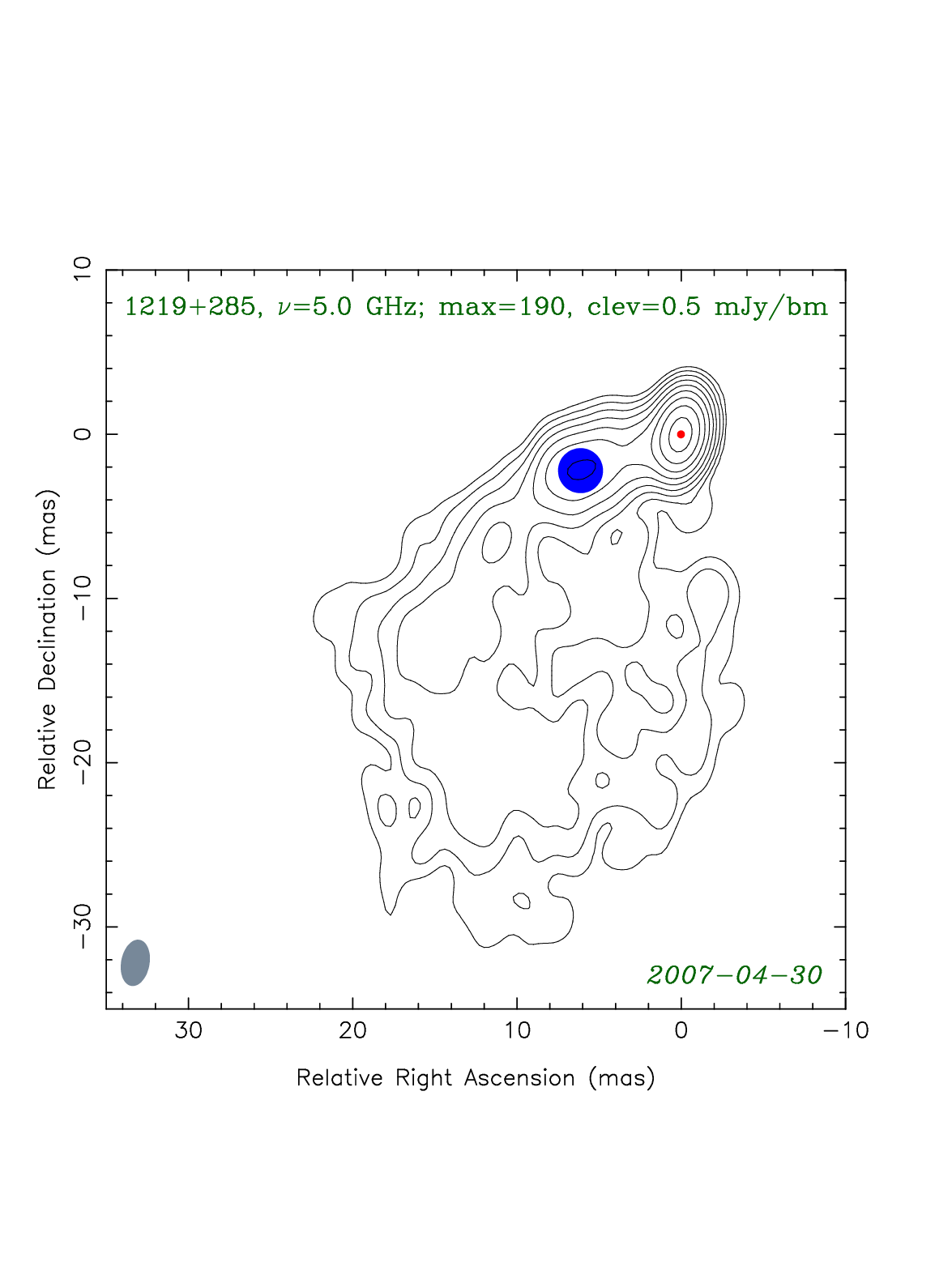}
 \includegraphics[width=0.33\textwidth,angle=0,trim=0.5cm 3cm 1.6cm 5cm,clip]{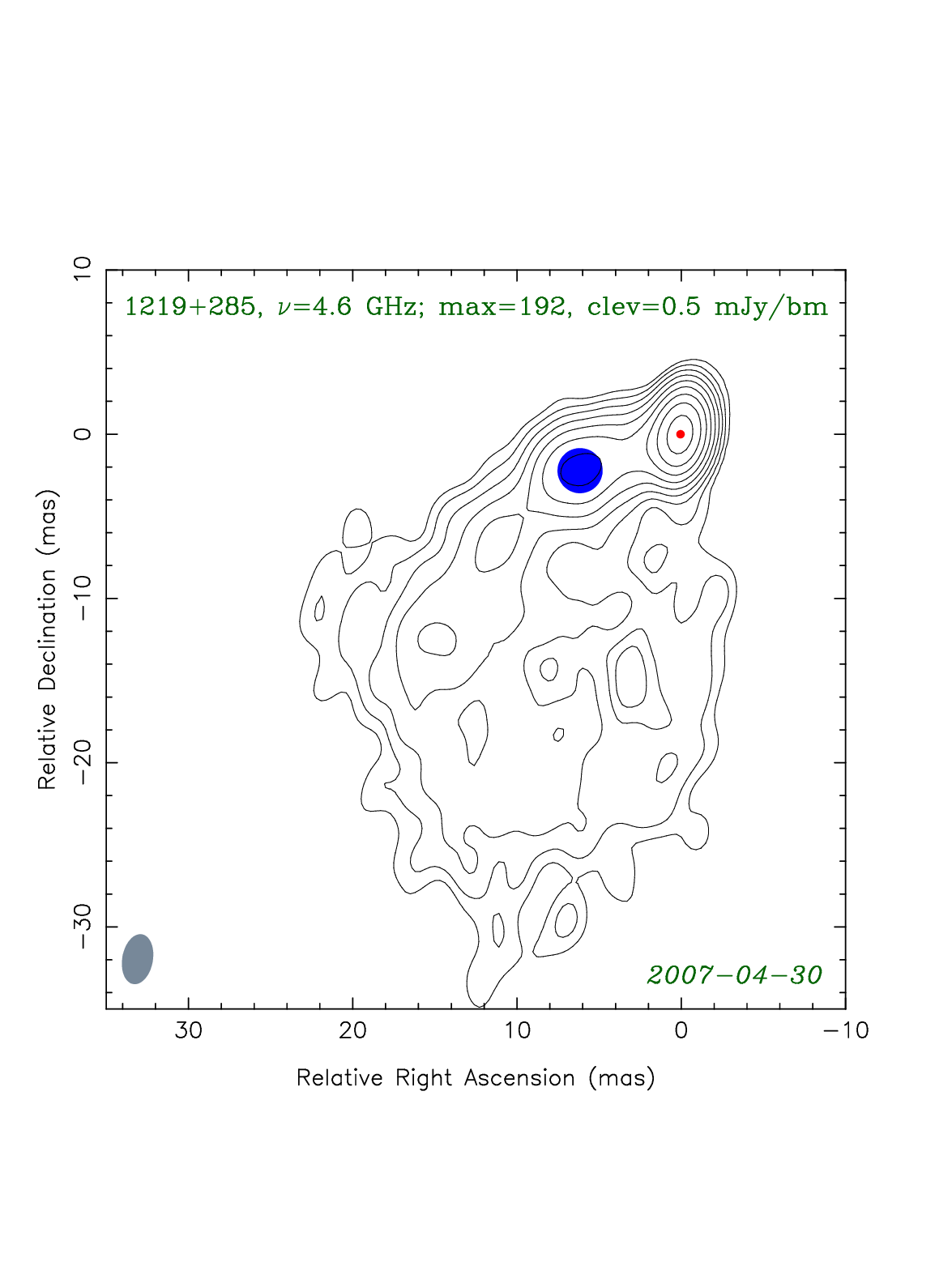}
 \includegraphics[width=0.33\textwidth,angle=0,trim=0.5cm 3cm 1.6cm 5cm,clip]{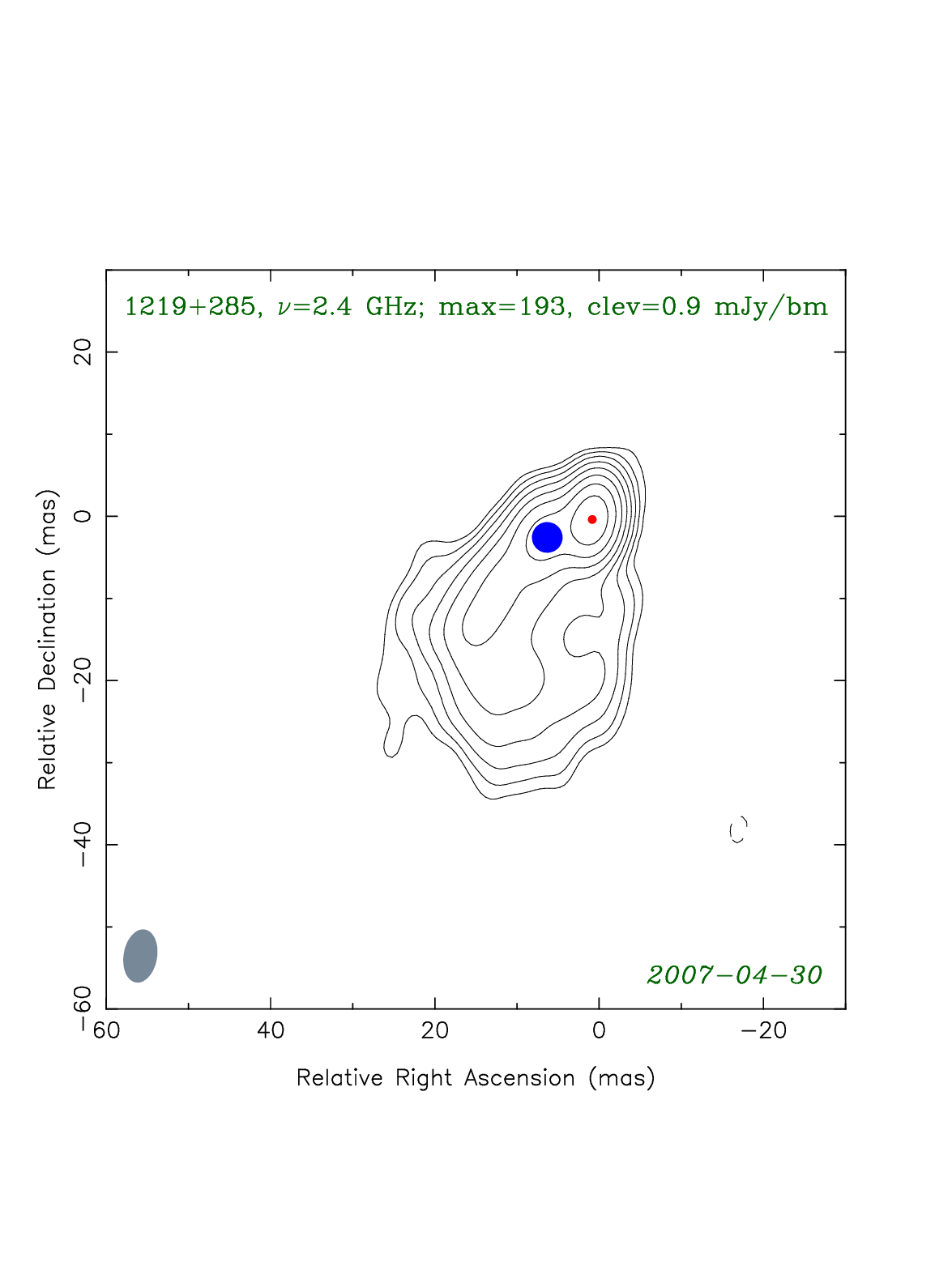}

 \includegraphics[width=0.33\textwidth,angle=0,trim=0.5cm 3cm 1.6cm 5cm,clip]{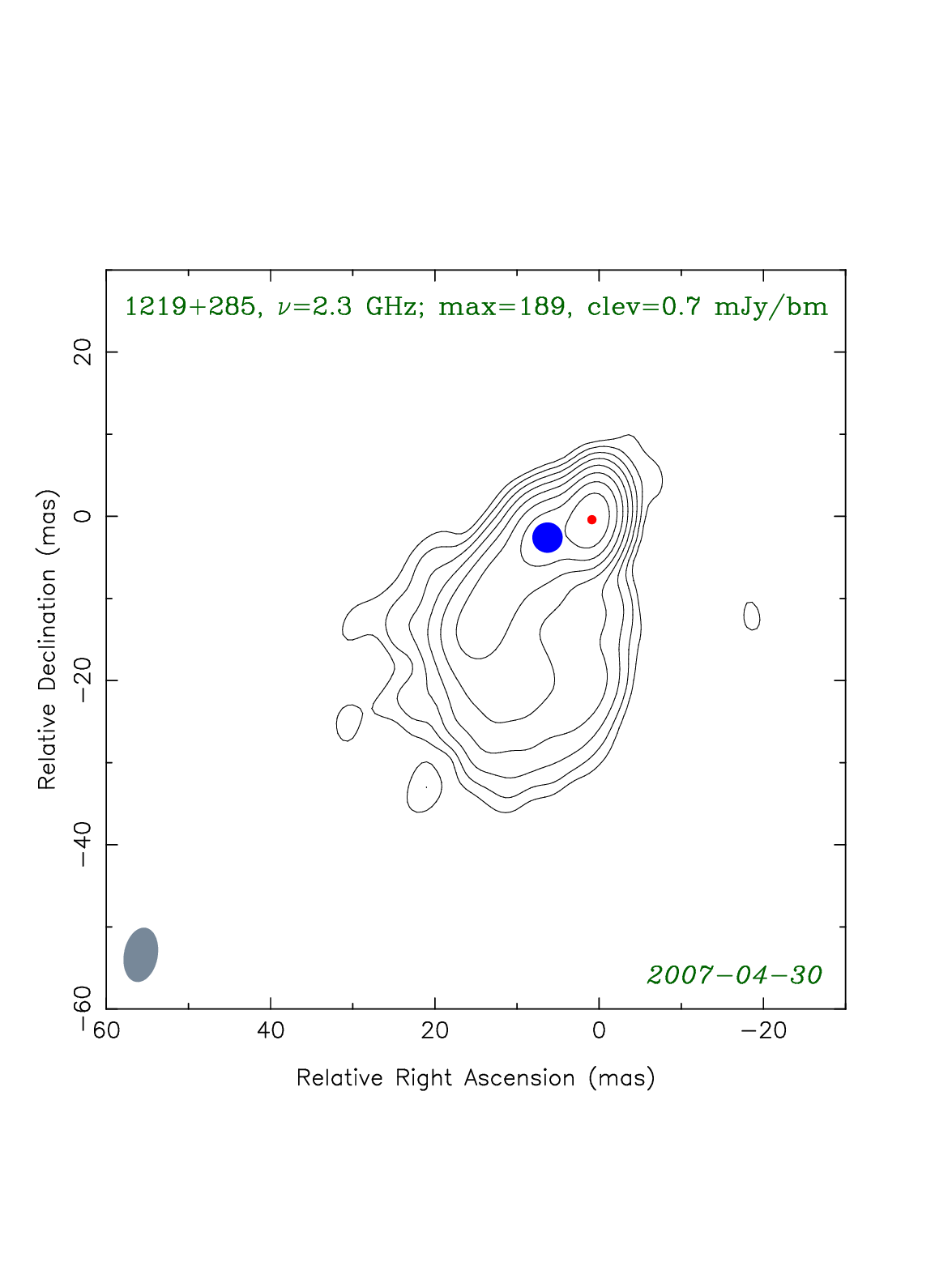}
 \includegraphics[width=0.33\textwidth,angle=0,trim=0.5cm 3cm 1.6cm 5cm,clip]{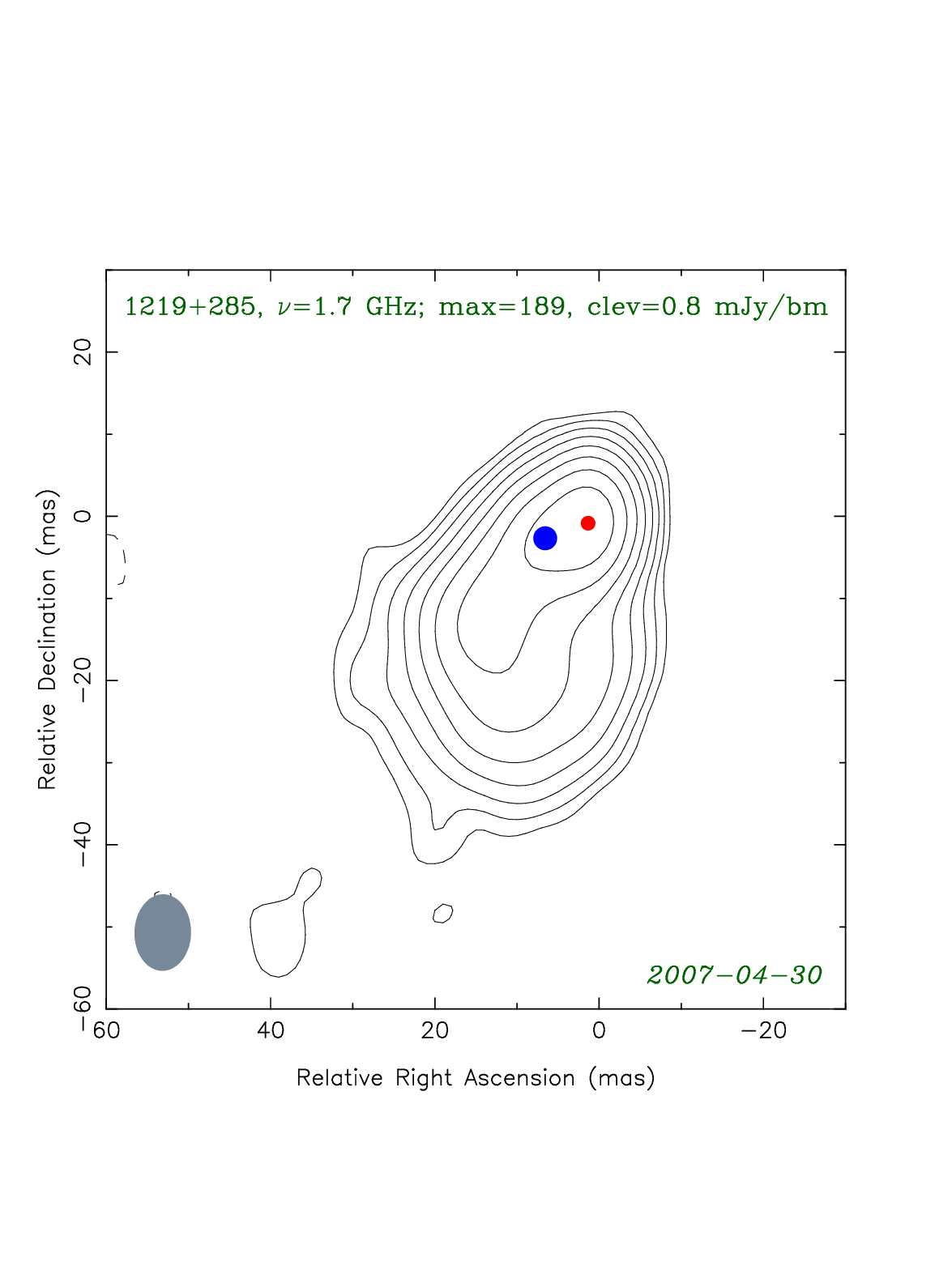}
 \includegraphics[width=0.33\textwidth,angle=0,trim=0.5cm 3cm 1.6cm 5cm,clip]{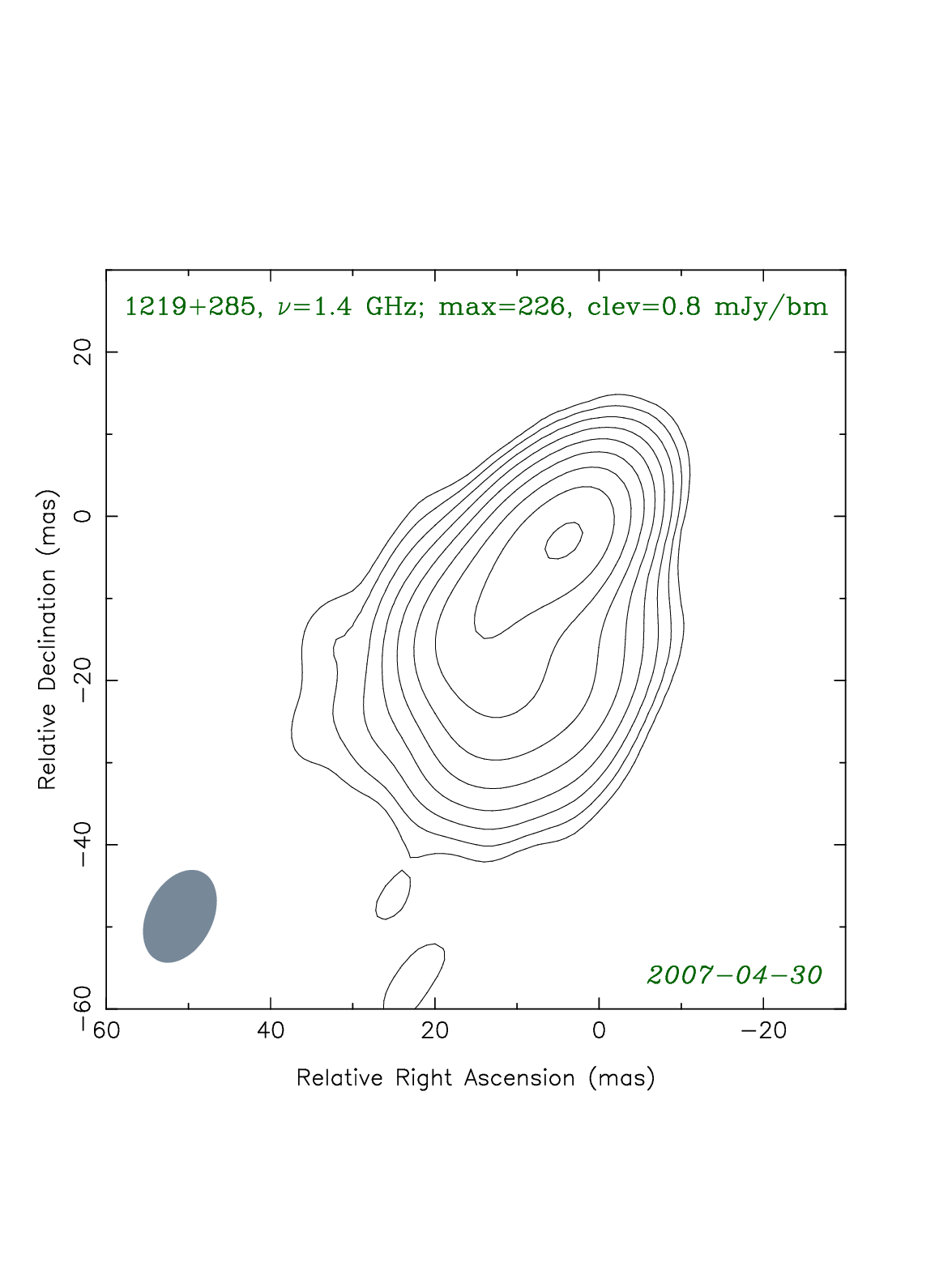}

 \caption{continued...}
\end{figure*}

\clearpage
\addtocounter{figure}{-1}
\begin{figure*}
 \centering

 \includegraphics[width=0.33\textwidth,angle=0,trim=0.5cm 3cm 1.6cm 5cm,clip]{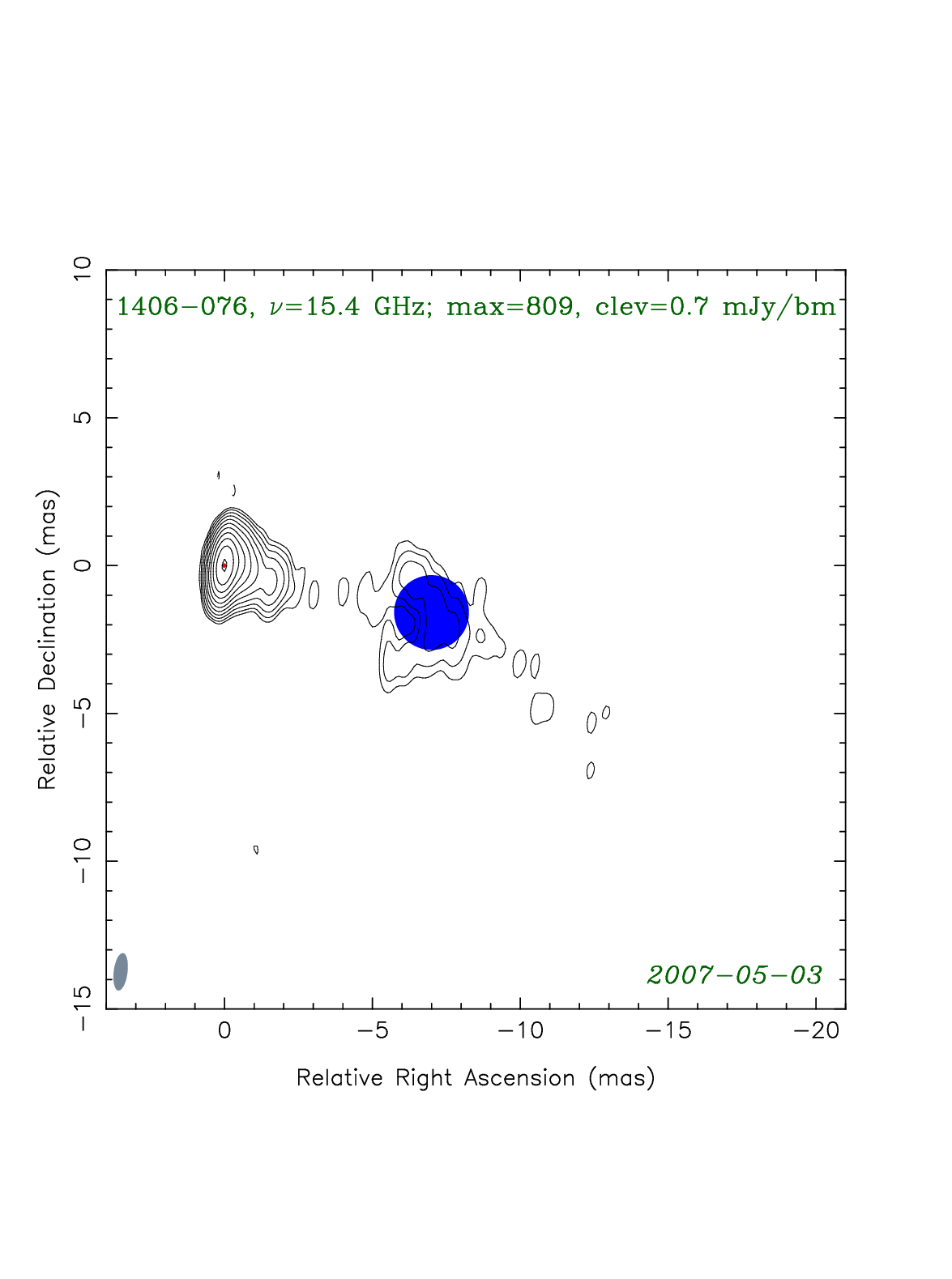}
 \includegraphics[width=0.33\textwidth,angle=0,trim=0.5cm 3cm 1.6cm 5cm,clip]{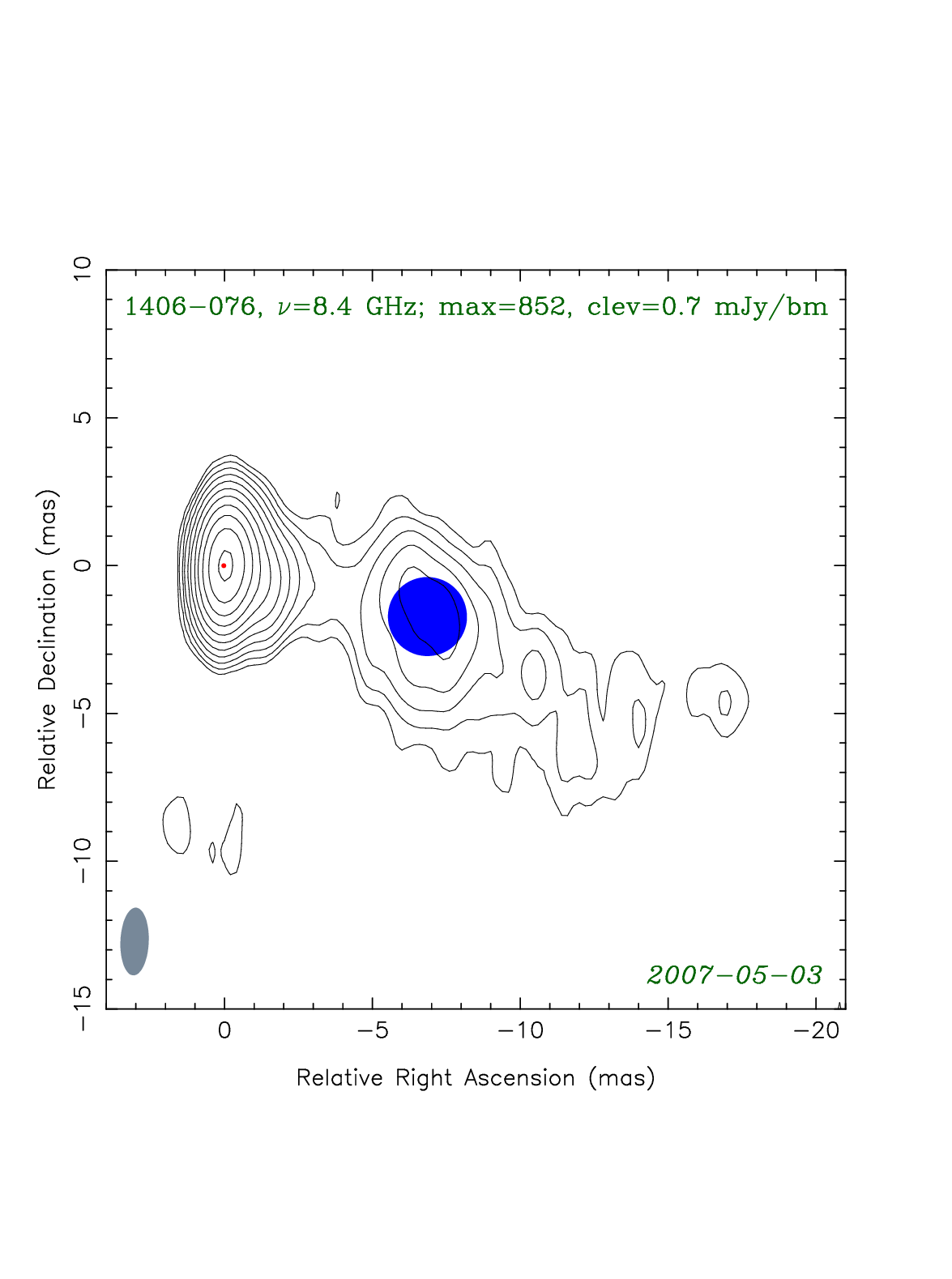}
 \includegraphics[width=0.33\textwidth,angle=0,trim=0.5cm 3cm 1.6cm 5cm,clip]{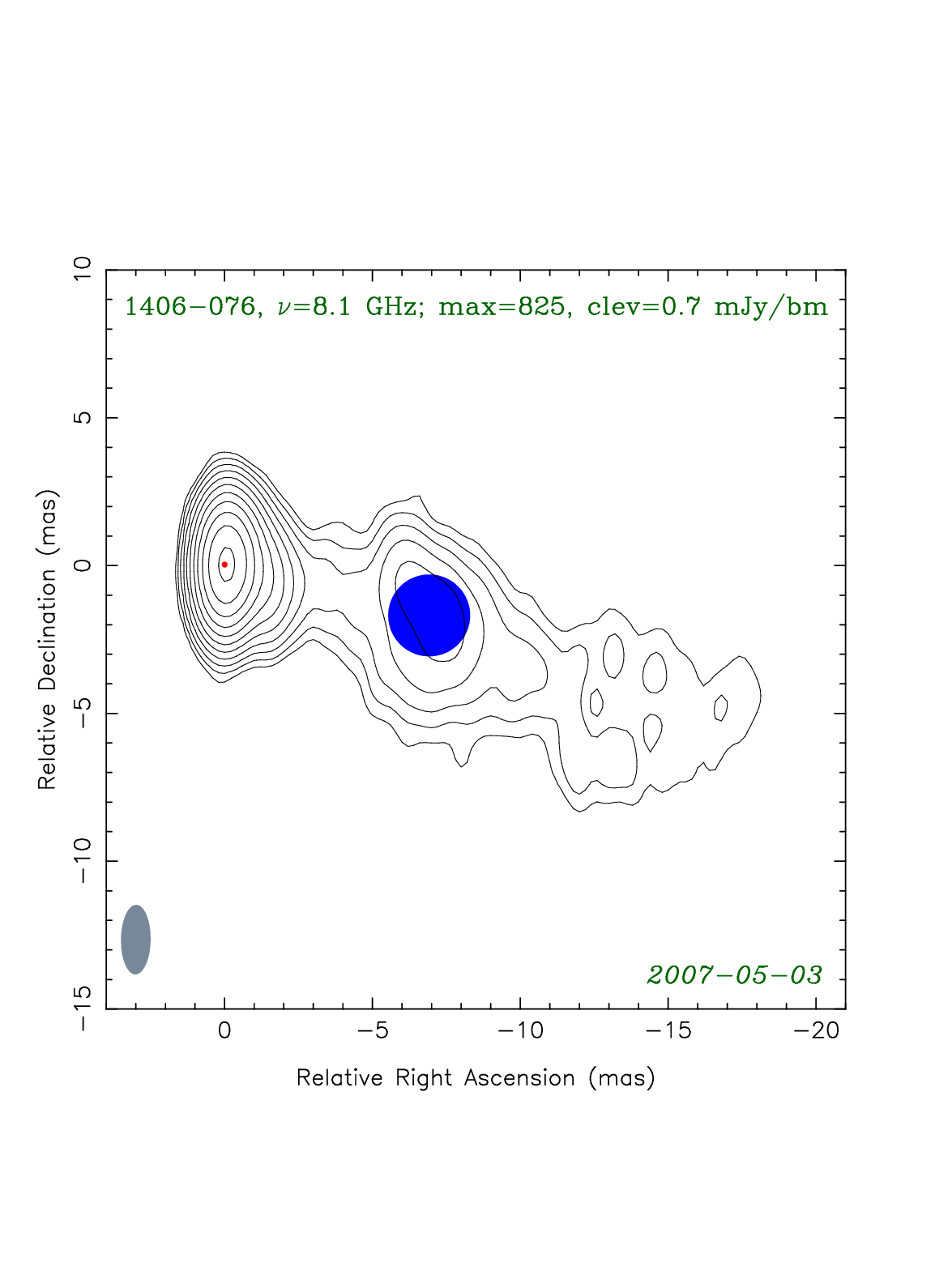}

 \includegraphics[width=0.33\textwidth,angle=0,trim=0.5cm 3cm 1.6cm 5cm,clip]{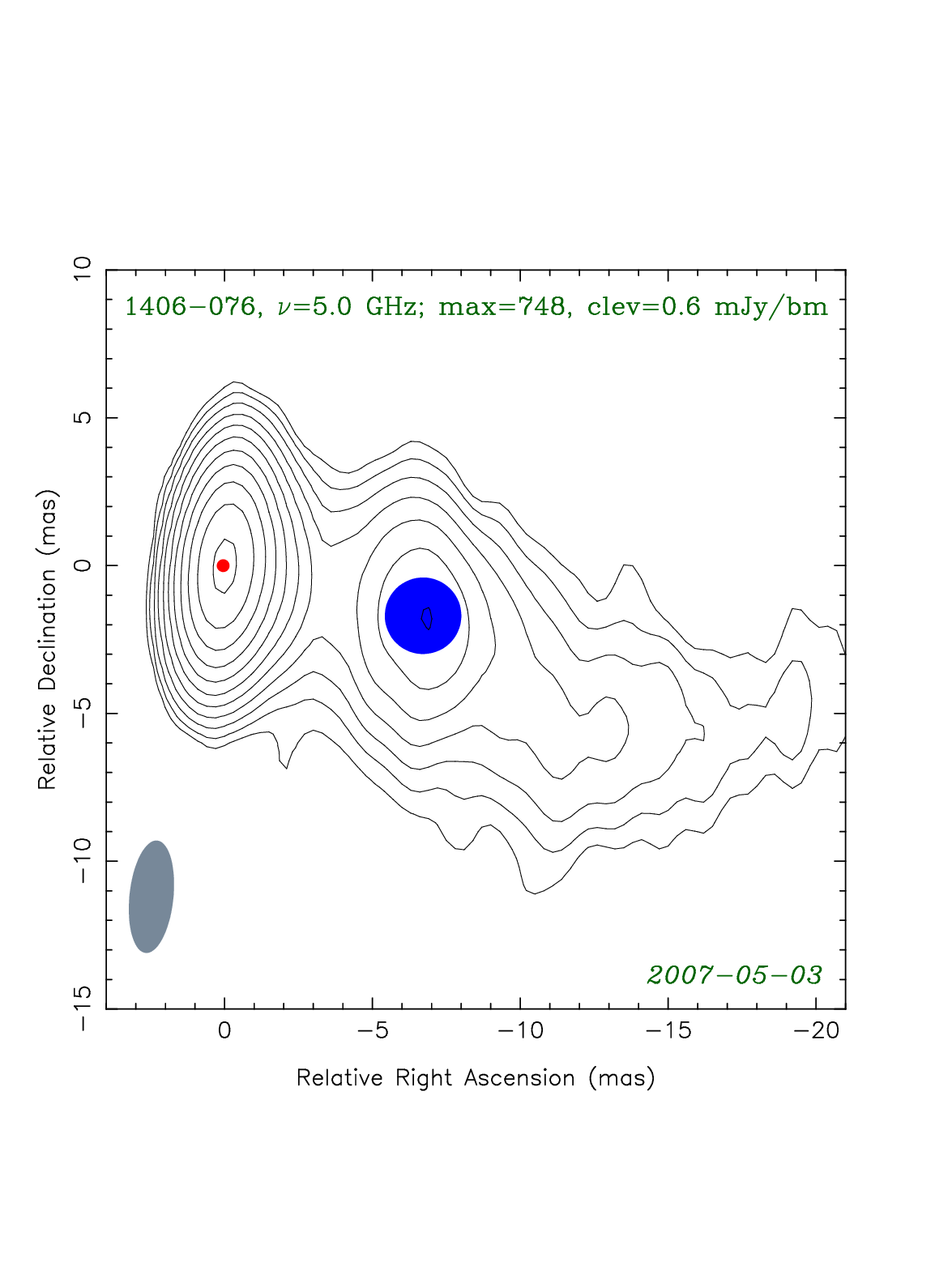}
 \includegraphics[width=0.33\textwidth,angle=0,trim=0.5cm 3cm 1.6cm 5cm,clip]{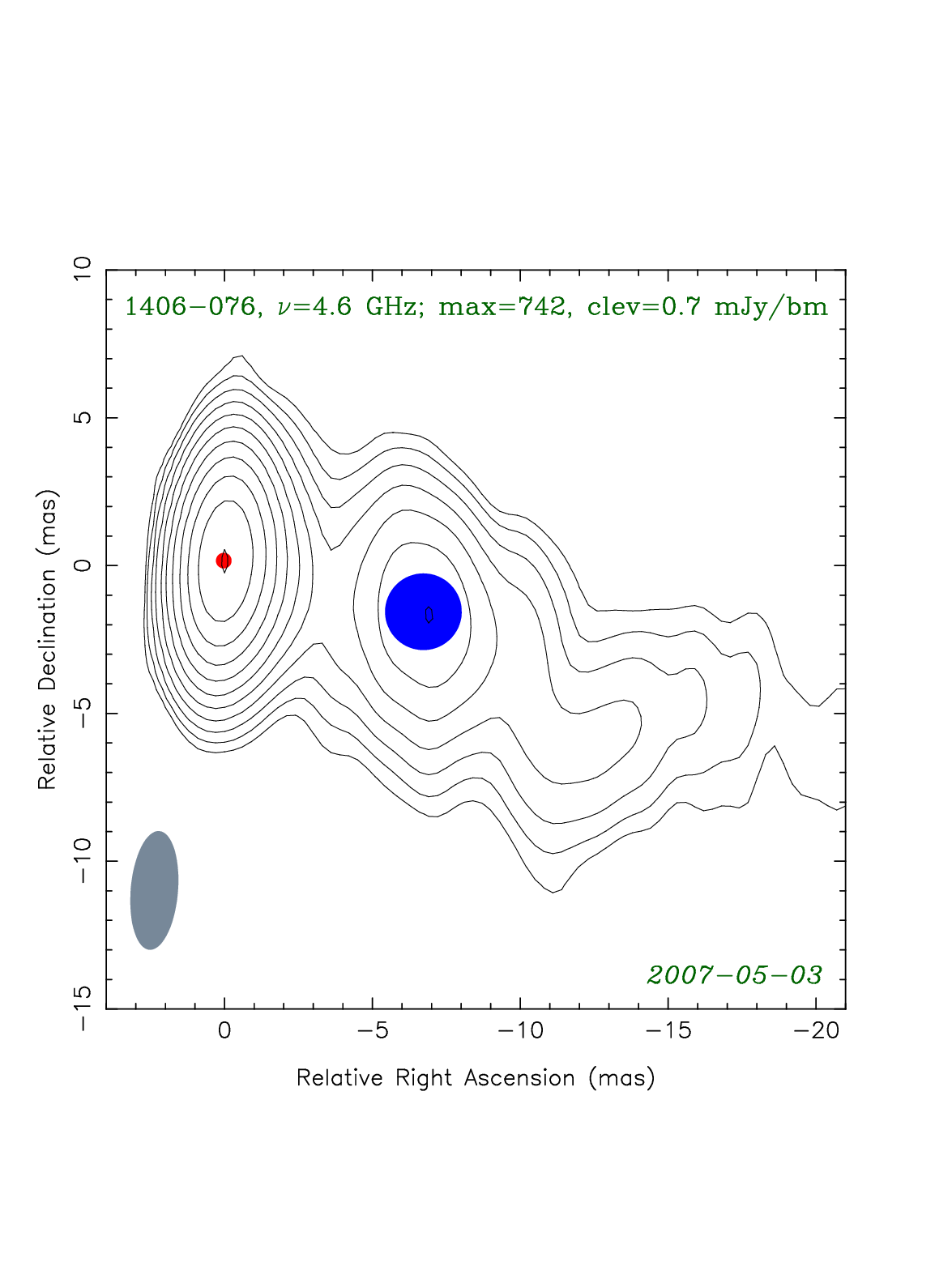}
 \includegraphics[width=0.33\textwidth,angle=0,trim=0.5cm 3cm 1.6cm 5cm,clip]{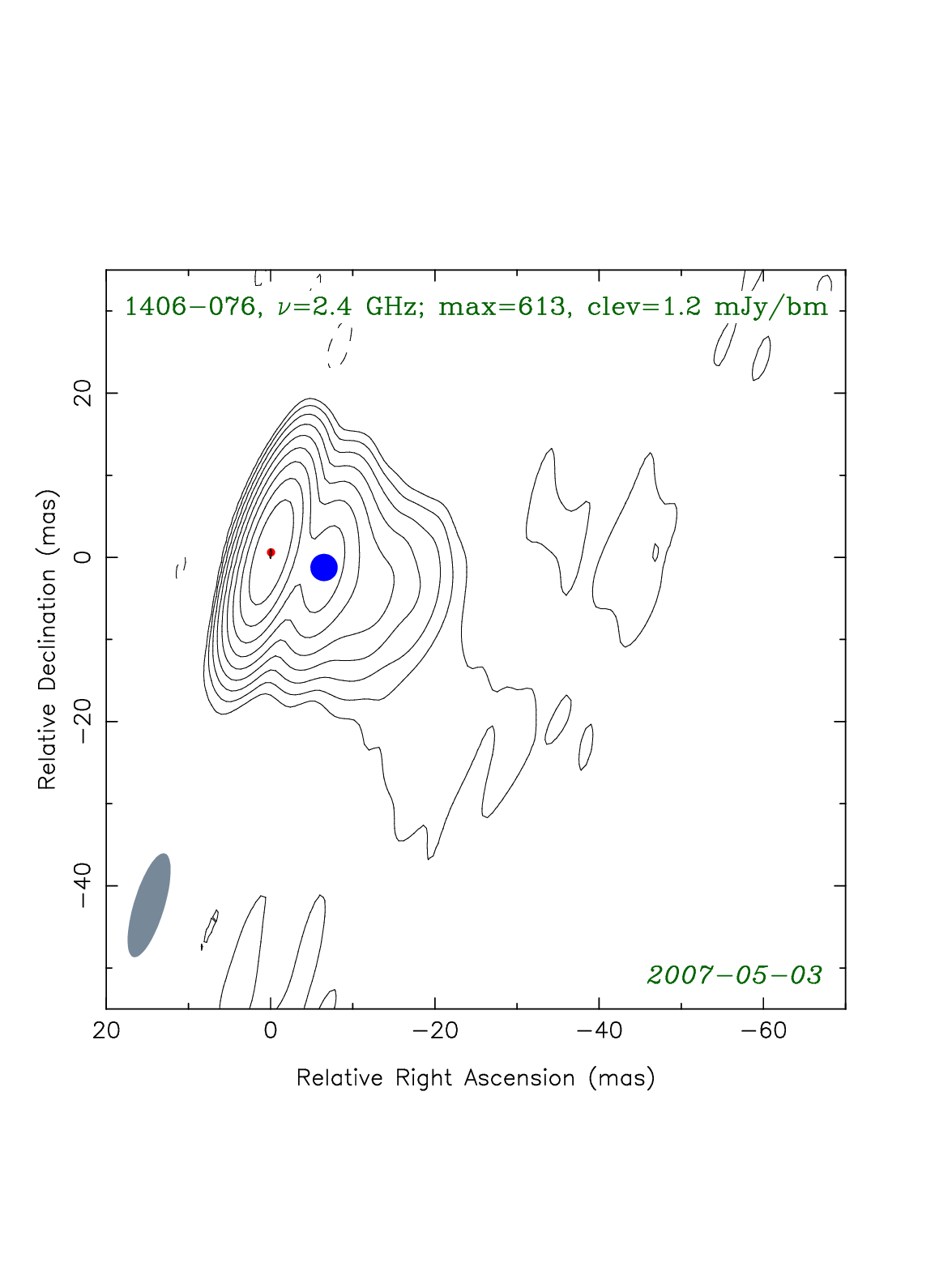}

 \includegraphics[width=0.33\textwidth,angle=0,trim=0.5cm 3cm 1.6cm 5cm,clip]{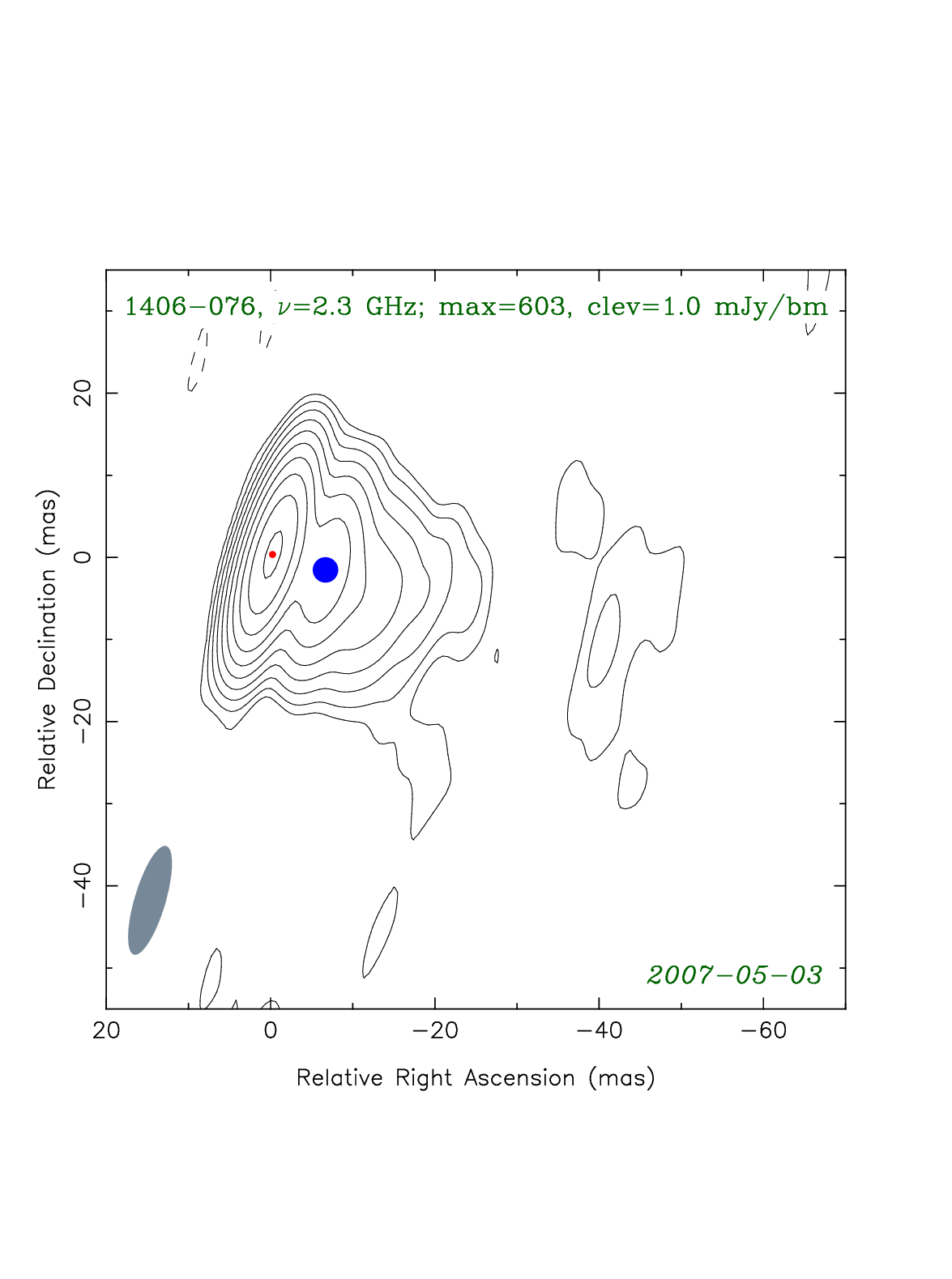}
 \includegraphics[width=0.33\textwidth,angle=0,trim=0.5cm 3cm 1.6cm 5cm,clip]{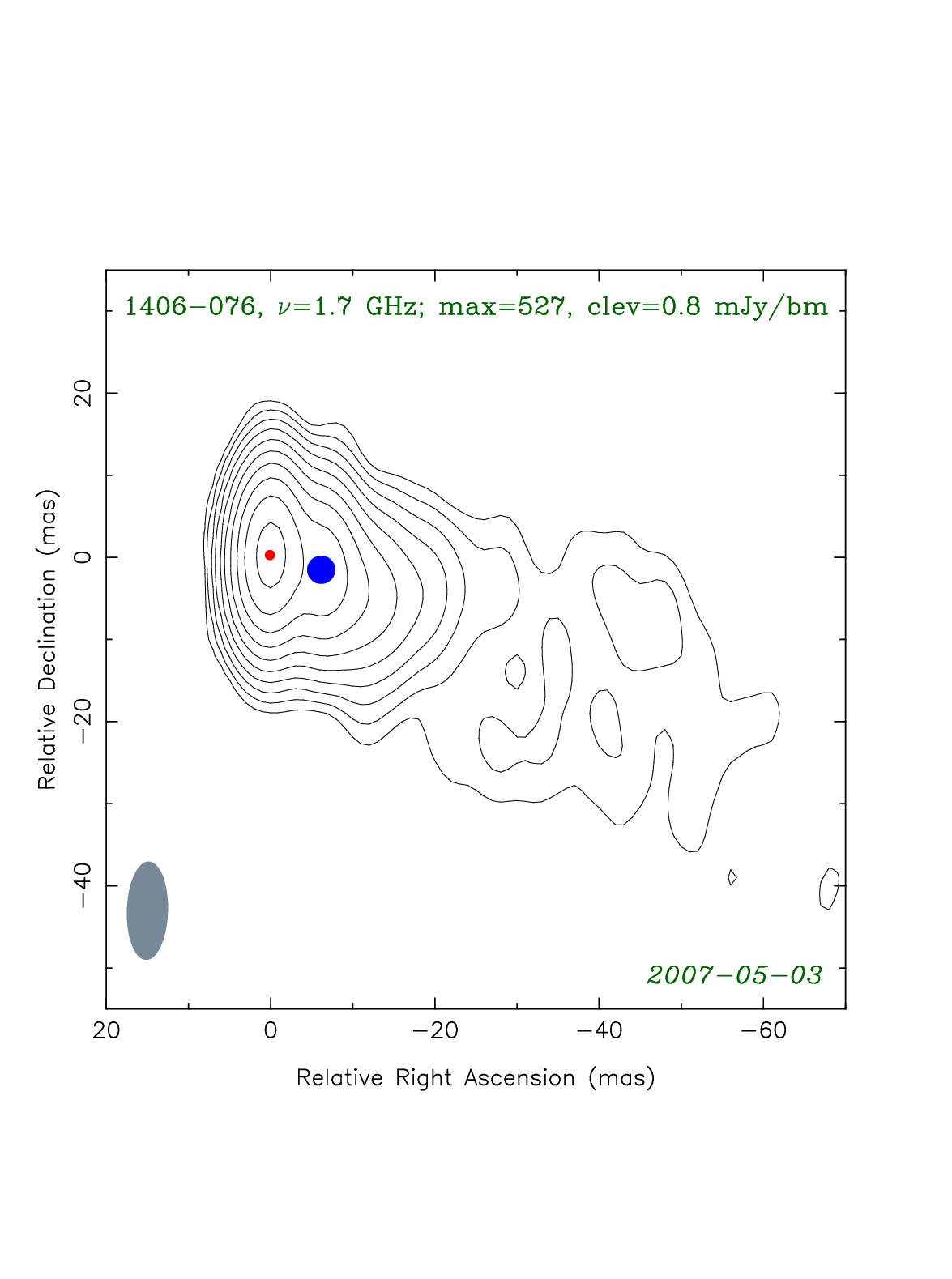}
 \includegraphics[width=0.33\textwidth,angle=0,trim=0.5cm 3cm 1.6cm 5cm,clip]{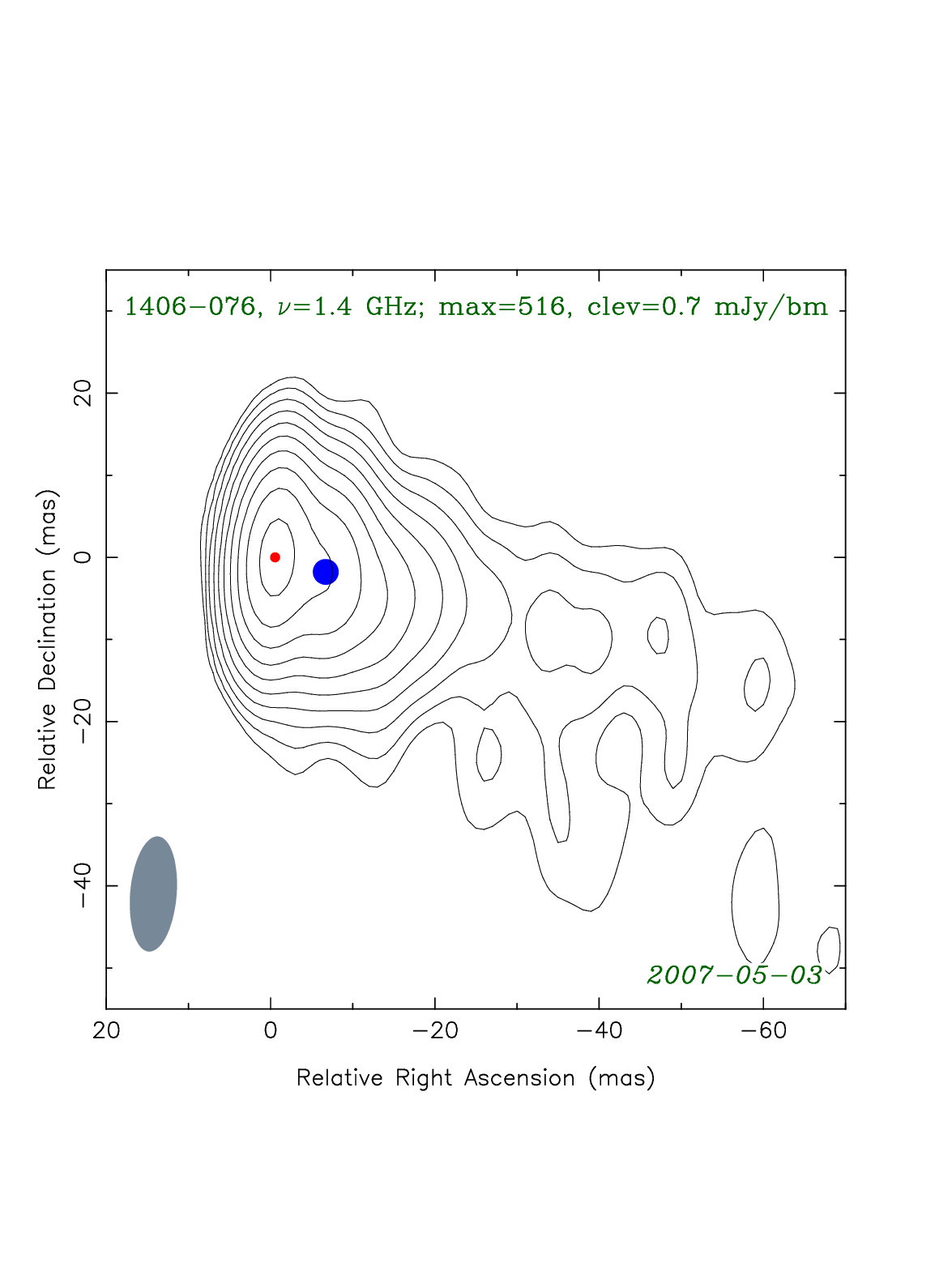}

 \caption{continued...}
\end{figure*}

\clearpage
\addtocounter{figure}{-1}
\begin{figure*}
 \centering

 \includegraphics[width=0.33\textwidth,angle=0,trim=0.5cm 3cm 1.6cm 5cm,clip]{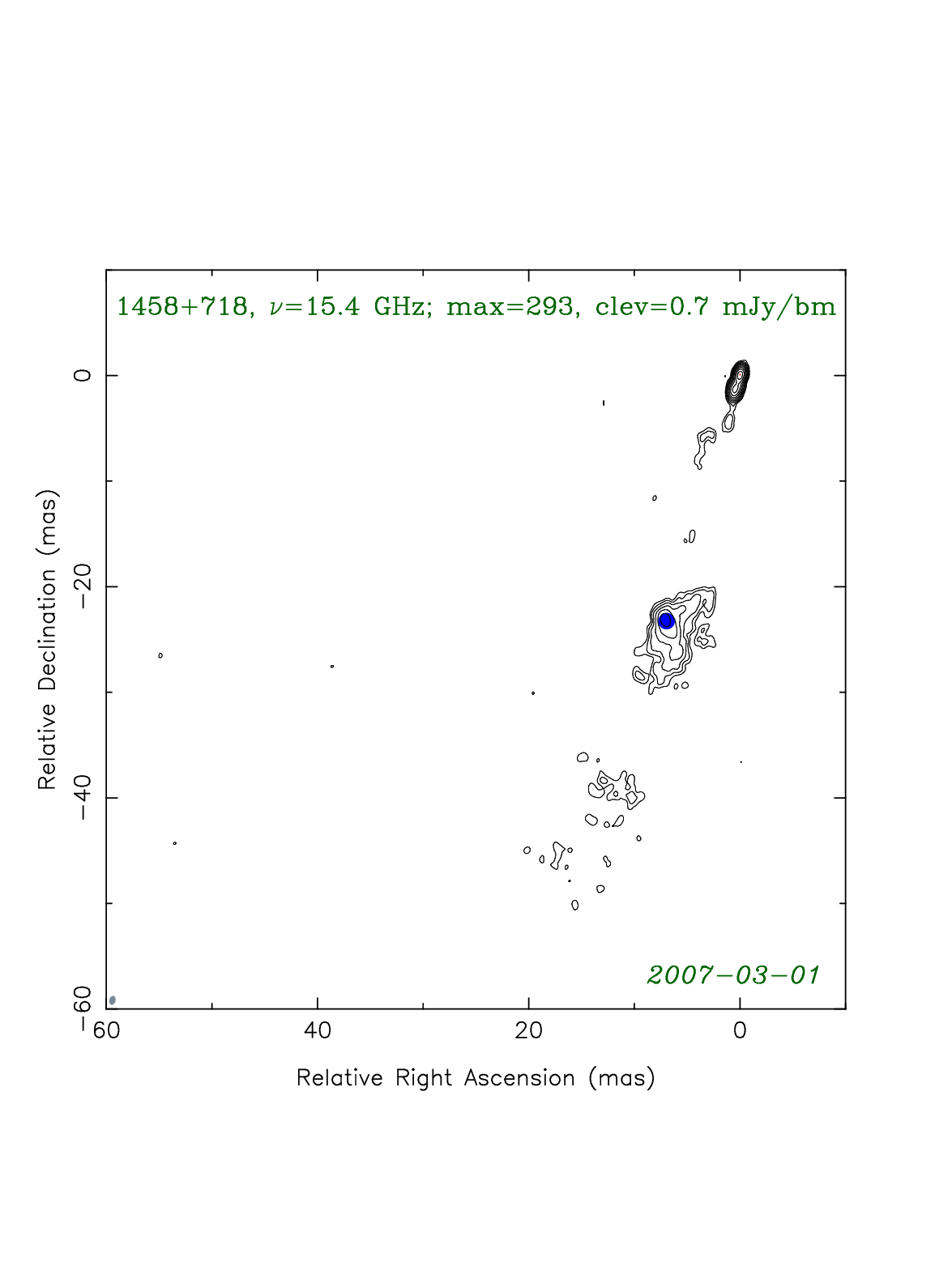}
 \includegraphics[width=0.33\textwidth,angle=0,trim=0.5cm 3cm 1.6cm 5cm,clip]{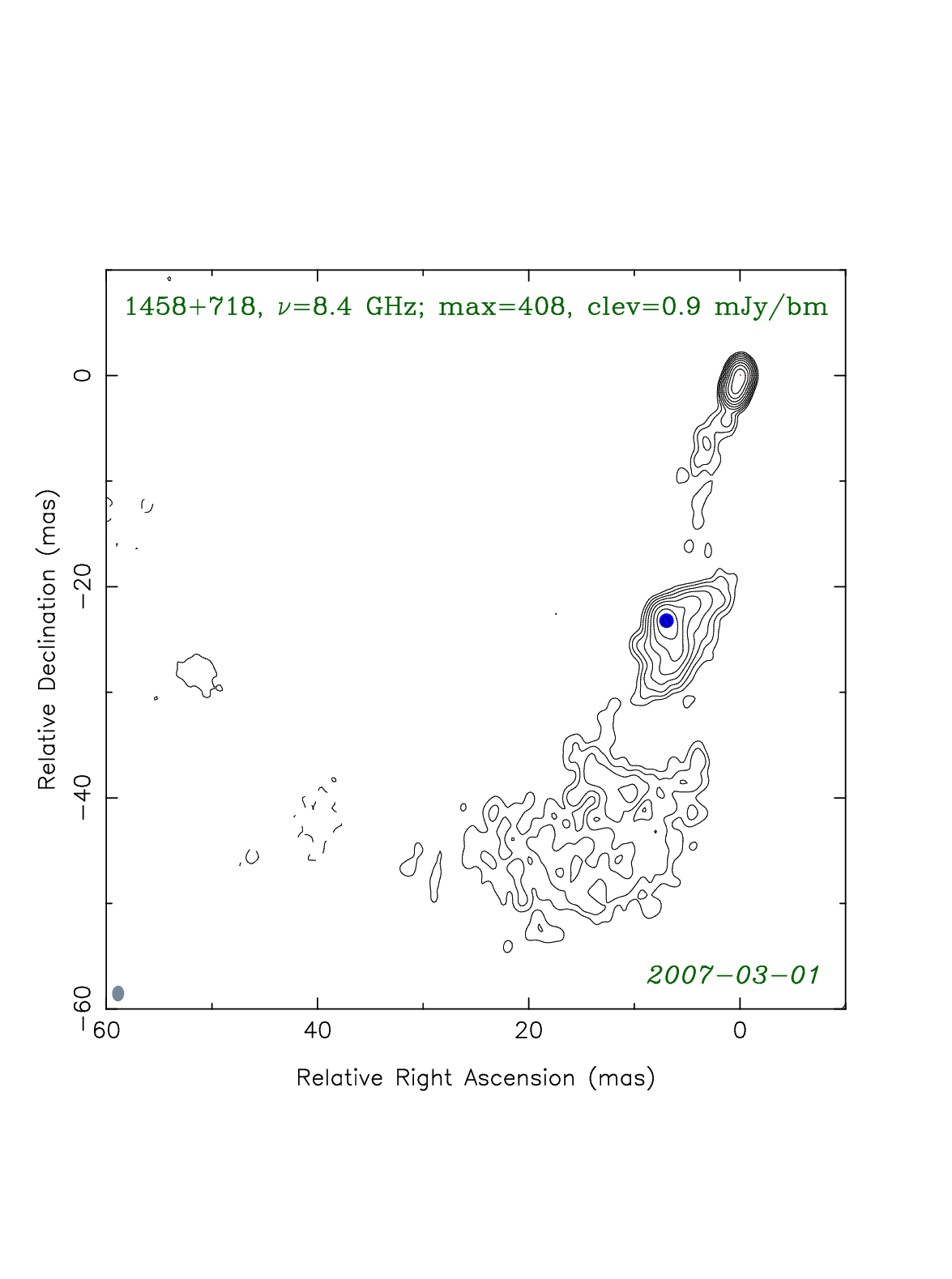}
 \includegraphics[width=0.33\textwidth,angle=0,trim=0.5cm 3cm 1.6cm 5cm,clip]{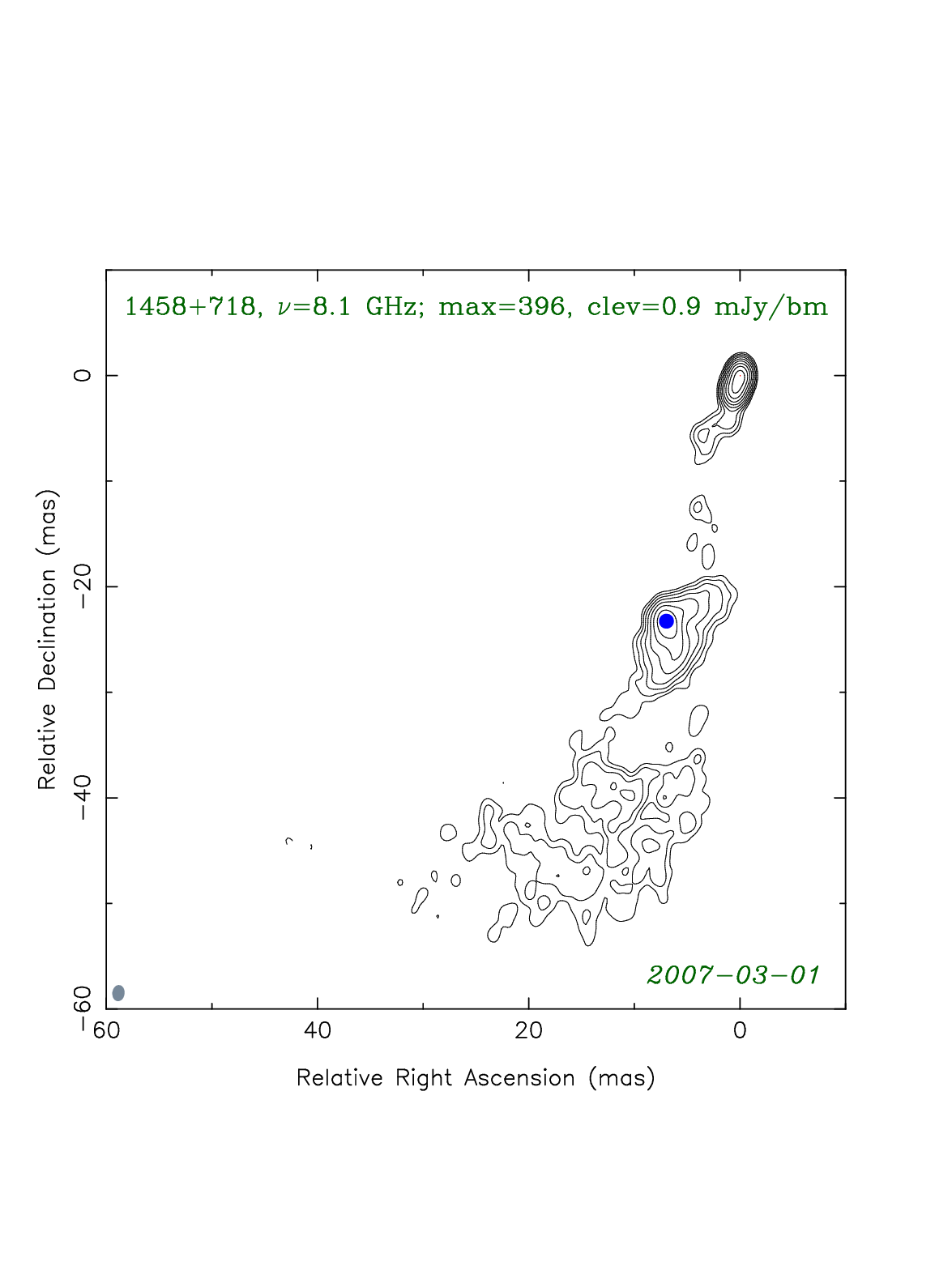}

 \includegraphics[width=0.33\textwidth,angle=0,trim=0.5cm 3cm 1.6cm 5cm,clip]{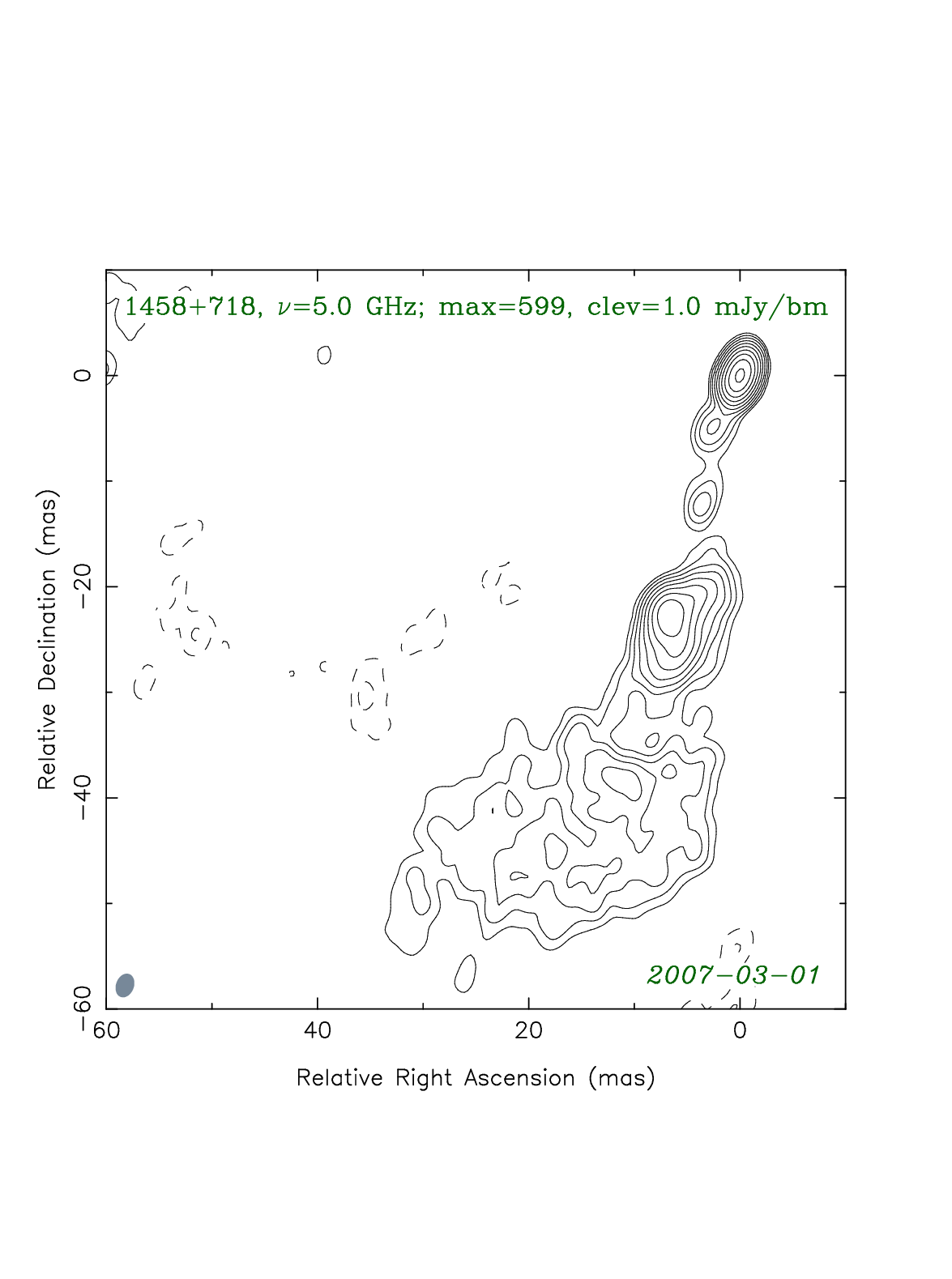}
 \includegraphics[width=0.33\textwidth,angle=0,trim=0.5cm 3cm 1.6cm 5cm,clip]{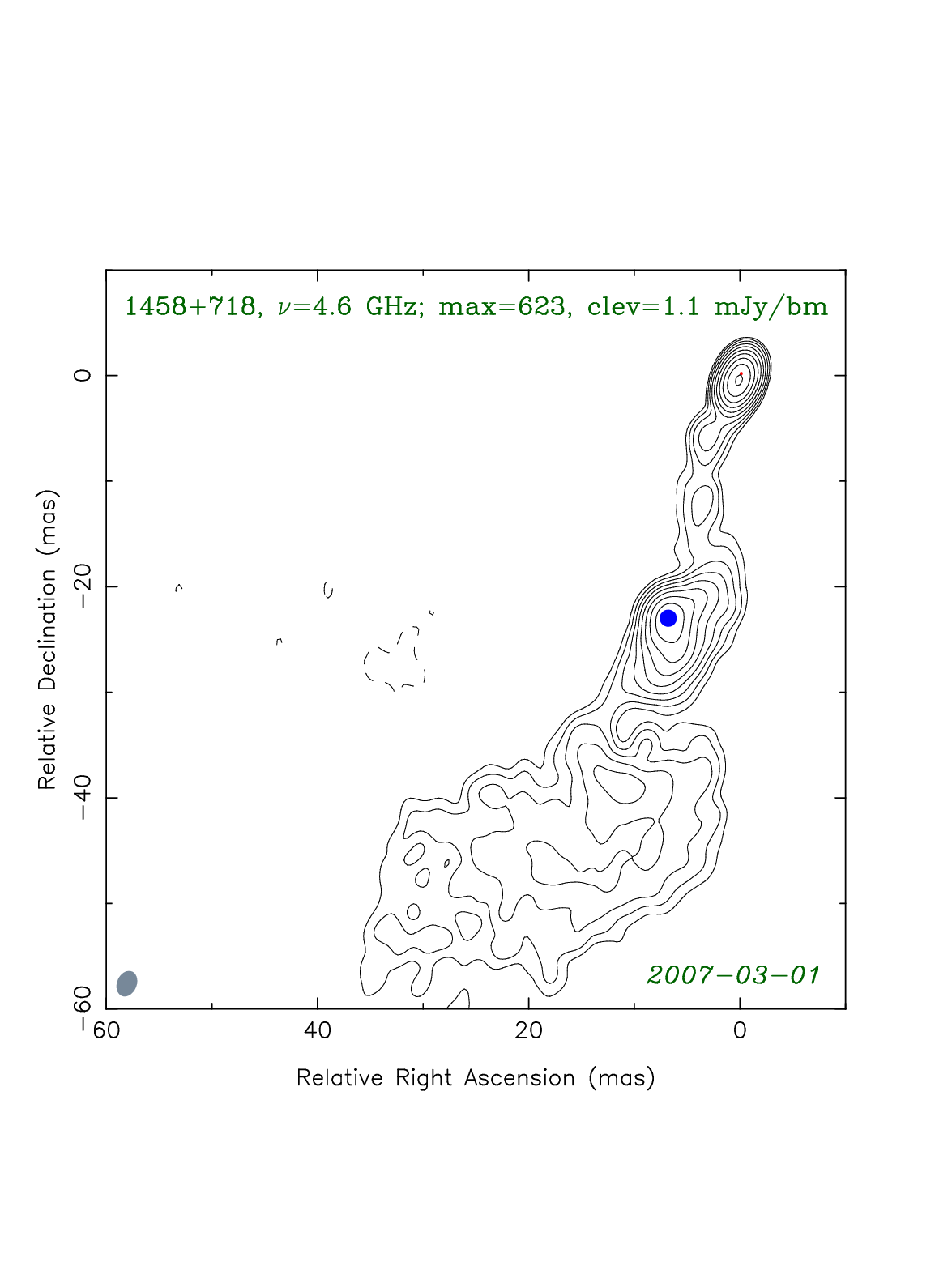}
 \includegraphics[width=0.33\textwidth,angle=0,trim=0.5cm 3cm 1.6cm 5cm,clip]{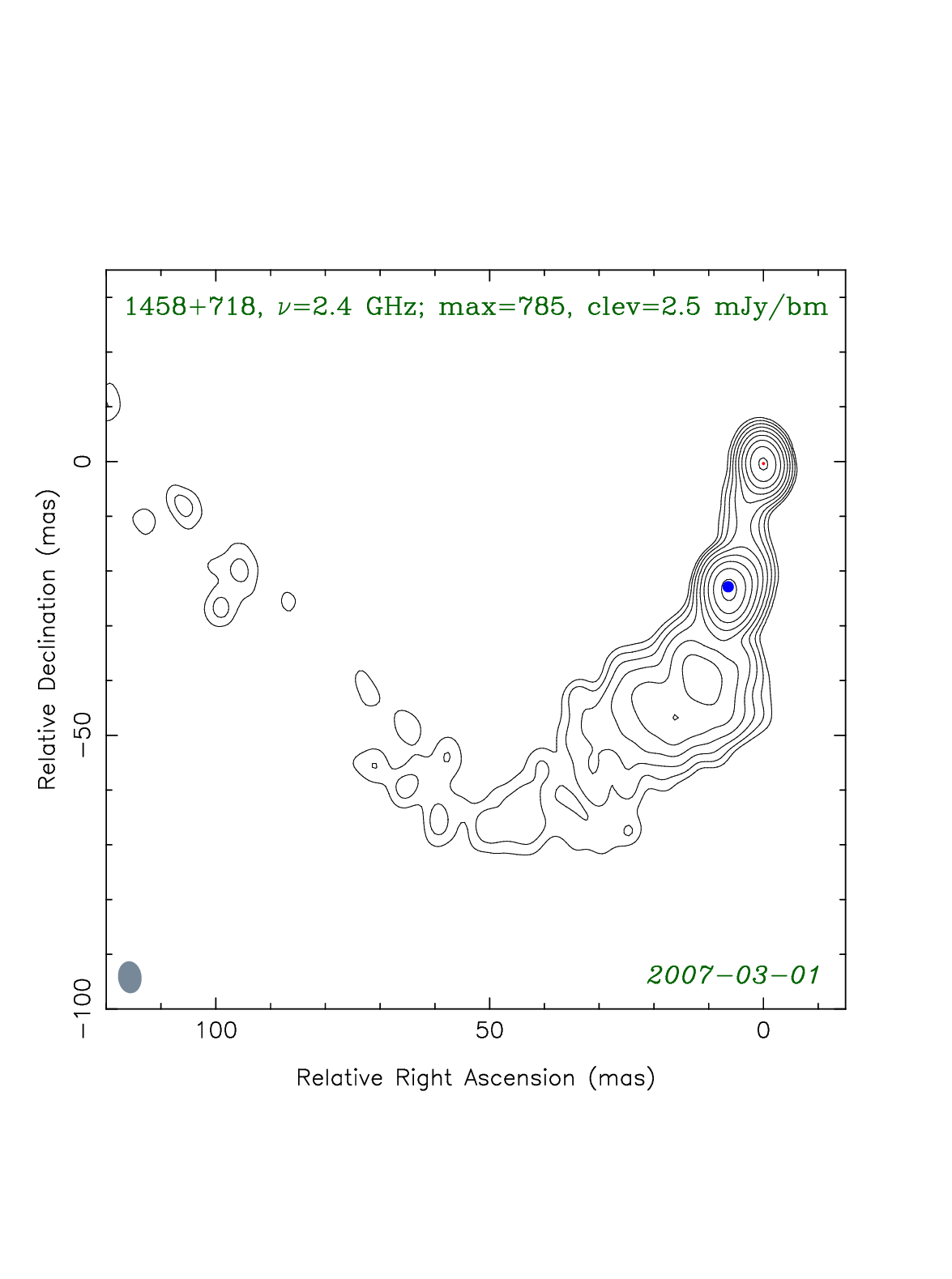}

 \includegraphics[width=0.33\textwidth,angle=0,trim=0.5cm 3cm 1.6cm 5cm,clip]{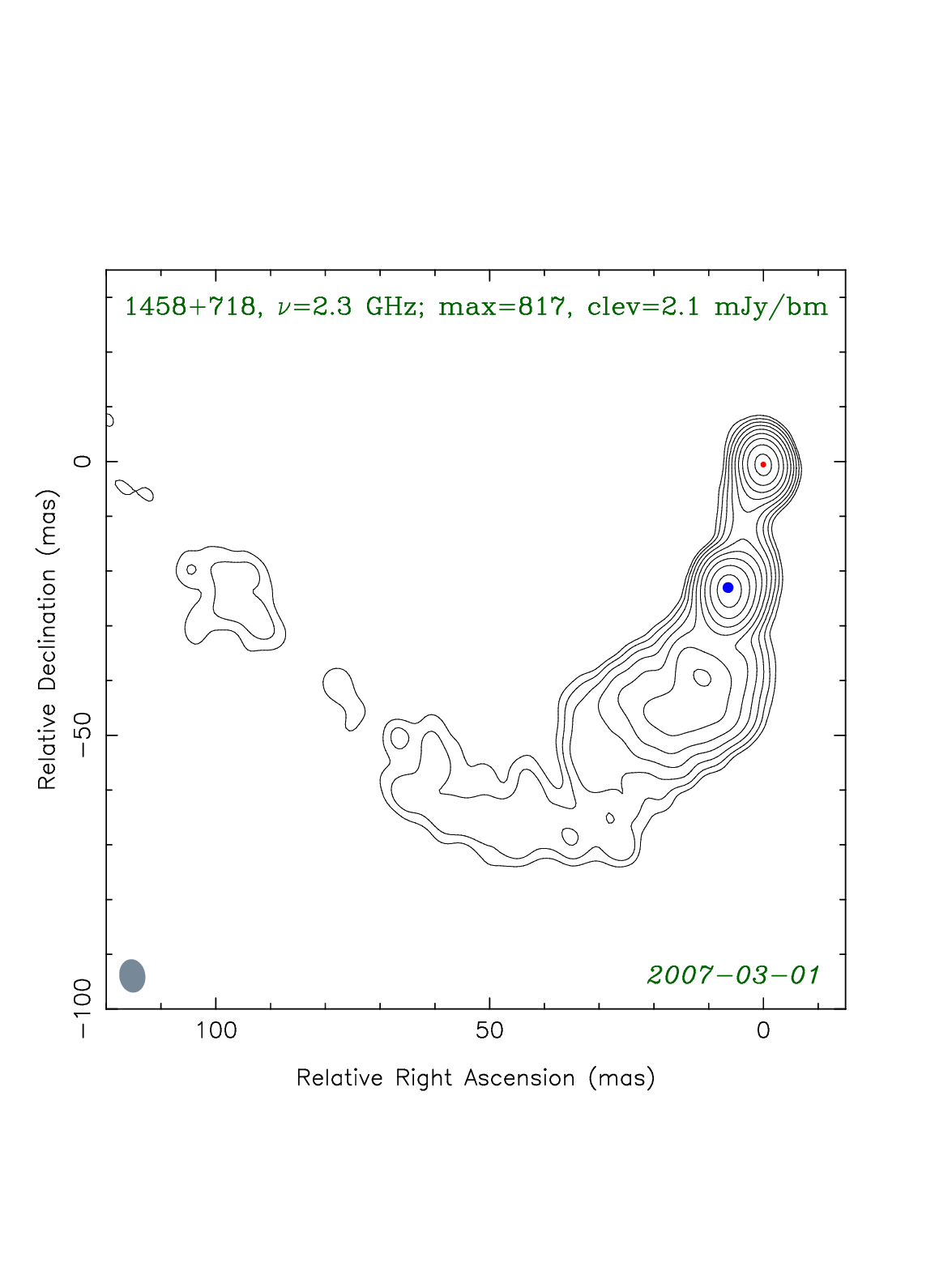}
 \includegraphics[width=0.33\textwidth,angle=0,trim=0.5cm 3cm 1.6cm 5cm,clip]{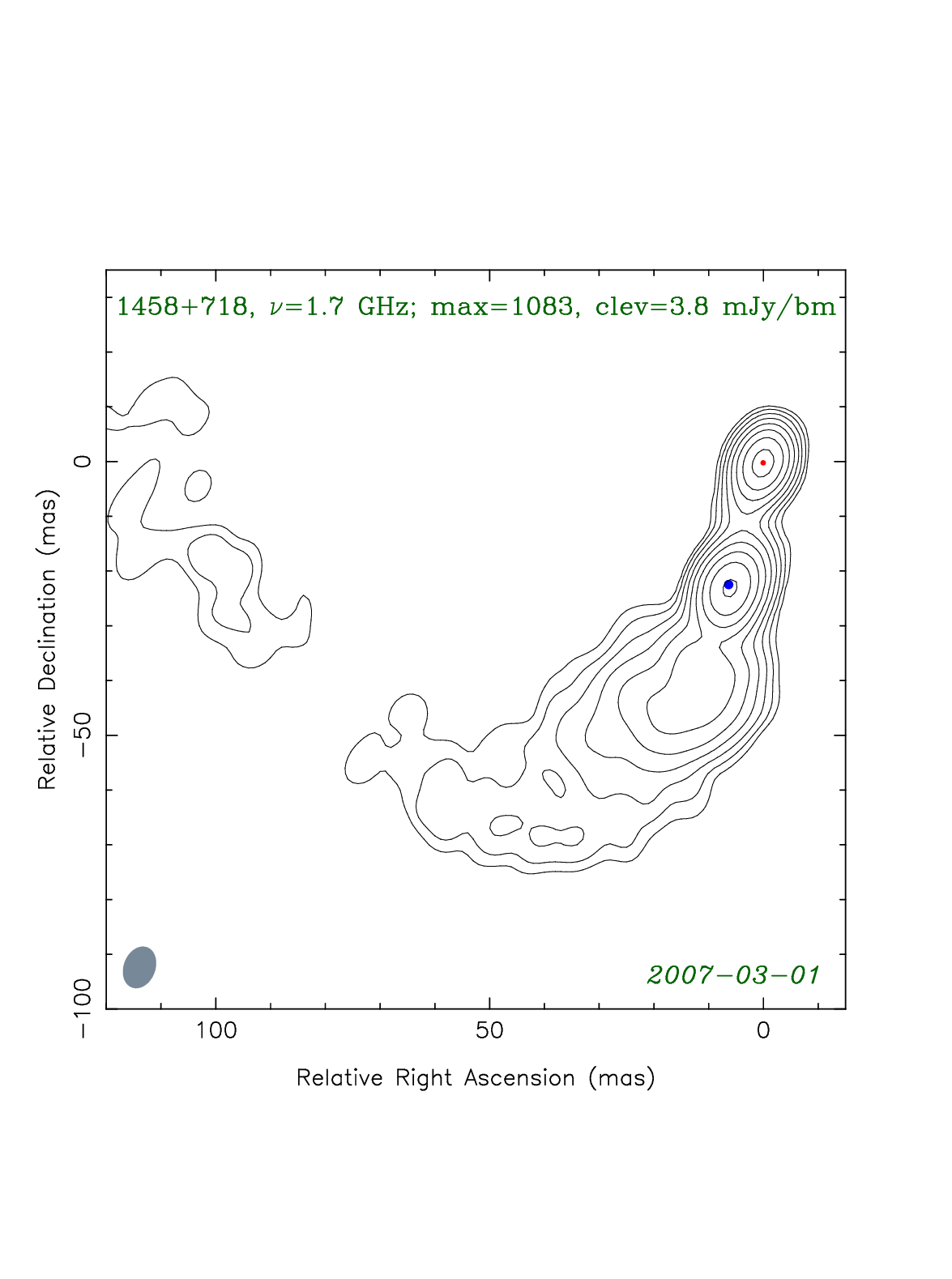}
 \includegraphics[width=0.33\textwidth,angle=0,trim=0.5cm 3cm 1.6cm 5cm,clip]{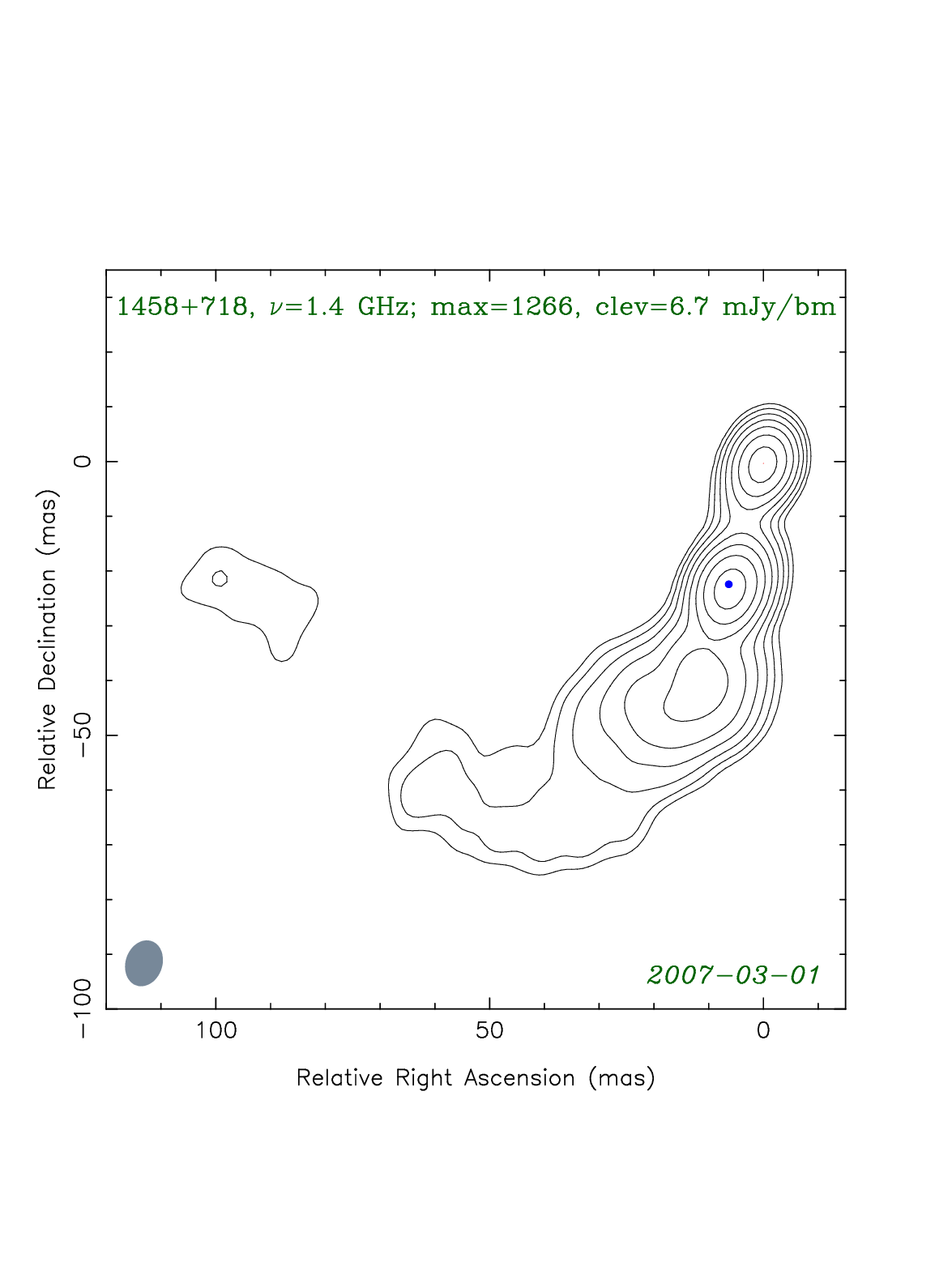}

 \caption{continued...}
\end{figure*}

\clearpage
\addtocounter{figure}{-1}
\begin{figure*}
 \centering

 \includegraphics[width=0.33\textwidth,angle=0,trim=0.5cm 3cm 1.6cm 5cm,clip]{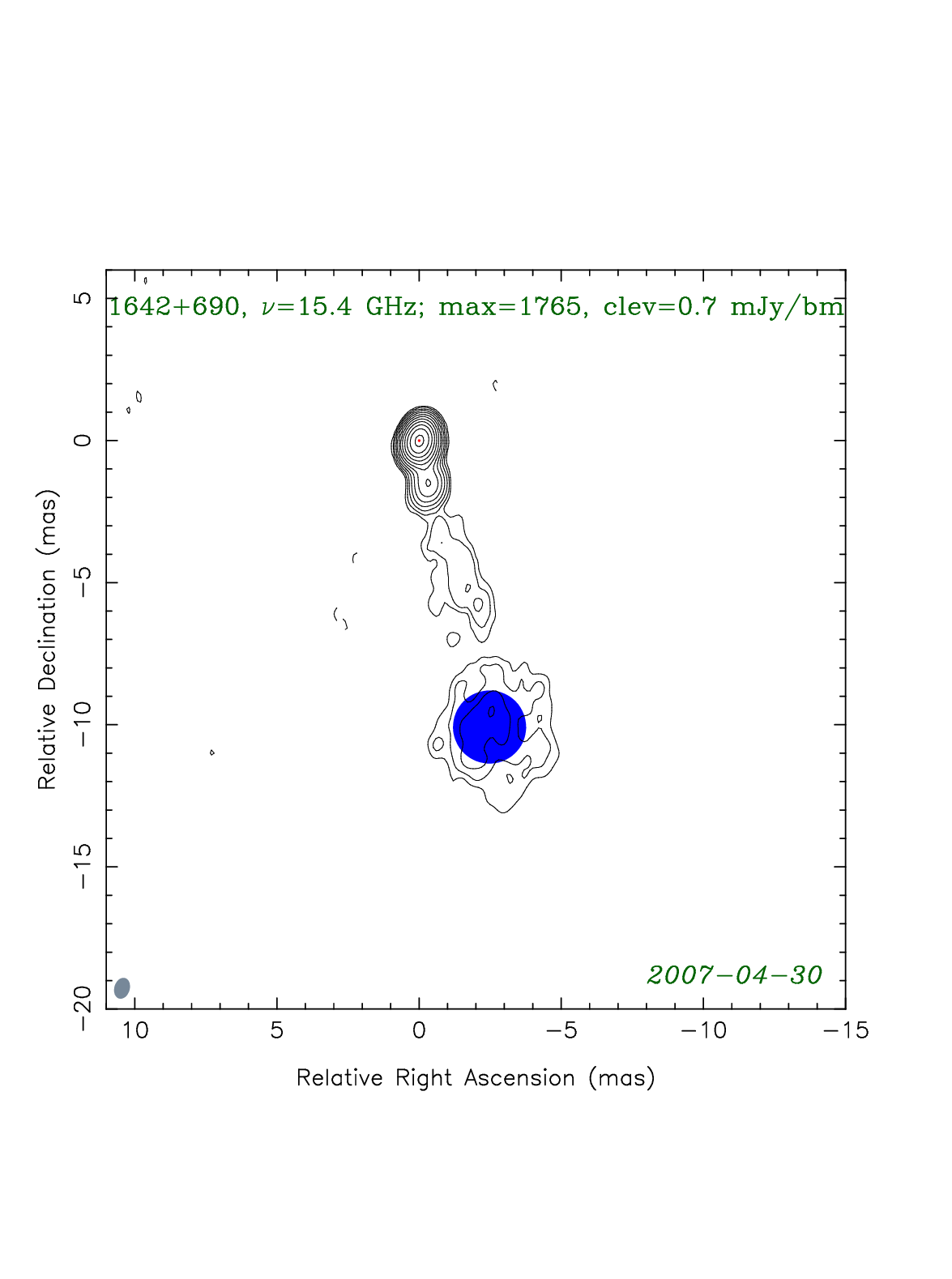}
 \includegraphics[width=0.33\textwidth,angle=0,trim=0.5cm 3cm 1.6cm 5cm,clip]{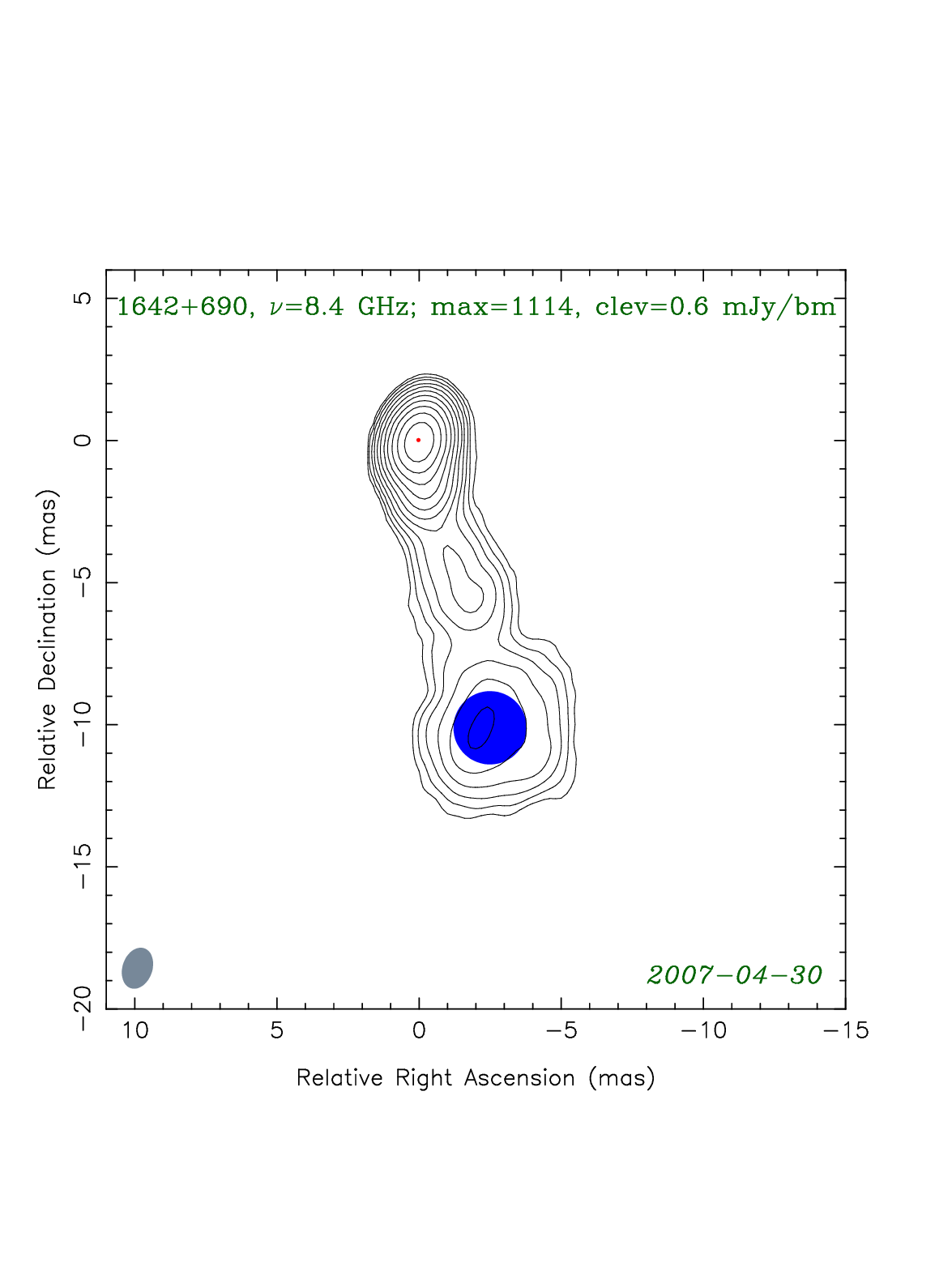}
 \includegraphics[width=0.33\textwidth,angle=0,trim=0.5cm 3cm 1.6cm 5cm,clip]{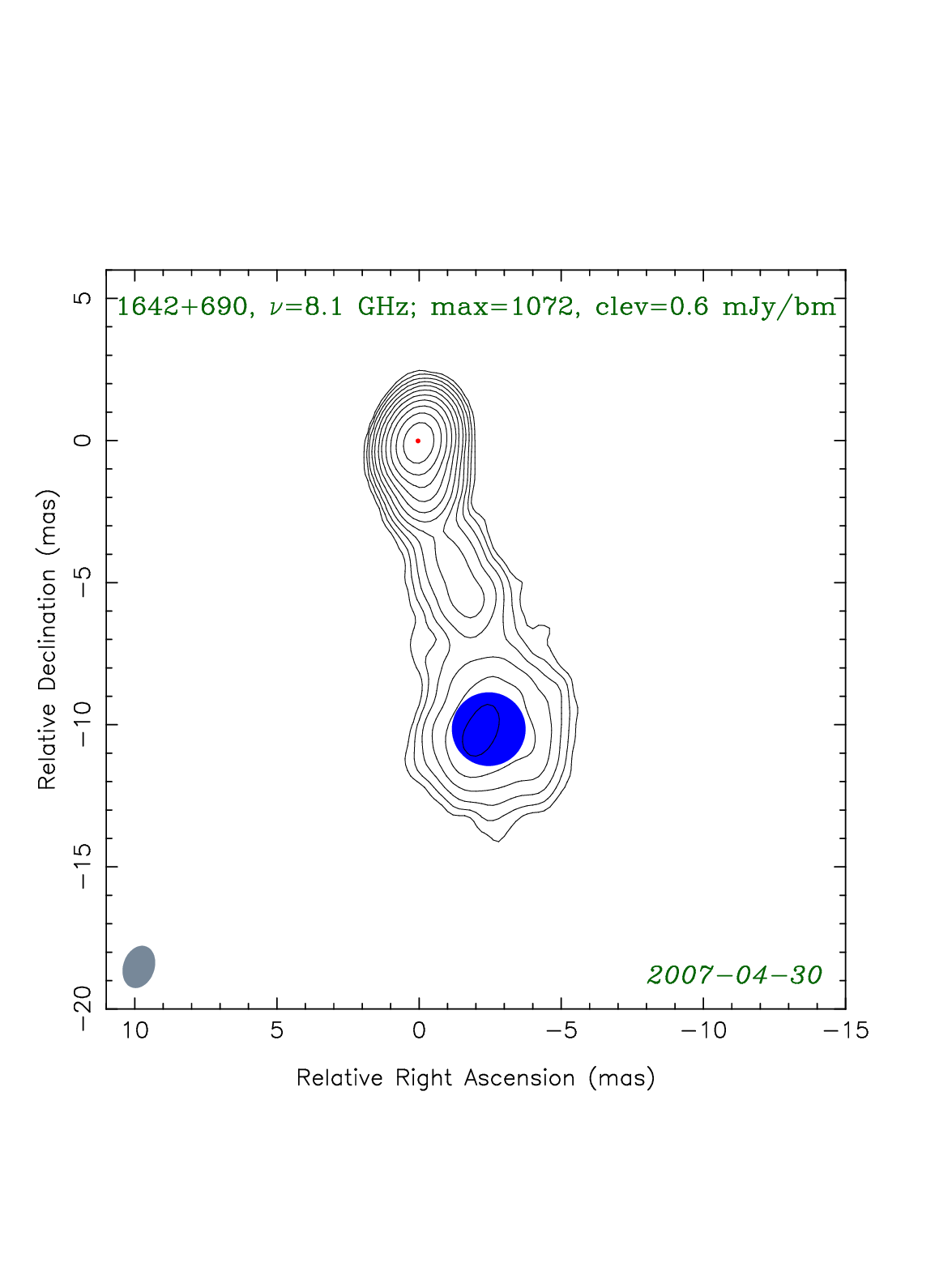}

 \includegraphics[width=0.33\textwidth,angle=0,trim=0.5cm 3cm 1.6cm 5cm,clip]{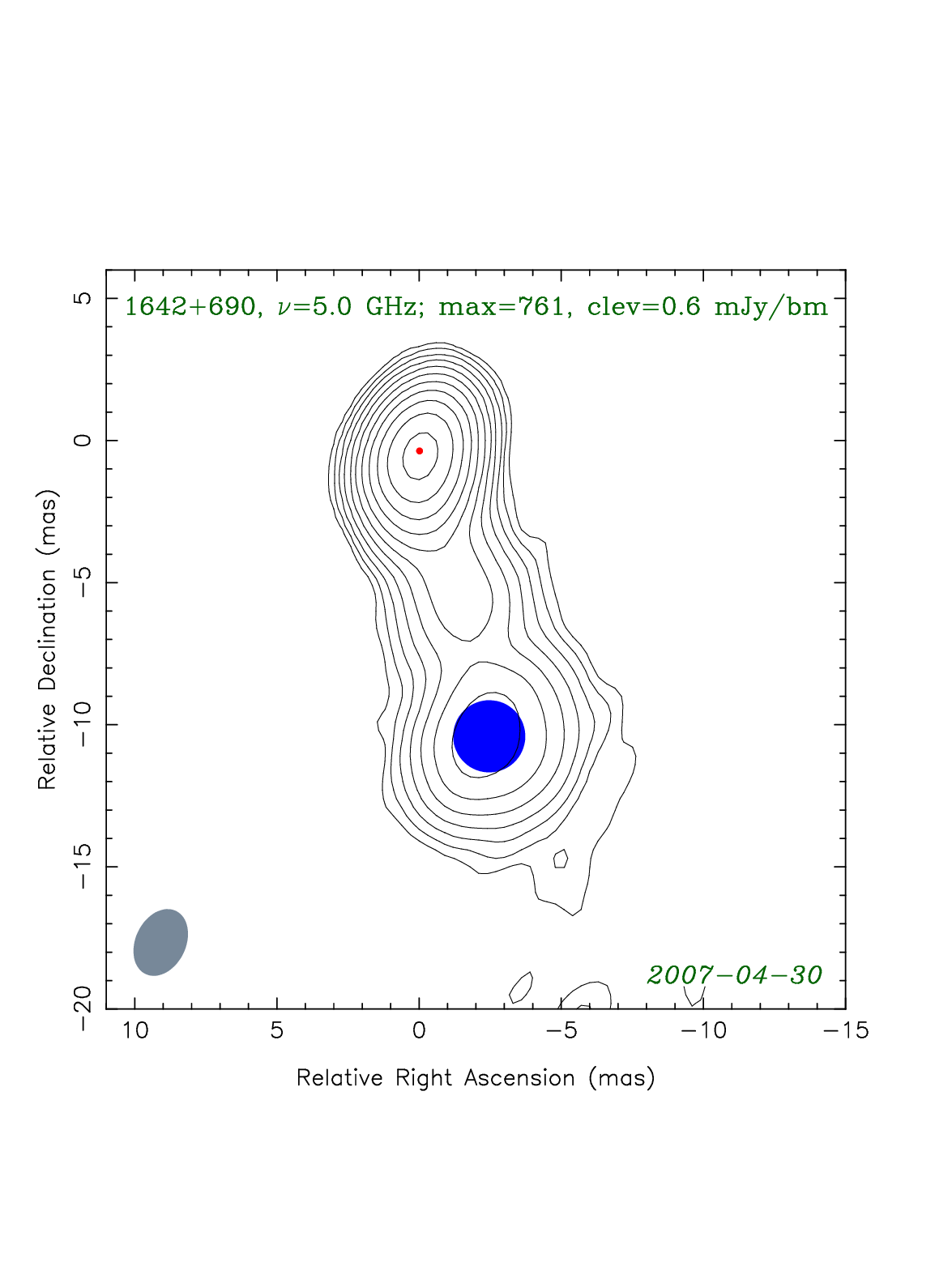}
 \includegraphics[width=0.33\textwidth,angle=0,trim=0.5cm 3cm 1.6cm 5cm,clip]{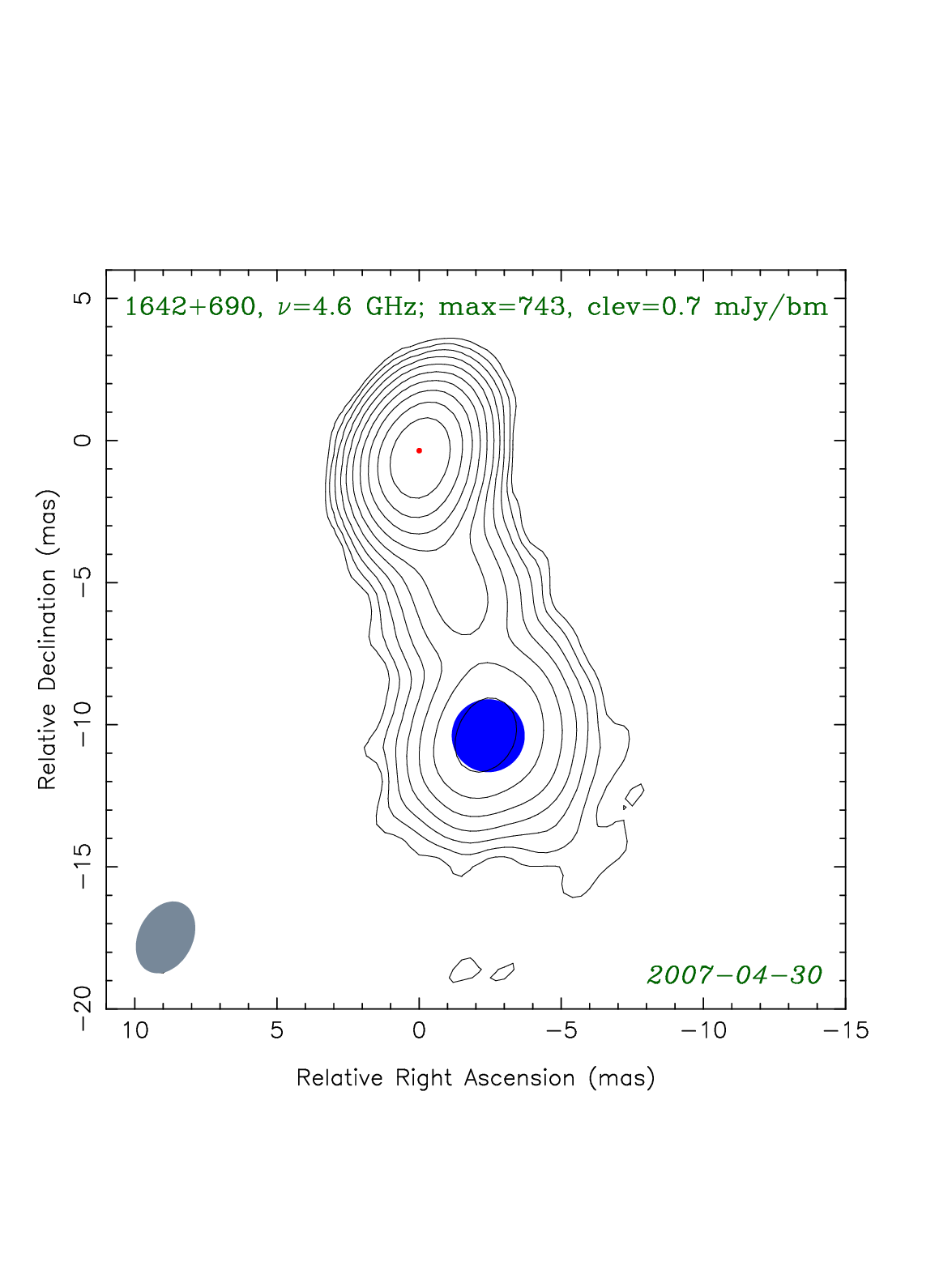}
 \includegraphics[width=0.33\textwidth,angle=0,trim=0.5cm 3cm 1.6cm 5cm,clip]{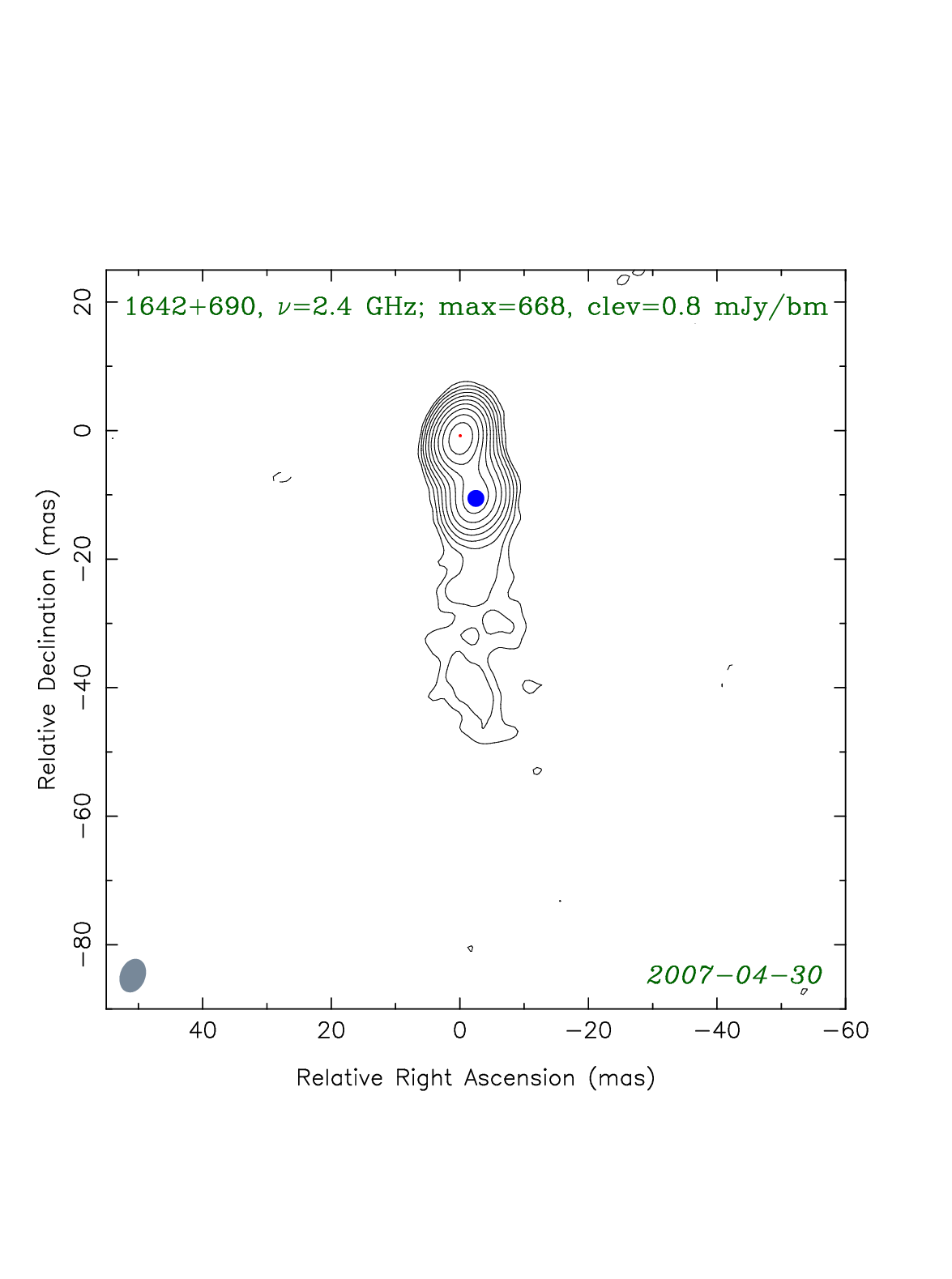}

 \includegraphics[width=0.33\textwidth,angle=0,trim=0.5cm 3cm 1.6cm 5cm,clip]{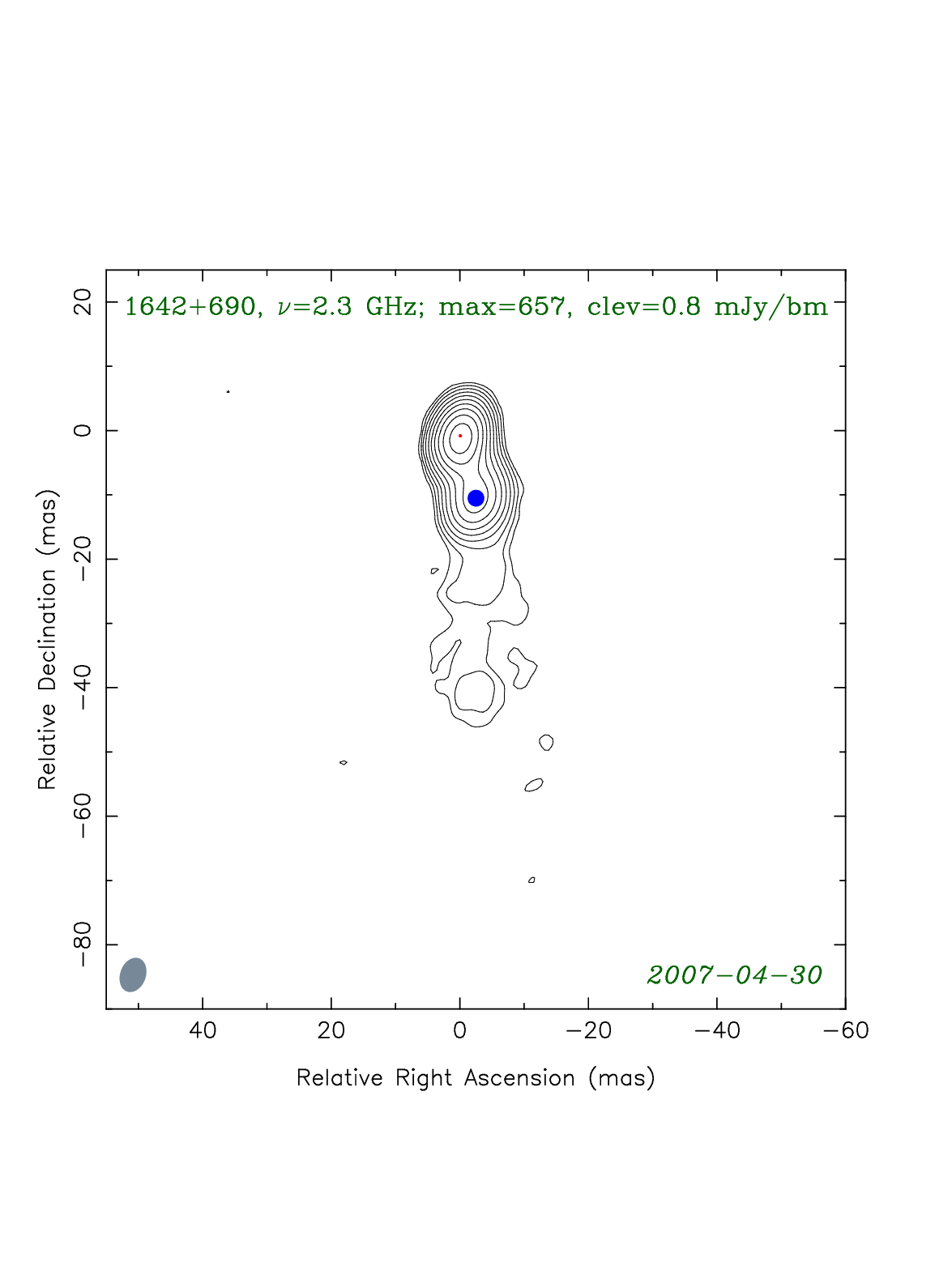}
 \includegraphics[width=0.33\textwidth,angle=0,trim=0.5cm 3cm 1.6cm 5cm,clip]{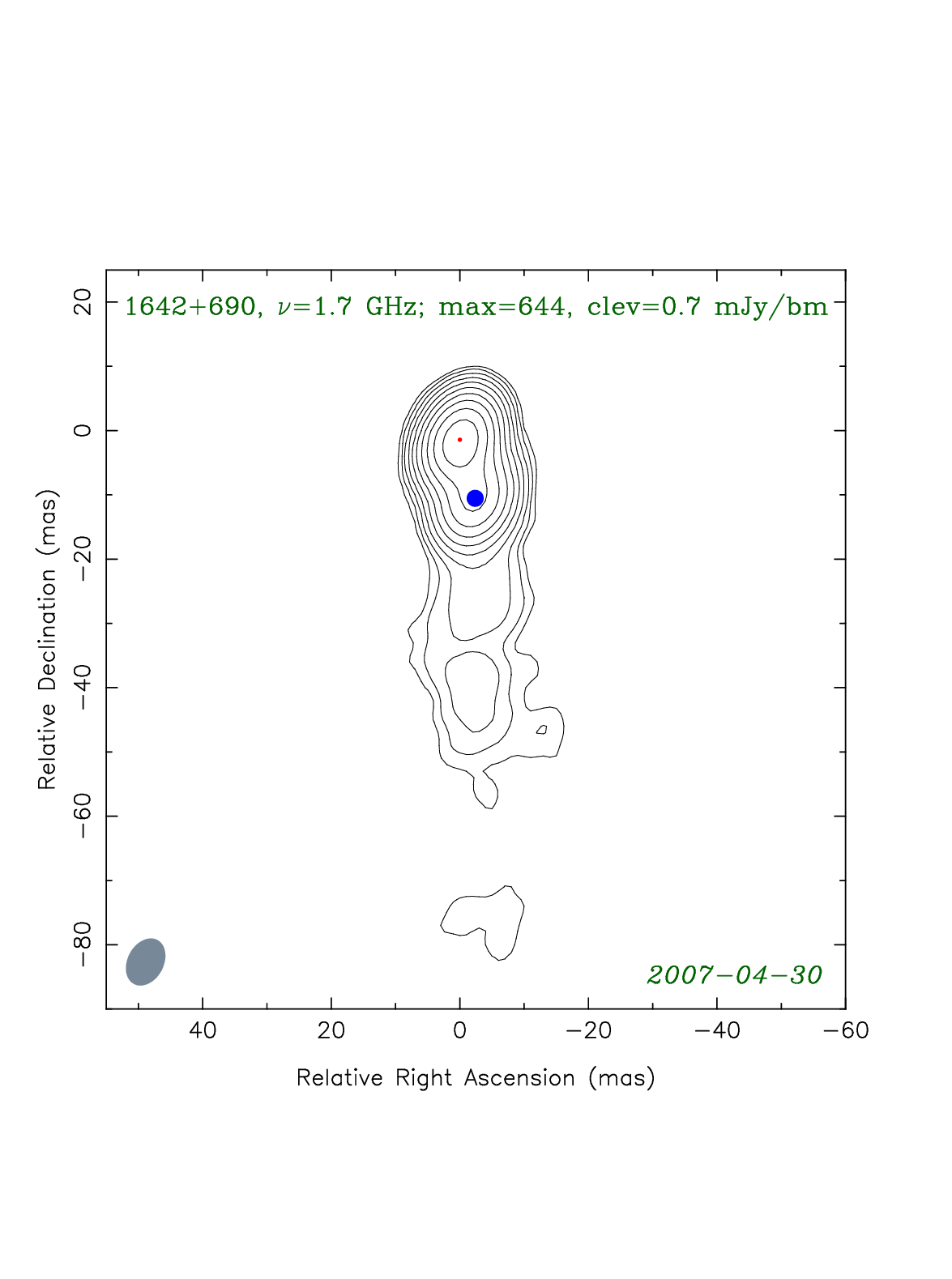}
 \includegraphics[width=0.33\textwidth,angle=0,trim=0.5cm 3cm 1.6cm 5cm,clip]{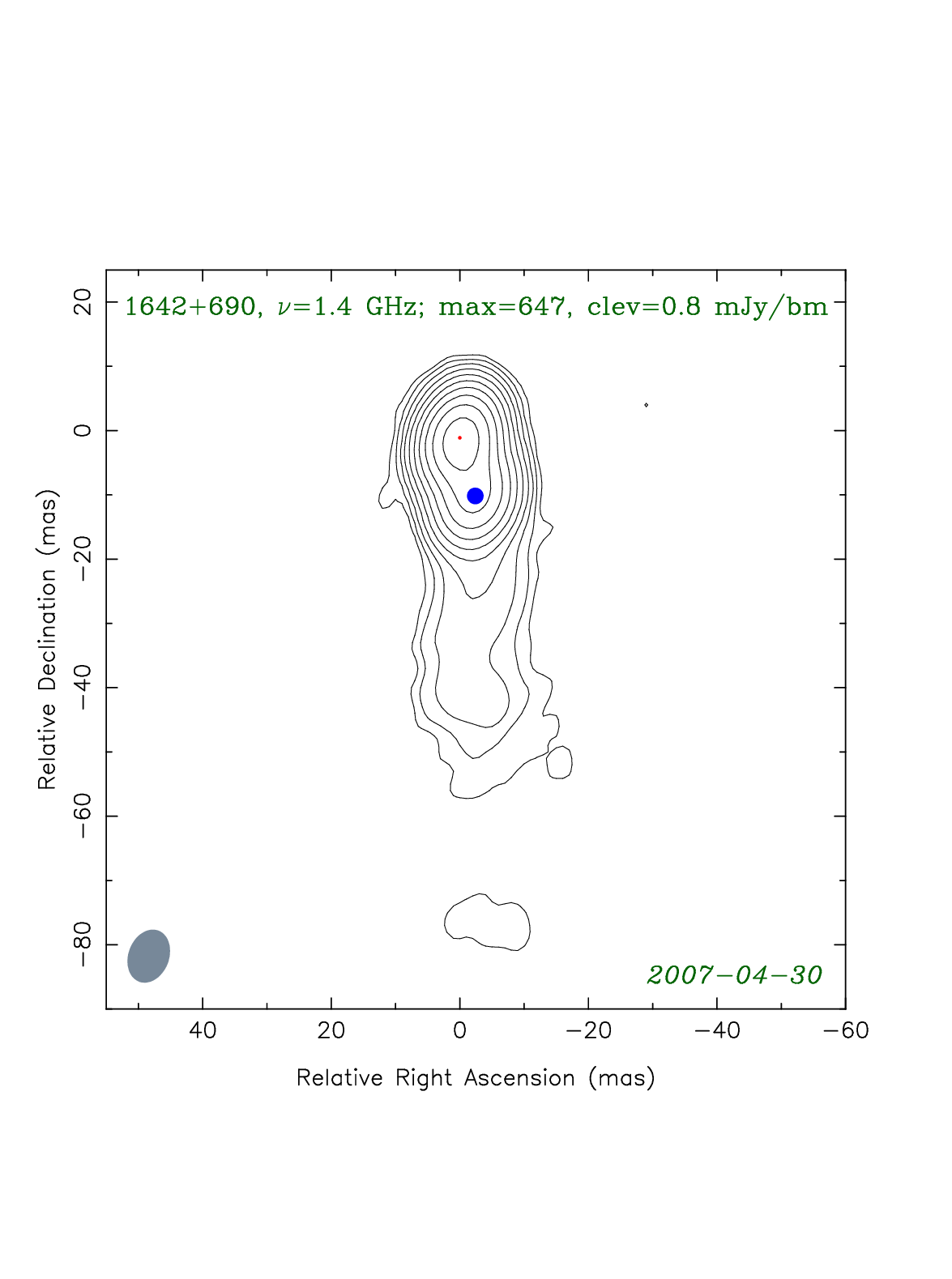}

 \caption{continued...}
\end{figure*}

\clearpage
\addtocounter{figure}{-1}
\begin{figure*}
 \centering

 \includegraphics[width=0.33\textwidth,angle=0,trim=0.5cm 3cm 1.6cm 5cm,clip]{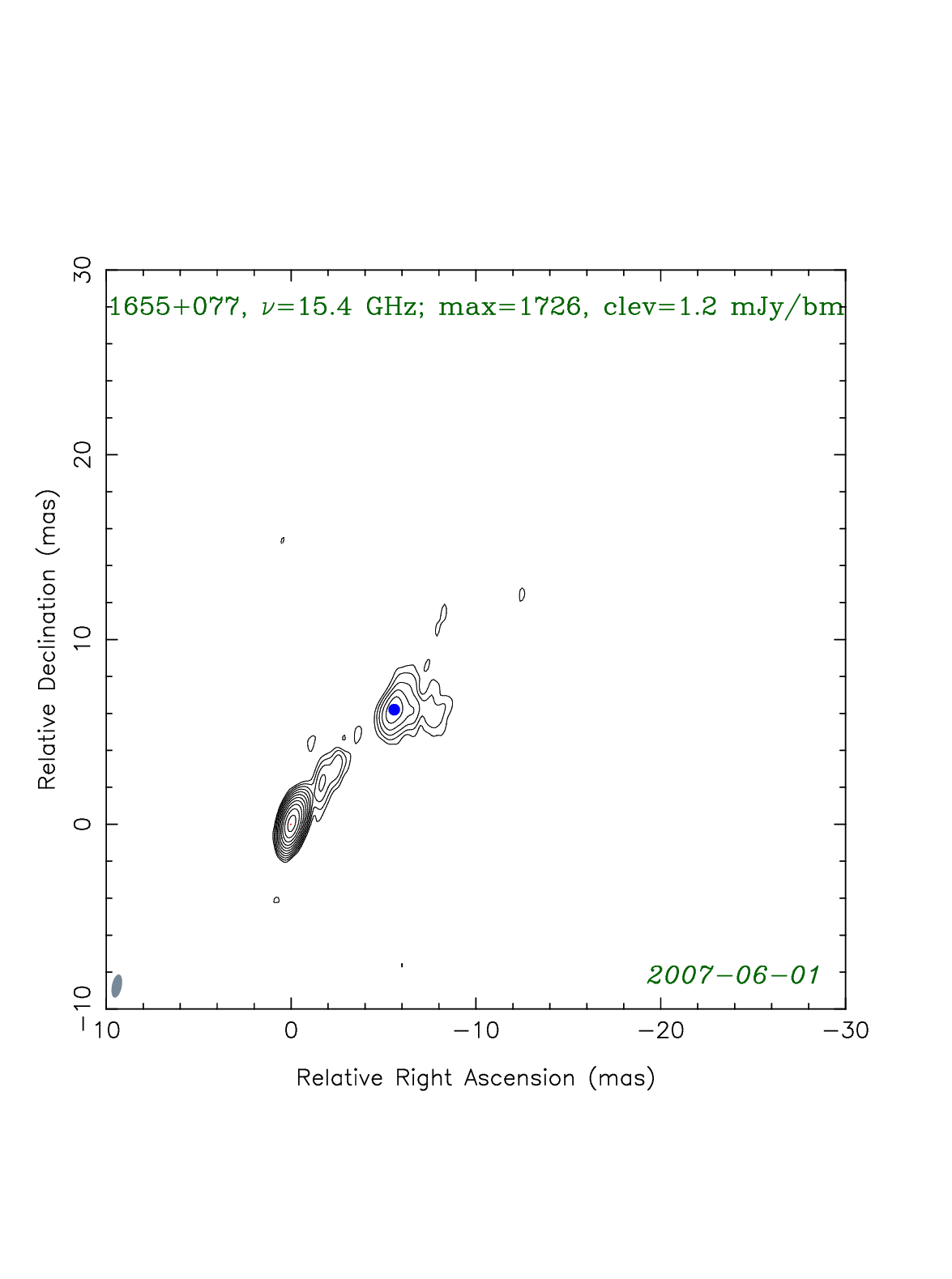}
 \includegraphics[width=0.33\textwidth,angle=0,trim=0.5cm 3cm 1.6cm 5cm,clip]{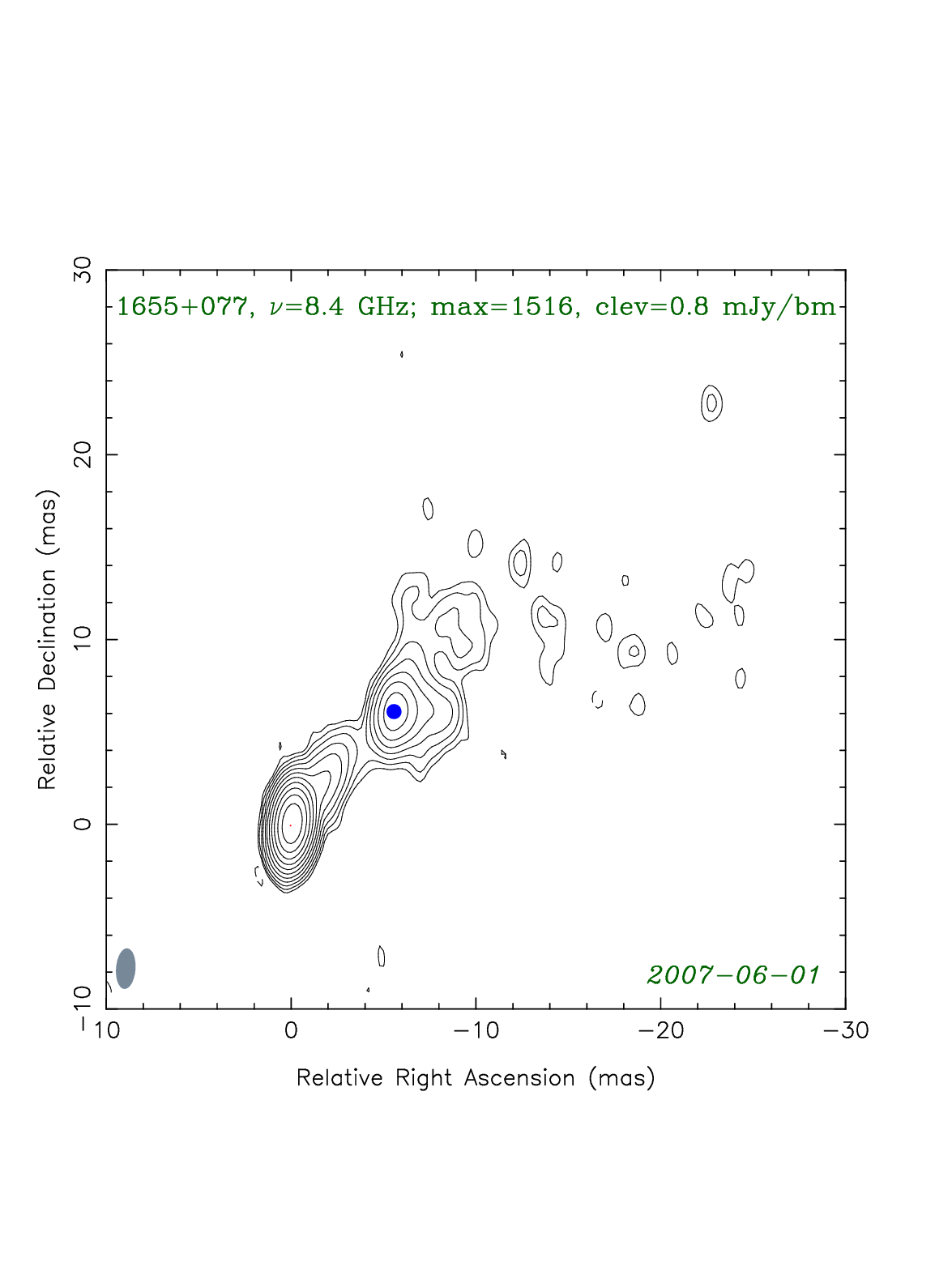}
 \includegraphics[width=0.33\textwidth,angle=0,trim=0.5cm 3cm 1.6cm 5cm,clip]{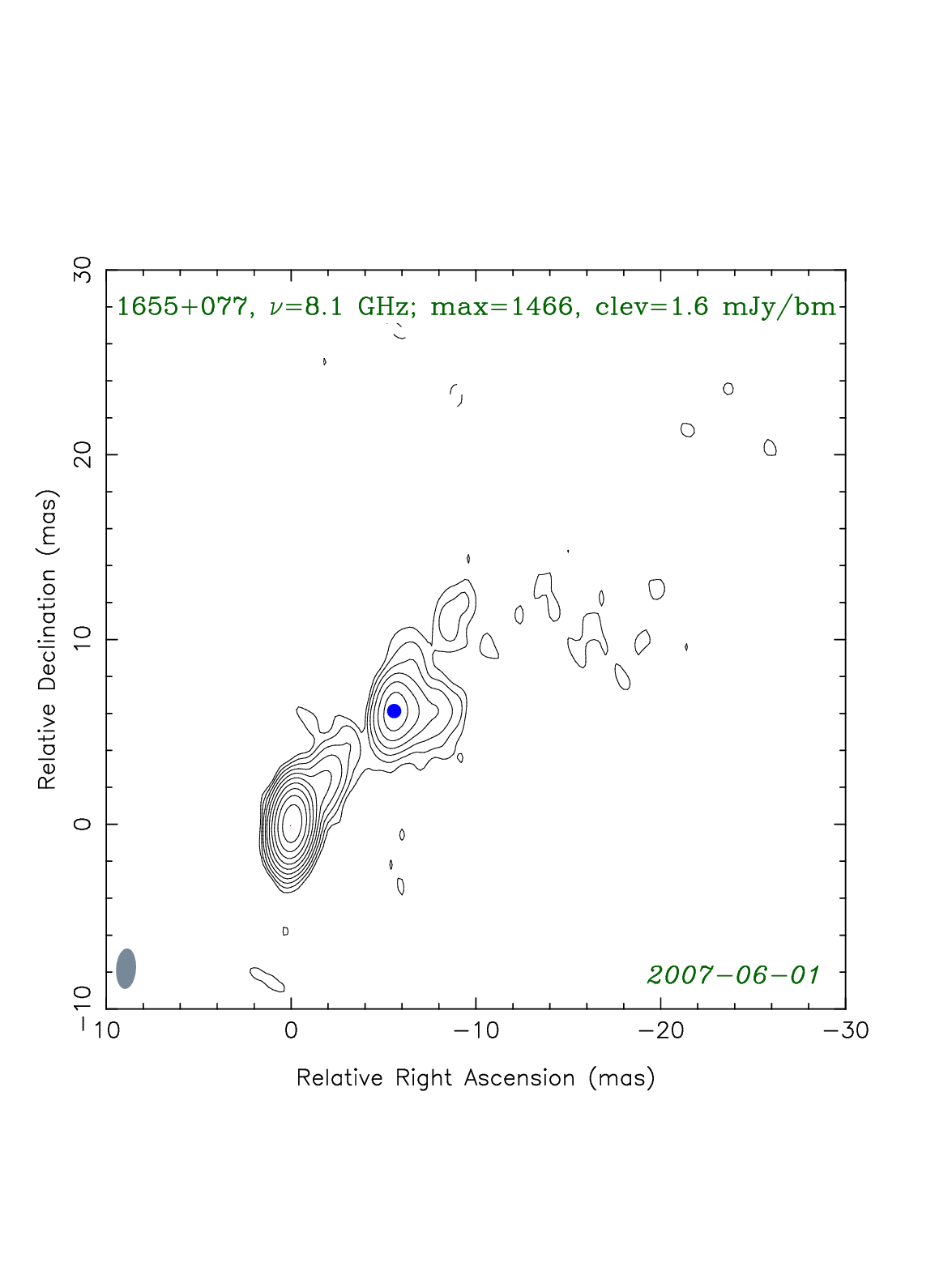}

 \includegraphics[width=0.33\textwidth,angle=0,trim=0.5cm 3cm 1.6cm 5cm,clip]{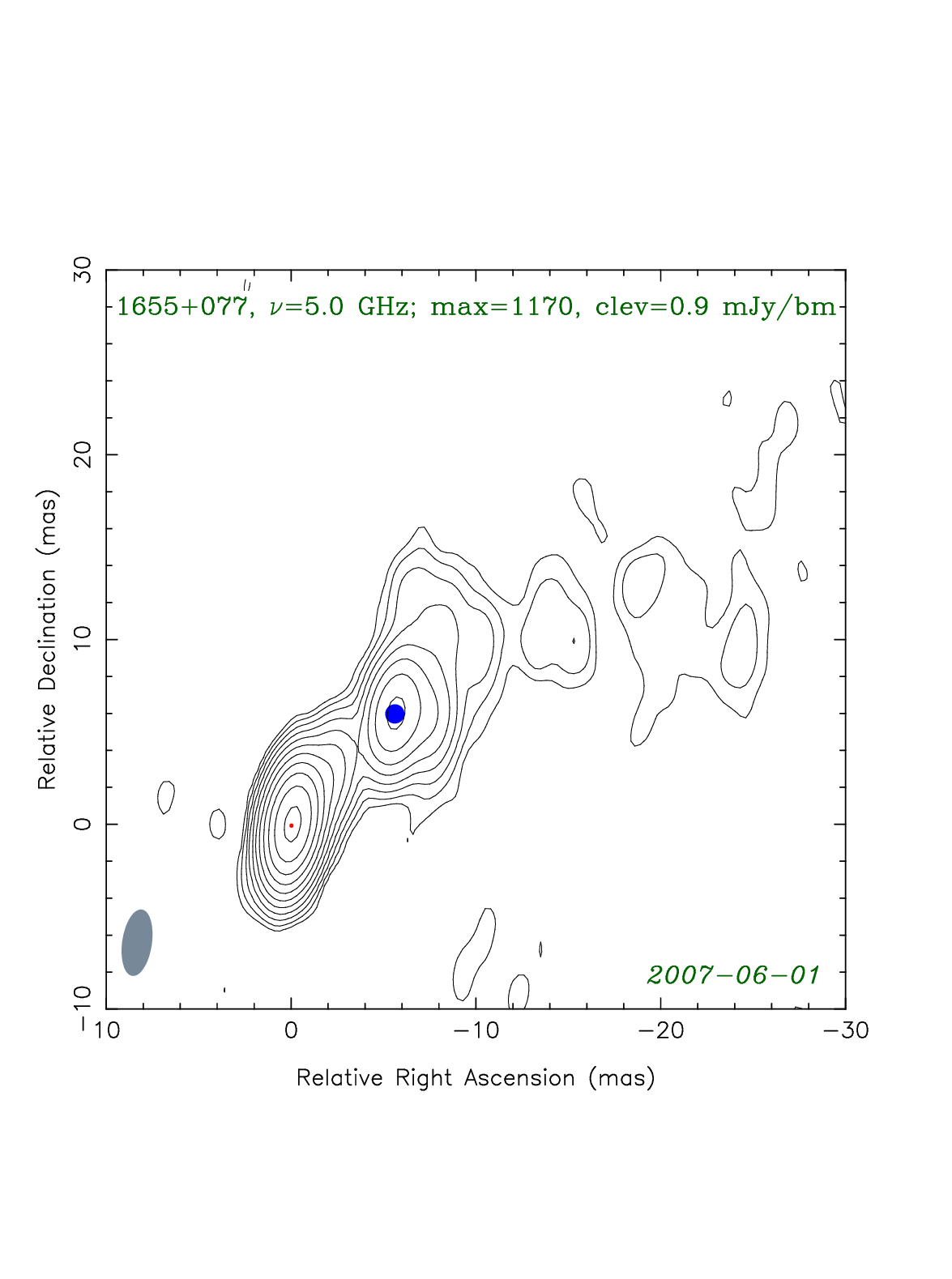}
 \includegraphics[width=0.33\textwidth,angle=0,trim=0.5cm 3cm 1.6cm 5cm,clip]{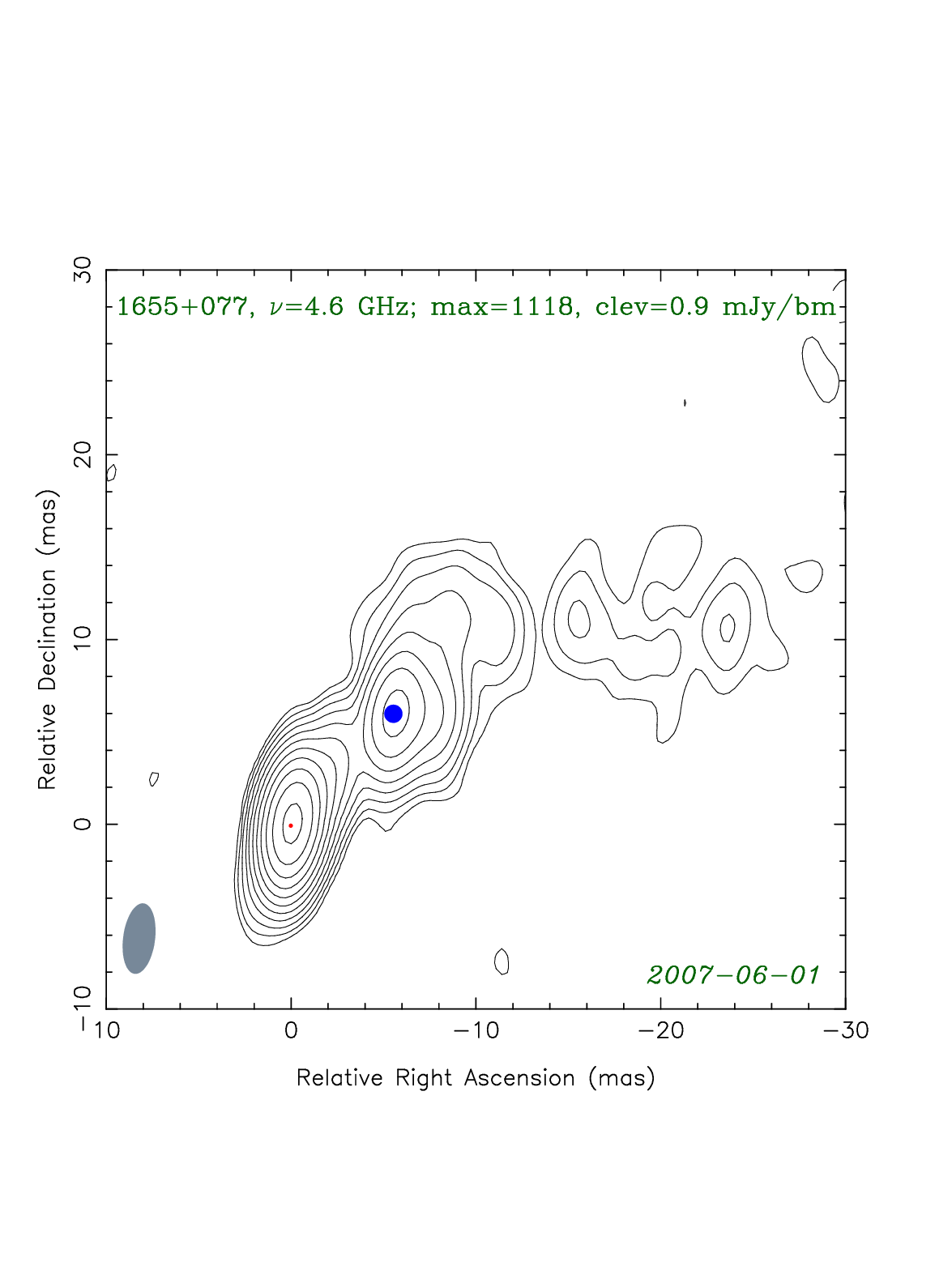}
 \includegraphics[width=0.33\textwidth,angle=0,trim=0.5cm 3cm 1.6cm 5cm,clip]{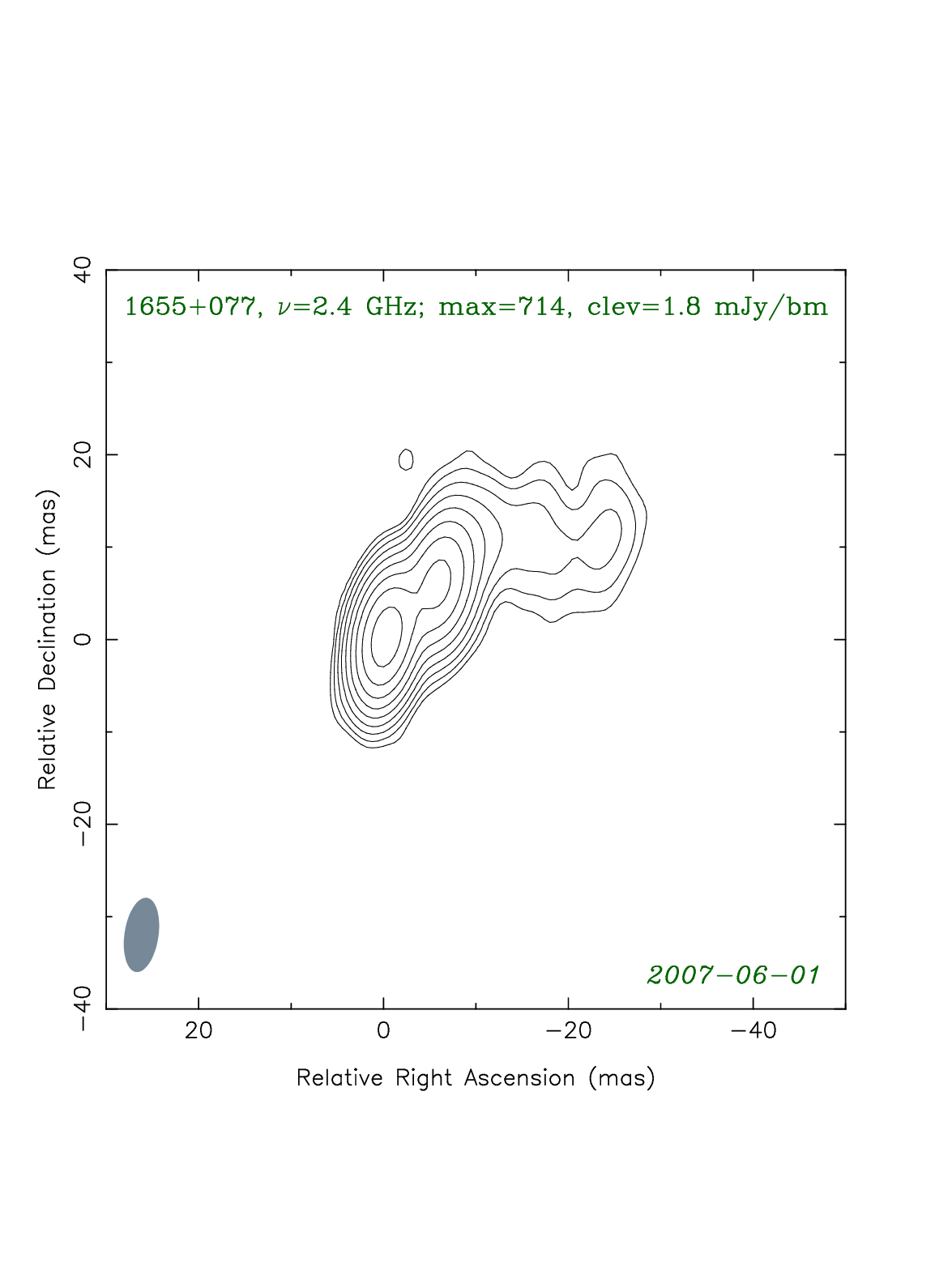}

 \includegraphics[width=0.33\textwidth,angle=0,trim=0.5cm 3cm 1.6cm 5cm,clip]{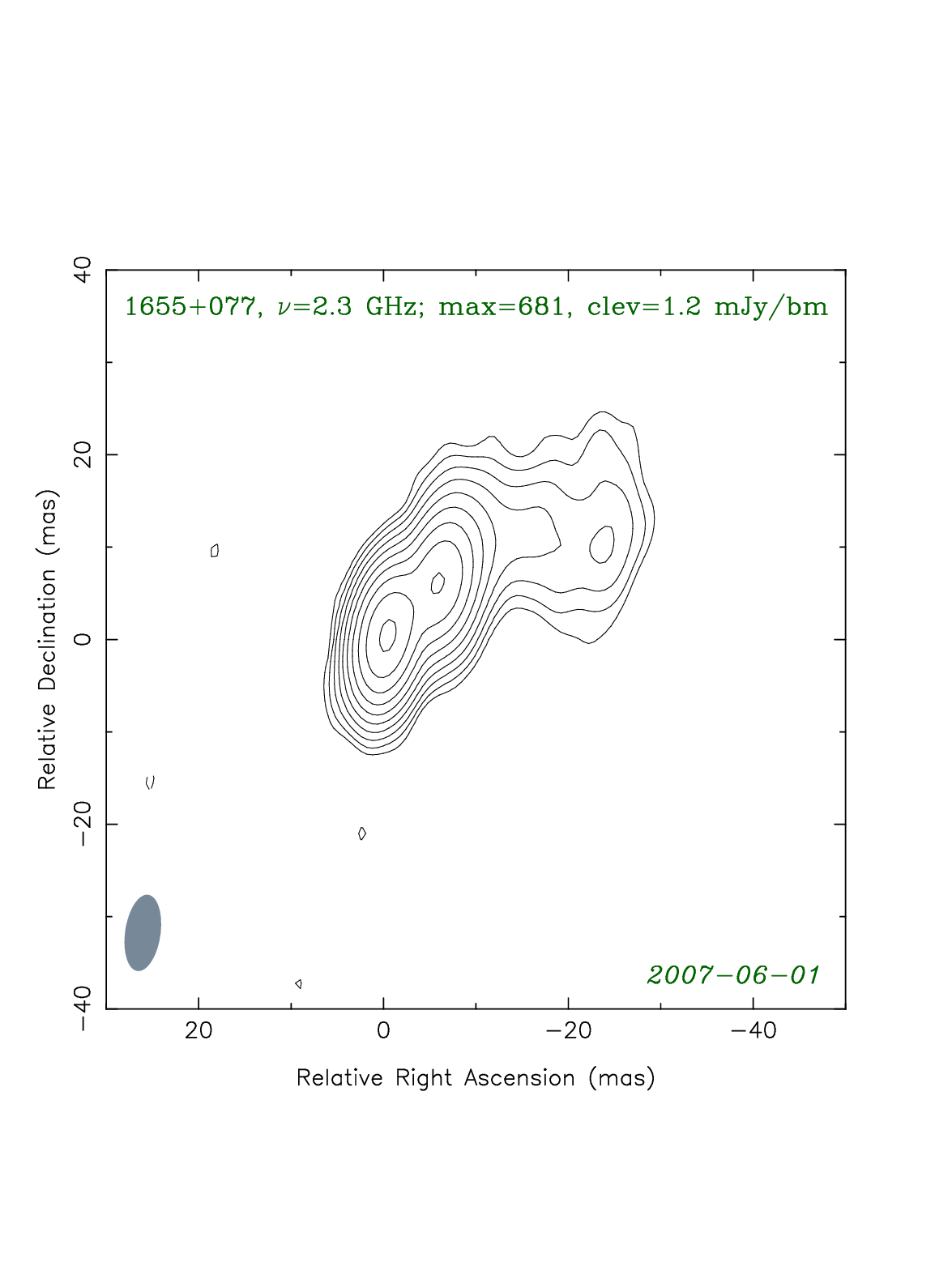}
 \includegraphics[width=0.33\textwidth,angle=0,trim=0.5cm 3cm 1.6cm 5cm,clip]{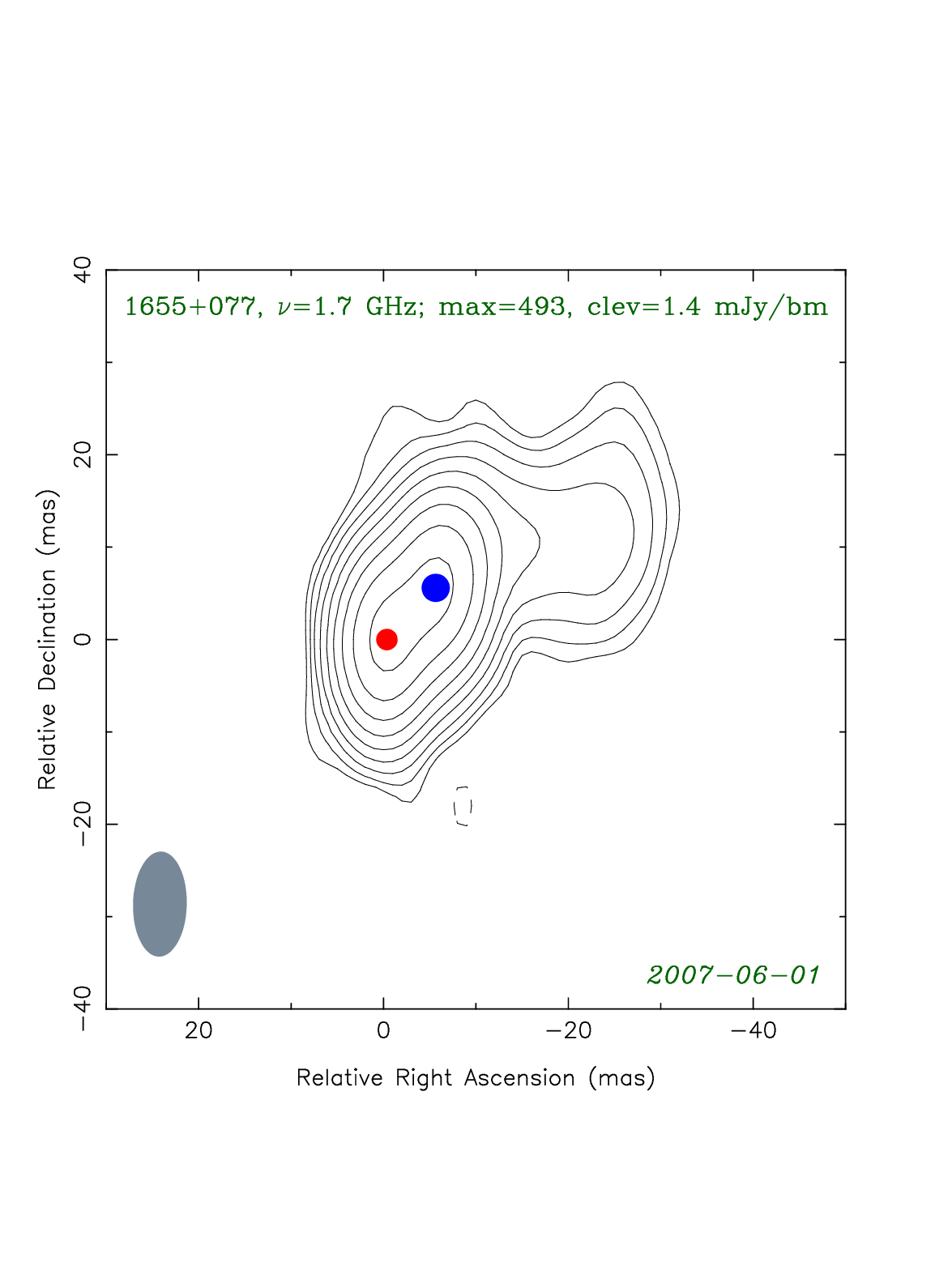}
 \includegraphics[width=0.33\textwidth,angle=0,trim=0.5cm 3cm 1.6cm 5cm,clip]{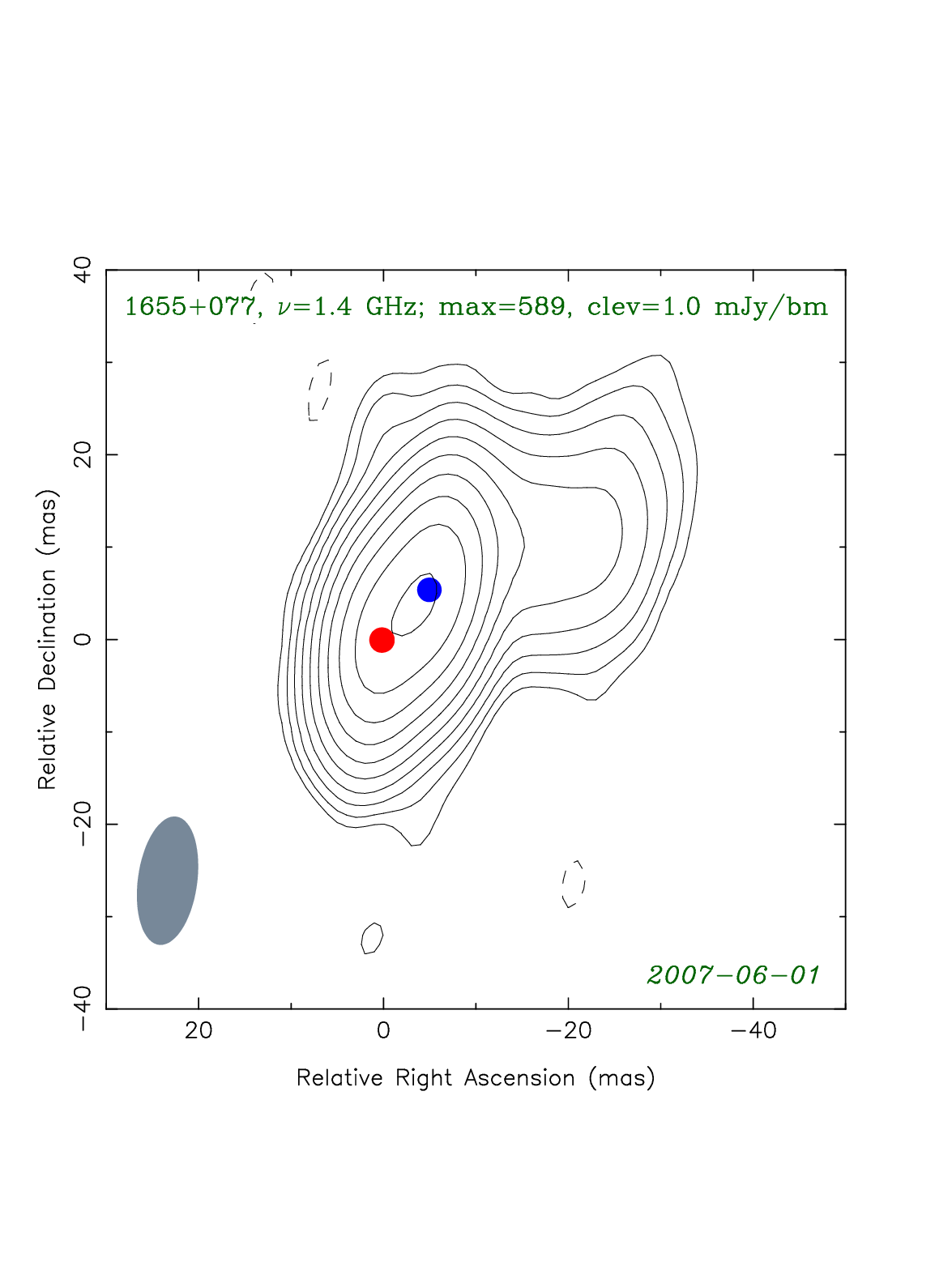}

 \caption{continued...}
\end{figure*}

\clearpage
\addtocounter{figure}{-1}
\begin{figure*}
 \centering

 \includegraphics[width=0.33\textwidth,angle=0,trim=0.5cm 3cm 1.6cm 5cm,clip]{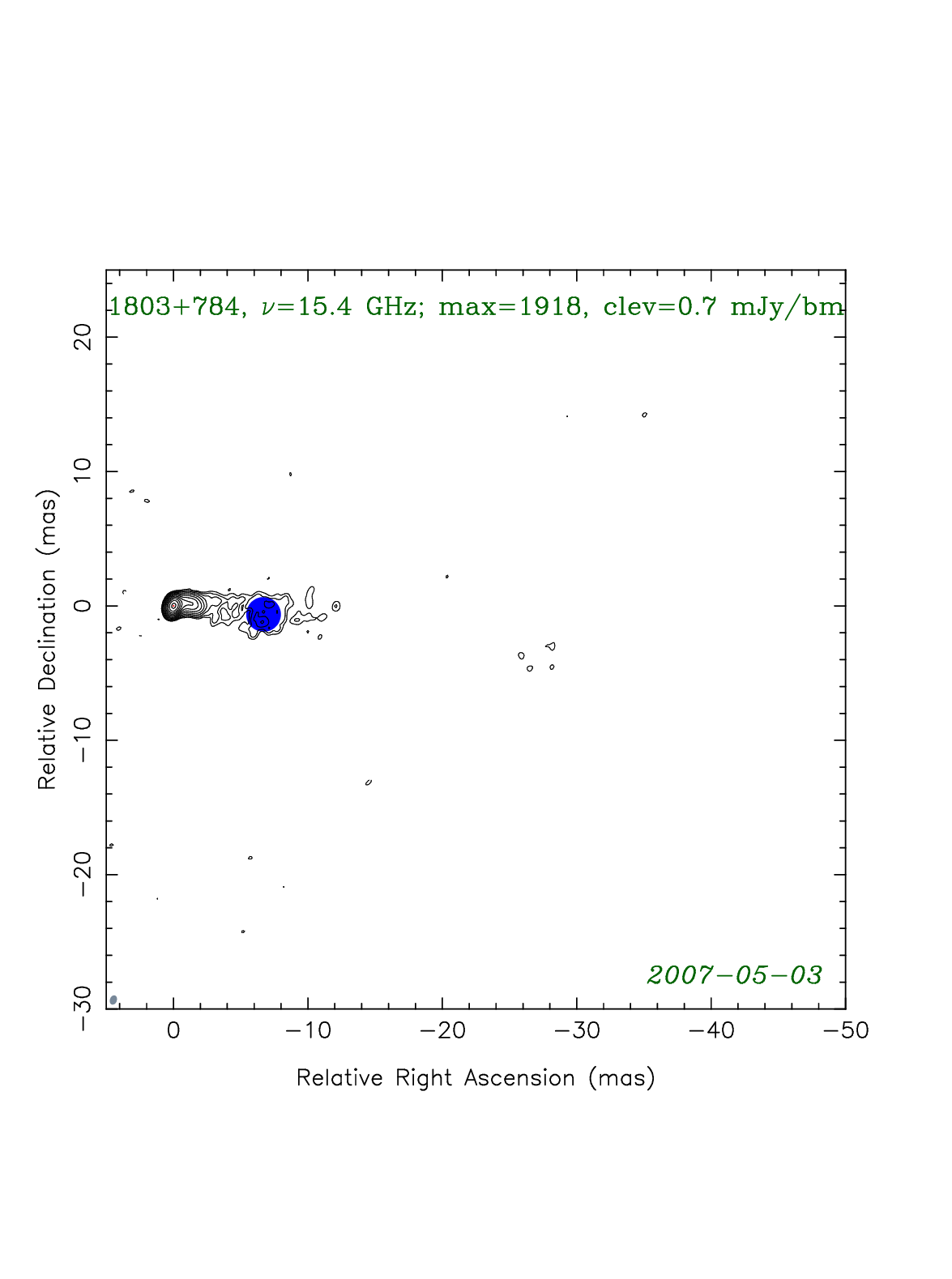}
 \includegraphics[width=0.33\textwidth,angle=0,trim=0.5cm 3cm 1.6cm 5cm,clip]{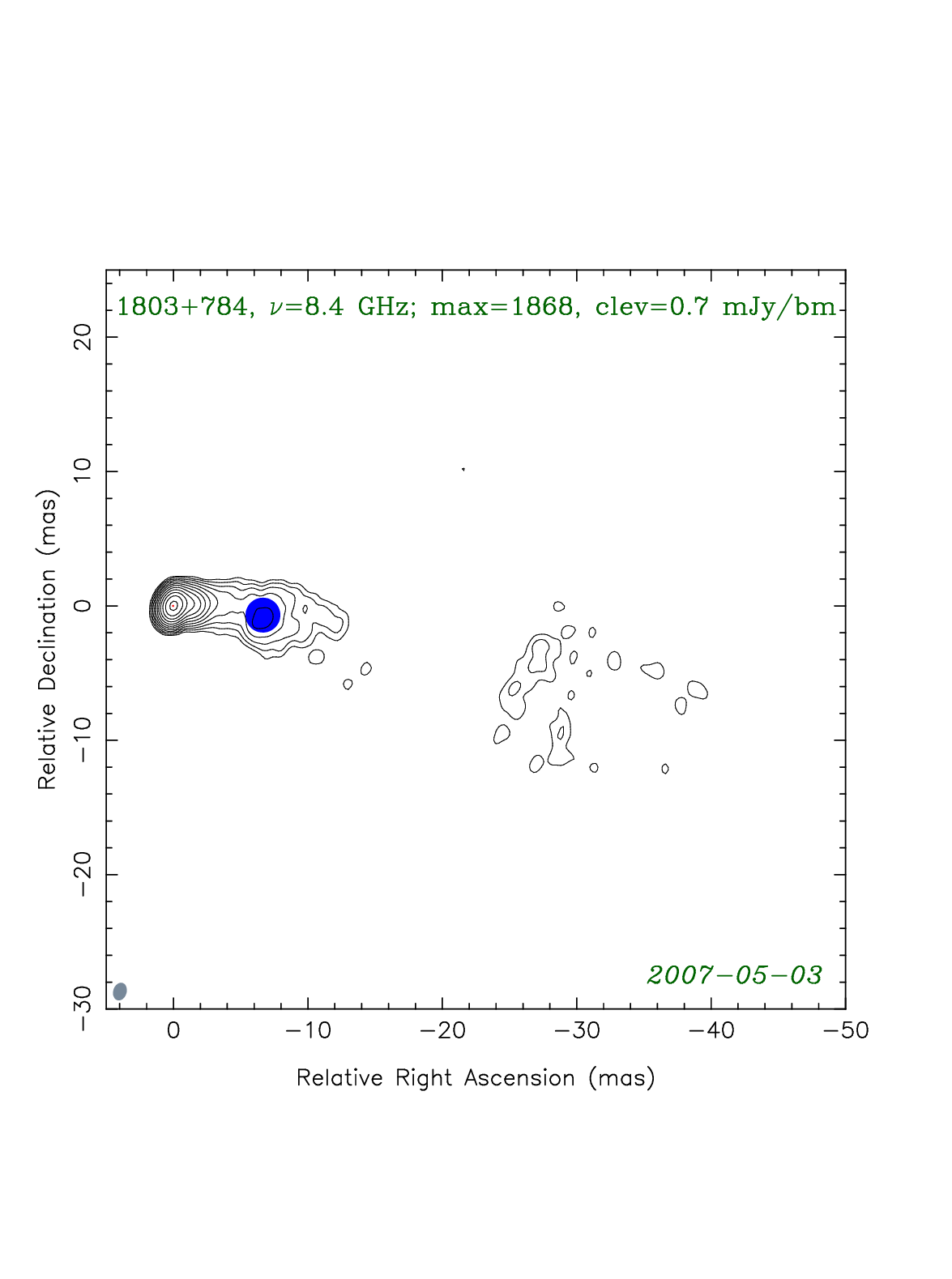}
 \includegraphics[width=0.33\textwidth,angle=0,trim=0.5cm 3cm 1.6cm 5cm,clip]{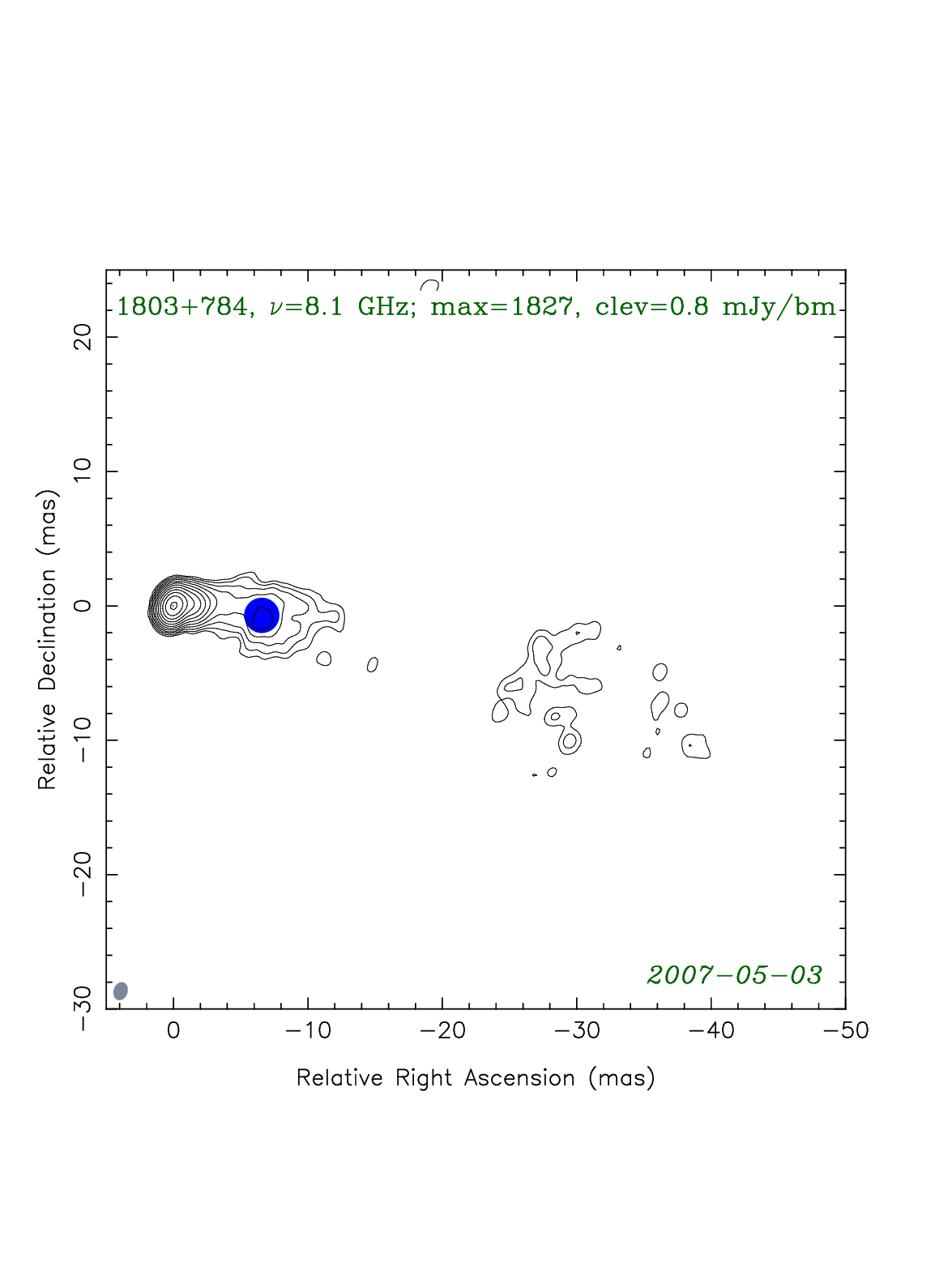}

 \includegraphics[width=0.33\textwidth,angle=0,trim=0.5cm 3cm 1.6cm 5cm,clip]{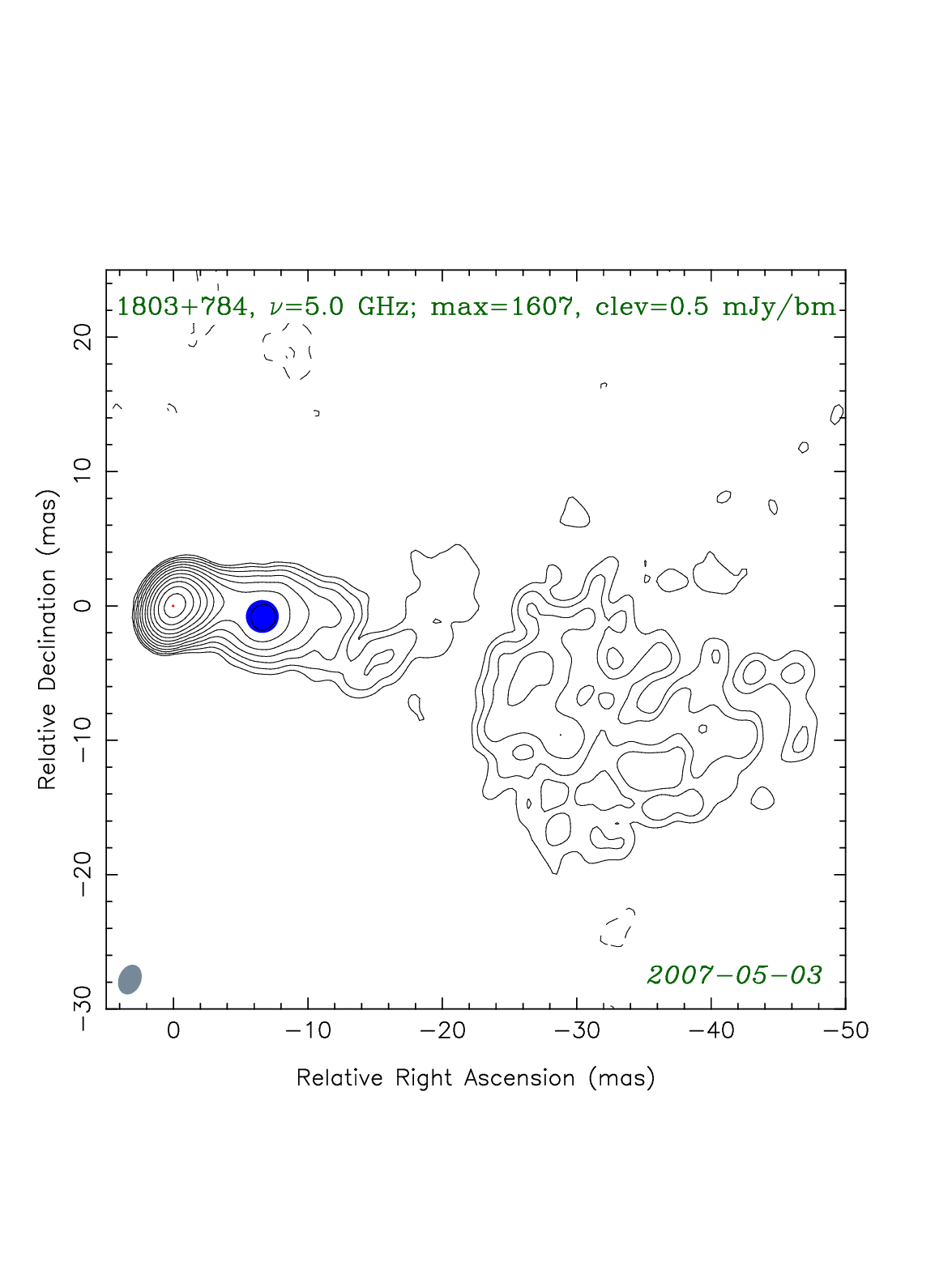}
 \includegraphics[width=0.33\textwidth,angle=0,trim=0.5cm 3cm 1.6cm 5cm,clip]{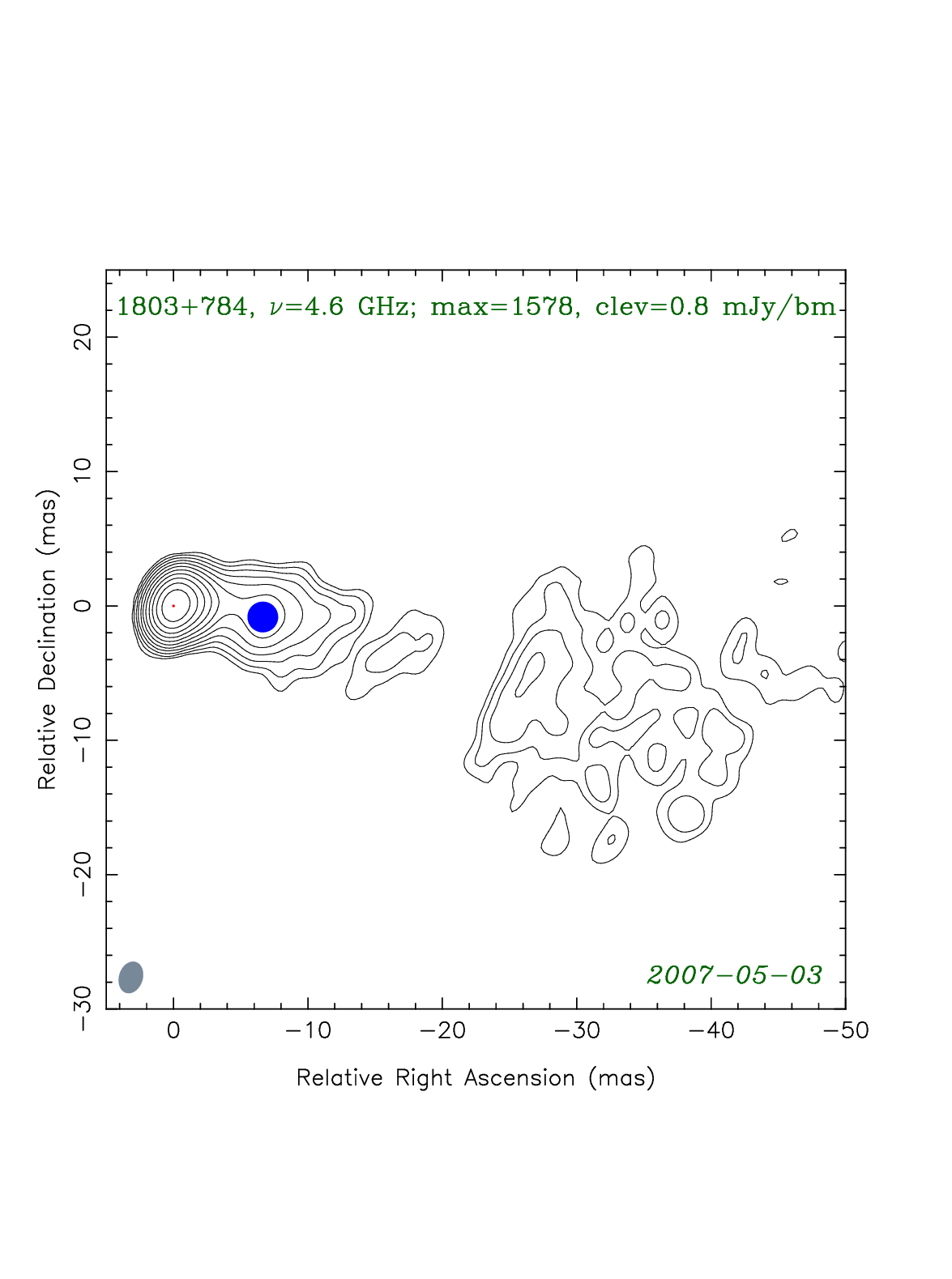}
 \includegraphics[width=0.33\textwidth,angle=0,trim=0.5cm 3cm 1.6cm 5cm,clip]{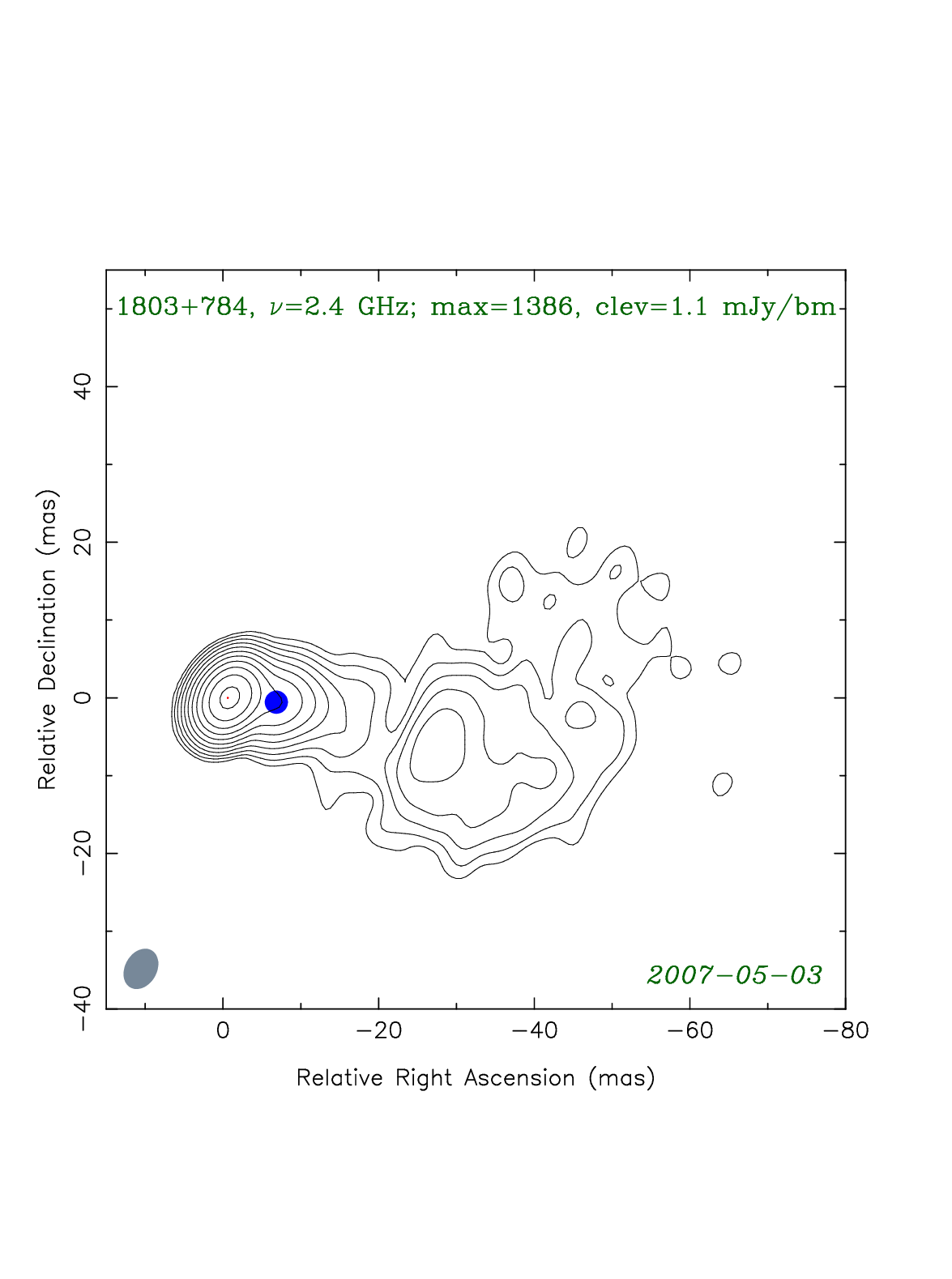}

 \includegraphics[width=0.33\textwidth,angle=0,trim=0.5cm 3cm 1.6cm 5cm,clip]{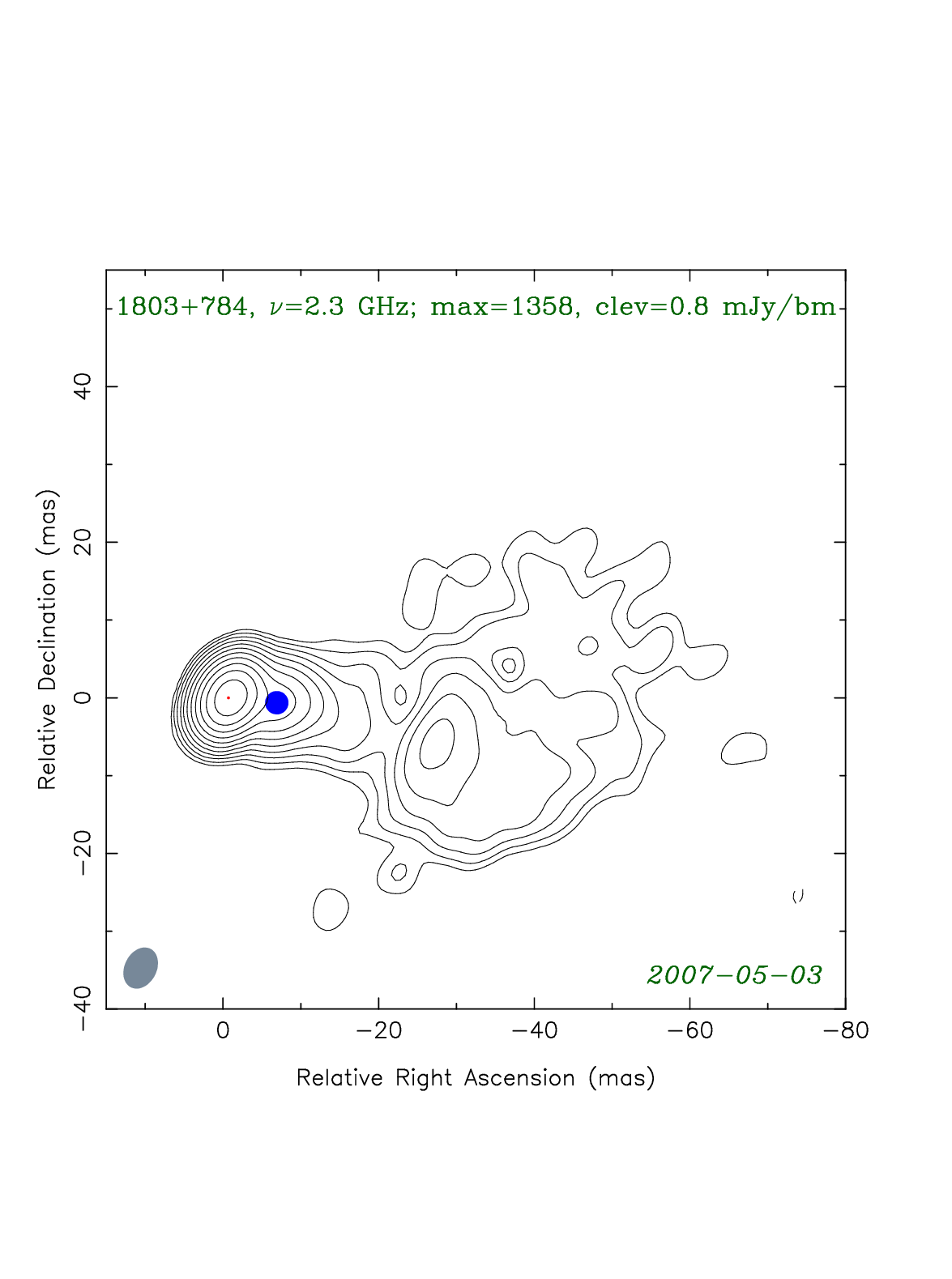}
 \includegraphics[width=0.33\textwidth,angle=0,trim=0.5cm 3cm 1.6cm 5cm,clip]{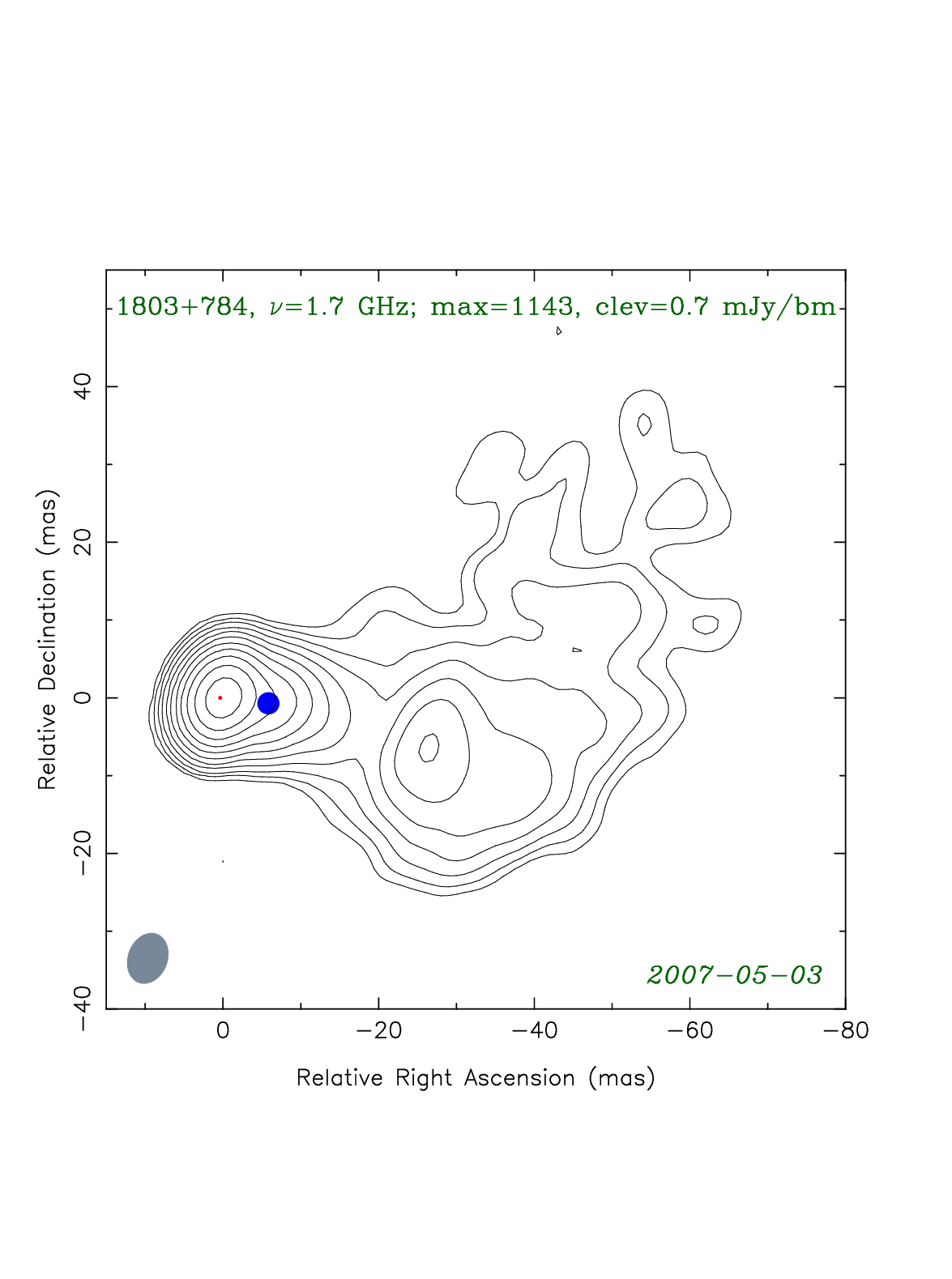}
 \includegraphics[width=0.33\textwidth,angle=0,trim=0.5cm 3cm 1.6cm 5cm,clip]{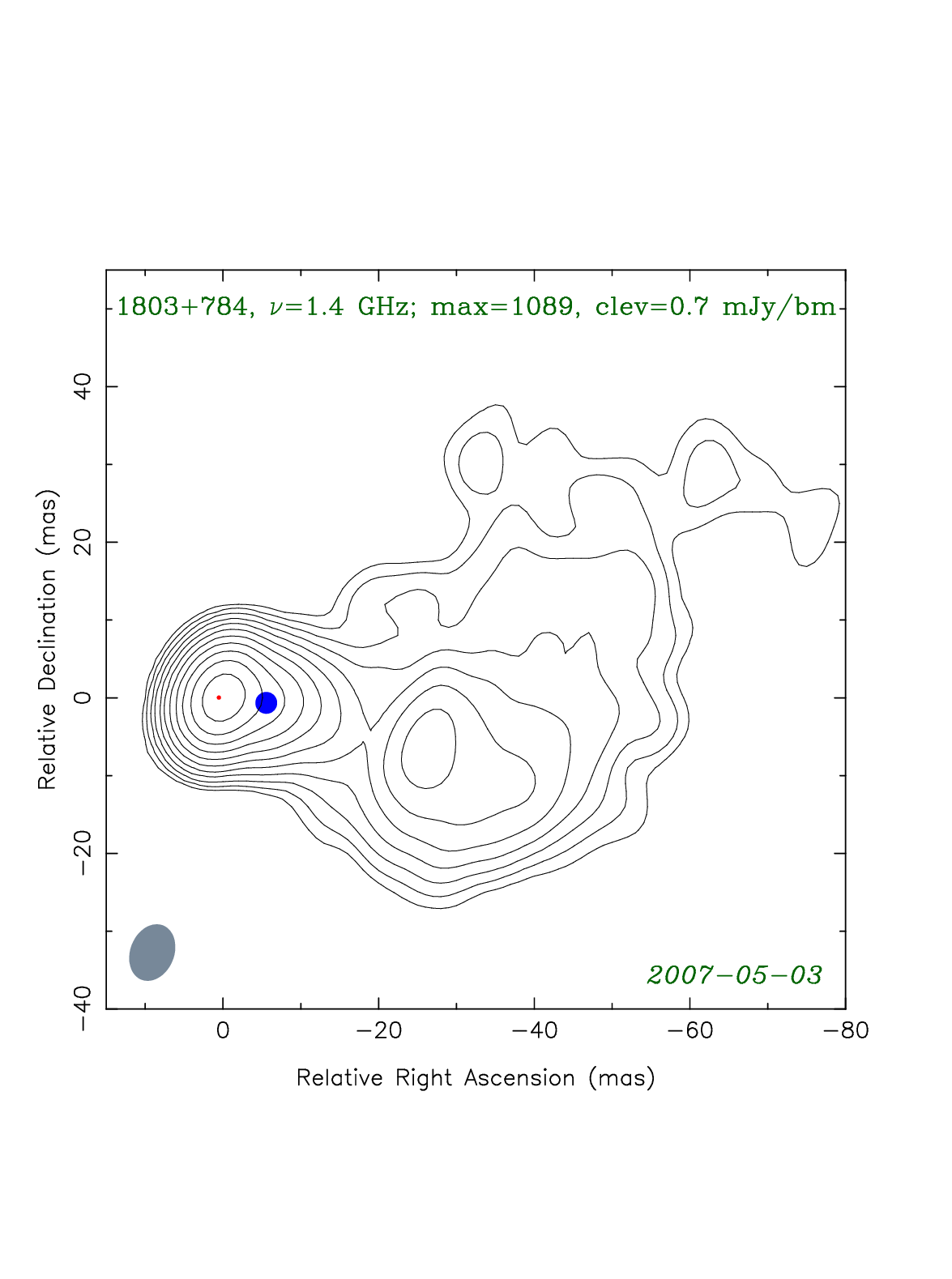}

 \caption{continued...}
\end{figure*}

\clearpage
\addtocounter{figure}{-1}
\begin{figure*}
 \centering

 \includegraphics[width=0.33\textwidth,angle=0,trim=0.5cm 3cm 1.6cm 5cm,clip]{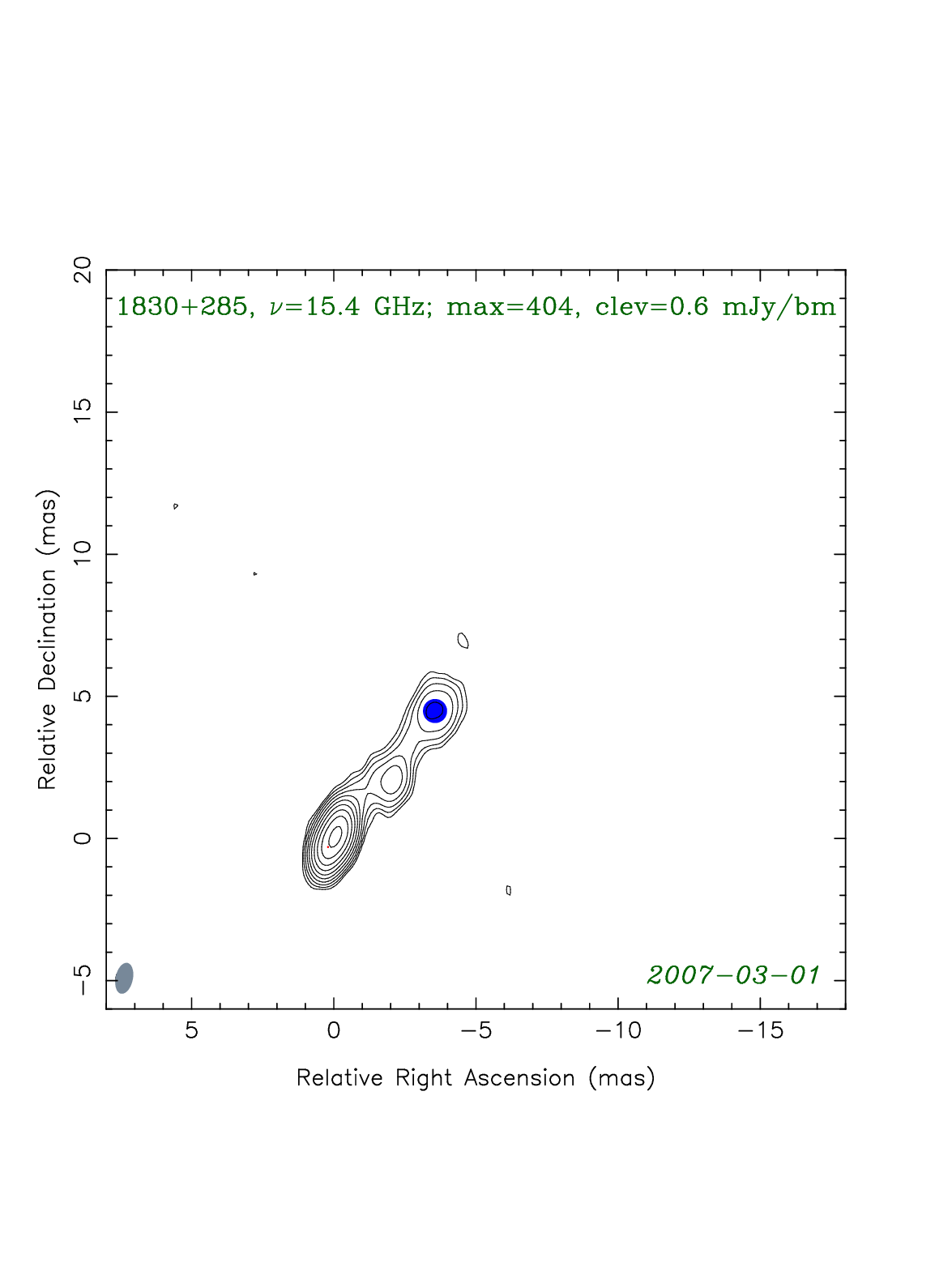}
 \includegraphics[width=0.33\textwidth,angle=0,trim=0.5cm 3cm 1.6cm 5cm,clip]{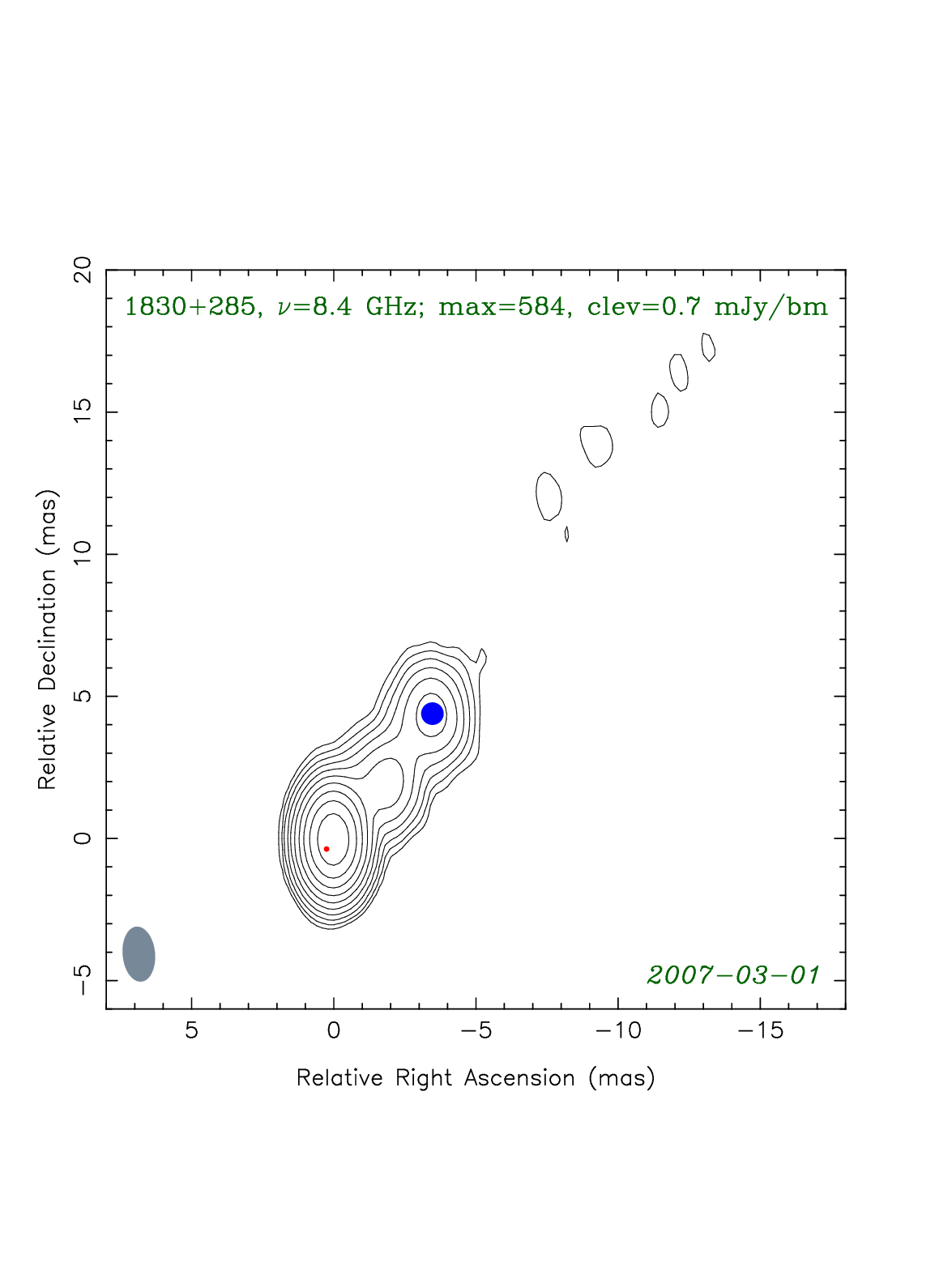}
 \includegraphics[width=0.33\textwidth,angle=0,trim=0.5cm 3cm 1.6cm 5cm,clip]{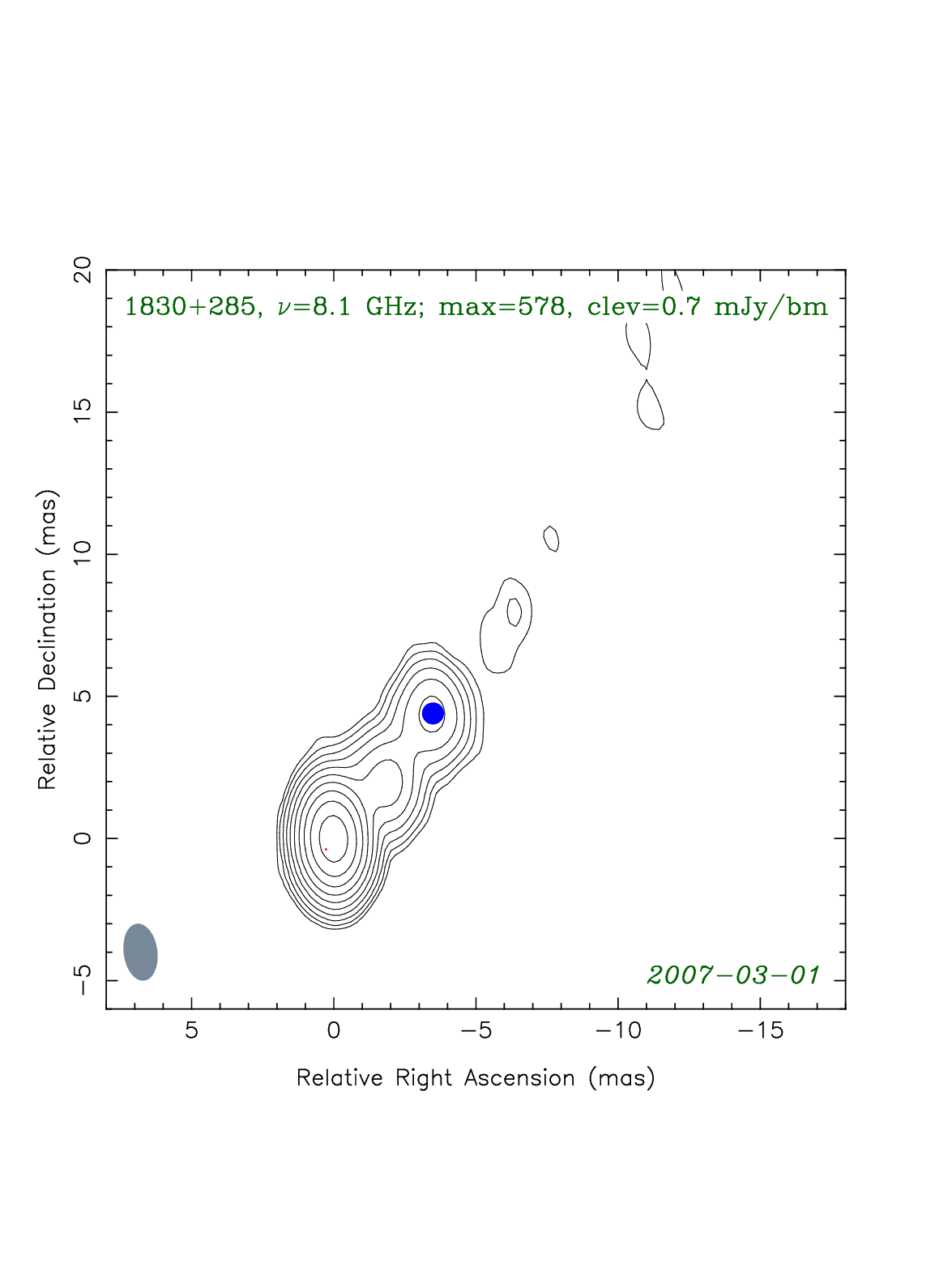}

 \includegraphics[width=0.33\textwidth,angle=0,trim=0.5cm 3cm 1.6cm 5cm,clip]{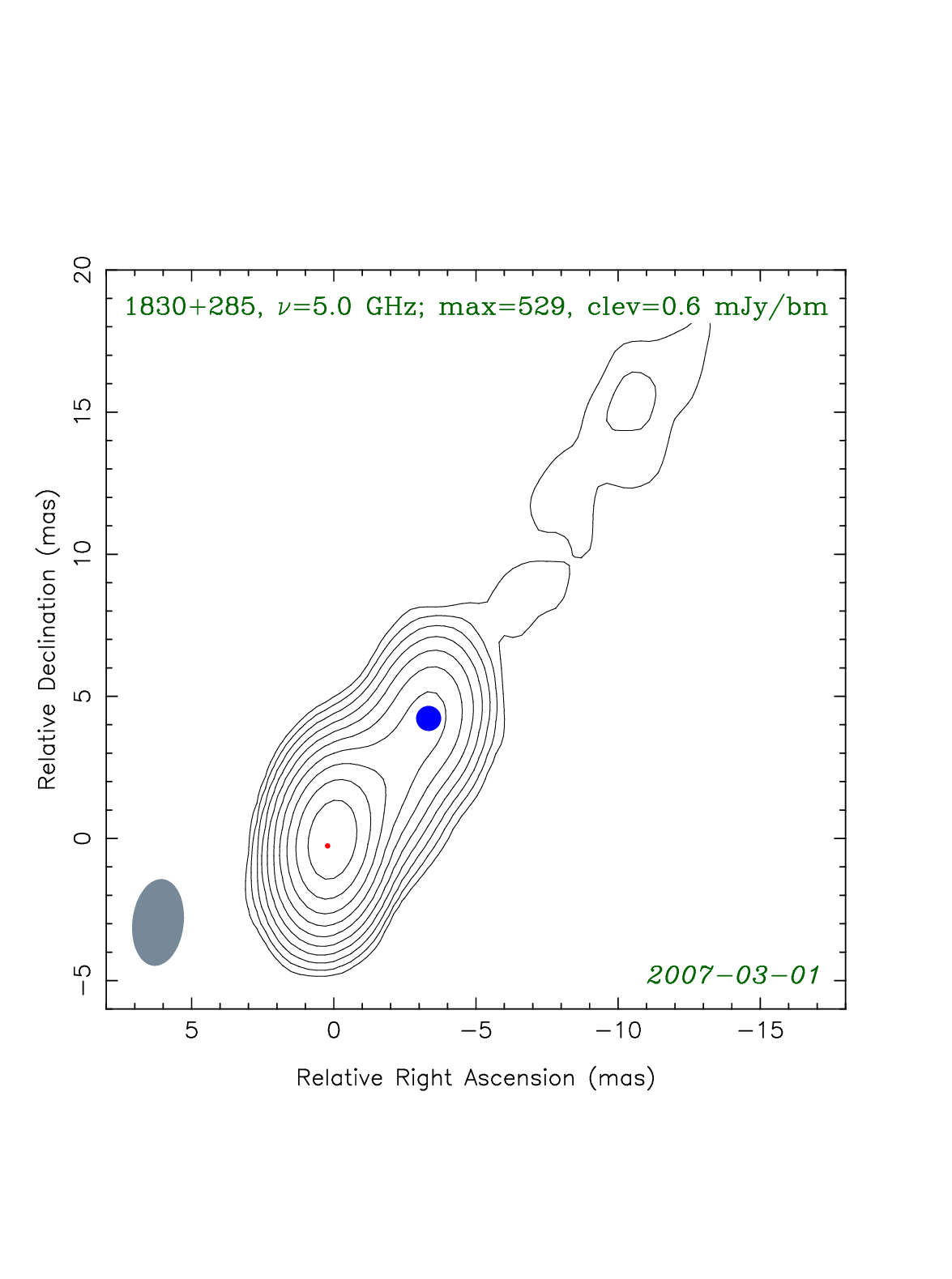}
 \includegraphics[width=0.33\textwidth,angle=0,trim=0.5cm 3cm 1.6cm 5cm,clip]{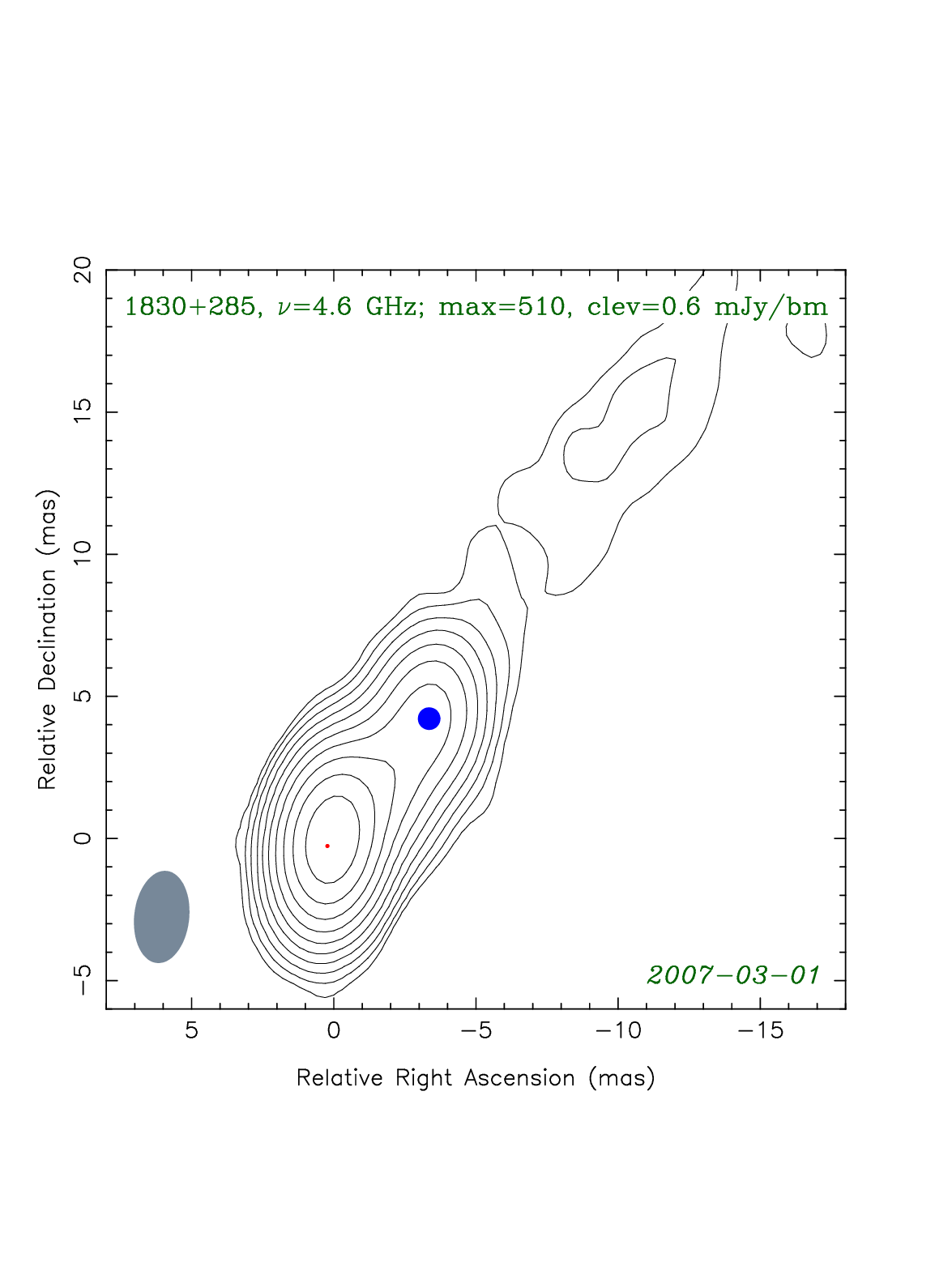}
 \includegraphics[width=0.33\textwidth,angle=0,trim=0.5cm 3cm 1.6cm 5cm,clip]{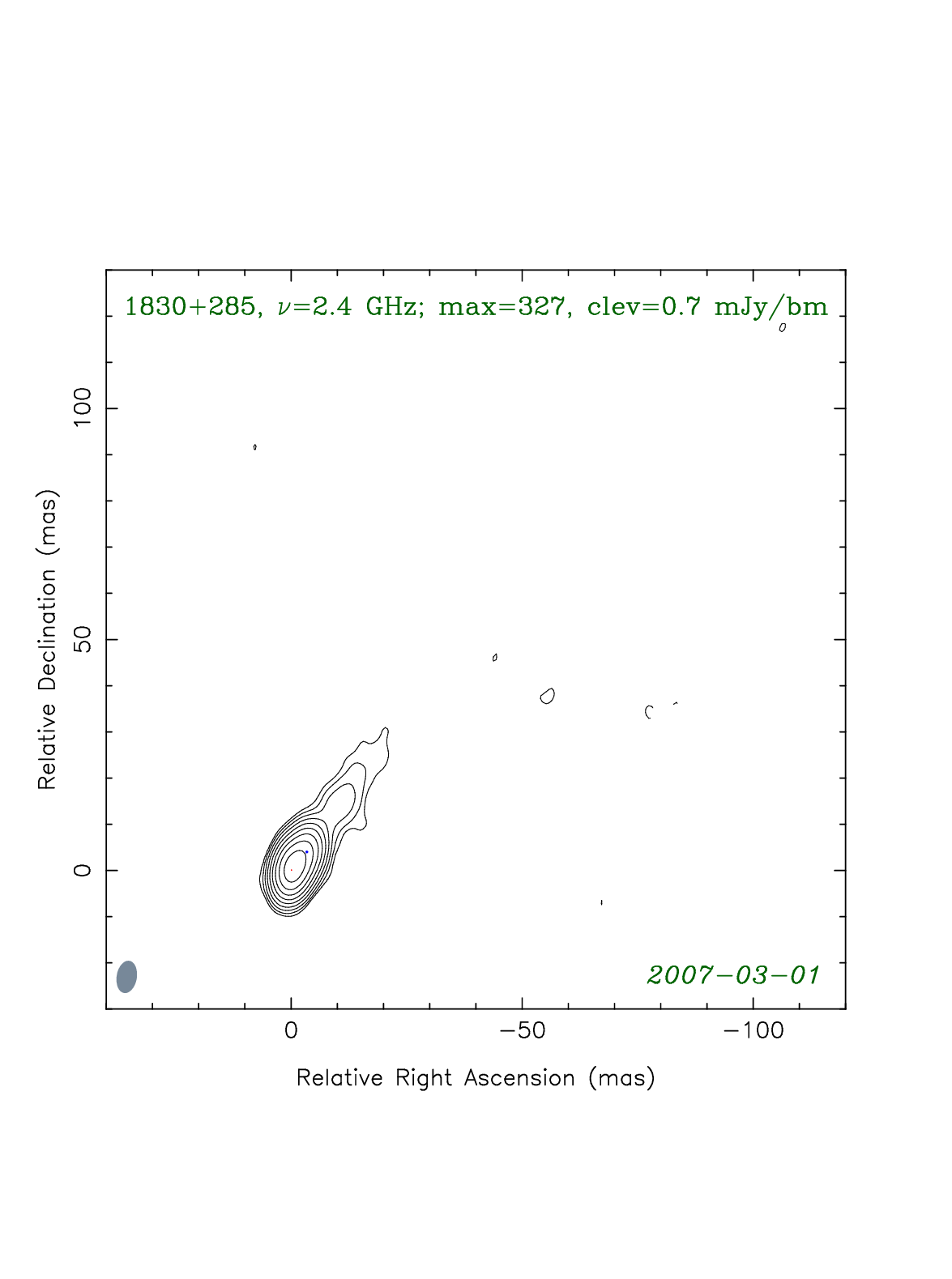}

 \includegraphics[width=0.33\textwidth,angle=0,trim=0.5cm 3cm 1.6cm 5cm,clip]{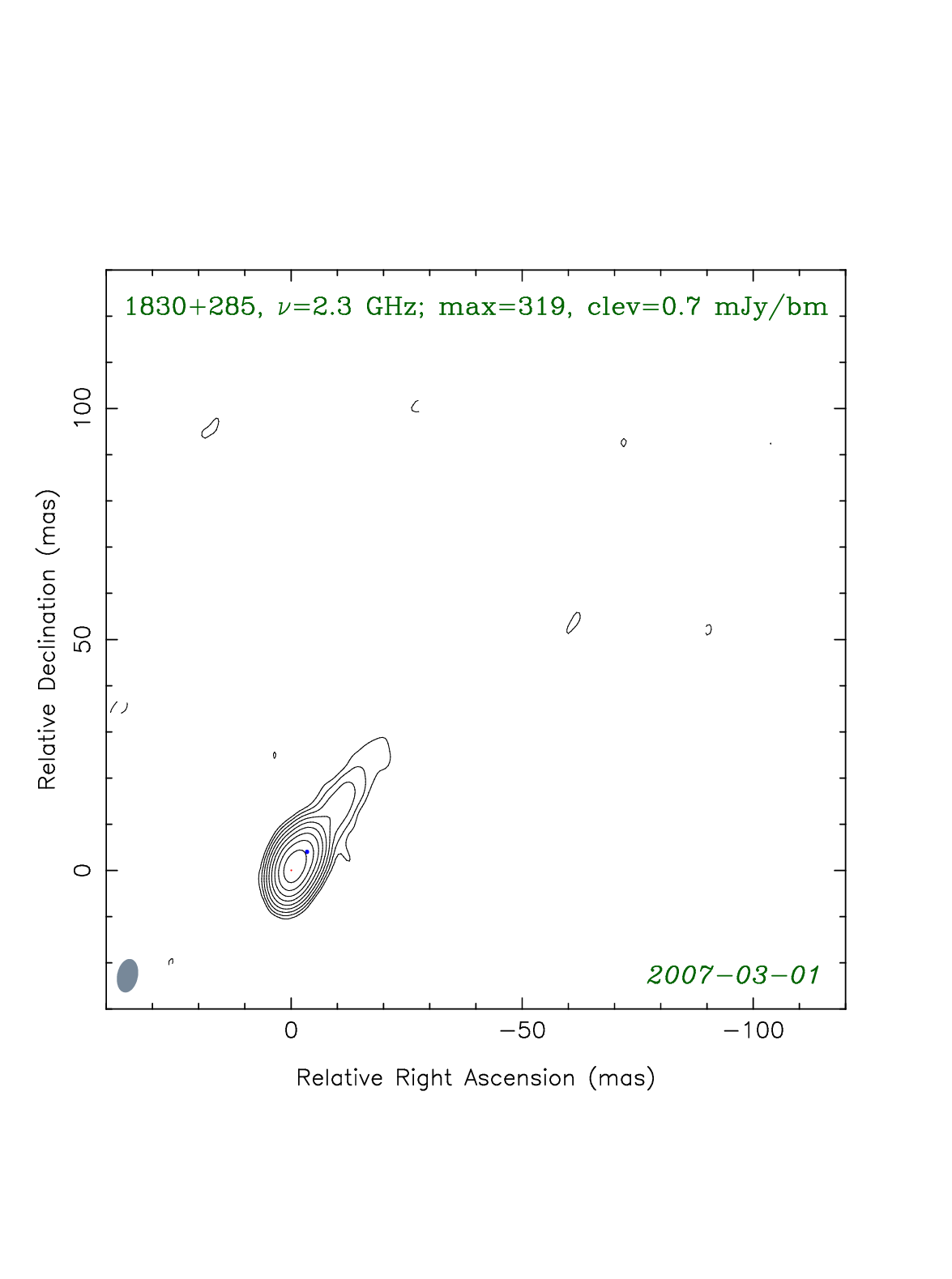}
 \includegraphics[width=0.33\textwidth,angle=0,trim=0.5cm 3cm 1.6cm 5cm,clip]{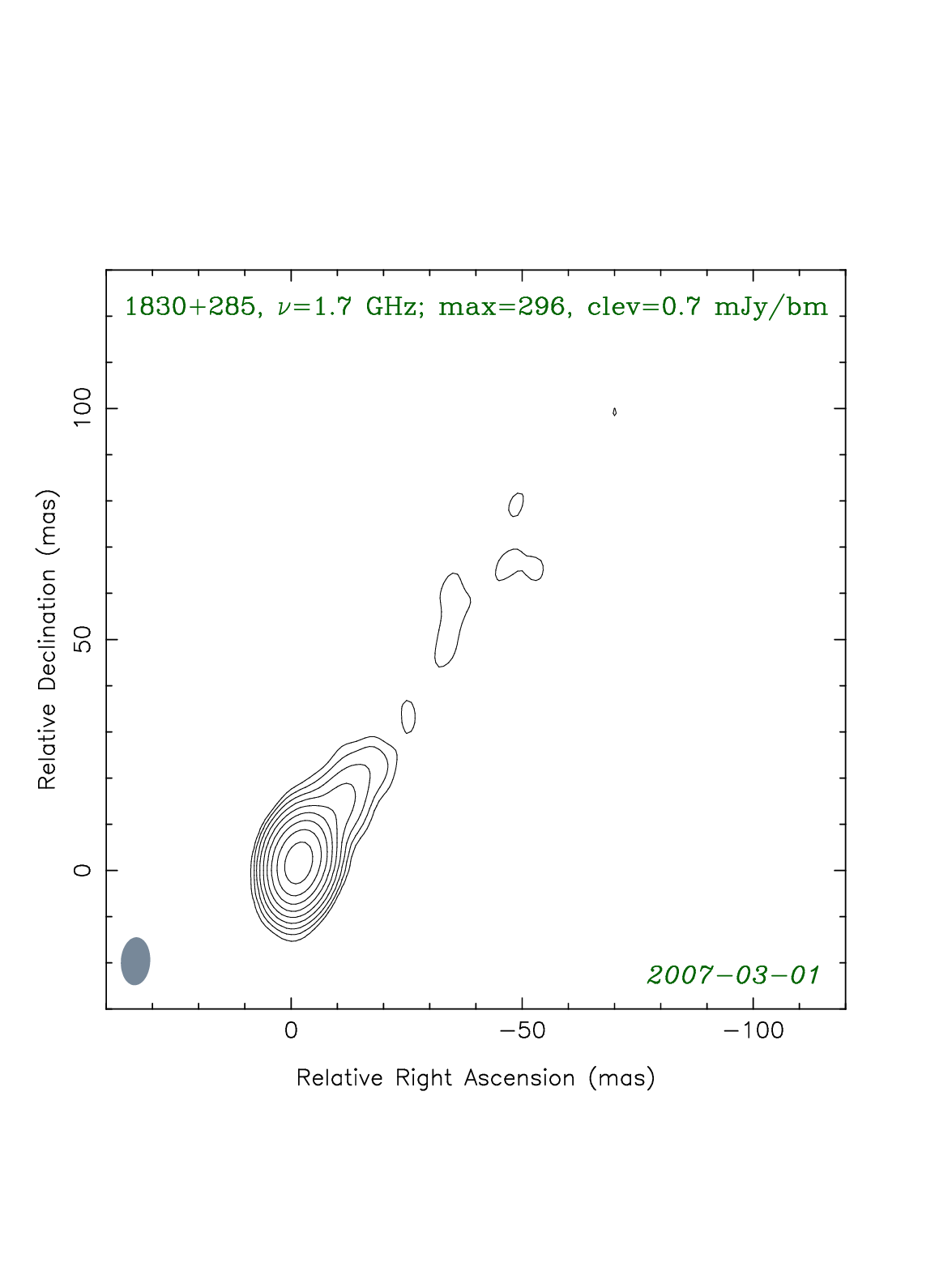}
 \includegraphics[width=0.33\textwidth,angle=0,trim=0.5cm 3cm 1.6cm 5cm,clip]{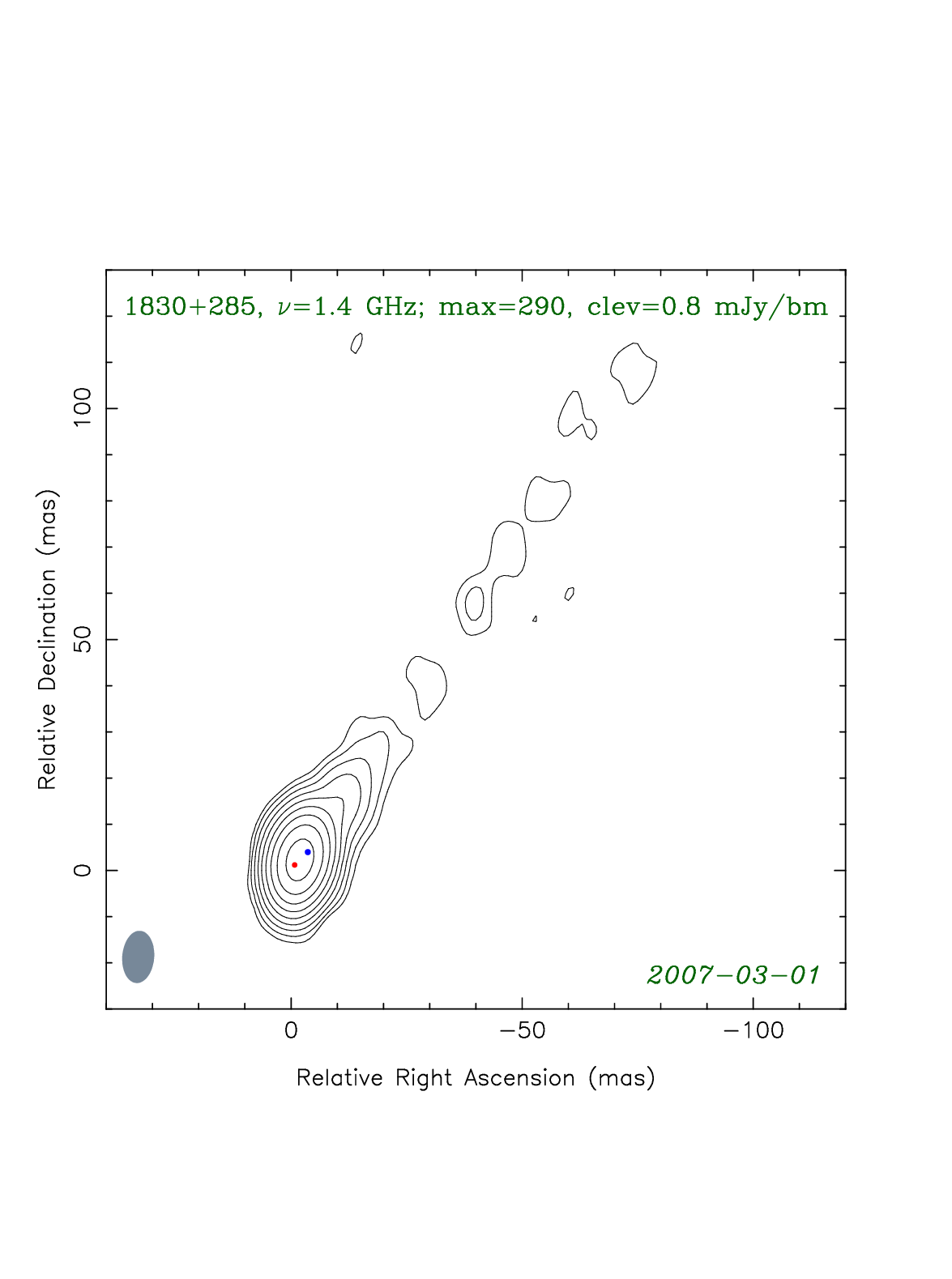}

 \caption{continued...}
\end{figure*}

\clearpage
\addtocounter{figure}{-1}
\begin{figure*}
 \centering

 \includegraphics[width=0.33\textwidth,angle=0,trim=0.5cm 3cm 1.6cm 5cm,clip]{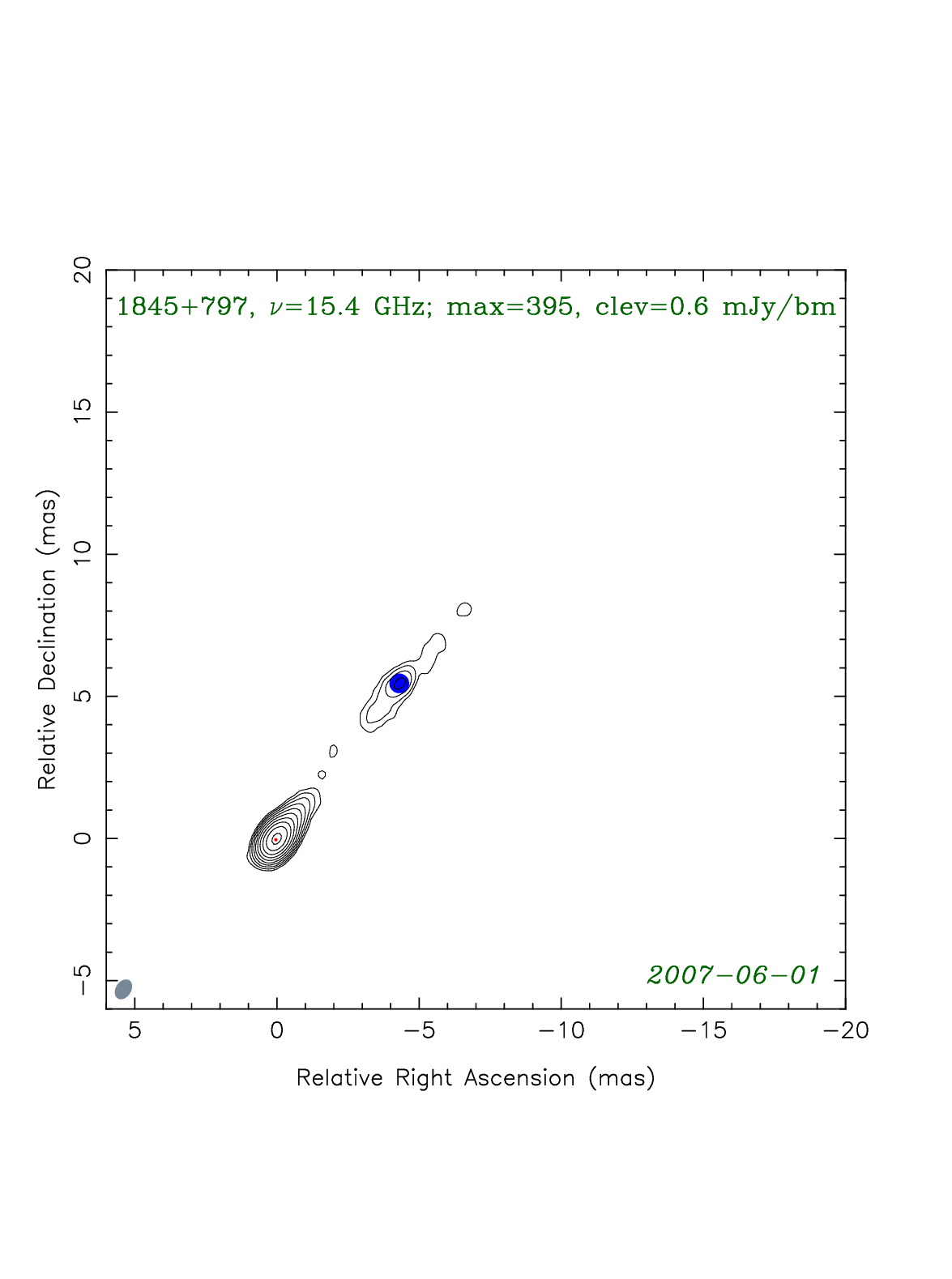}
 \includegraphics[width=0.33\textwidth,angle=0,trim=0.5cm 3cm 1.6cm 5cm,clip]{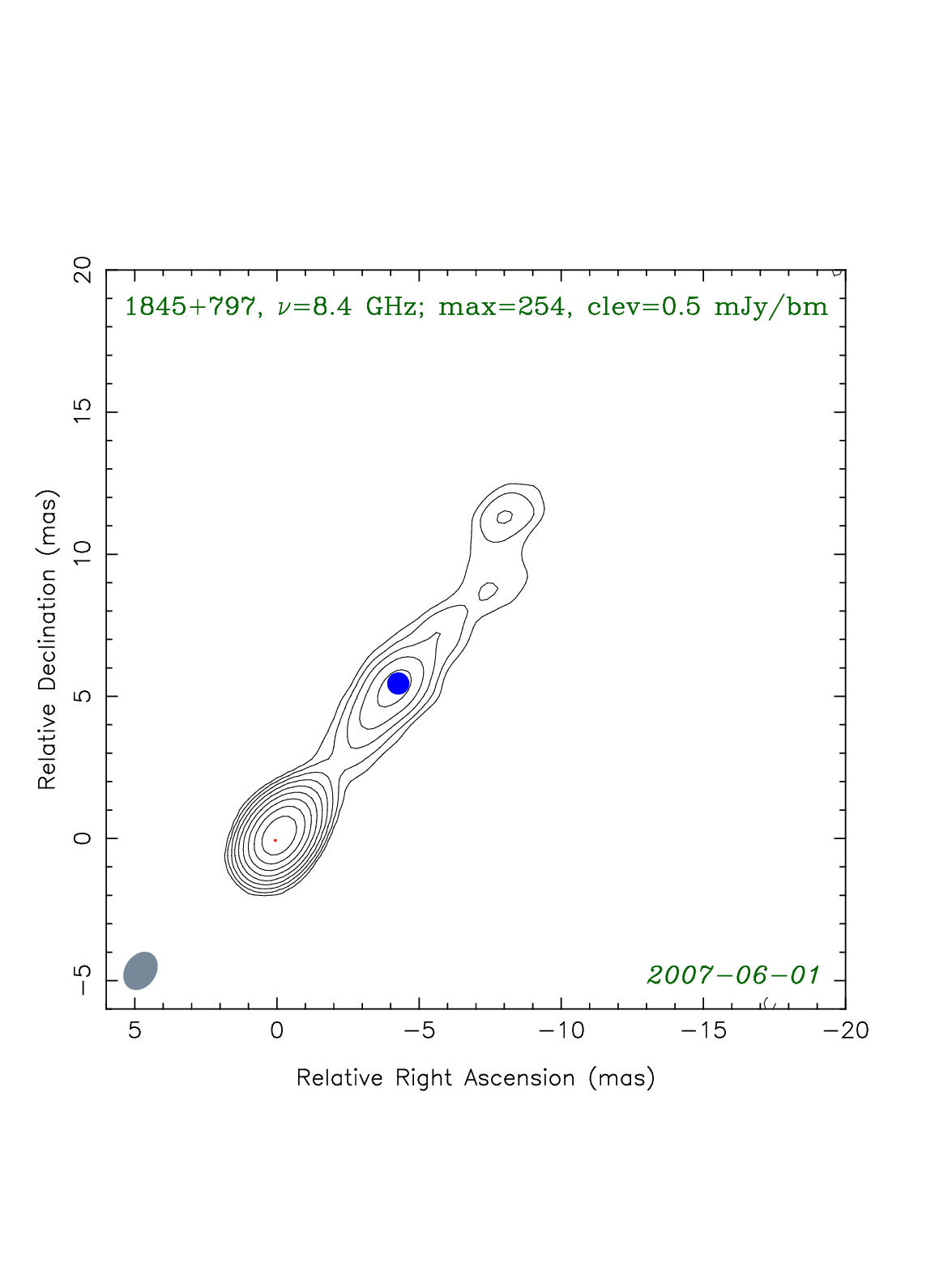}
 \includegraphics[width=0.33\textwidth,angle=0,trim=0.5cm 3cm 1.6cm 5cm,clip]{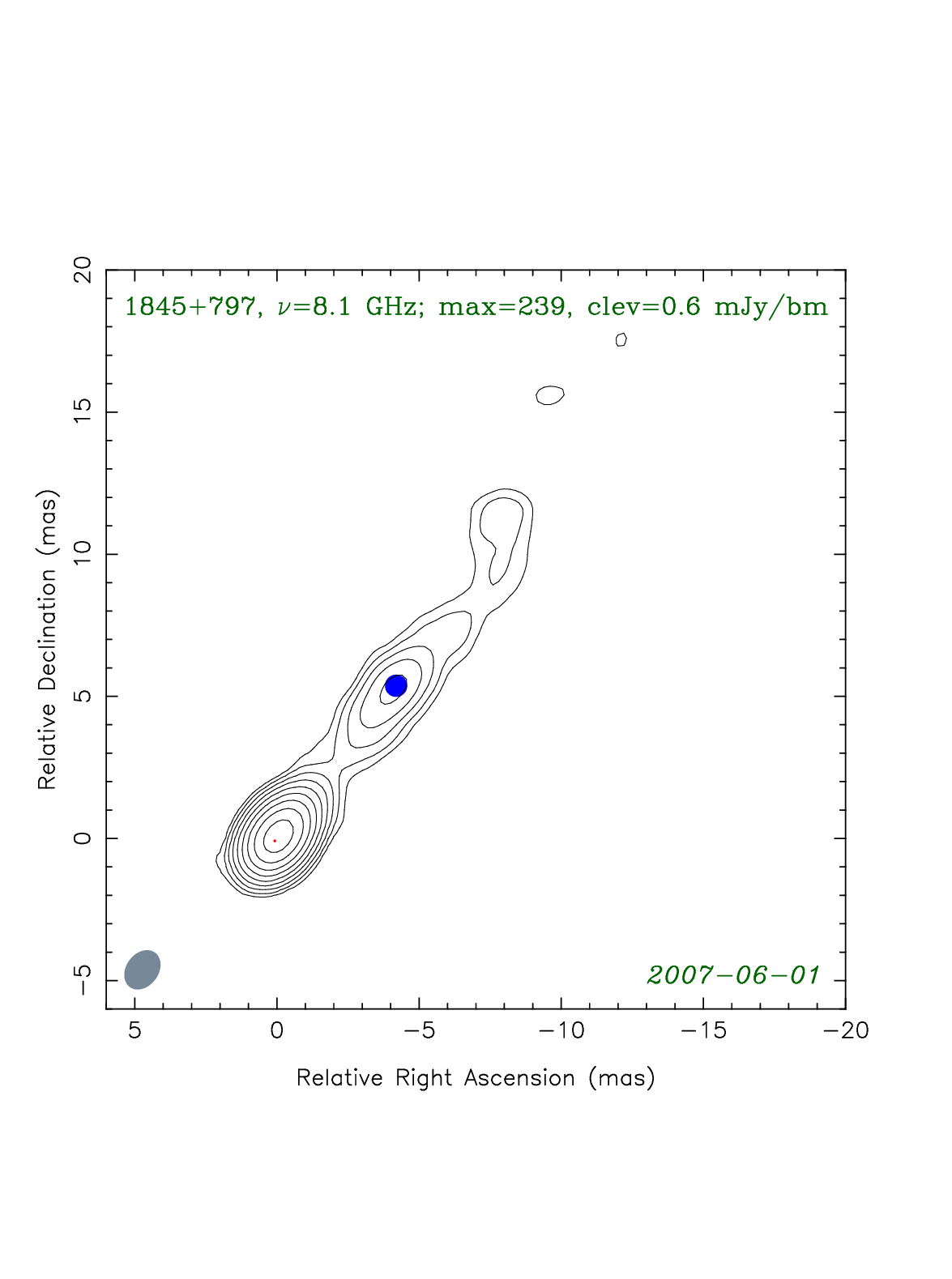}

 \includegraphics[width=0.33\textwidth,angle=0,trim=0.5cm 3cm 1.6cm 5cm,clip]{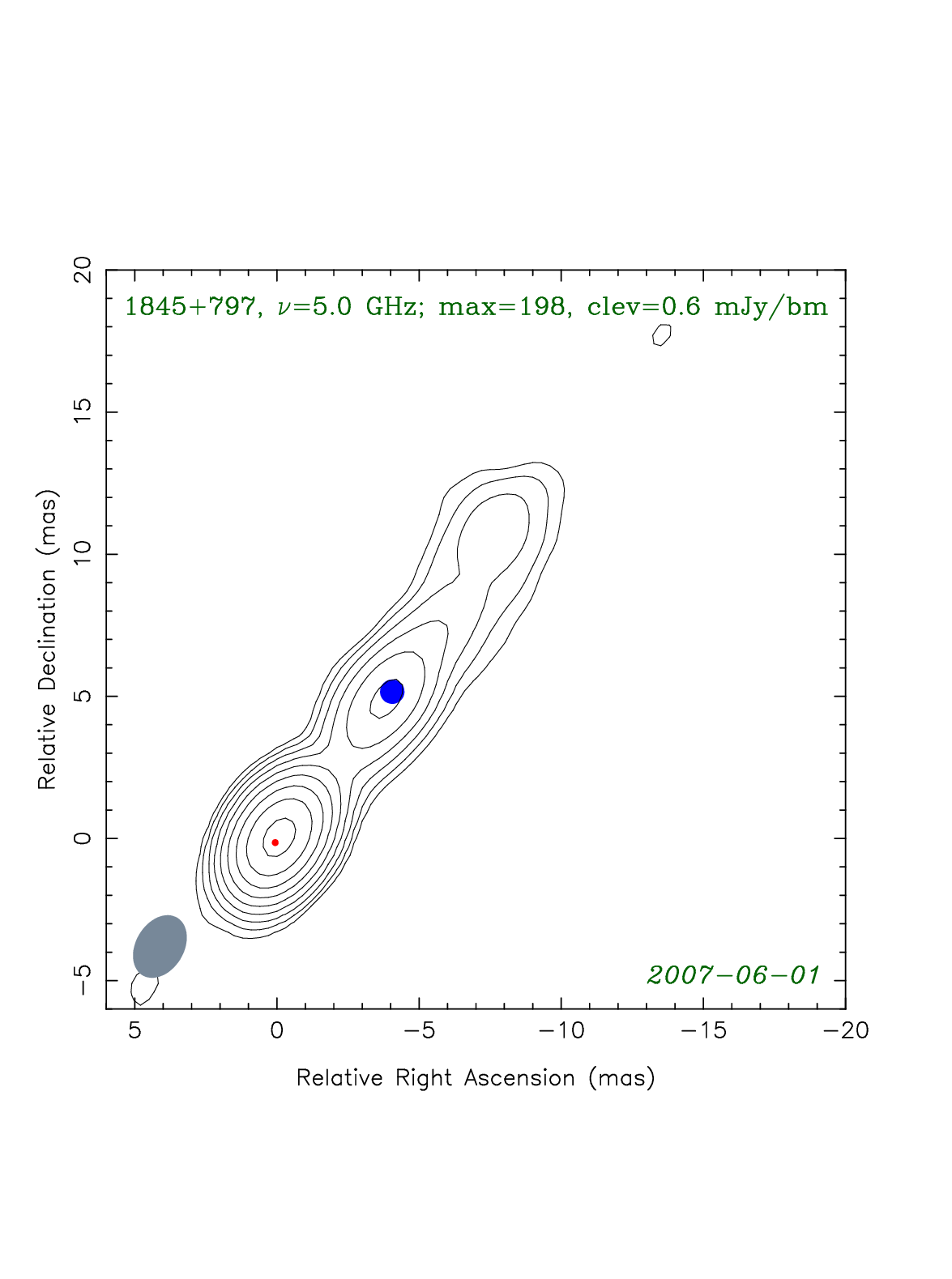}
 \includegraphics[width=0.33\textwidth,angle=0,trim=0.5cm 3cm 1.6cm 5cm,clip]{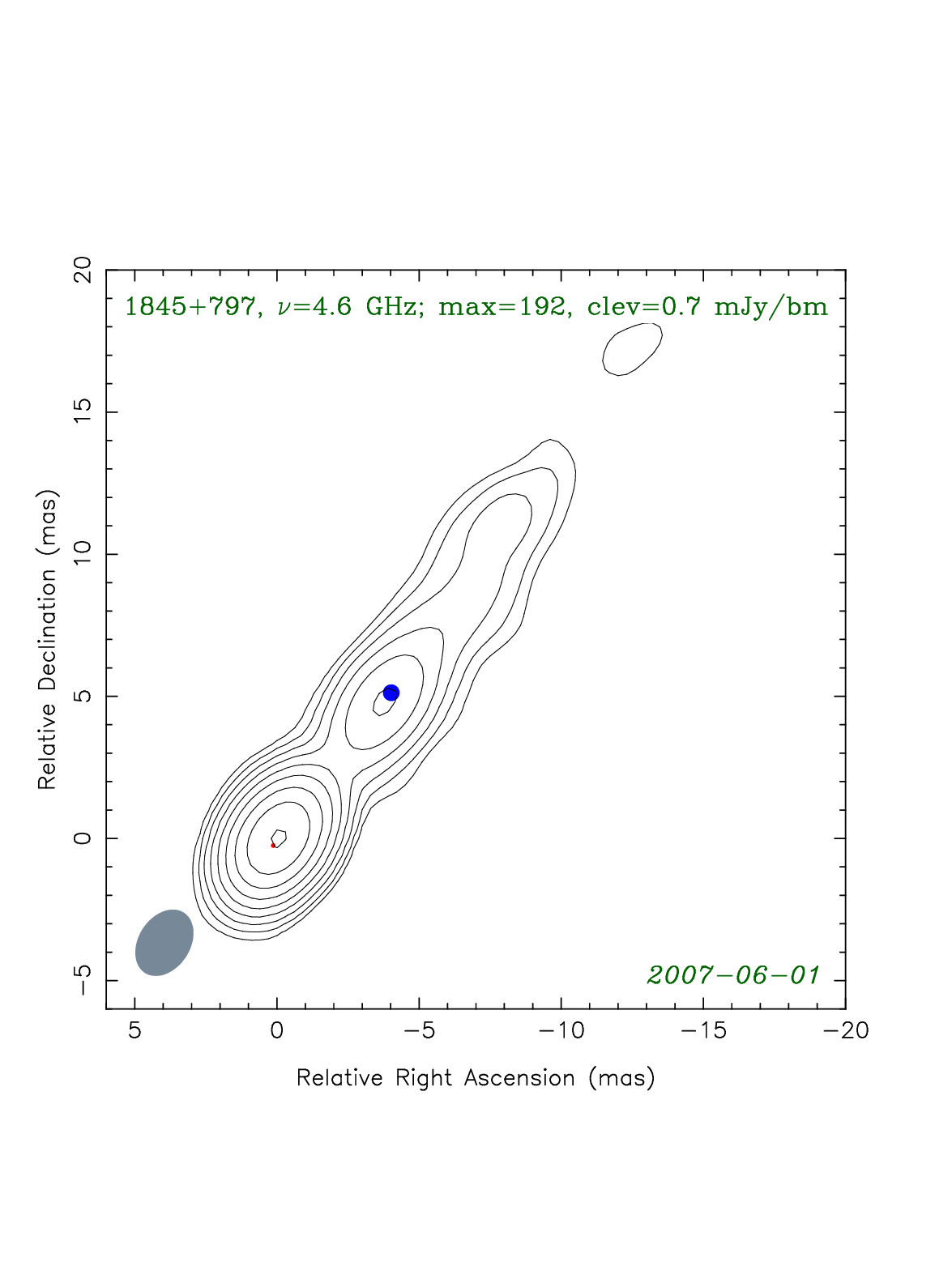}
 \includegraphics[width=0.33\textwidth,angle=0,trim=0.5cm 3cm 1.6cm 5cm,clip]{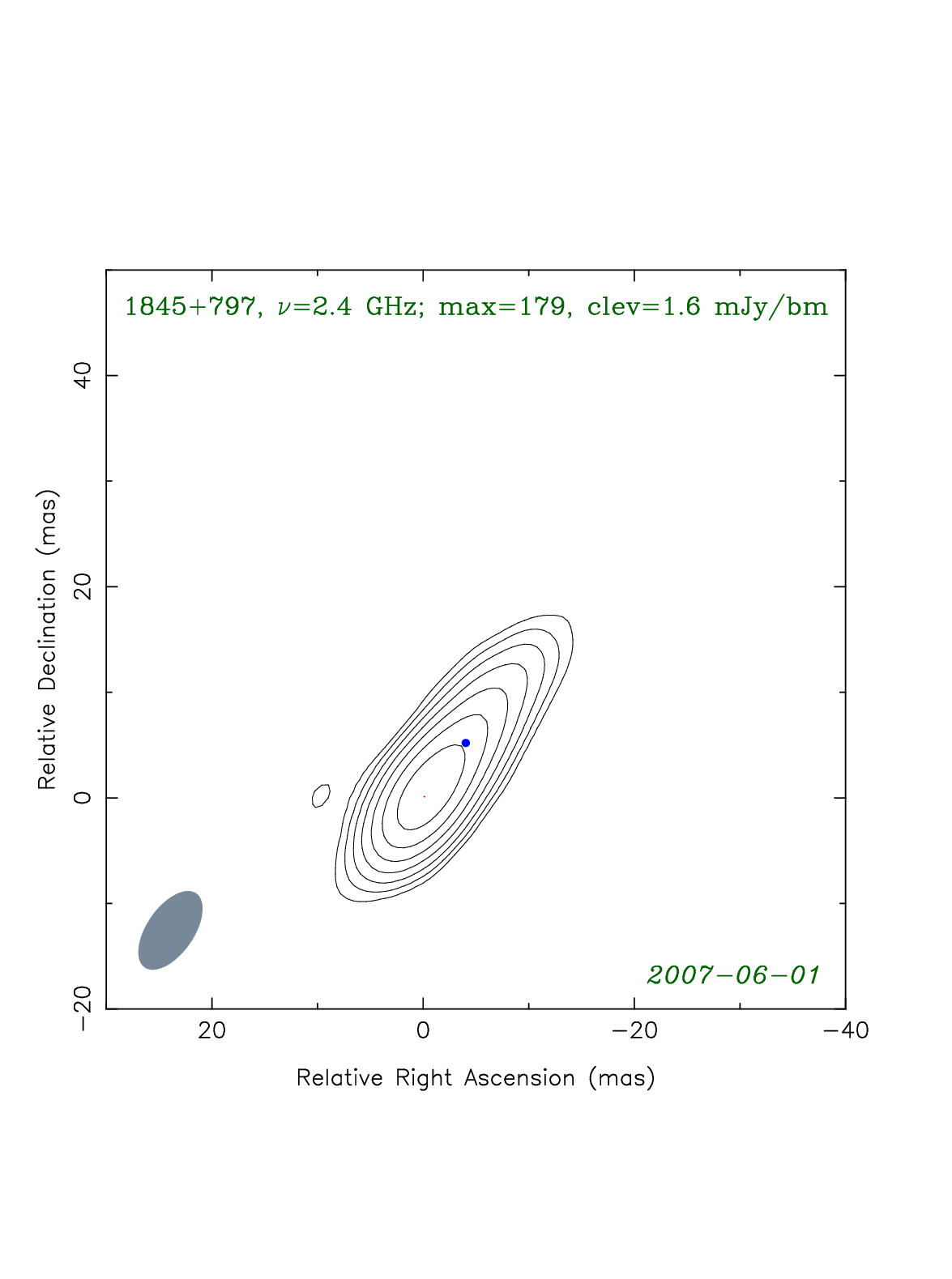}

 \includegraphics[width=0.33\textwidth,angle=0,trim=0.5cm 3cm 1.6cm 5cm,clip]{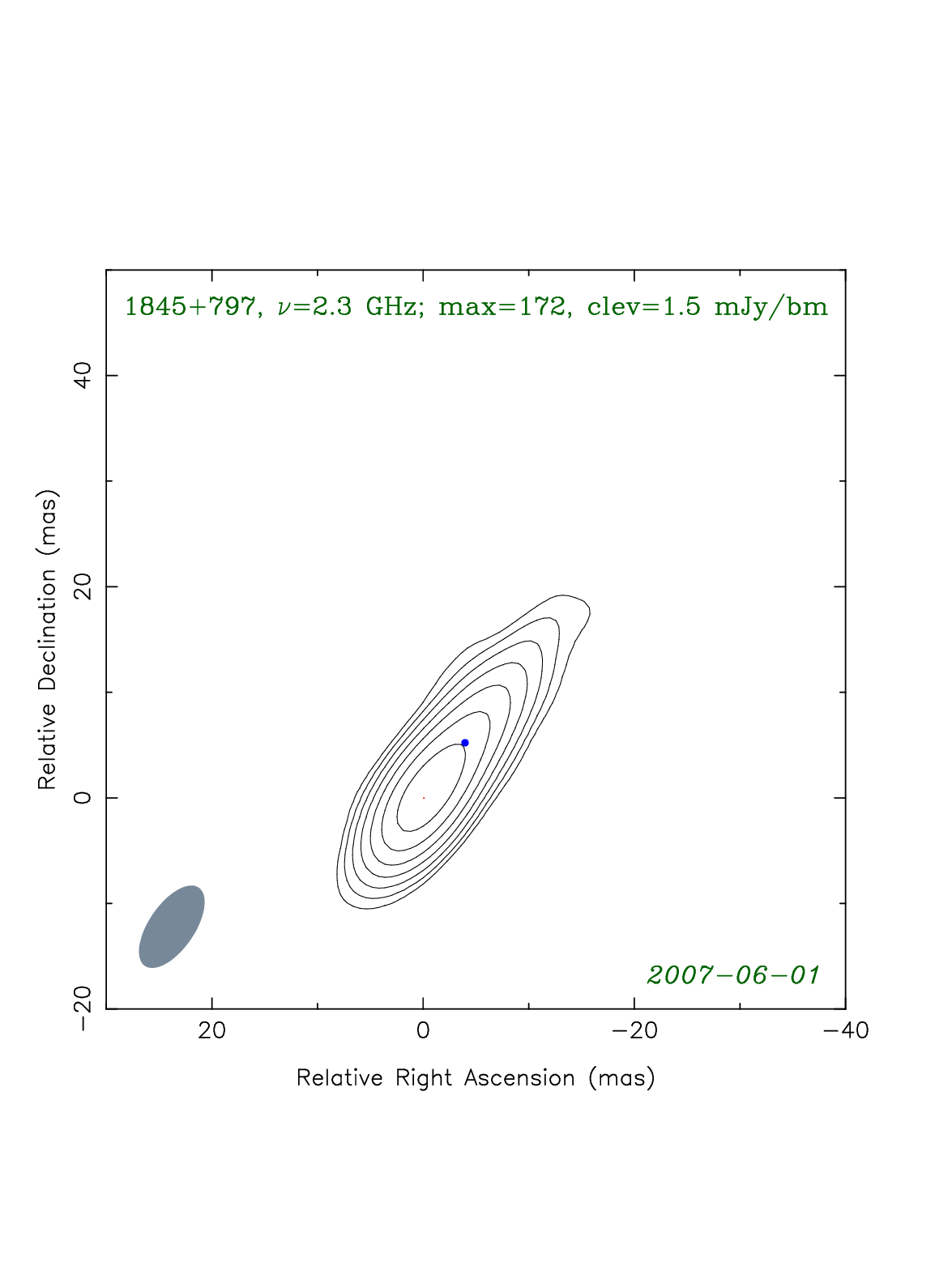}
 \includegraphics[width=0.33\textwidth,angle=0,trim=0.5cm 3cm 1.6cm 5cm,clip]{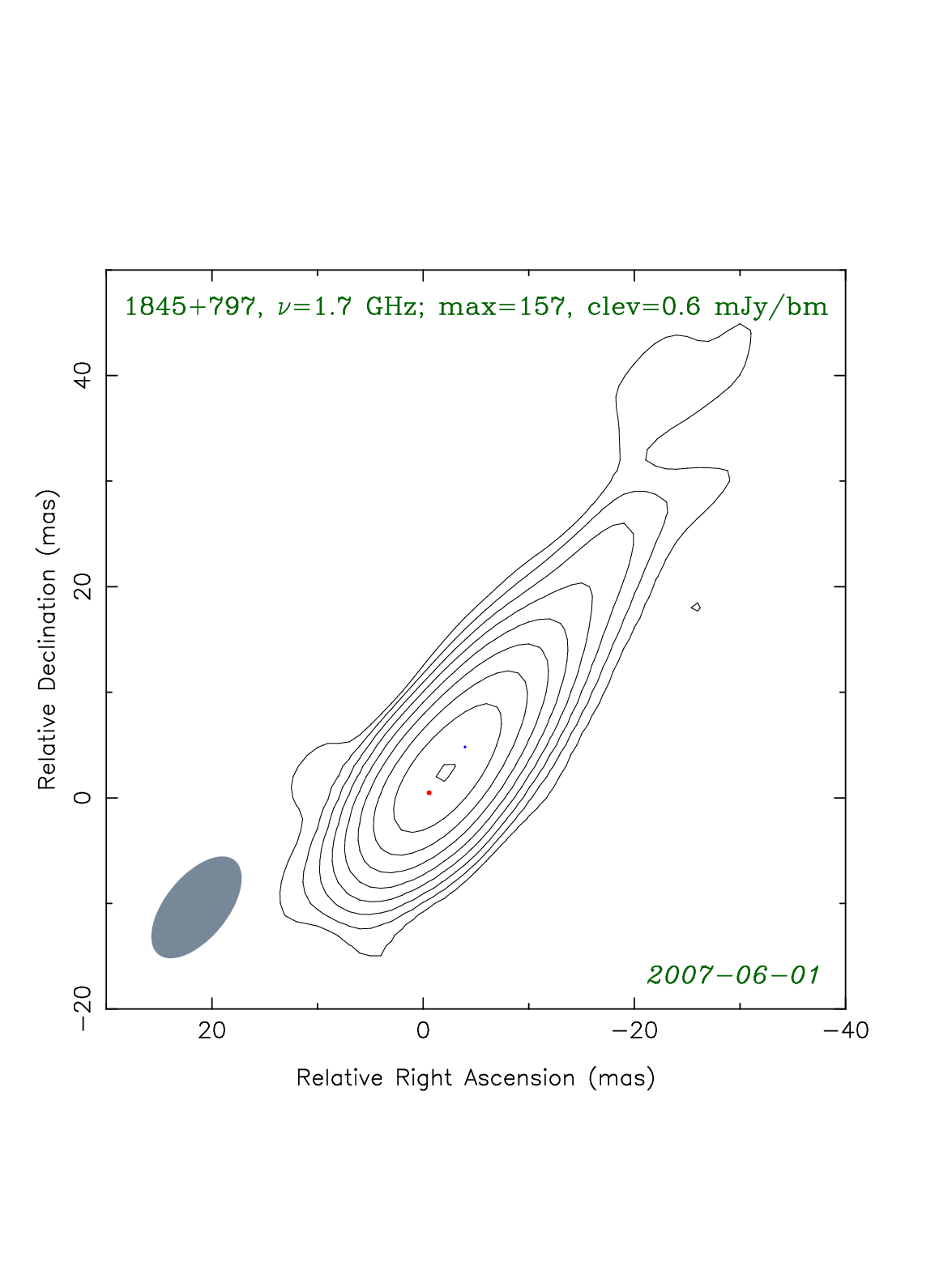}
 \includegraphics[width=0.33\textwidth,angle=0,trim=0.5cm 3cm 1.6cm 5cm,clip]{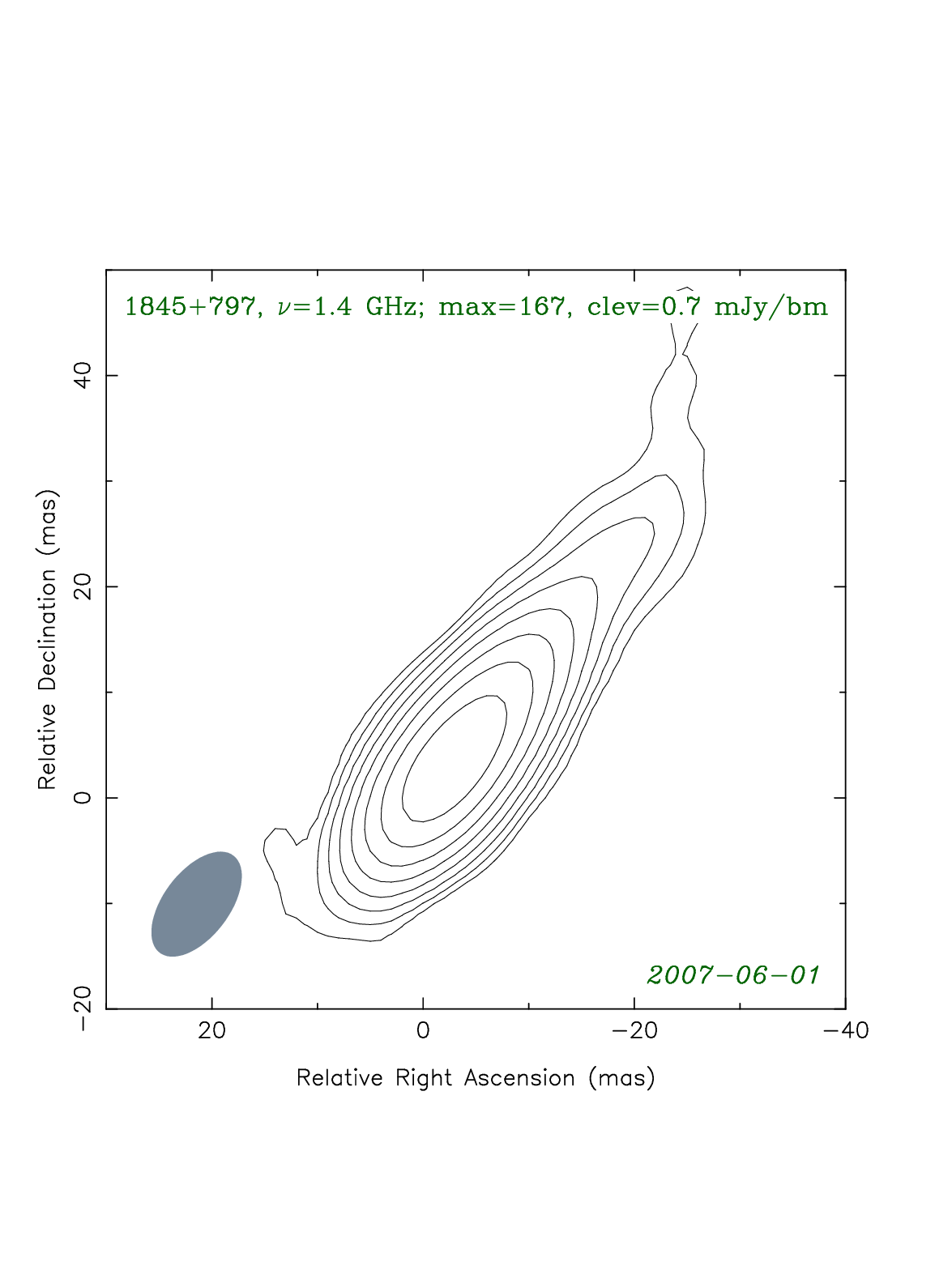}

 \caption{continued...}
\end{figure*}

\clearpage
\addtocounter{figure}{-1}
\begin{figure*}
 \centering

 \includegraphics[width=0.33\textwidth,angle=0,trim=0.5cm 3cm 1.6cm 5cm,clip]{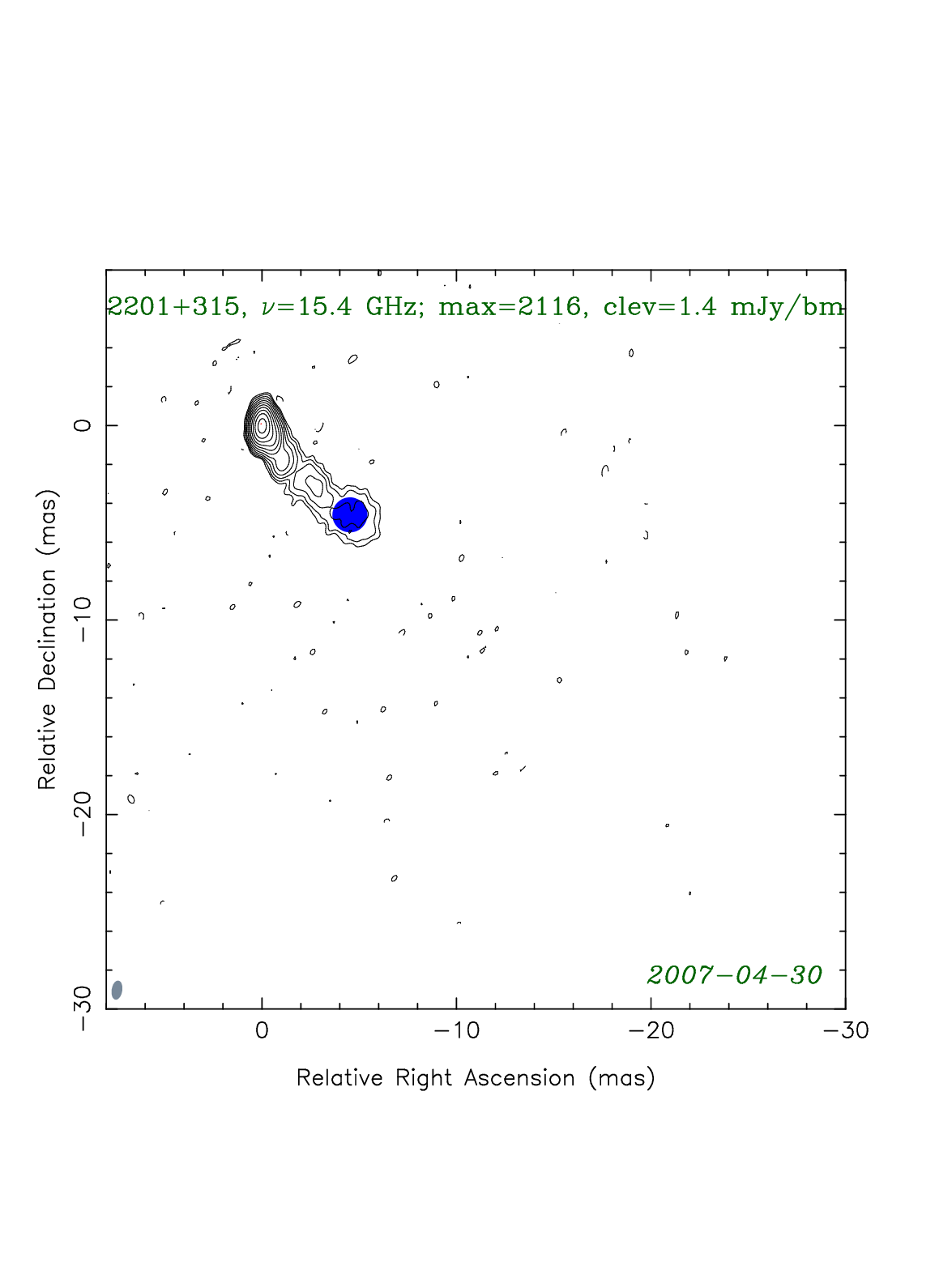}
 \includegraphics[width=0.33\textwidth,angle=0,trim=0.5cm 3cm 1.6cm 5cm,clip]{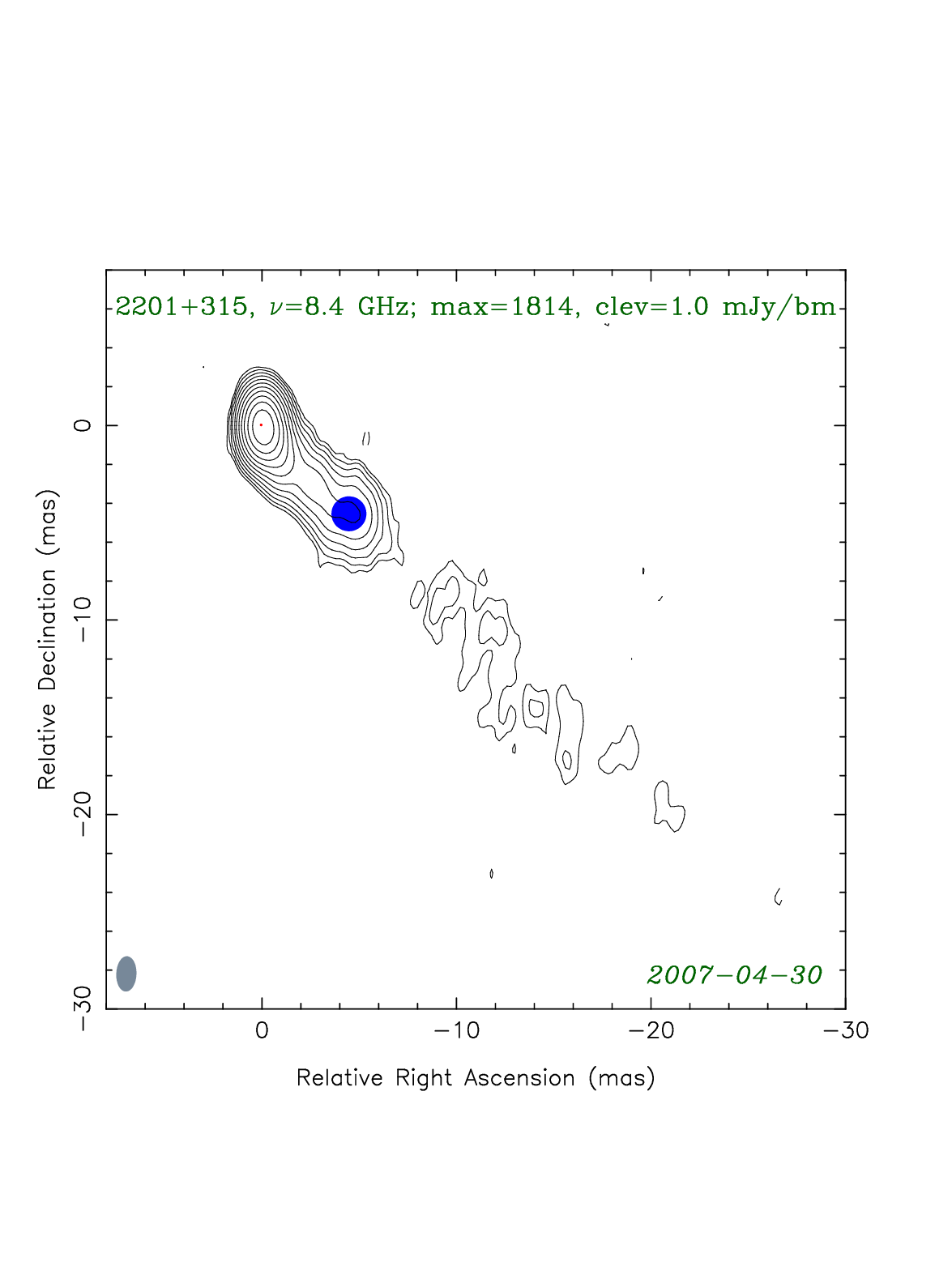}
 \includegraphics[width=0.33\textwidth,angle=0,trim=0.5cm 3cm 1.6cm 5cm,clip]{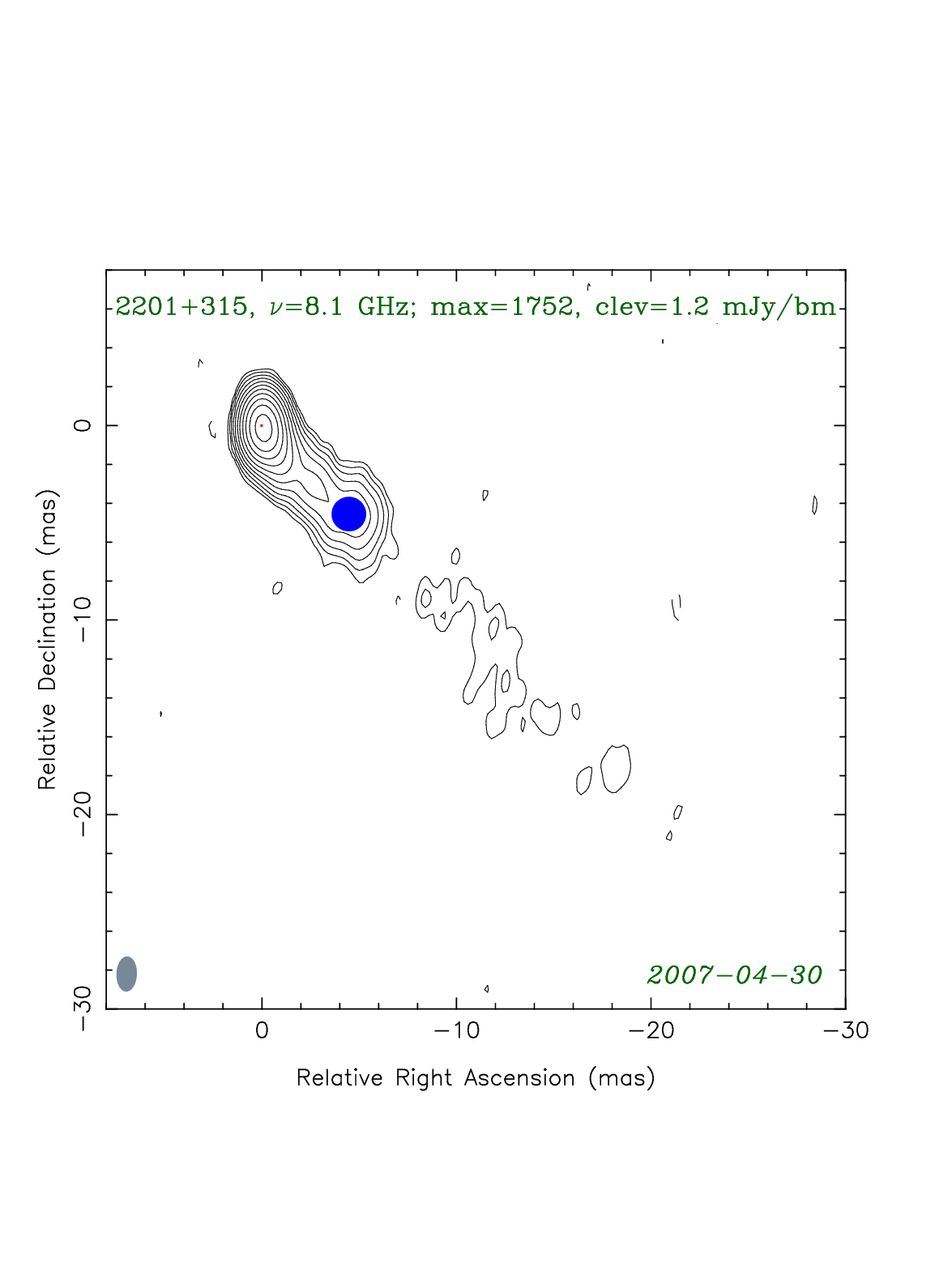}

 \includegraphics[width=0.33\textwidth,angle=0,trim=0.5cm 3cm 1.6cm 5cm,clip]{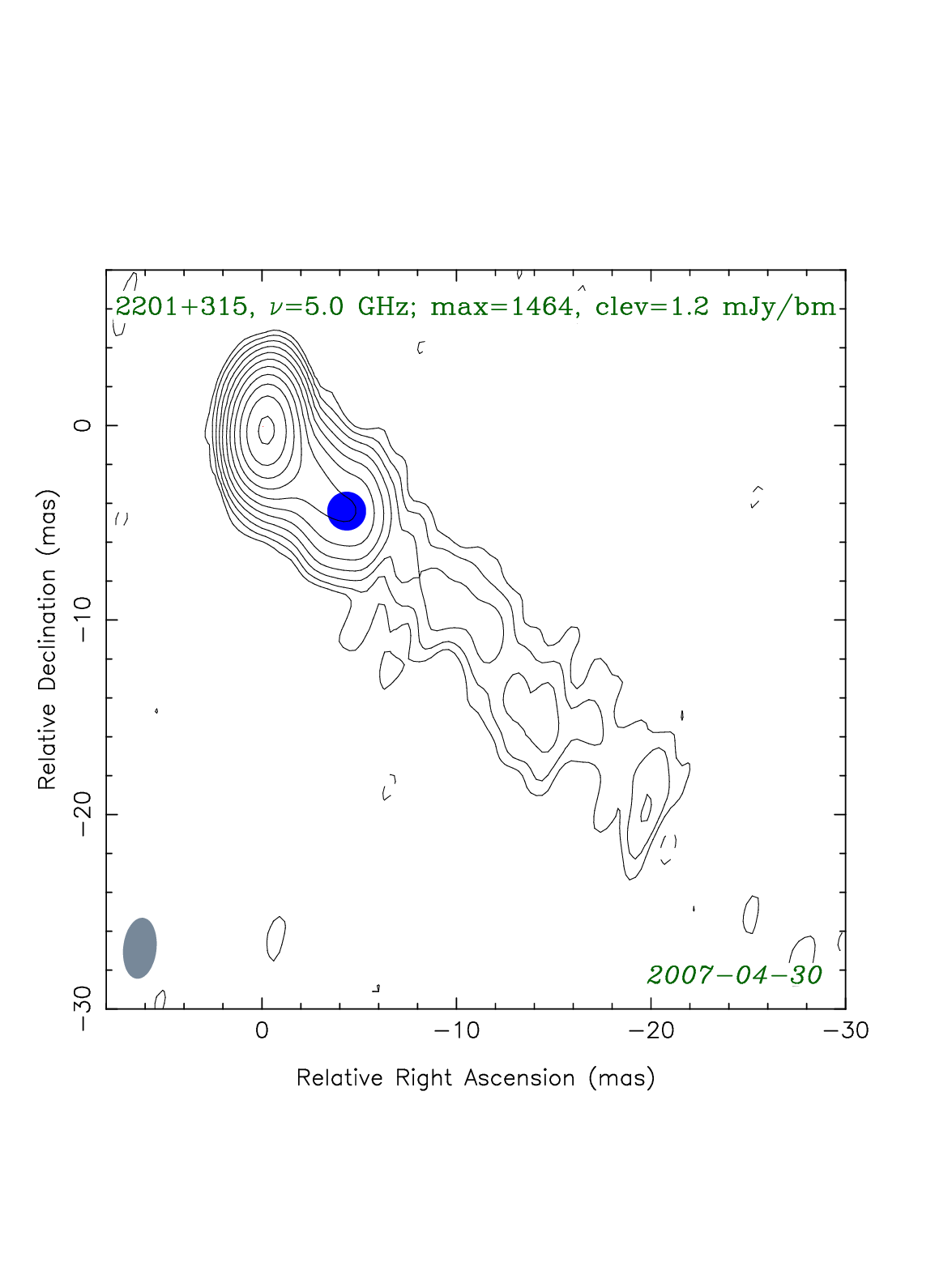}
 \includegraphics[width=0.33\textwidth,angle=0,trim=0.5cm 3cm 1.6cm 5cm,clip]{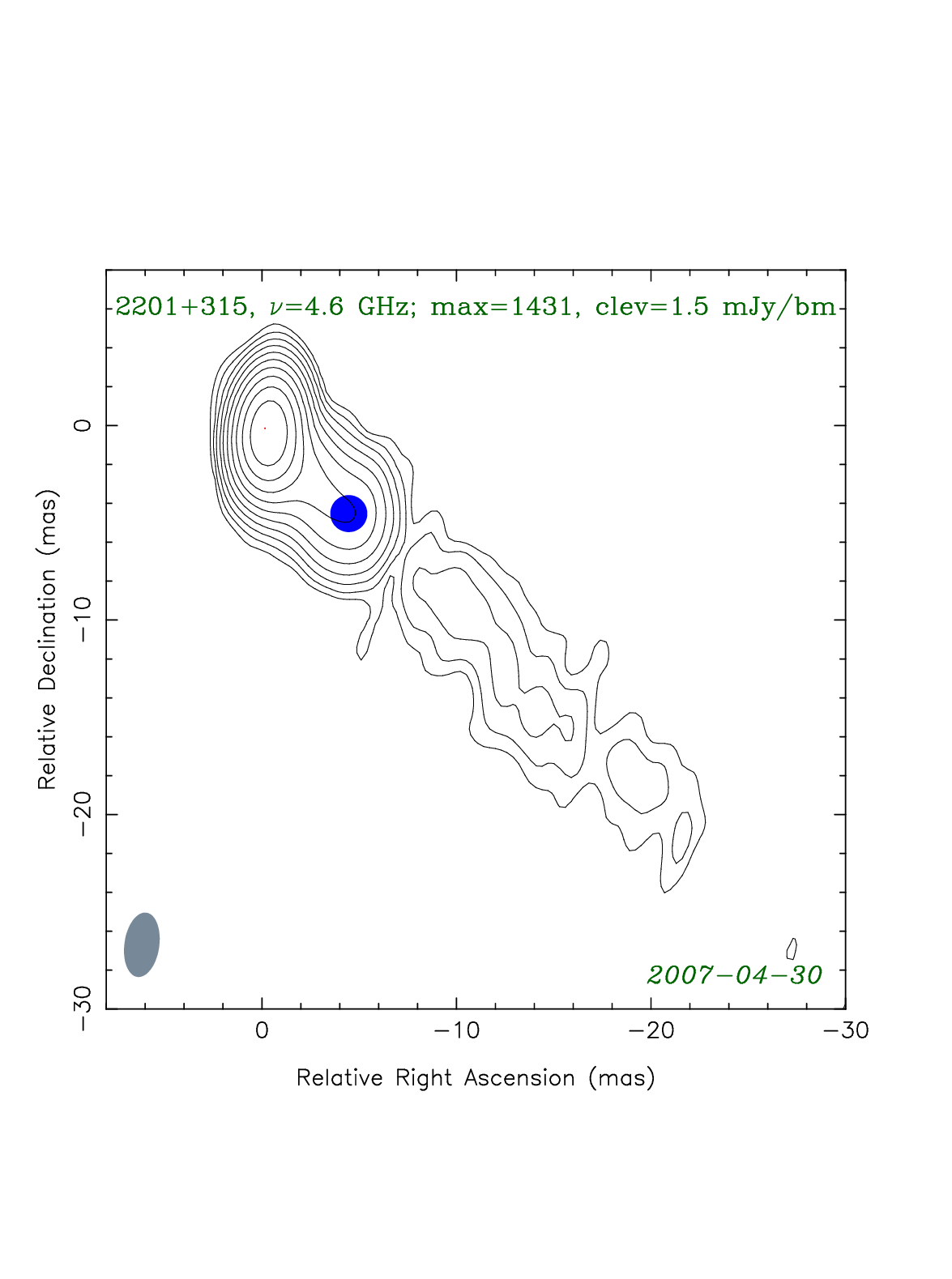}
 \includegraphics[width=0.33\textwidth,angle=0,trim=0.5cm 3cm 1.6cm 5cm,clip]{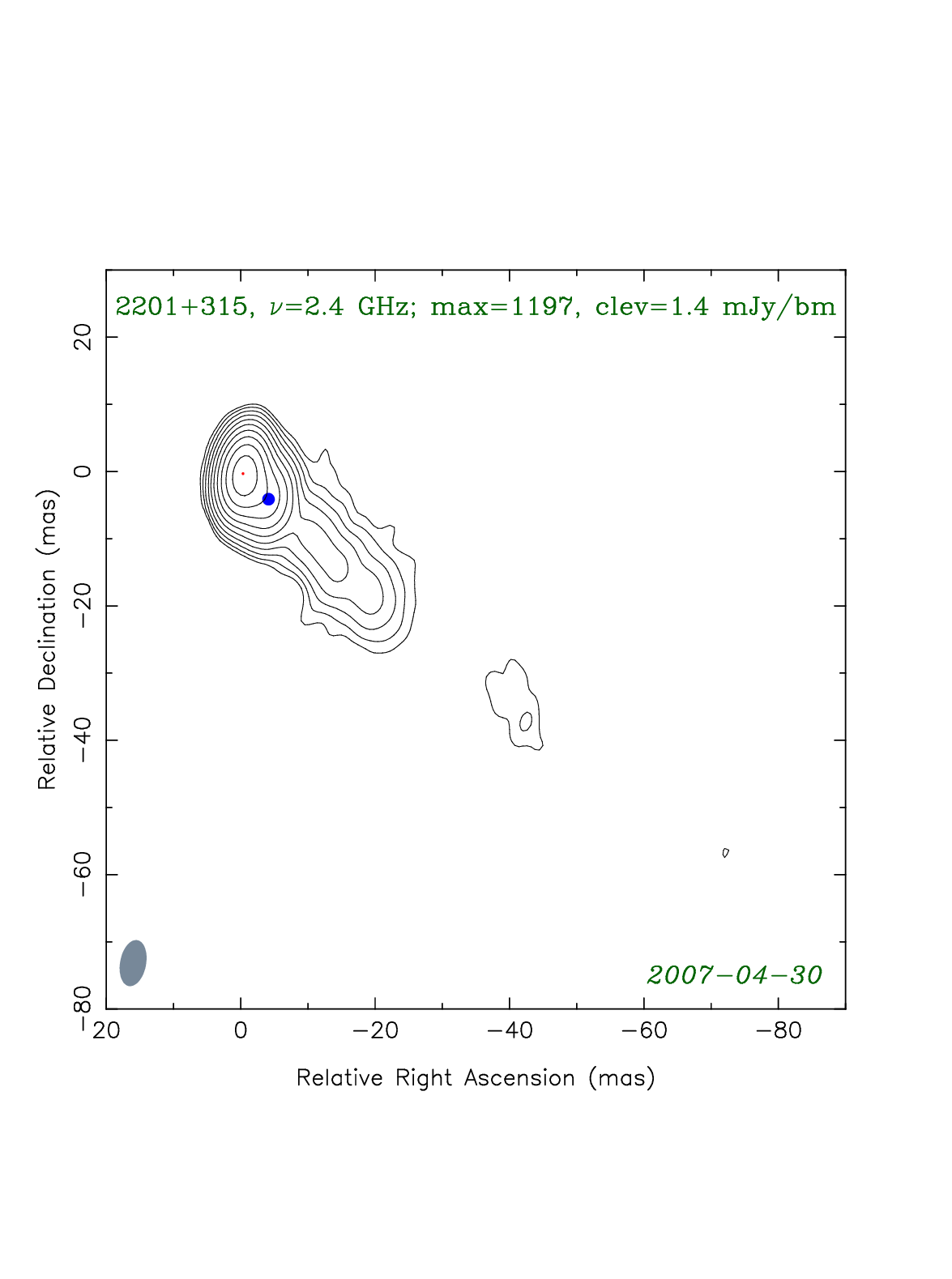}

 \includegraphics[width=0.33\textwidth,angle=0,trim=0.5cm 3cm 1.6cm 5cm,clip]{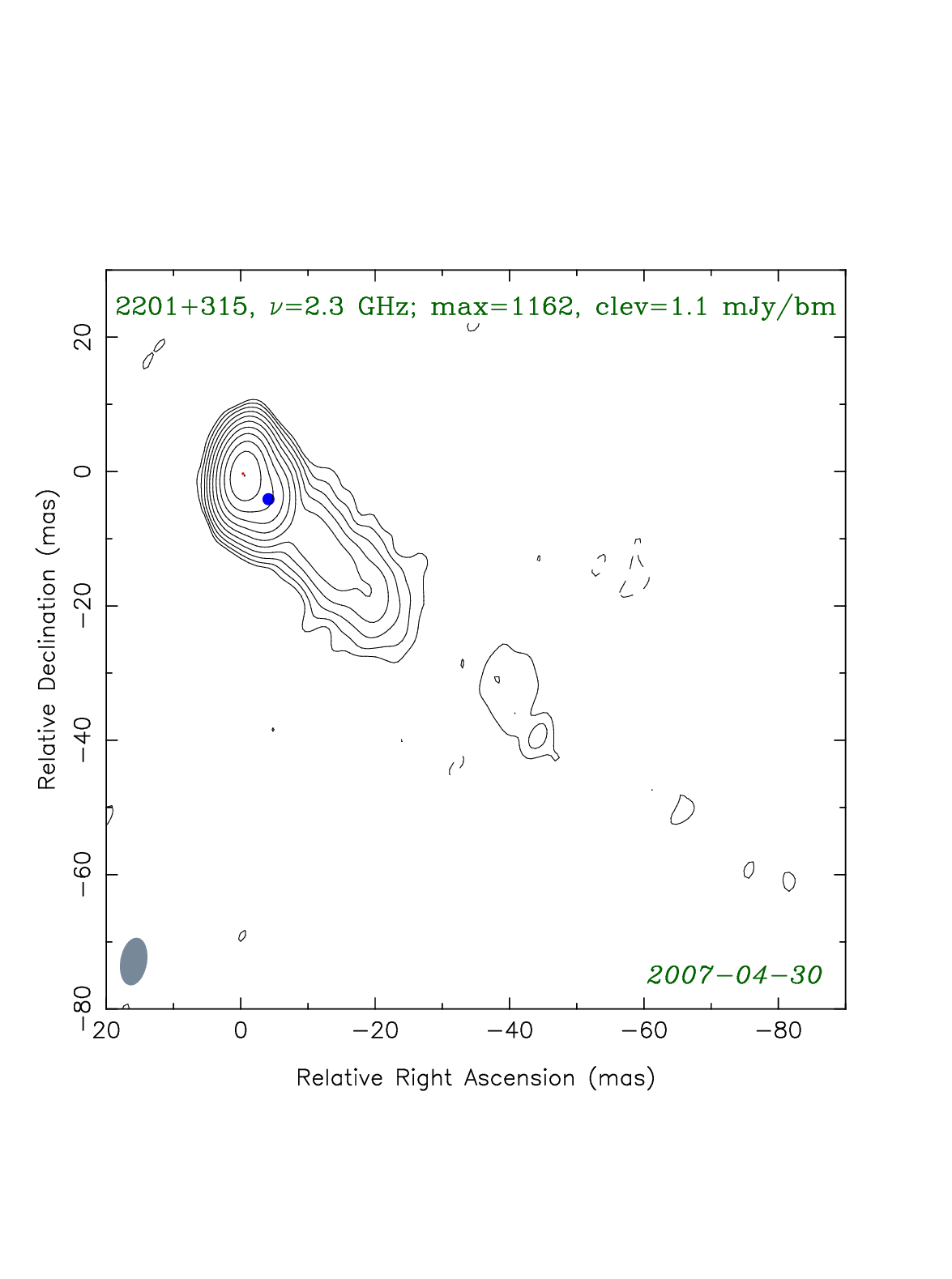}
 \includegraphics[width=0.33\textwidth,angle=0,trim=0.5cm 3cm 1.6cm 5cm,clip]{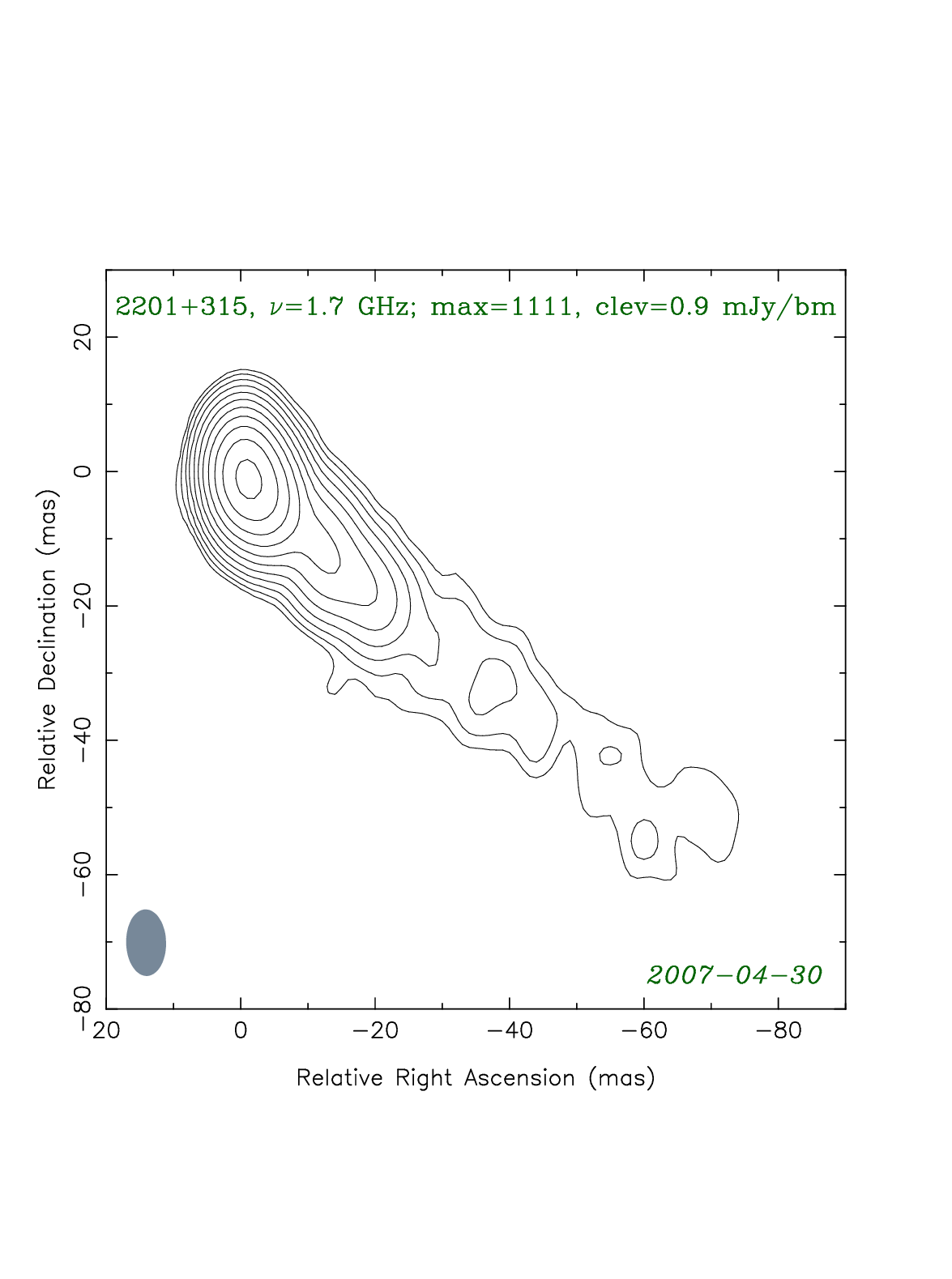}
 \includegraphics[width=0.33\textwidth,angle=0,trim=0.5cm 3cm 1.6cm 5cm,clip]{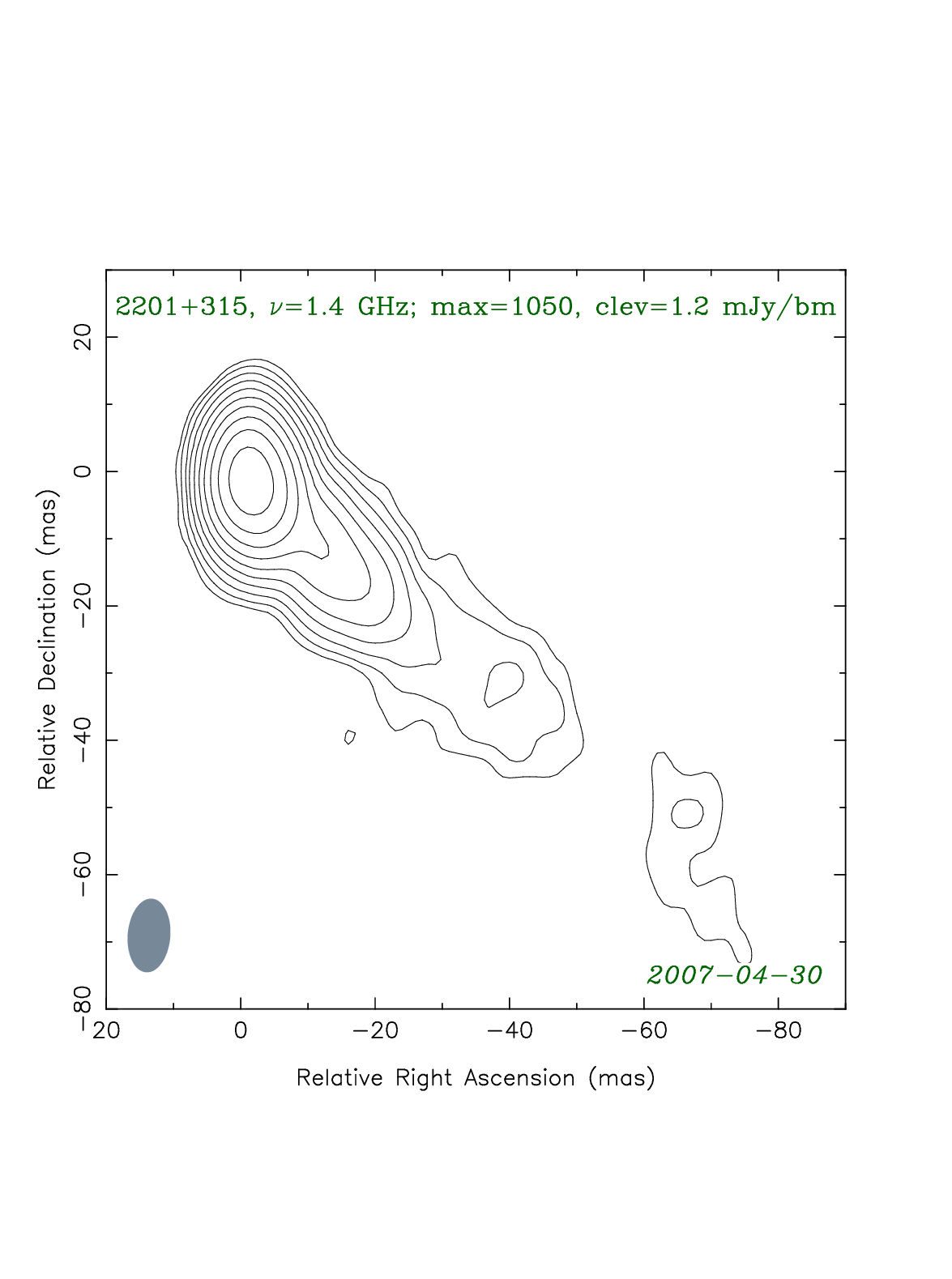}

 \caption{continued...}
\end{figure*}

\clearpage
\addtocounter{figure}{-1}
\begin{figure*}
 \centering

 \includegraphics[width=0.33\textwidth,angle=0,trim=0.5cm 3cm 1.6cm 5cm,clip]{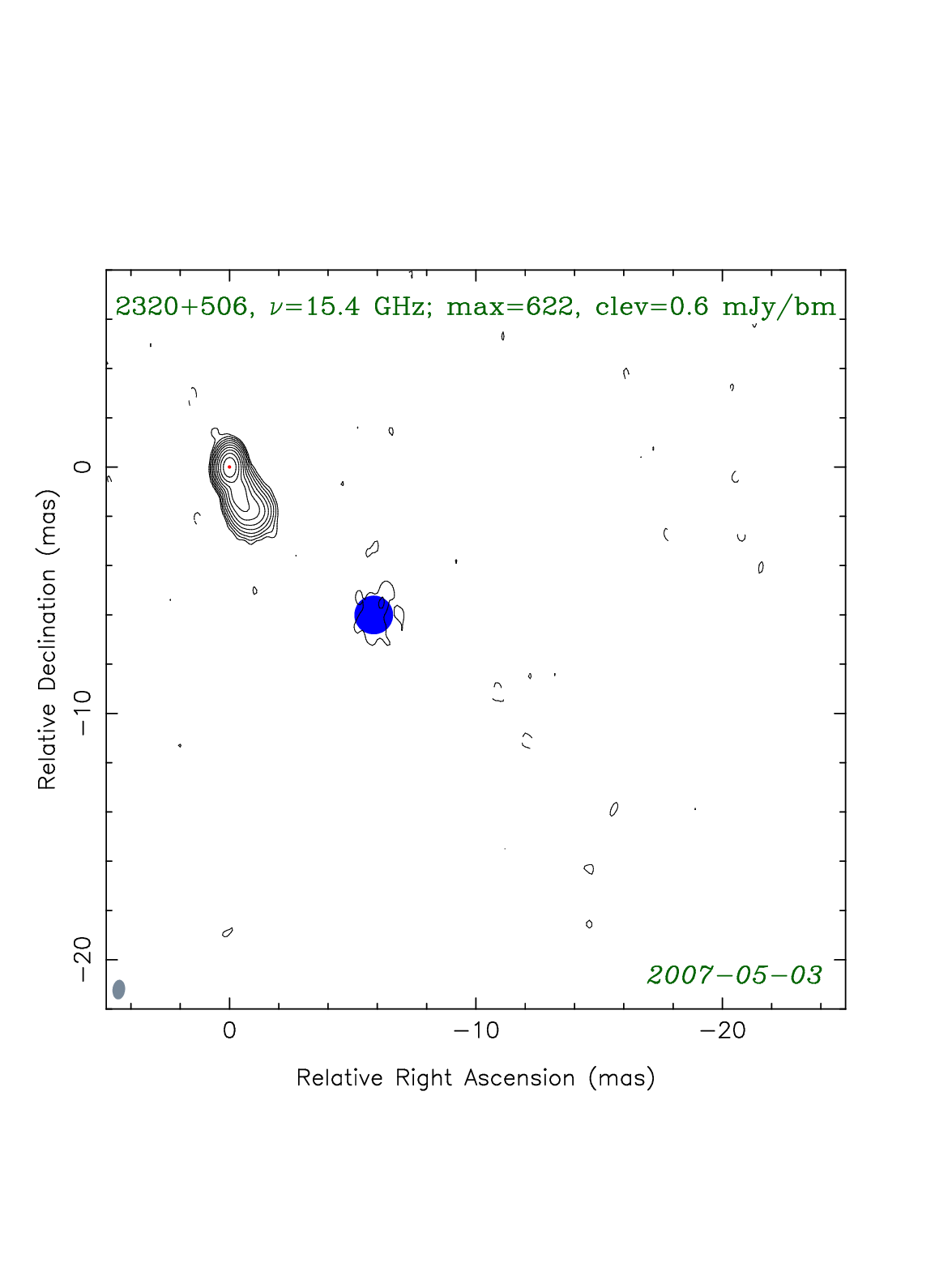}
 \includegraphics[width=0.33\textwidth,angle=0,trim=0.5cm 3cm 1.6cm 5cm,clip]{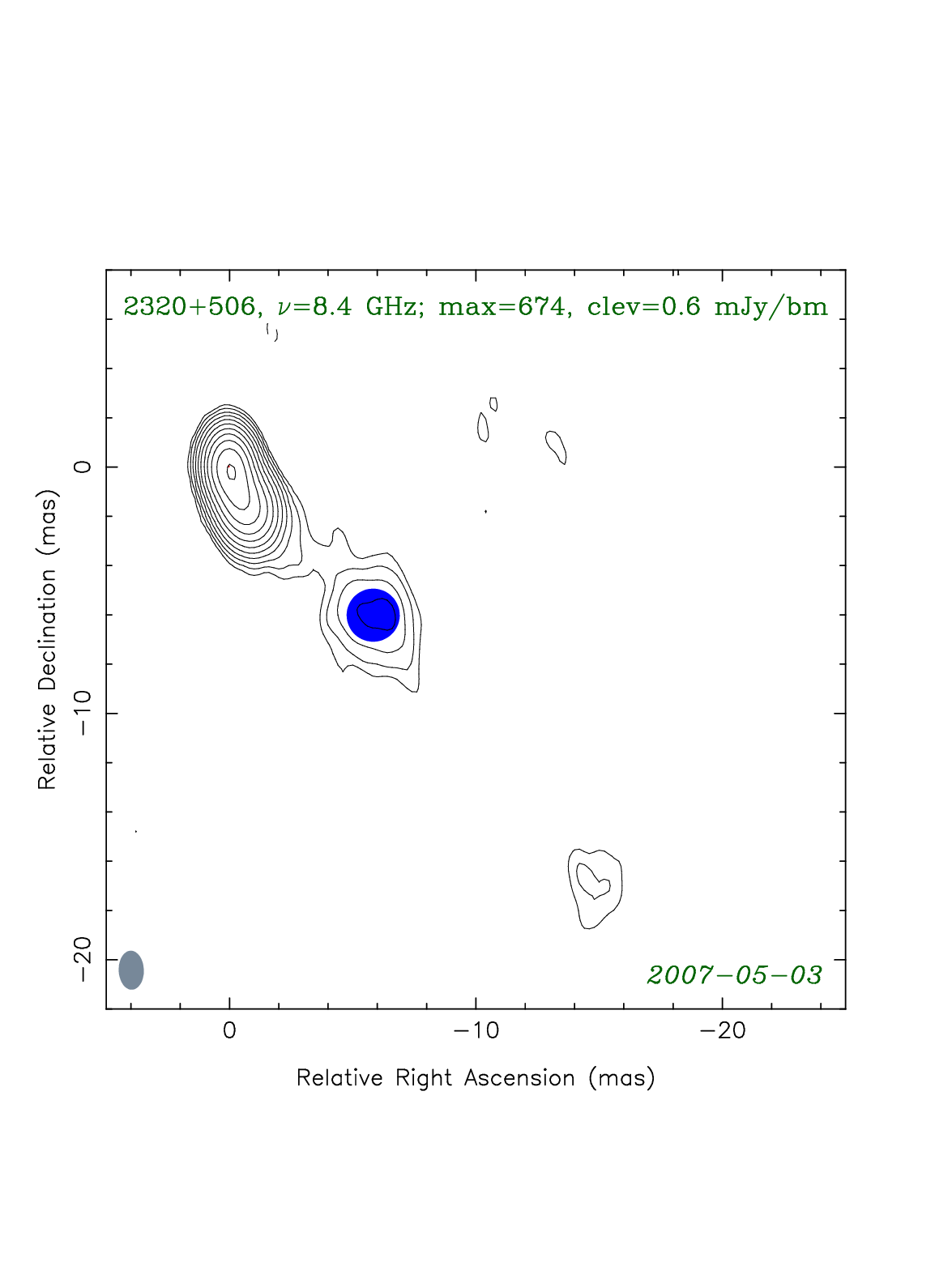}
 \includegraphics[width=0.33\textwidth,angle=0,trim=0.5cm 3cm 1.6cm 5cm,clip]{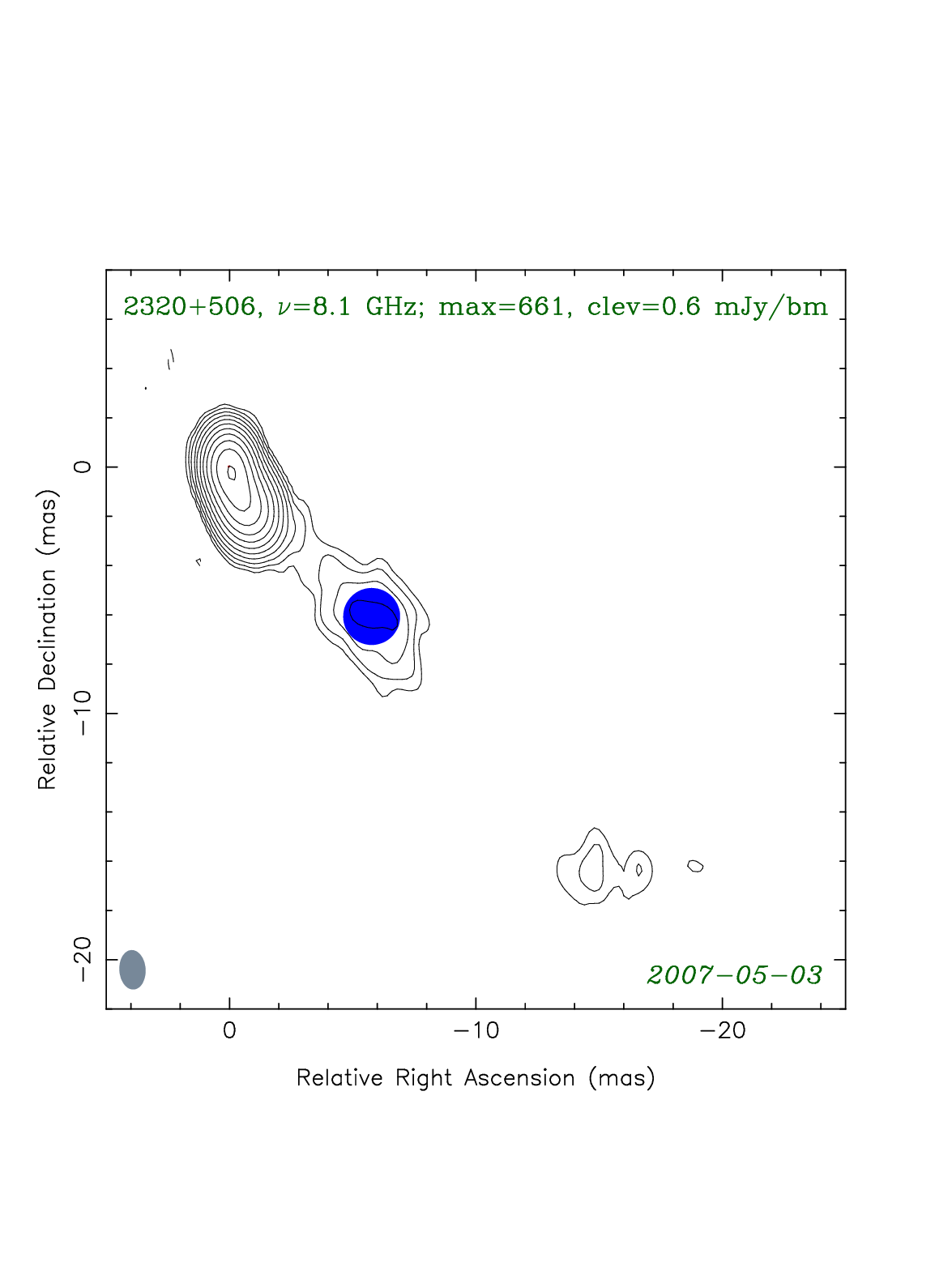}

 \includegraphics[width=0.33\textwidth,angle=0,trim=0.5cm 3cm 1.6cm 5cm,clip]{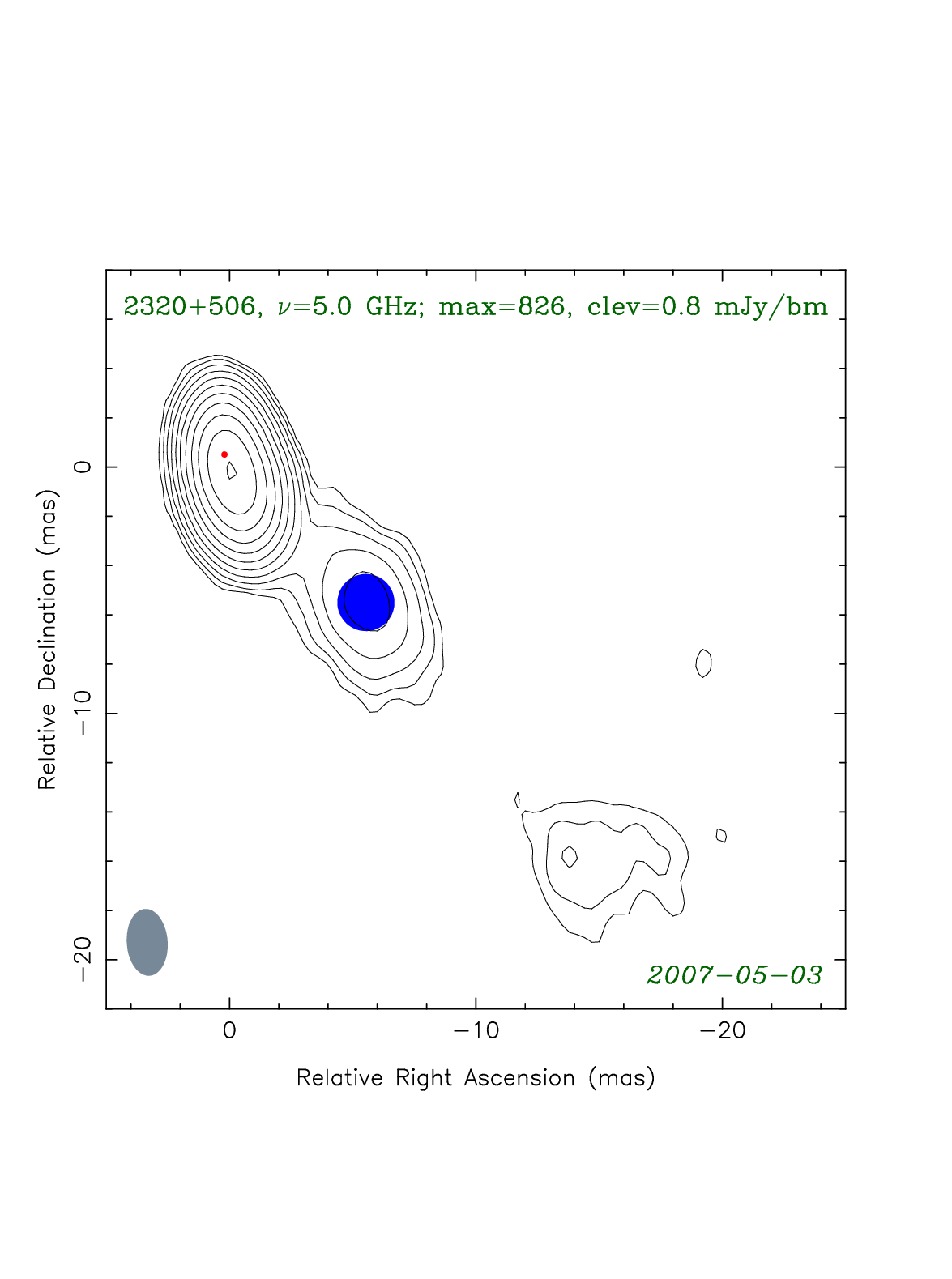}
 \includegraphics[width=0.33\textwidth,angle=0,trim=0.5cm 3cm 1.6cm 5cm,clip]{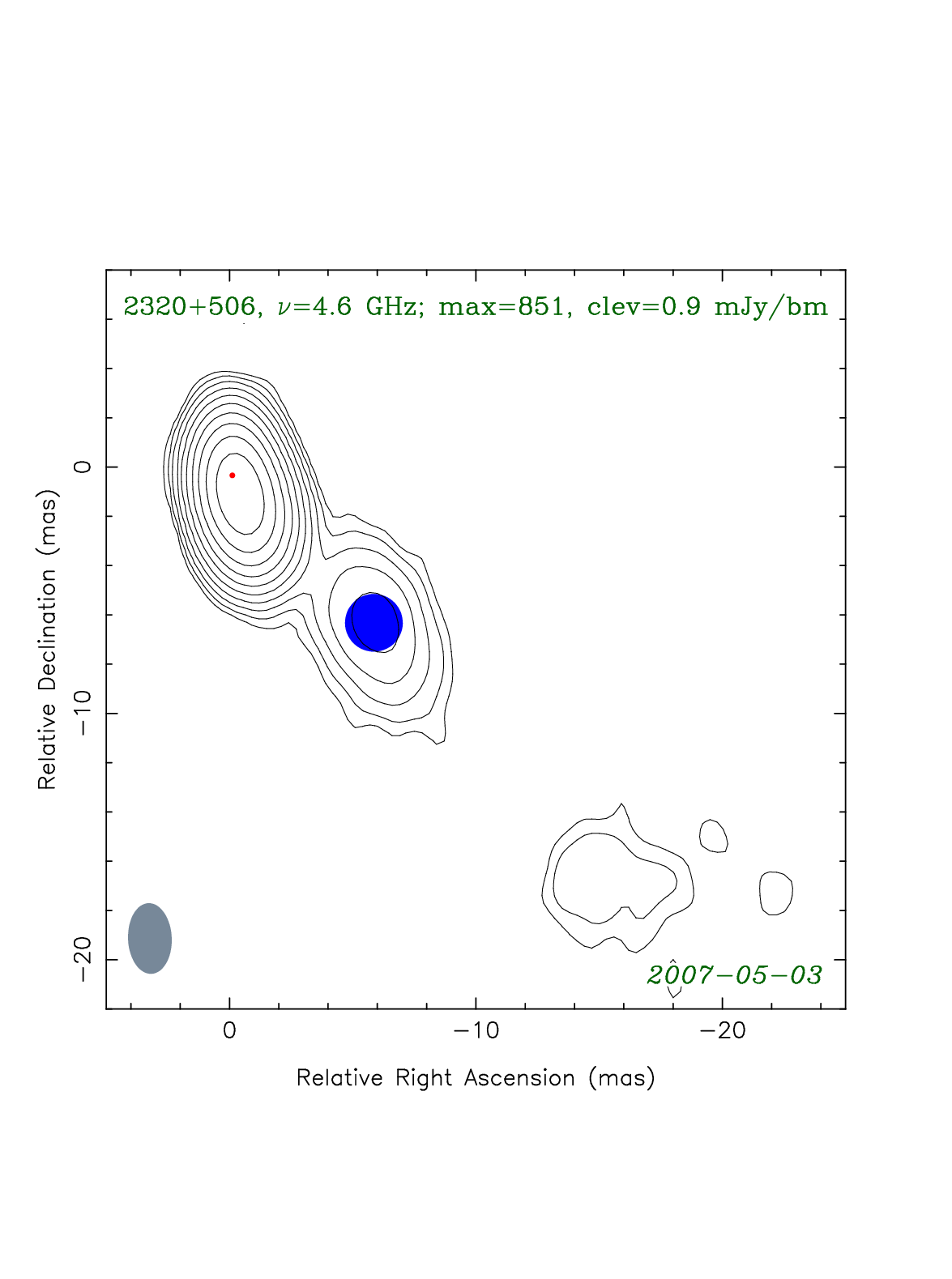}
 \includegraphics[width=0.33\textwidth,angle=0,trim=0.5cm 3cm 1.6cm 5cm,clip]{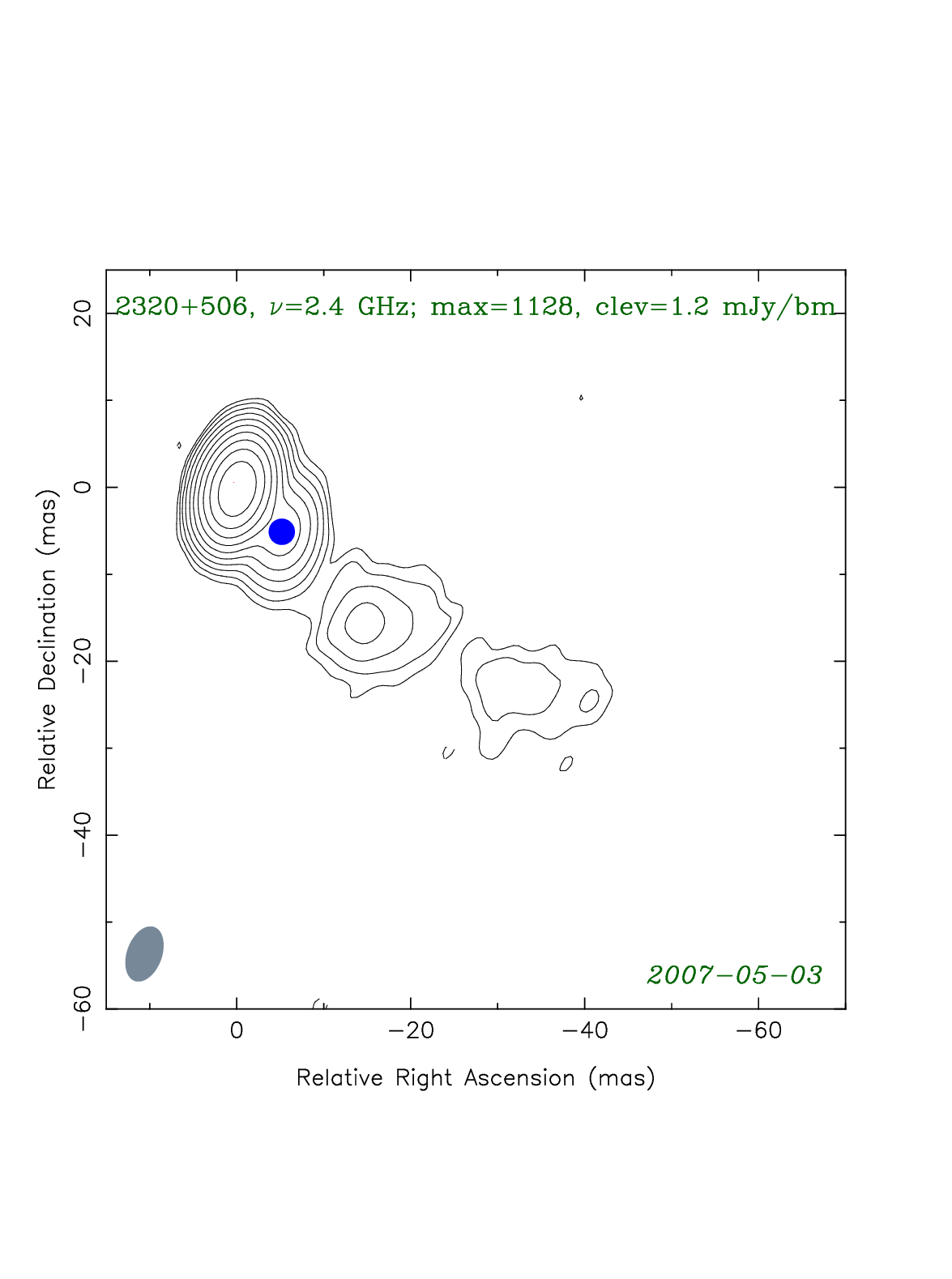}

 \includegraphics[width=0.33\textwidth,angle=0,trim=0.5cm 3cm 1.6cm 5cm,clip]{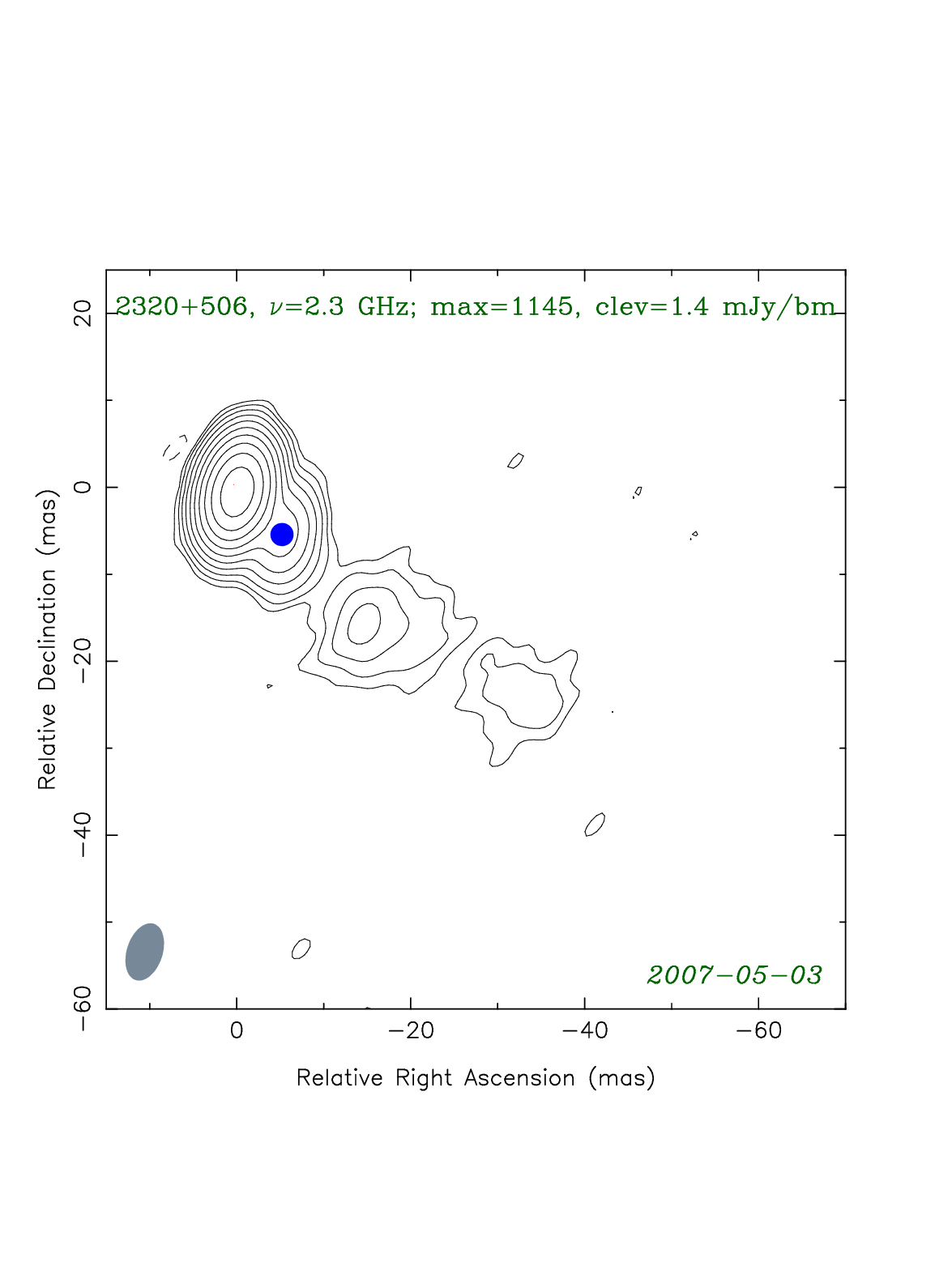}
 \includegraphics[width=0.33\textwidth,angle=0,trim=0.5cm 3cm 1.6cm 5cm,clip]{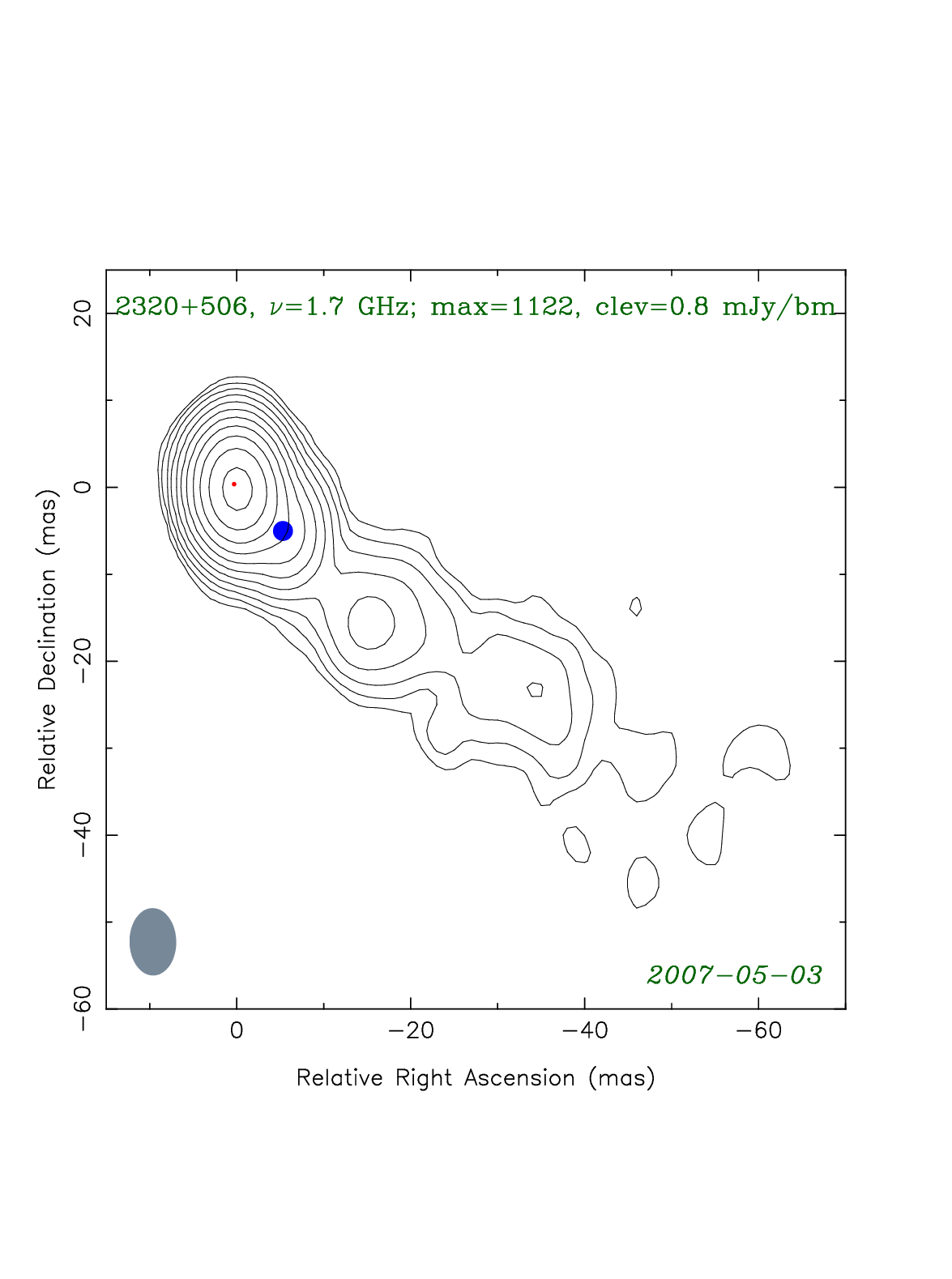}
 \includegraphics[width=0.33\textwidth,angle=0,trim=0.5cm 3cm 1.6cm 5cm,clip]{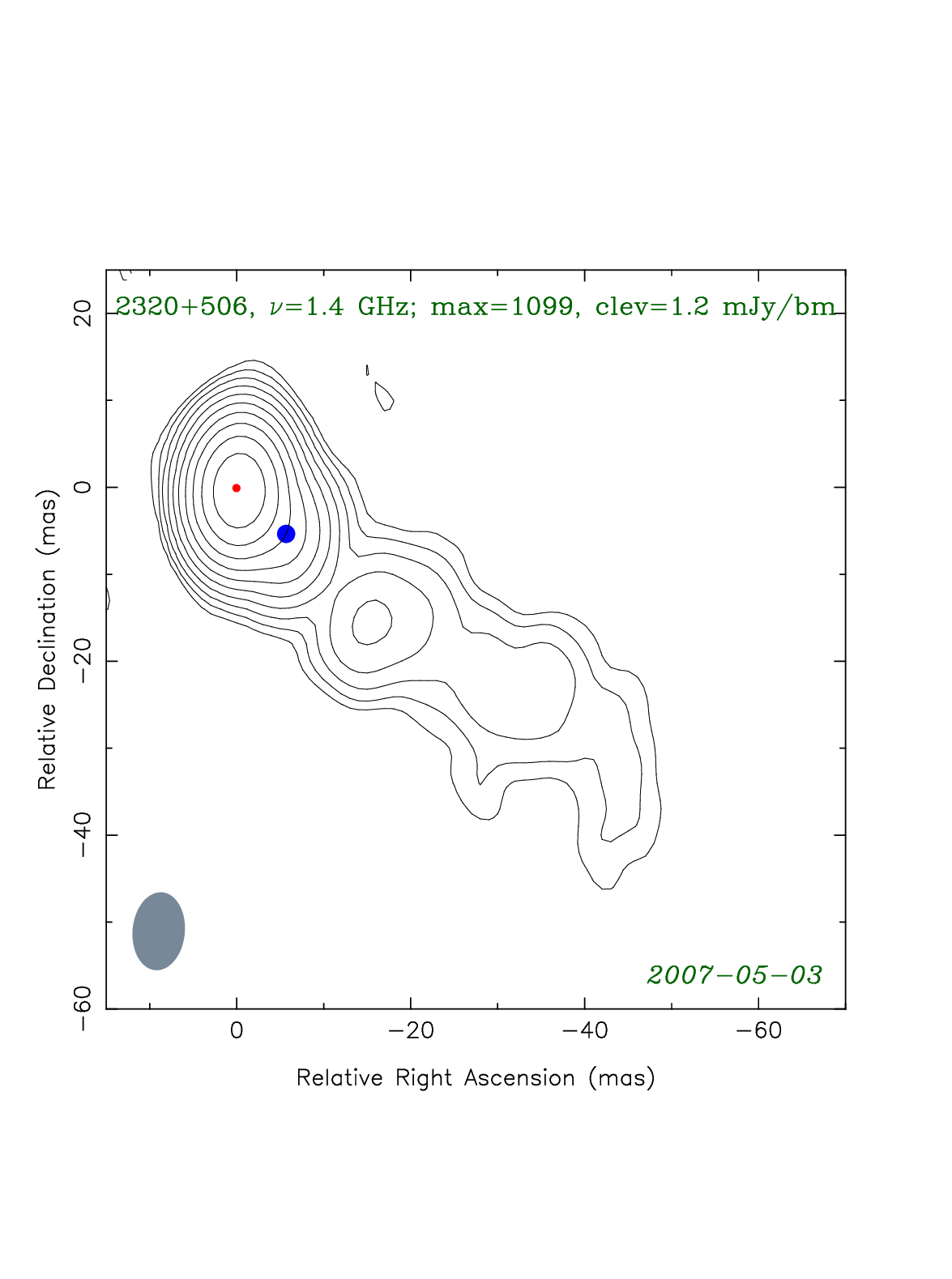}

 \caption{continued...}
\end{figure*}
}

%__________________________________________________________________

\section{Model fitting and core shift measurements}
\label{sec:core_shift_measurement}

The structure of each source at each frequency was modeled in the 
visibility ($uv$) plane with a number of circular Gaussian components using the 
\textit{Difmap} software \citep{difmap}. 
A suitable number of components (typically $\sim 6$) was used to account for all
significantly detected features of the source structure at a given frequency.
The components were visually cross-identified between frequencies on
the basis of their position  with respect to the core.
The VLBI core is identified at each frequency as the bright component at the apparent jet
base. As pointed out in Section~\ref{sec:sample_selection}, the target
sources were preselected in a way that such an identification would be possible in
a wide frequency range for the brightest jet components.

\begin{figure}[tb]            
 \centering
 \includegraphics[width=0.5\textwidth,angle=0,trim=0cm 0cm 0cm 0cm,clip]{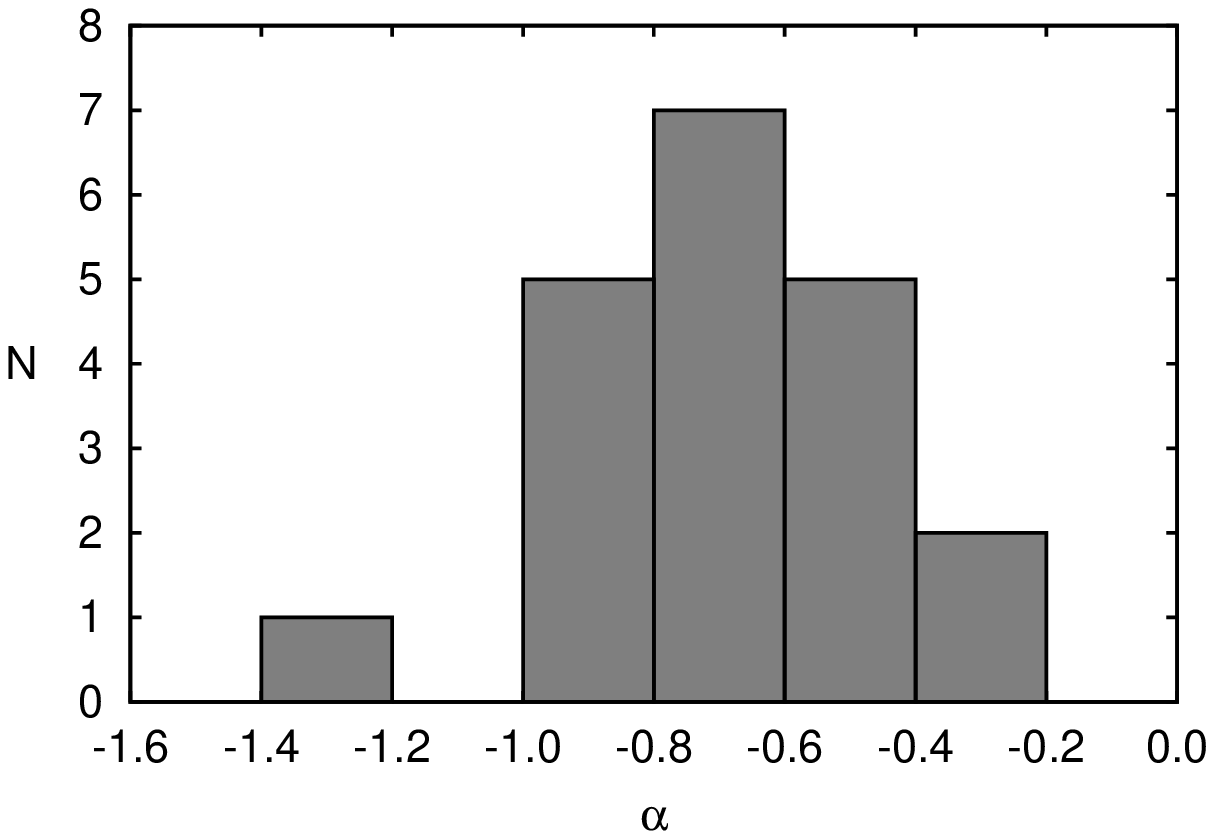}
 \caption{Distribution of the spectral index ($\alpha$, flux density 
 $S_\nu \propto \nu^\alpha$) of reference jet
 components. The indexes were computed by fitting a power law to the
 $1.4$--$15.4$~GHz spectrum. The component flux at each frequency was
 obtained from the $uv$-model. The source with the steepest reference
 component spectrum ($\alpha=-1.29$) is 1011$+$250.}
 \label{fig:a}
\end{figure}

One bright, optically-thin jet component, which could be
identified across all (or most of) the observing frequency range, was
chosen, and
its distance from the core was measured at each frequency. 
Model component positions and sizes for the core, as well as the jet
feature used for this analysis, are indicated in Fig.~\ref{images}.
In some rare cases we were unable to construct a consistent model to
allow for accurate and robust component cross-identification with other
frequencies. In these cases no model components are shown in
Fig.~\ref{images}, and no data are used in the analysis.
Figure~\ref{fig:a} shows the distribution of spectral indexes of the
reference components, indicating that they are optically thin in the
considered frequency range.

 The technique of using an optically thin
jet component as a reference point for the core position measurement has
previously been successfully tested on $S$/$X$ band RDV data by
\cite{2008A&A...483..759K} in several ways: {\em (i)}~similar core shift
values were obtained referencing to different jet features of the same
source; {\em (ii)}~the core shift for the quasar 1655$+$077 agreed
within the errors with previous phase-referenced measurements \citep{BH103};
{\em (iii)}~two RDV epochs of the quasar 1642$+$690 resulted in values
being in good agreement.
\cite{2008A&A...483..759K} have also investigated the effect of
the different $uv$-coverage resulting in a different resolution (and amount
of blending between the core and nearby jet regions) at $S$ and $X$
bands on core shift measurements. The effect was found to be small
compared to the core position shift between these bands (see Fig.~5 in
\citealt{2008A&A...483..759K}).

In addition to that, we performed the following test to look for 
indications of blending in our $1.4$--$15.4$~GHz results. 
If significant blending is present, we should measure systematically larger
core shifts for sources with jet directions aligned
with the position angle (P.A.) of the elongated VLBA beam. However, no such dependence is
found in the data.
 
To assess the quality of image alignment using a reference
jet component, we inspected spectral index maps of the sources. 
The spectral index maps were computed by constructing a
data cube from the images at different frequencies convolved with the same beam
(naturally weighted beam corresponding to the lowest frequency) and
fitting a power law to the spectrum in each individual pixel of the data cube.
The power law index (corresponding to the slope of a straight line if plotted
in logarithmic scale) is the spectral index. A map with a smooth spectral 
index gradient along the jet and an absent/weak gradient across the jet
would indicate, in general, the correct alignment of images 
obtained at different frequencies (see, e.g., 
\citealt{1984ApJ...276...56M,2008A&A...483..759K}).
In all cases the spectral index maps revealed no major problems with image
alignment, and the core was found to be the only optically thick feature in
the jet. The detailed discussion of the spectral imaging results will be
presented elsewhere.

\begin{figure*}[p!]
 \centering

 \includegraphics[width=0.32\textwidth,angle=0,trim=0.15cm 0cm 0cm 0.3cm,clip]{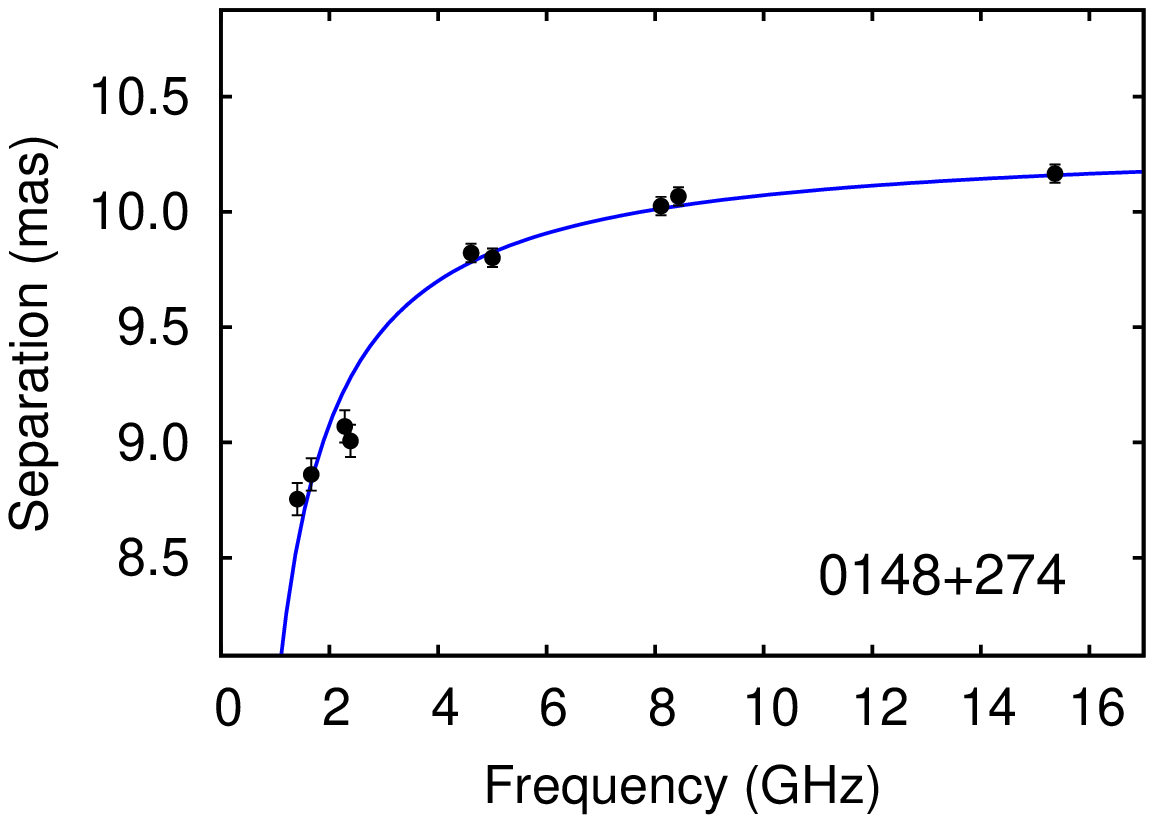}
 \includegraphics[width=0.32\textwidth,angle=0,trim=0.15cm 0cm 0cm 0.3cm,clip]{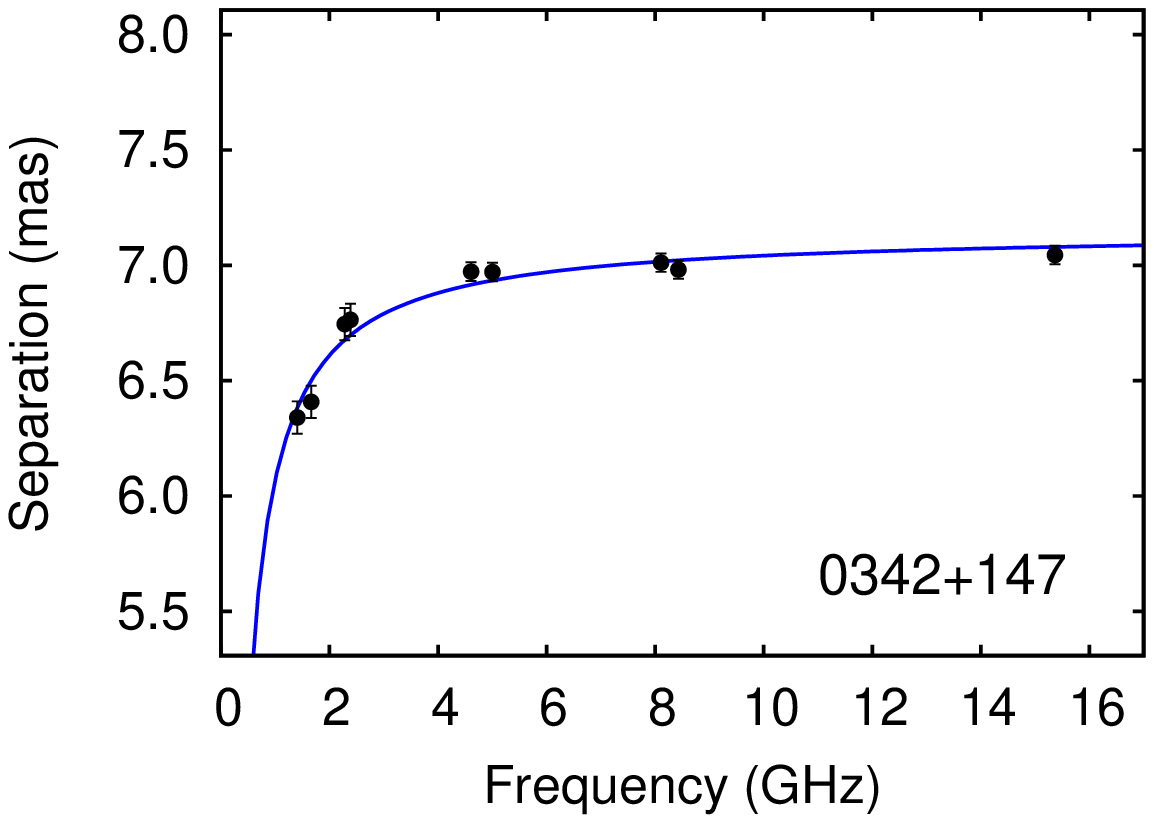}
 \includegraphics[width=0.32\textwidth,angle=0,trim=0.15cm 0cm 0cm 0.3cm,clip]{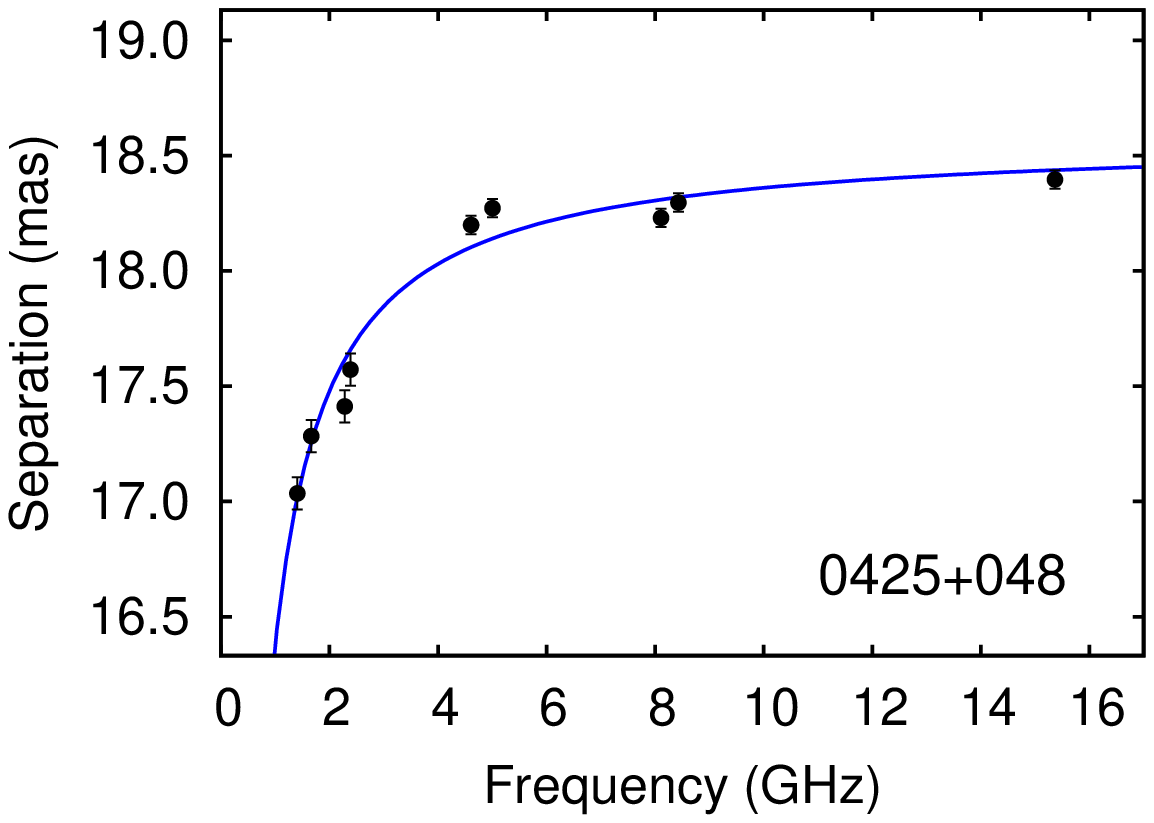}
\\
 \includegraphics[width=0.32\textwidth,angle=0,trim=0.15cm 0cm 0cm 0.3cm,clip]{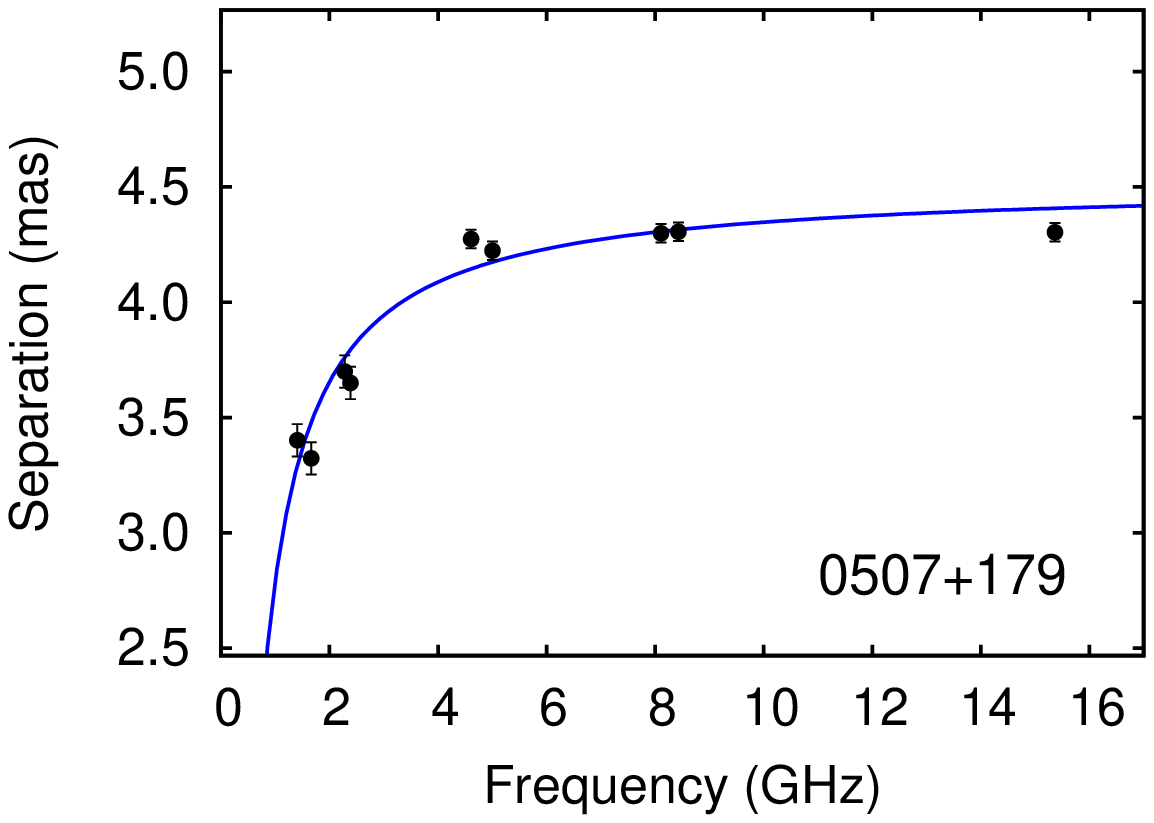}
 \includegraphics[width=0.32\textwidth,angle=0,trim=0.15cm 0cm 0cm 0.3cm,clip]{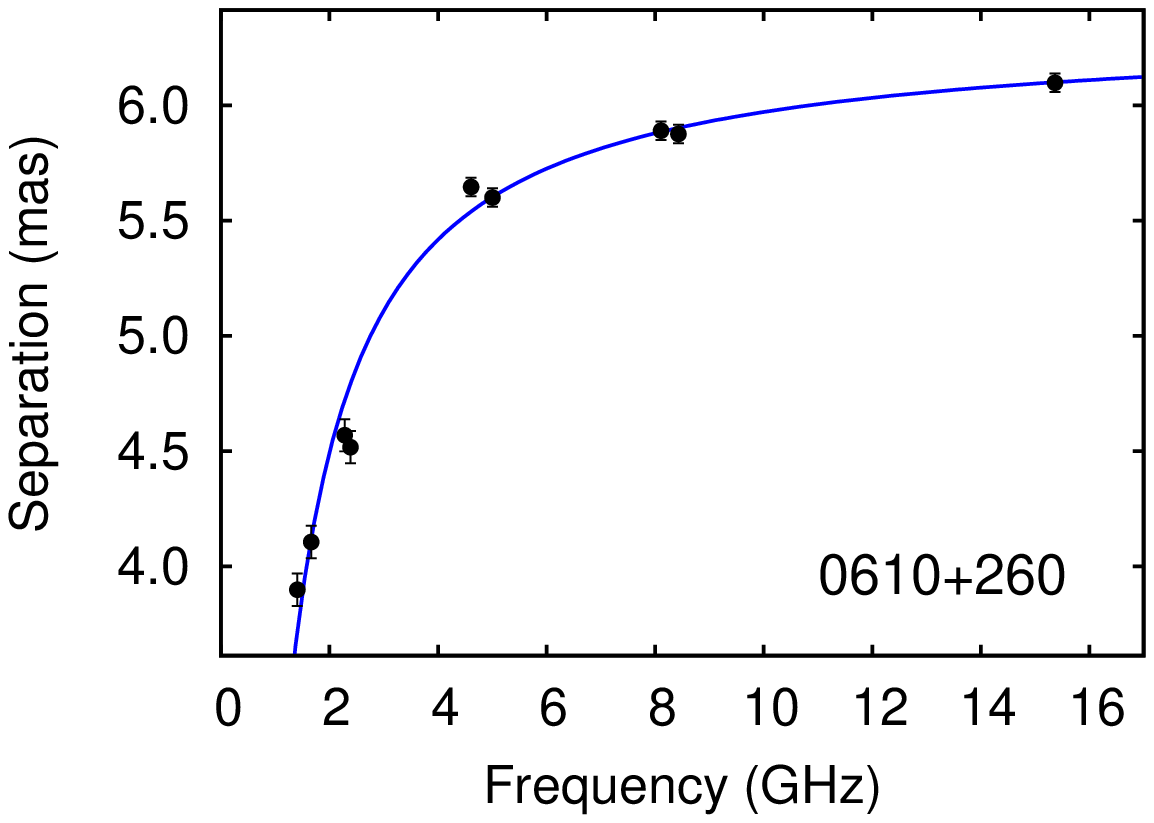}
 \includegraphics[width=0.32\textwidth,angle=0,trim=0.15cm 0cm 0cm 0.3cm,clip]{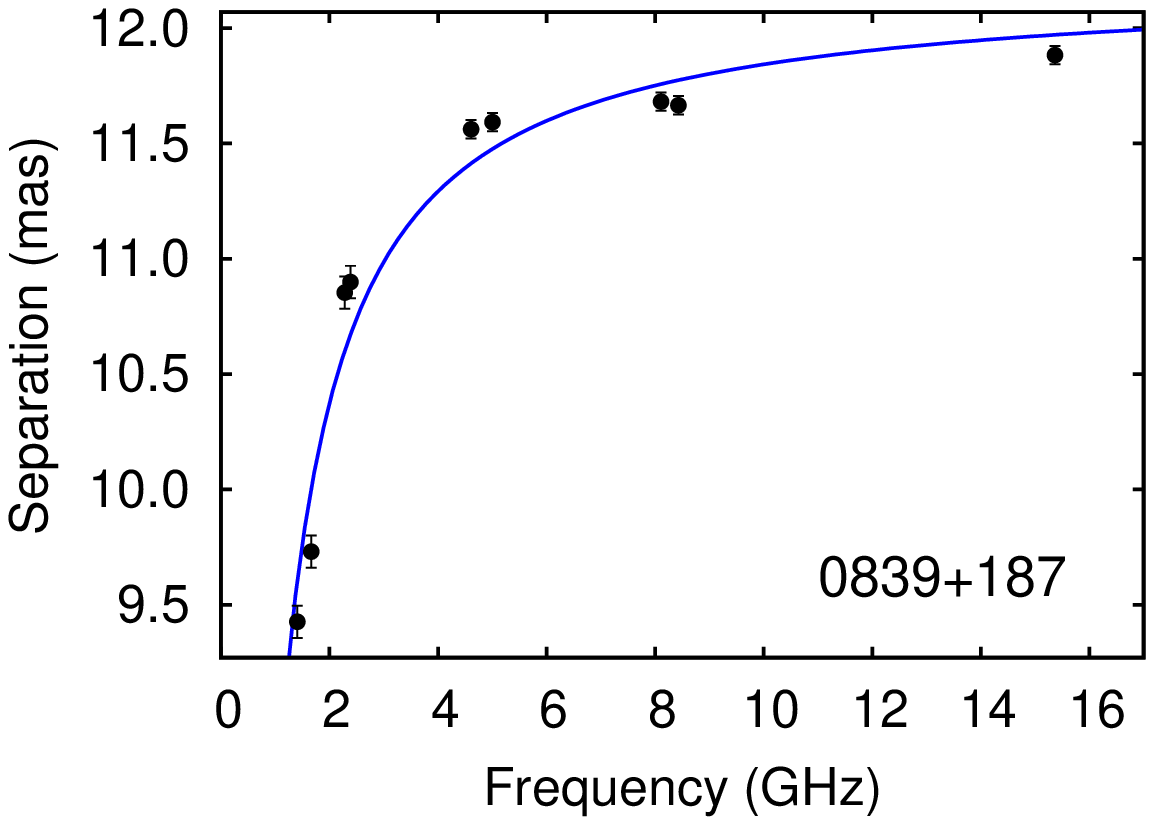}
\\
 \includegraphics[width=0.32\textwidth,angle=0,trim=0.15cm 0cm 0cm 0.3cm,clip]{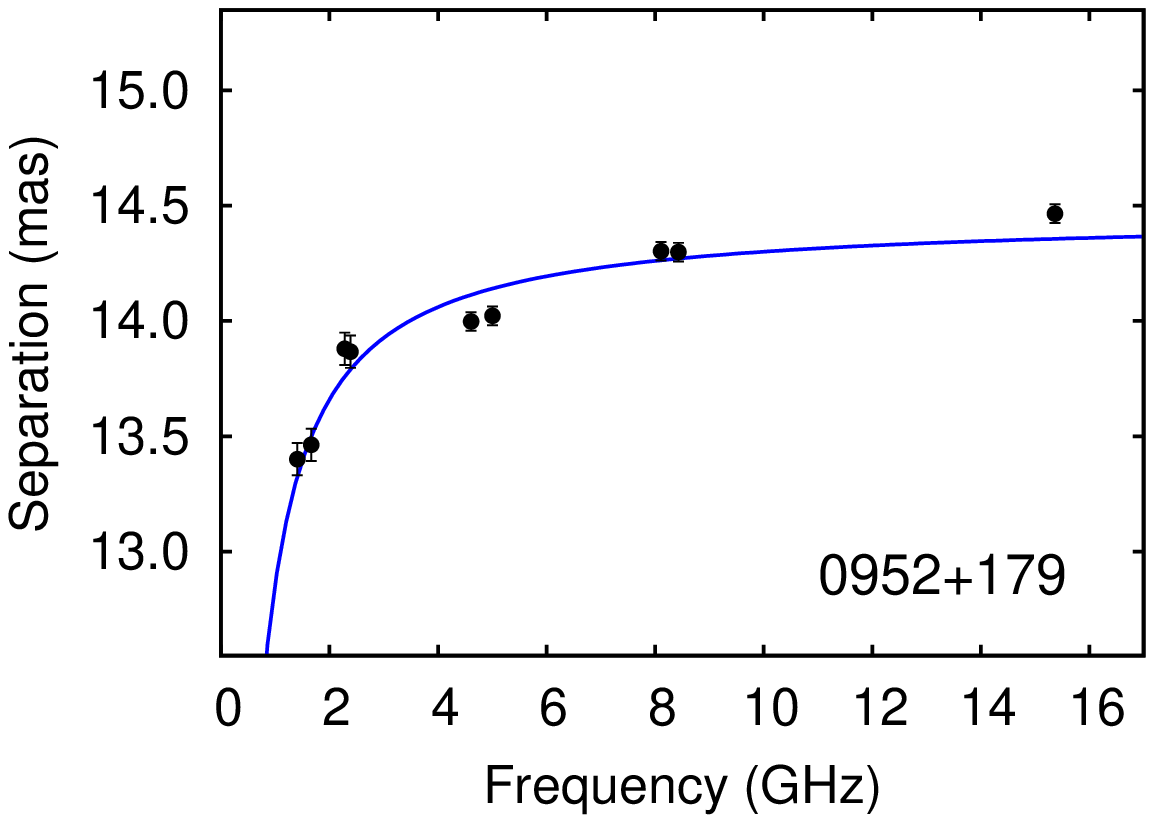} 
 \includegraphics[width=0.32\textwidth,angle=0,trim=0.15cm 0cm 0cm 0.3cm,clip]{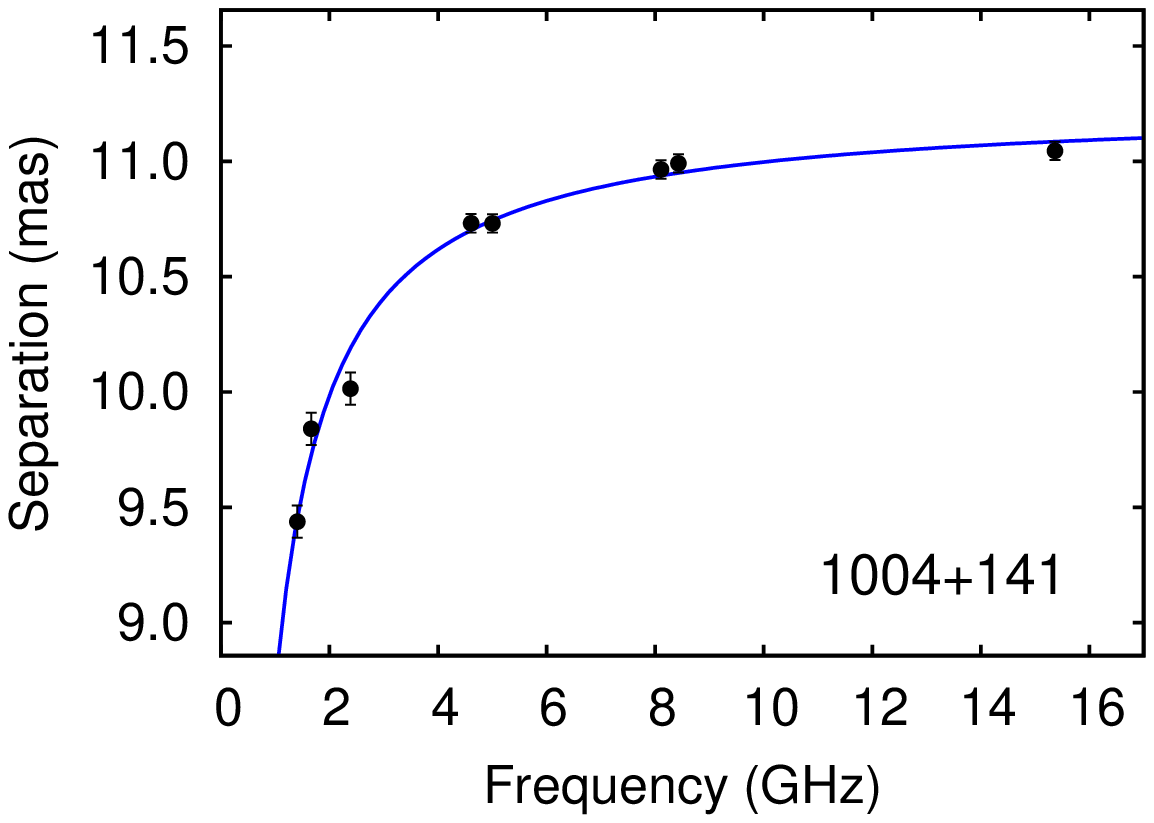} 
 \includegraphics[width=0.32\textwidth,angle=0,trim=0.15cm 0cm 0cm 0.3cm,clip]{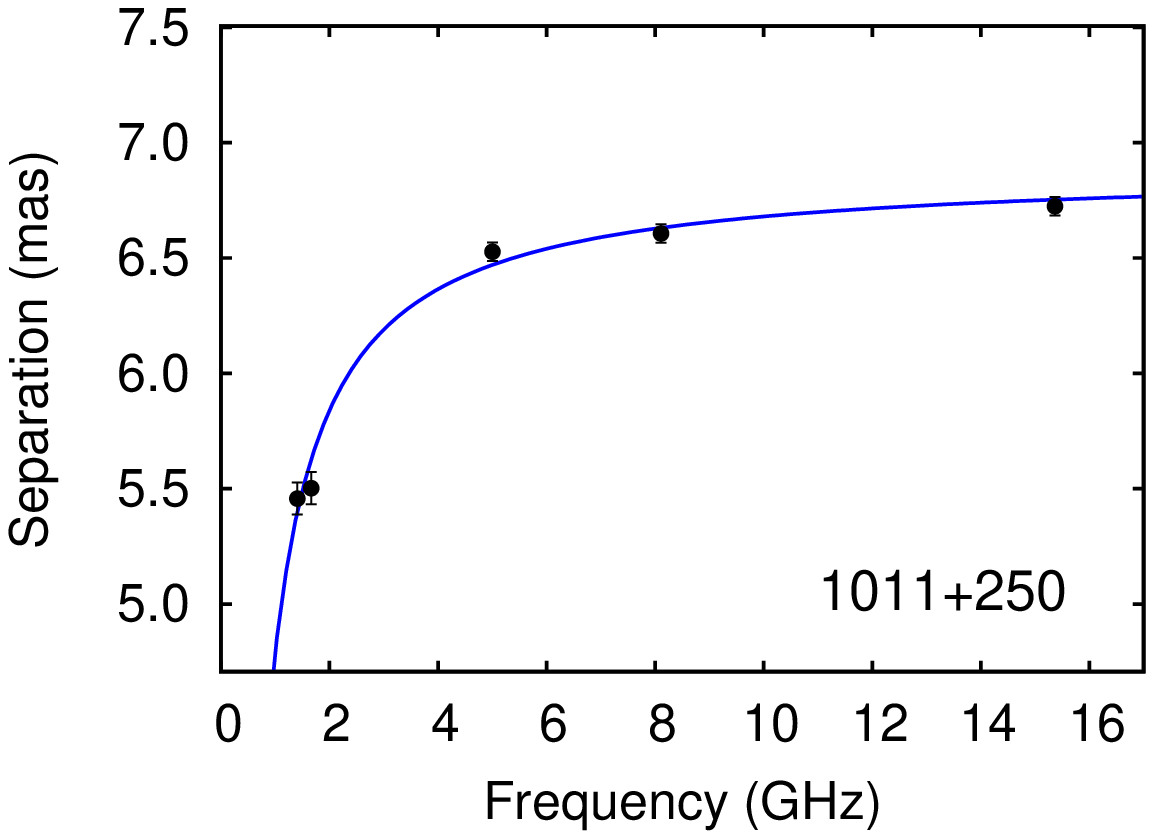}
\\
 \includegraphics[width=0.32\textwidth,angle=0,trim=0.15cm 0cm 0cm 0.3cm,clip]{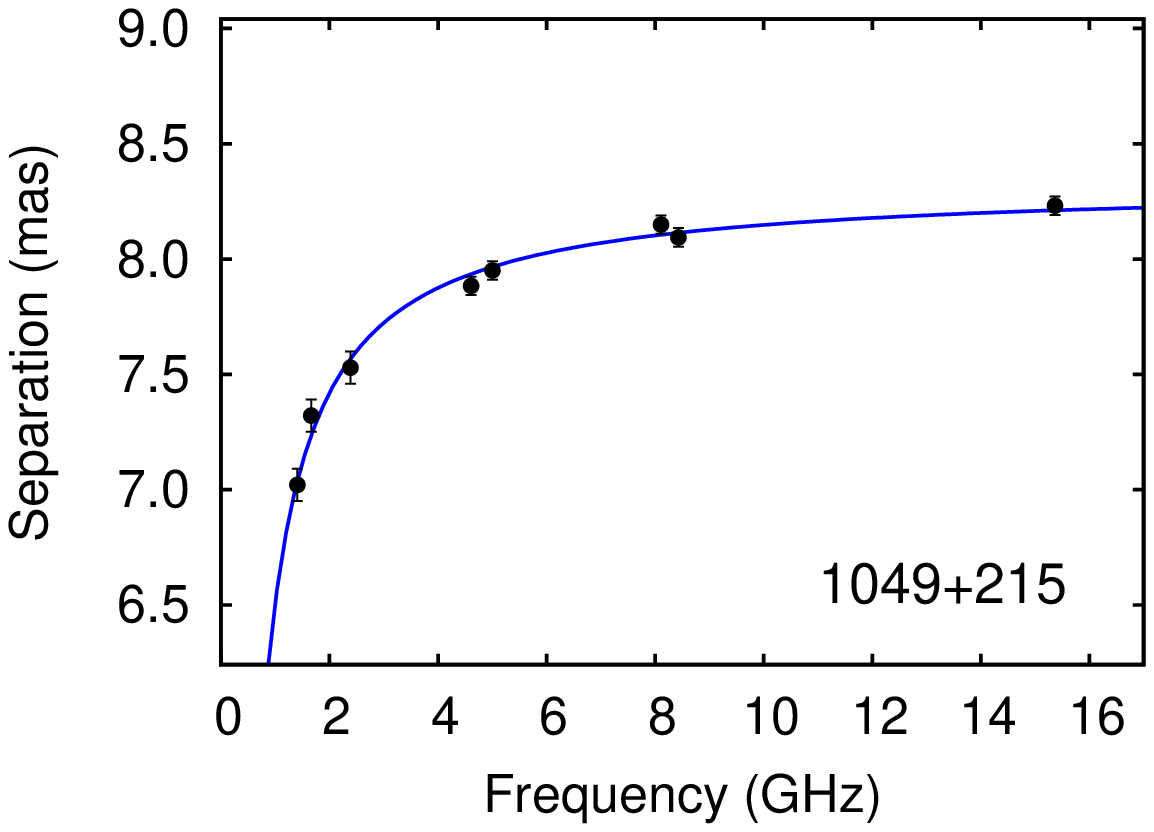} 
 \includegraphics[width=0.32\textwidth,angle=0,trim=0.15cm 0cm 0cm 0.3cm,clip]{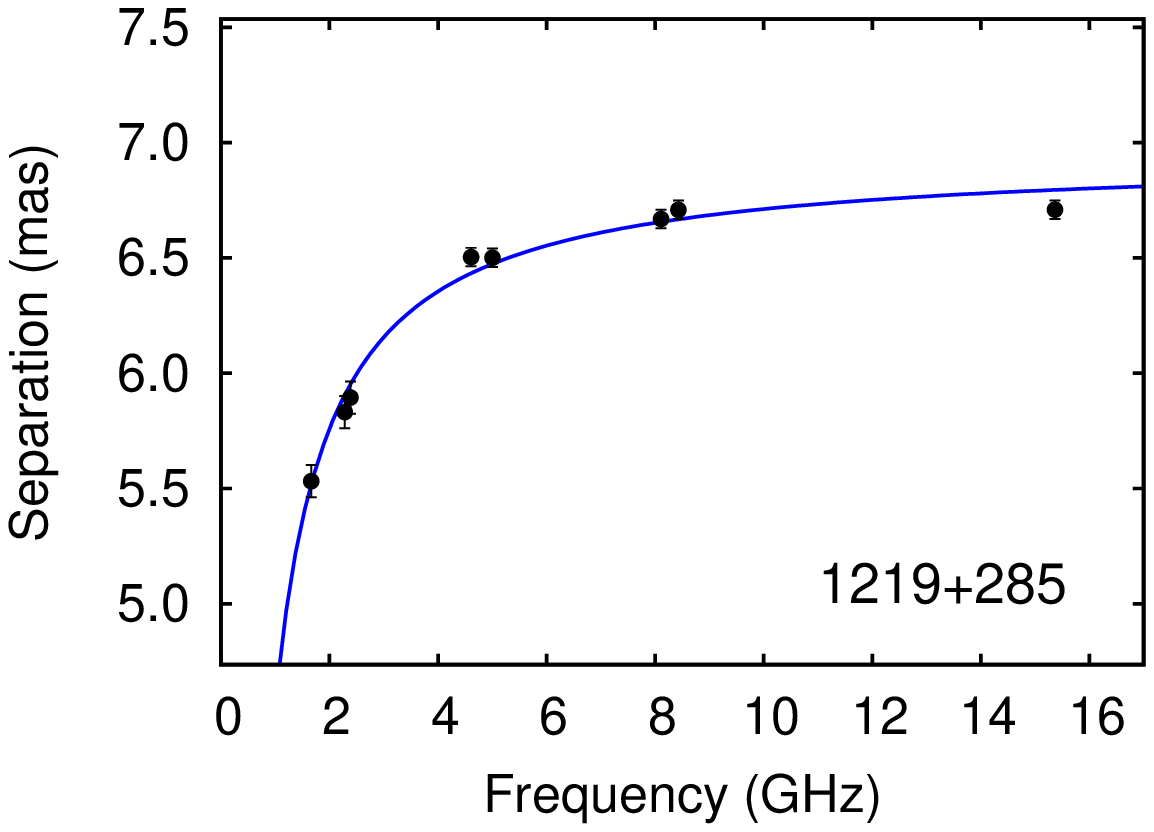} 
 \includegraphics[width=0.32\textwidth,angle=0,trim=0.15cm 0cm 0cm 0.3cm,clip]{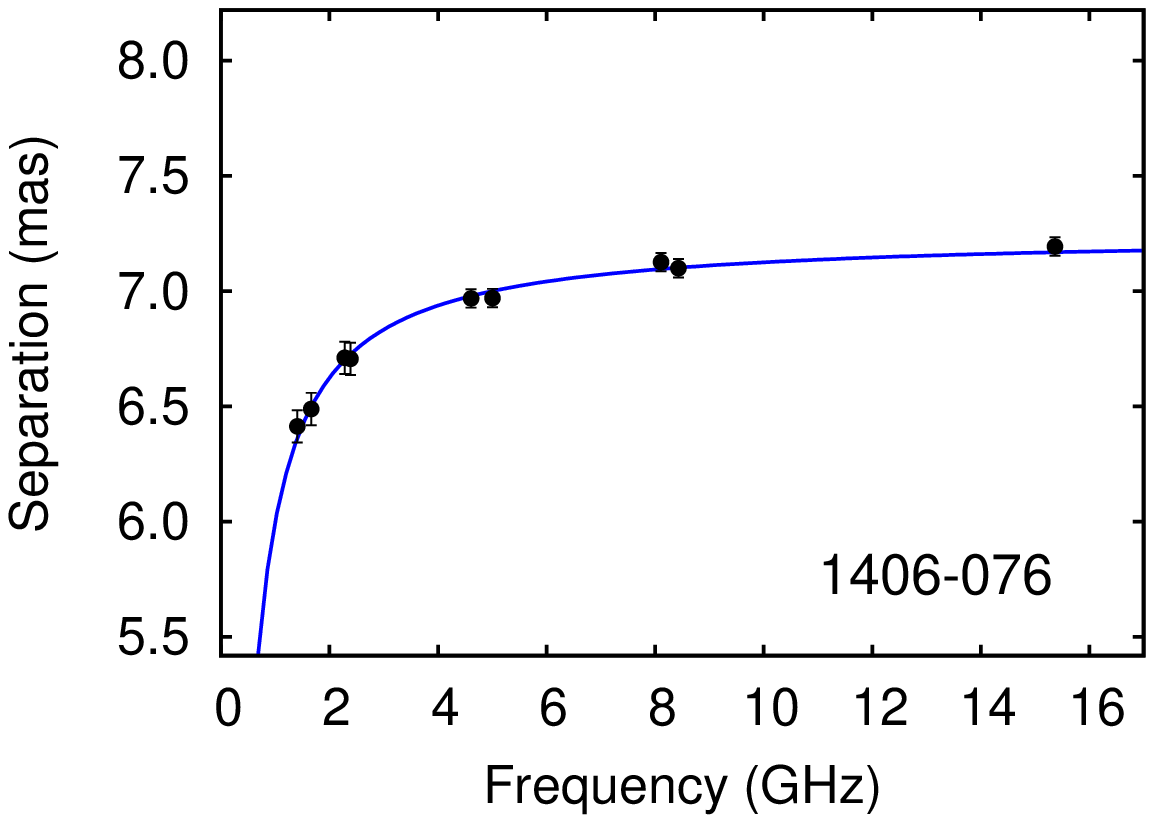}
\\
 \includegraphics[width=0.32\textwidth,angle=0,trim=0.15cm 0cm 0cm 0.3cm,clip]{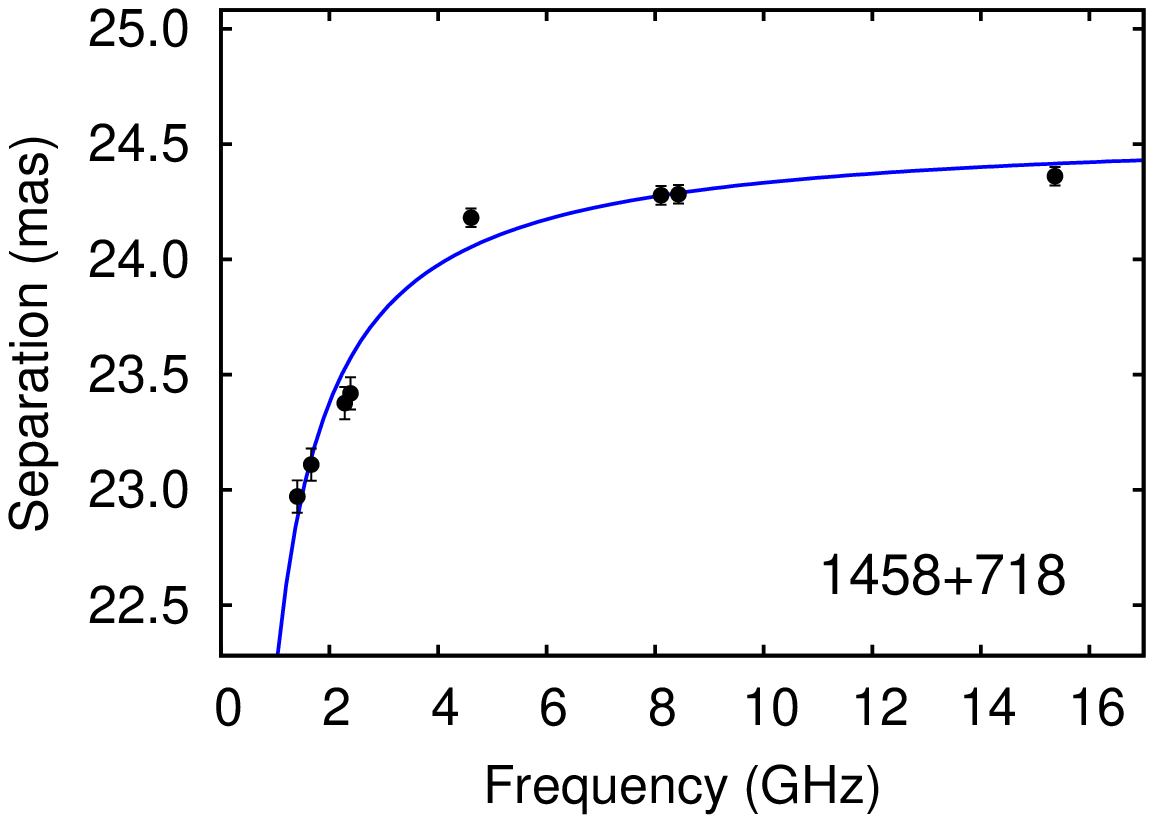}
 \includegraphics[width=0.32\textwidth,angle=0,trim=0.15cm 0cm 0cm 0.3cm,clip]{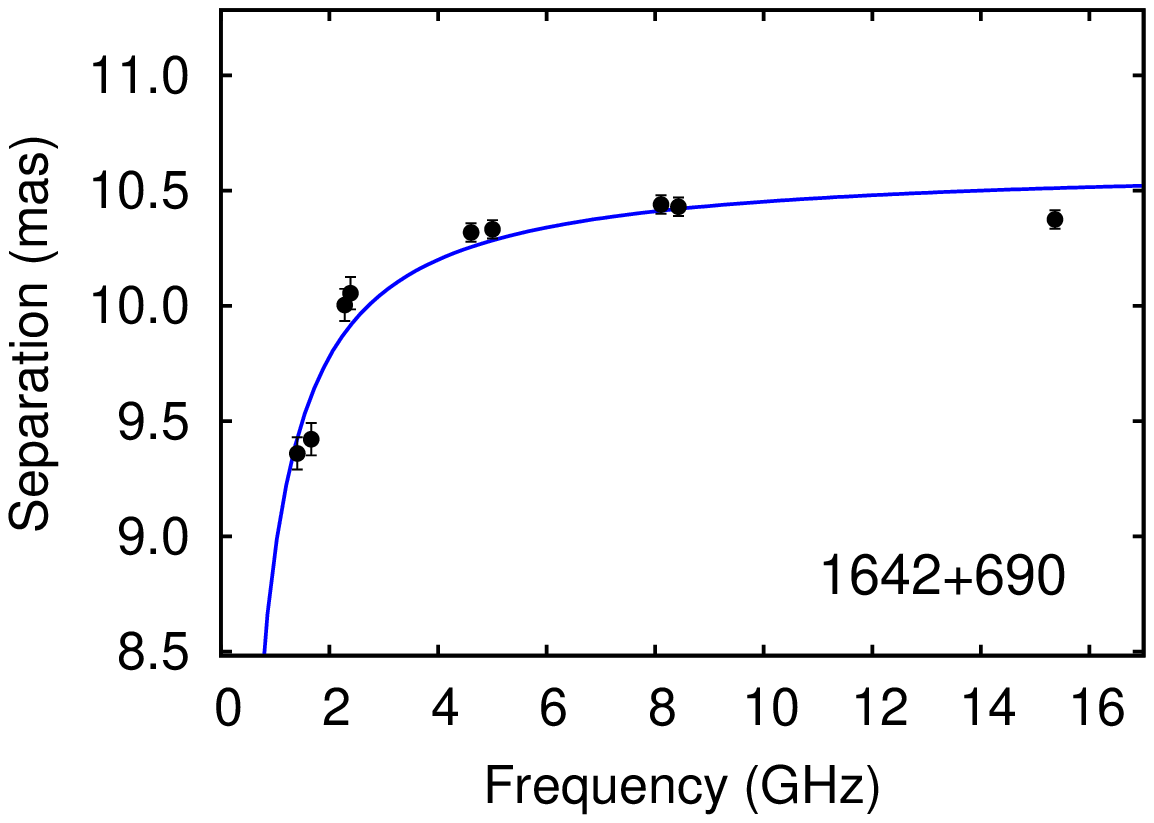}
 \includegraphics[width=0.32\textwidth,angle=0,trim=0.15cm 0cm 0cm 0.3cm,clip]{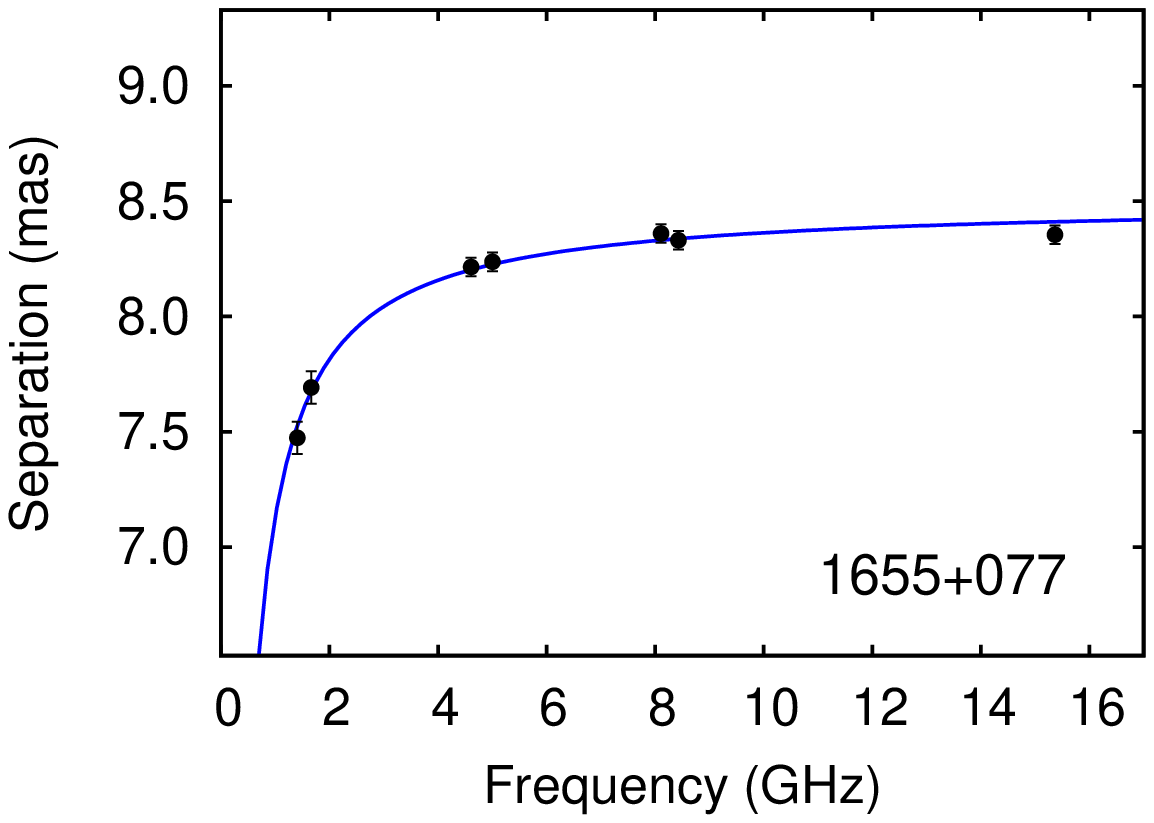}
\\
\includegraphics[width=0.32\textwidth,angle=0,trim=0.15cm 0cm 0cm 0.3cm,clip]{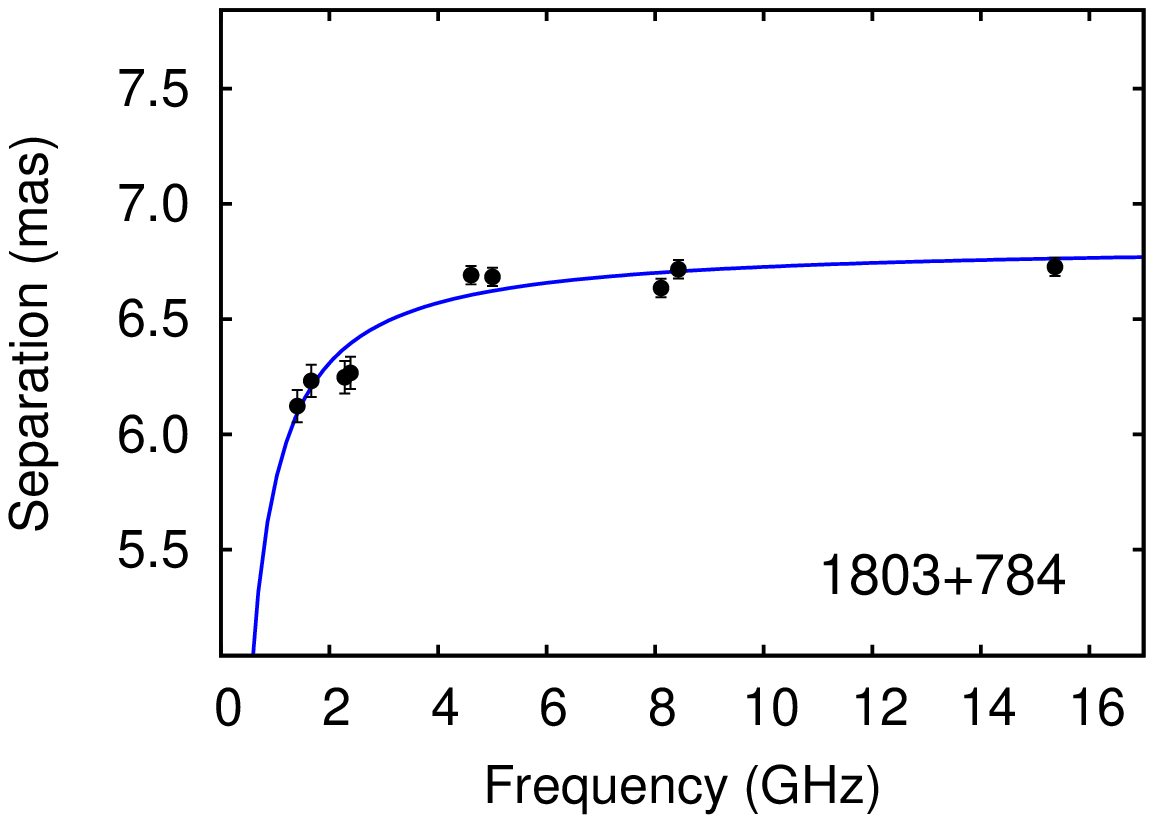}
\includegraphics[width=0.32\textwidth,angle=0,trim=0.15cm 0cm 0cm 0.3cm,clip]{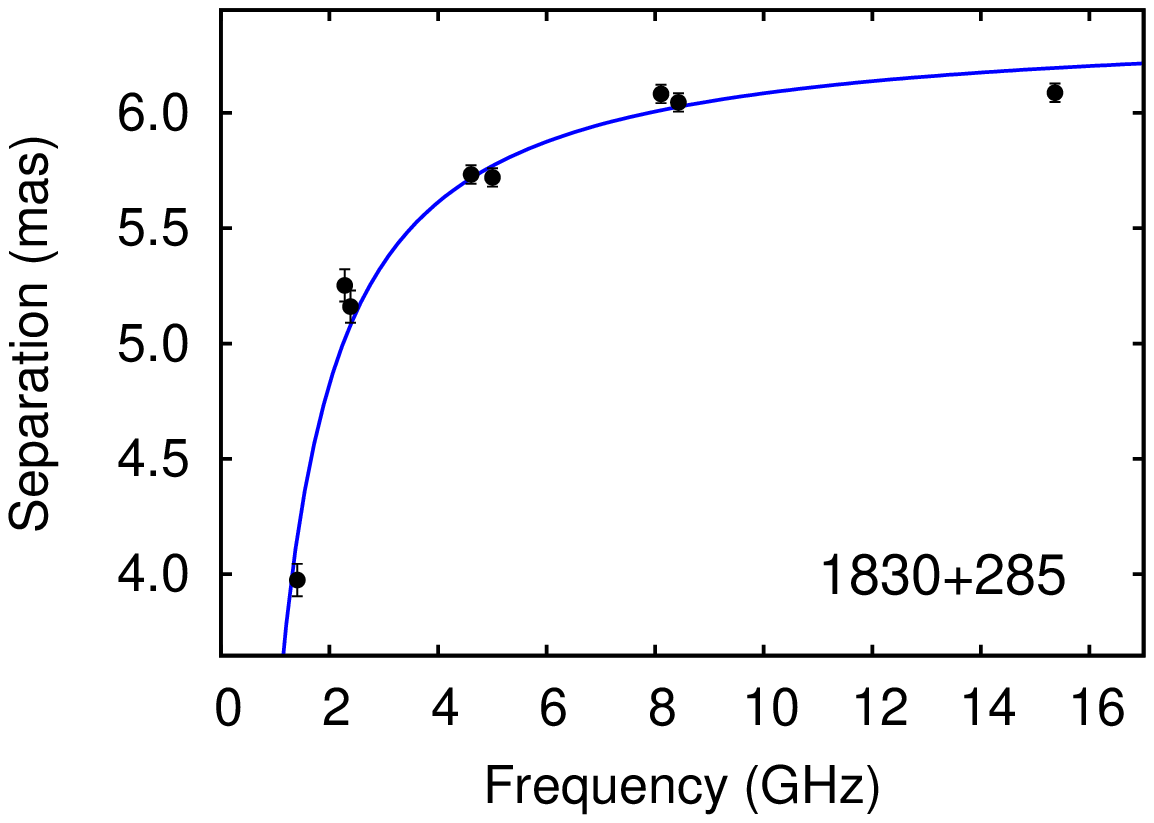}
\includegraphics[width=0.32\textwidth,angle=0,trim=0.15cm 0cm 0cm 0.3cm,clip]{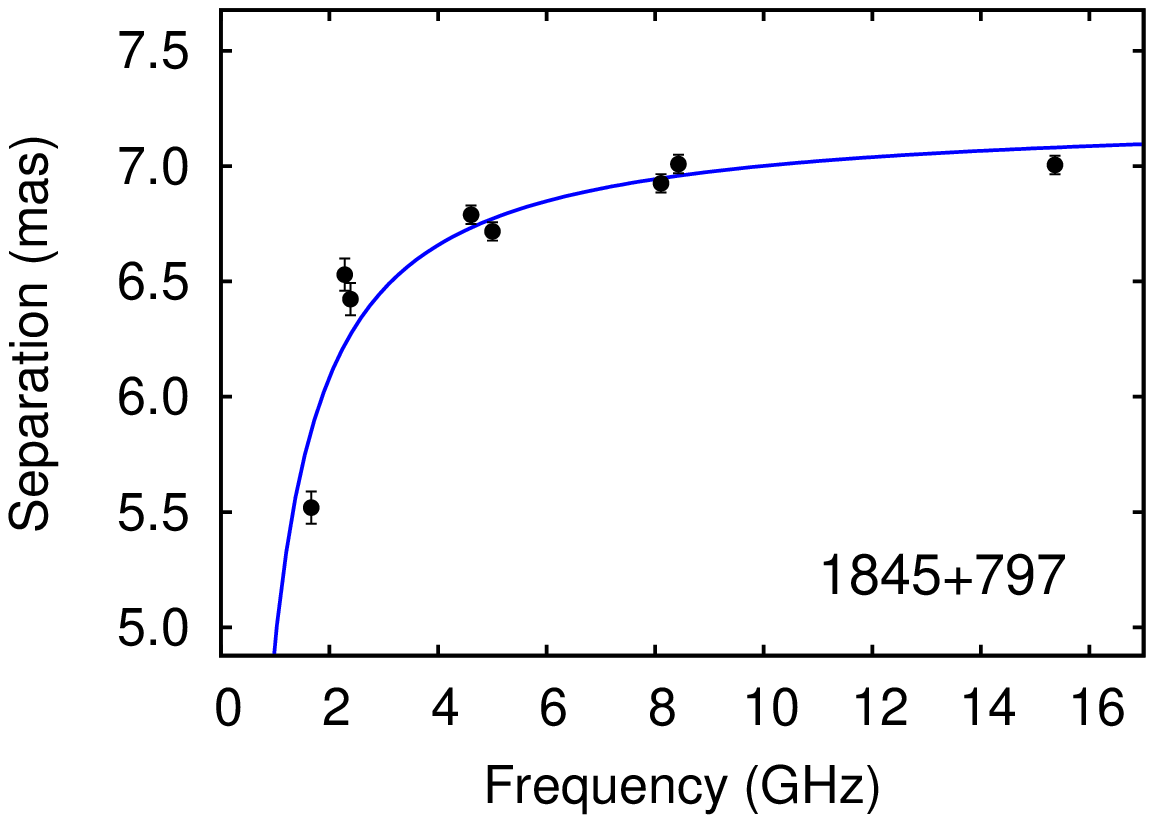}

\caption{Separation of the core from a reference optically-thin jet component as a function of frequency.
The curve represents the best-fit function $r_c(\nu) = a + b/\nu$.
The best-fit parameters are presented in Table~\ref{tab:fit_results}.}
\label{fig:coreshiftplots}
\end{figure*}

\addtocounter{figure}{-1}
\begin{figure*}[t!]
 \centering
\includegraphics[width=0.32\textwidth,angle=0,trim=0.15cm 0cm 0cm 0cm,clip]{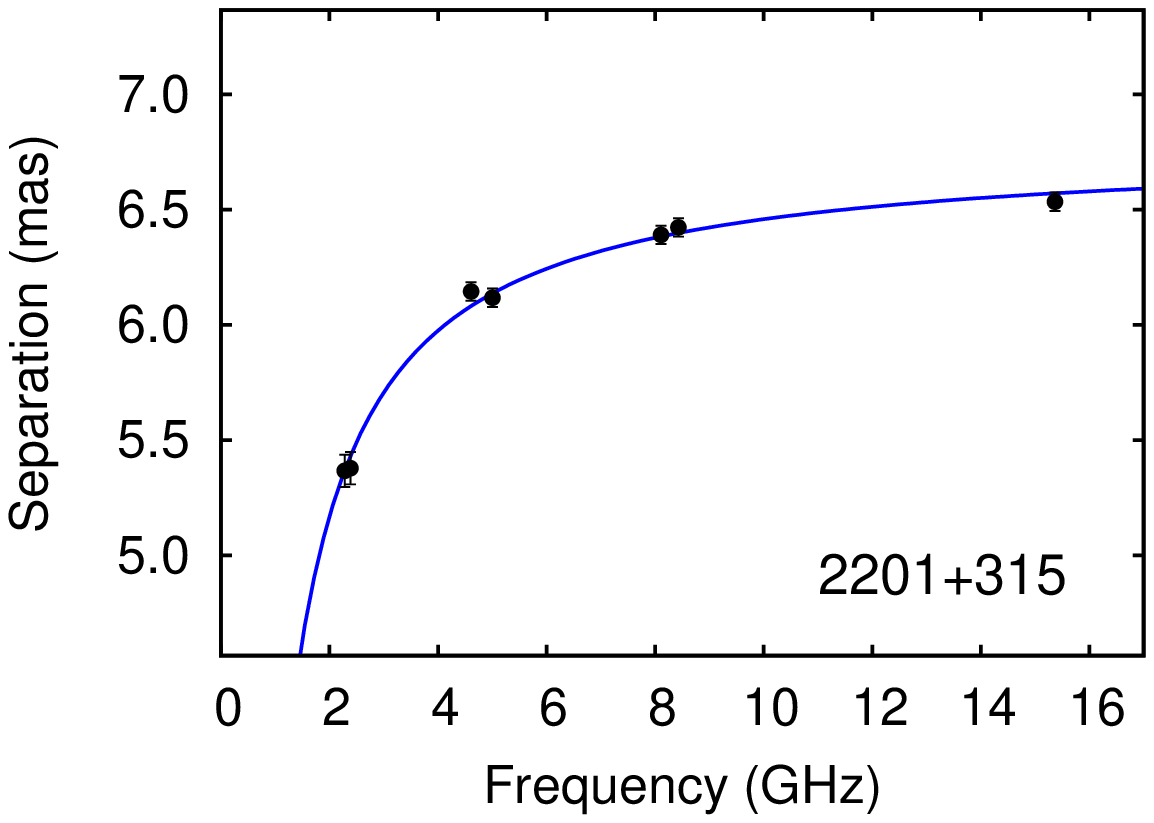}
\includegraphics[width=0.32\textwidth,angle=0,trim=0.15cm 0cm 0cm 0cm,clip]{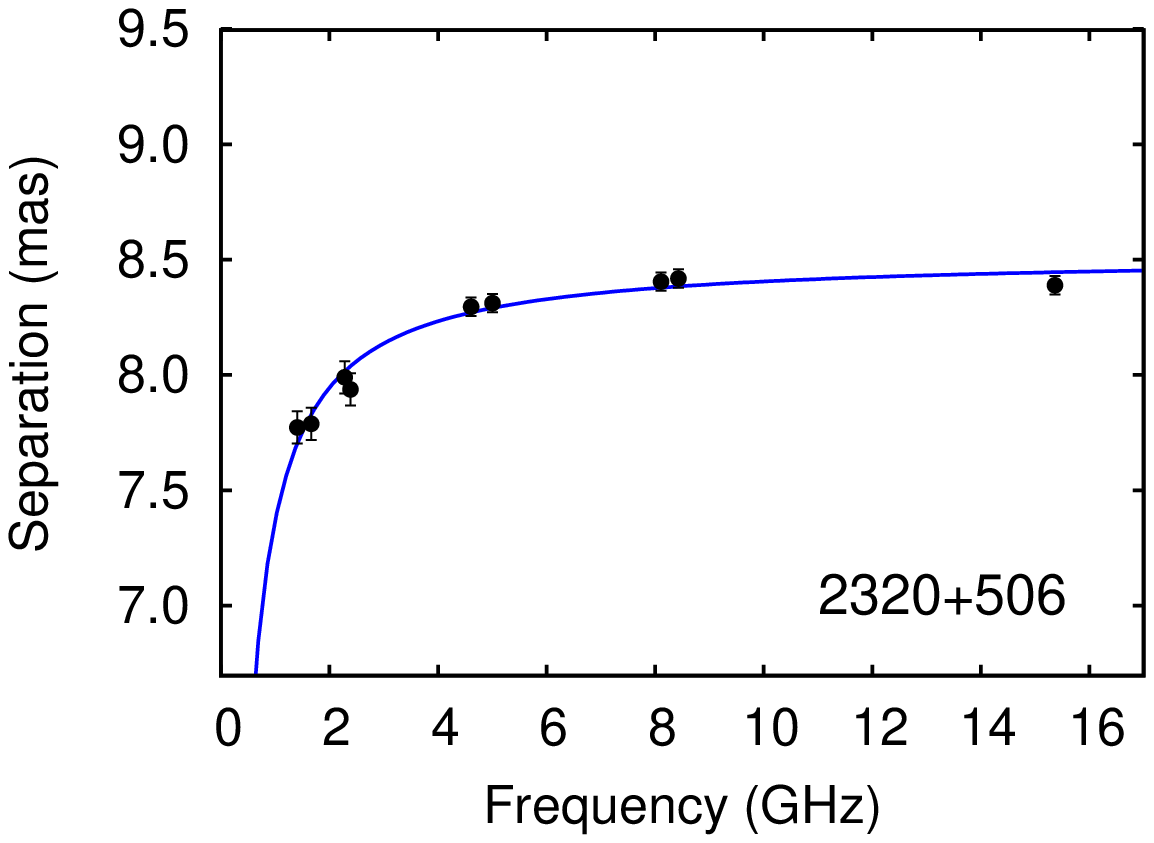}
\caption{continued...}
\end{figure*}

\begin{table*}[tb]
\renewcommand{\thefootnote}{\alph{footnote}}
\begin{center}
\caption{Core shift in milliarcseconds with respect to its position measured at 15.37~GHz}
\label{tab:measuredshifts}
\begin{tabular}{c|rrrrrrrr}
\hline
\hline
                    & \multicolumn{8}{c}{Frequency (GHz)} \\
%\hline
                    &  1.41  &  1.66  &  2.28  &  2.39  &  4.61  &  5.00  &  8.11   &  8.43   \\
\hline
Name                &        &        &        &        &        &        &         &         \\
%\hline
\object{0148$+$274} & $1.41$ & $1.30$ & $1.10$ & $1.16$ & $0.34$ & $0.37$ & $ 0.14$ & $ 0.10$ \\
\object{0342$+$147} & $0.71$ & $0.64$ & $0.30$ & $0.28$ & $0.07$ & $0.07$ & $ 0.03$ & $ 0.06$ \\
\object{0425$+$048} & $1.36$ & $1.11$ & $0.98$ & $0.82$ & $0.20$ & $0.12$ & $ 0.17$ & $ 0.10$ \\
\object{0507$+$179} & $0.90$ & $0.98$ & $0.60$ & $0.65$ & $0.03$ & $0.08$ & $ 0.00$ & $ 0.00$ \\
\object{0610$+$260} & $2.20$ & $1.99$ & $1.53$ & $1.58$ & $0.45$ & $0.50$ & $ 0.21$ & $ 0.22$ \\
\object{0839$+$187} & $2.46$ & $2.15$ & $1.03$ & $0.98$ & $0.32$ & $0.29$ & $ 0.20$ & $ 0.22$ \\
\object{0952$+$179} & $1.07$ & $1.00$ & $0.59$ & $0.60$ & $0.47$ & $0.44$ & $ 0.16$ & $ 0.17$ \\
\object{1004$+$141} & $1.61$ & $1.21$ & $    $ & $1.03$ & $0.31$ & $0.31$ & $ 0.08$ & $ 0.05$ \\
\object{1011$+$250} & $1.27$ & $1.22$ & $    $ & $    $ & $    $ & $0.20$ & $ 0.12$ & $     $ \\
\object{1049$+$215} & $1.21$ & $0.91$ & $    $ & $0.70$ & $0.35$ & $0.28$ & $ 0.08$ & $ 0.14$ \\
\object{1219$+$285} & $    $ & $1.18$ & $0.88$ & $0.81$ & $0.21$ & $0.21$ & $ 0.04$ & $ 0.00$ \\
\object{1406$-$076} & $0.78$ & $0.70$ & $0.48$ & $0.49$ & $0.23$ & $0.22$ & $ 0.07$ & $ 0.09$ \\
\object{1458$+$718} & $1.39$ & $1.25$ & $0.98$ & $0.94$ & $0.18$ & $    $ & $ 0.08$ & $ 0.08$ \\
\object{1642$+$690} & $1.01$ & $0.95$ & $0.37$ & $0.32$ & $0.06$ & $0.04$ & $-0.06$ & $-0.06$ \\
\object{1655$+$077} & $0.88$ & $0.66$ & $    $ & $    $ & $0.14$ & $0.12$ & $-0.01$ & $ 0.02$ \\
\object{1803$+$784} & $0.60$ & $0.49$ & $0.48$ & $0.46$ & $0.04$ & $0.04$ & $ 0.09$ & $ 0.01$ \\
\object{1830$+$285} & $2.11$ & $    $ & $0.84$ & $0.93$ & $0.35$ & $0.37$ & $ 0.01$ & $ 0.04$ \\
\object{1845$+$797} & $    $ & $1.49$ & $0.48$ & $0.58$ & $0.22$ & $0.29$ & $ 0.08$ & $-0.00$ \\
\object{2201$+$315} & $    $ & $    $ & $1.17$ & $1.16$ & $0.39$ & $0.42$ & $ 0.14$ & $ 0.11$ \\
\object{2320$+$506} & $0.62$ & $0.60$ & $0.40$ & $0.45$ & $0.09$ & $0.08$ & $-0.02$ & $-0.03$ \\
\hline
            Adopted &        &        &        &        &        &        &         &        \\
        uncertainty & $0.07$ & $0.07$ & $0.07$ & $0.07$ & $0.04$ & $0.04$ &  $0.04$ &  $0.04$ \\
\hline
\end{tabular}
\end{center}

{\bf Designation:}
Col.~1~-- IAU source name (B1950),
Cols.~2~to~9-- core shift measured at the given frequency (in milliarcseconds).
The last row of the table presents the adopted measurement uncertainty at
the given frequency (in milliarcseconds). The adopted measurement uncertainty at 15.37~GHz is 0.04~mas.
\renewcommand{\thefootnote}{\arabic{footnote}}
\end{table*}

An alternative approach to multifrequency VLBI image alignment based on
two-dimensional cross-correlation has been proposed by
\cite{2000ApJ...530..233W}. It has been successfully applied for core
shift measurements by \cite{2009MNRAS.400...26O} in sources with smooth jets 
where referencing to a distinct bright jet component was not possible. A potential disadvantage of
this method is that, in principle, it may introduce artificial offsets between
frequencies in the presence of spectral gradients along the jet.
A similar problem may affect the model component-based method used here, if
there is a large spectral gradient across a resolved reference component.
However, the spectral index gradient is smaller for an
individual component then for a large section of parsec-scale jet used in
cross-correlation analysis. 
Early observations of the frequency dependent difference in distance between the core
and jet components were actually interpreted by \cite{1986ApJ...308...93B} as
a sign of spectral gradients across the components. This interpretation is
disfavored by the detection of core position shift
with frequency in phase-referenced experiments 
\citep{1985A&A...142...71M,1997A&A...325..383R,BH103} and the fact that
strong spectral gradients across individual jet components are not routinely
observed if images are aligned correctly.

A significant core shift has been detected in each of the twenty sources
observed within our VLBA program (Table~\ref{tab:measuredshifts}).
The effect is especially pronounced 
at frequencies below 5~GHz. 
The results are discussed in detail in Section~\ref{sec:core_shift_discussion} and
presented visually in Fig.~\ref{fig:coreshiftplots}.

\begin{figure}[bt]
 \centering
 \includegraphics[width=0.5\textwidth,angle=0,trim=2.5cm 0.57cm 2.5cm 0cm,clip]{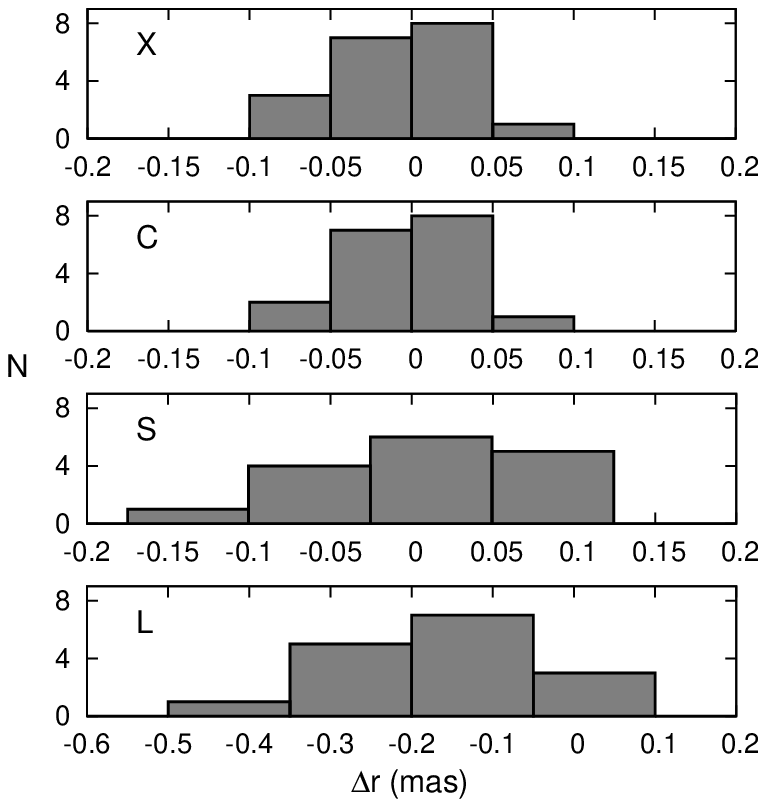}
 \caption{Distribution of the observed difference in distance between the core and reference
 component measured at two sub-bands of $X$, $C$, $S$, and $L$ bands.}
 \label{fig:diffs}
\end{figure}

To statistically estimate typical uncertainty $\sigma_r$
of the measured distance $r$ between the core and
reference jet component, we used pairs of sub-bands of $X$, $C$, $S$, and
$L$ bands, assuming that the source structure at these closely separated 
frequencies is essentially the same and the non-zero difference $\delta r$ 
between sub-bans in the model fit 
parameters is determined purely by errors (Fig.~\ref{fig:diffs}).
This assumption turns out to be correct for $X$, $C$, and $S$ bands,
while at $L$ band a significant core shift between the sub-bands becomes evident. 
We underline that this approach is possible since data processing of these
sub-bands was done completely independently.
It should be also noted that the core-shift effect is found to be
not significant between the two sub-bands at $S$-band most probably
due to the relatively narrow frequency range (Table~\ref{tab:IFs})
achievable at this particular VLBA band.
We have also estimated individual errors $\sigma_r$
for every source and frequency following
two approaches: the one suggested by \cite{1999ASPC..180..301F} and 
its modification by \cite{2008AJ....136..159L}.
Then we compared the measured values of $\delta r$ with estimations
of its error $\sigma_{\delta r}$ for every sub-band pair.
If the $\sigma_{\delta r}$ are estimated correctly following
\cite{1999ASPC..180..301F} or \cite{2008AJ....136..159L},
we expect that for about 67\% of the sources
$\sigma_{\delta r} > \delta r$.
We have found out that the  \cite{1999ASPC..180..301F} method
has provided meaningful error estimates for
S, C, and X, while the L-bans errors were significantly
underestimated due to a very high signal-to-noise ratio
of jet components in this band.
At the same time, the \cite{2008AJ....136..159L}
method has significantly overestimated the errors in all the bands
by 3 -- 40 times. On the basis of these results, we made a decision to
use a statistical method and estimate 
only a typical $\sigma_r$ error for every band.
An approach to adopting the same error for each target at a given frequency
has an obvious weakness. However, we believe that this turns out to be
the most reliable estimate of $\sigma_r$ in view of the general problem
of calculating the model-dependent error of a component position in VLBI.

% so, what errors do we finally use?
In the following, for $X$, $C$, and $S$ bands we adopt a single typical uncertainty of the
distance between the core and reference component, which is the standard deviation of the distance
difference measurements made in the two sub-bands of each band, computed using all the sources
in our experiment (Fig.~\ref{fig:diffs}). For the $X$ band the adopted distance error is $\sigma_X
= 0.04$~mas, for $C$ band $\sigma_C = 0.04$~mas, and for $S$ band
$\sigma_S = 0.07$~mas. The $K_u$ band has not been divided into two
independent sub-bands and for that band we adopt the same position
uncertainty as for $X$ band: $\sigma_{K_u} = \sigma_X = 0.04$~mas. In
the $L$ band the mean difference between the distance measured at the lower
sub-band ($1.4$~GHz) and the higher sub-band ($1.7$~GHz) is $-0.14 \pm 0.03$~mas,
which is significantly different from zero. We interpret this as
a sign of the core shift effect, which should be more pronounced at lower
frequencies, and use the uncertainty derived from the scatter of $S$ band
measurements as an estimate of the $L$ band error instead:
$\sigma_L = \sigma_S = 0.07$~mas. We note that $\sigma_L$ may be
underestimated.

%__________________________________________________________________

\section{Discussion}
\label{sec:core_shift_discussion}

\subsection{Typically measured values of the shift}

% Measured values and how typical are they?
In the 20 sources observed in our experiment we found the following
median values of the core shift with respect to its position at $15.4$~GHz:
$0.06$~mas for $8.4$~GHz, $0.08$~mas for $8.1$~GHz, $0.22$~mas for
$5.0$~GHz, $0.22$ for $4.6$~GHz, $0.76$~mas for $2.4$~GHz, $0.72$~mas for
$2.3$~GHz, $1.06$~mas for $1.7$~GHz, and $1.21$~mas for $1.4$~GHz measurements.
See Table~\ref{tab:measuredshifts} for more details.
It is difficult to say how applicable these values are to the population of
compact extragalactic radio sources as a whole. It is likely that the above
values represent a higher extreme of the possible range of values, since the
sources with a high core shift were selected for our observations from the
list of \cite{2008A&A...483..759K}.
The median S/X band core shift value of $0.69$~mas obtained for our
sample in this study may be compared to the median value of $0.44$~mas from
\cite{2008A&A...483..759K} for a sample of 29 sources. These
samples have 15 sources in common;
see more detailed discussion in Section~\ref{s:var}.

\cite{2009MNRAS.400...26O} have measured
the median core shift of $0.22$~mas between $4.6$ and $15.4$~GHz using a
small sample of three objects, none of which is present in our sample. 
This value is the same as the median $4.6$/$15.4$~GHz shift in our sample ($0.22$~mas).
A detailed investigation of the core shift effect in a large complete sample of 
sources selected not on the basis of previous core shift 
measurements is needed to derive typical magnitude of the effect 
in a compact extragalactic radio source.

Finally, we note that misalignment between the parsec-scale jet
(resolved in our VLBI observations) and the subparsec scale jet (which we
observe as the core) will reduce the frequency-dependent distance difference
between the core and a reference feature in the parsec-scale jet.
Therefore, the sources with the highest measured core shift may be those
reflecting the true intrinsic magnitude of the effect.

\subsection{Core position as a function of frequency}

\begin{figure}[tb]
 \centering
 \includegraphics[width=0.5\textwidth,angle=0,trim=0cm 0cm 0cm 0cm,clip]{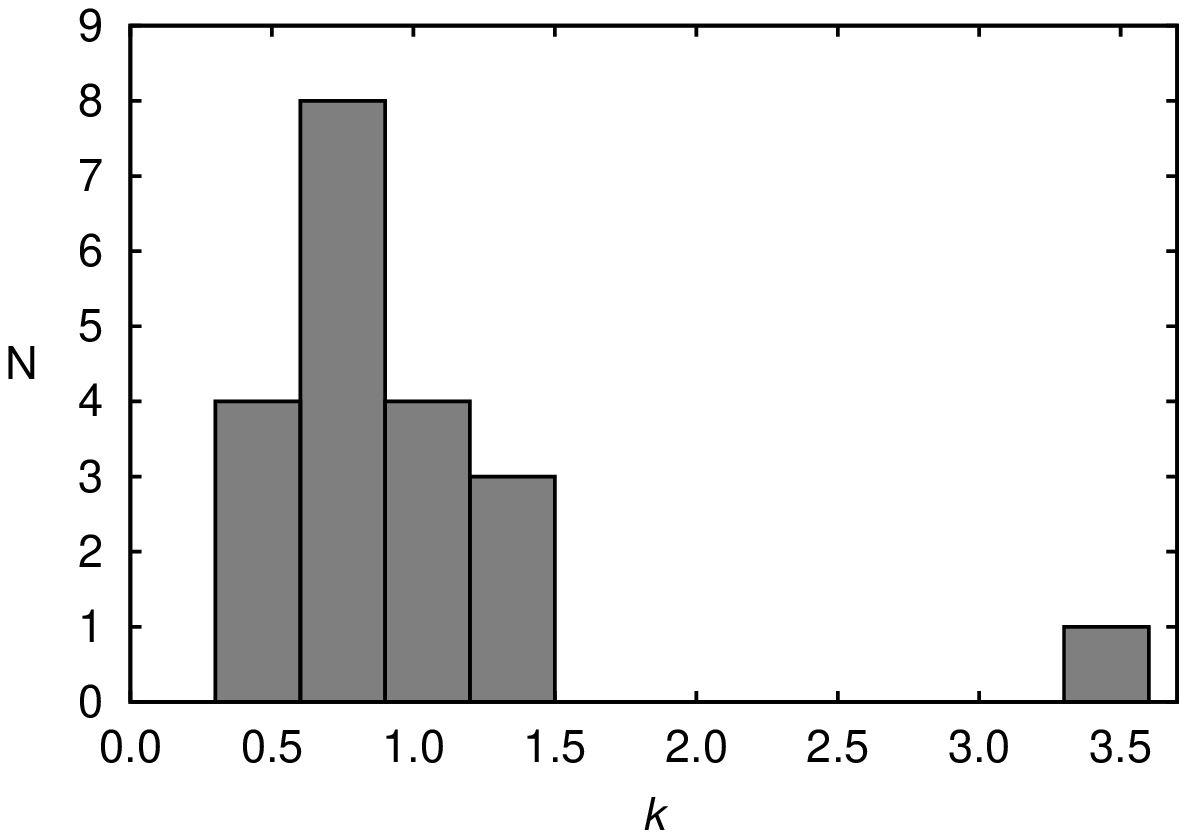}
 \caption{Distribution of the $k$ parameter in the core position as a
 function of frequency fit: $r_c(\nu) \propto \nu^{-1/k}$. The mean value is 
 $k = 0.99 \pm 0.14$, while the median is $0.82$. One discrepant measurement on the
 histogram corresponds to the source 0952$+$179 where the value of $k$ is not
 well constrained: $k = 3.4 \pm2.7$.}
 \label{fig:k}
\end{figure}

% In this subsection we test if k=1
According to \cite{1998A&A...330...79L}, the position of the core $r_c(\nu) \propto \nu^{-1/k}$
where the coefficient $k = 1$ if {\em (i)} the dominating absorption mechanism is 
synchrotron self-absorption, {\em (ii)} the jet has a conical shape and {\em (iii)} is in
equipartition. These assumptions must hold for the ultracompact jet region
that we observe as the VLBI core and does not need to be correct
all the way along the extended parsec-scale jet.

To test the above assumptions, we fitted the observed distances between
the reference jet component and the core in each source with the function
$r_c(\nu) = a + b \nu^{-1/k}$ leaving the coefficients $a$, $b$, and $k$ as free parameters. 
The Levenberg--Marquardt algorithm \citep[e.g.,][]{NumRec} was used to perform the nonlinear
least-squares fit.
Figure~\ref{fig:k} shows the distribution of the estimated $k$ values.
The mean value of the distribution is $k = 0.99 \pm 0.14$. This suggests
that, for a typical source in the sample in the frequency range of our
observations, the above assumptions about the structure of the inner jet
(which we observe as the VLBI core) are generally correct.
Specifically, we conclude that our observations
are consistent with synchrotron self-absorption being the dominating
opacity mechanism acting in parsec-scale core regions in the GHz frequency
range.

It should be noted that, since the core shift effect is most pronounced at
lower frequencies, the best-fit values of $k$ obtained heavily depend
on the $L$-band measurements. At the same time, $L$-band data are 
also more likely to be affected by blending, an effect that could interfere with the
core shift measurements \citep{2008A&A...483..759K}. To test how
sensitive our conclusions are to the $L$-band results, we have repeated the
above fitting procedure excluding all $L$-band data. The resulting values of
$k$ are on average 30\,\% lower than to those obtained with $L$-band data
included. Naturally, a larger uncertainty is associated with the values of
$k$ obtained without $L$-band data. The mean values of the two distributions
of best-fit $k$ values are consistent at the $2 \sigma$ level.

\begin{table*}[t!]
\begin{center}
\caption{ Distance of the reference jet component from the core as a function of
frequency: fit results ($r = a + b/\nu$)}
\label{tab:fit_results}
\begin{tabular}{crcrrr}
\hline
\hline
Source     &     $a$ (mas)~~~    & $b$ ($\text{mas} \cdot \text{GHz}$)  & $a$ (pc)~~~~~~  & $b$ ($\text{pc} \cdot \text{GHz}$)~ & P.A.$^\circ$ \\
\hline
%%%%%%%%%%%%%%%%%%%%%%%%%%%%%%%%%%%%%%%%%%%%%%%%%%%%%%%%%%%%%%%%%%%%%%%%%%%%%%%%%%%%%%%%%%%%%%%%%%%%%%%%%%%%%%%%%       z     D_A[Mpc]
0148$+$274  & $ 10.32 \pm 0.05 $ &  $ -2.48 \pm 0.19 $ &  $  86.75 \pm  0.42 $ & $ -47.11 \pm  3.61 $  & $-40.9$  \\ % 1.26   1733.8
0342$+$147  & $  7.15 \pm 0.03 $ &  $ -1.08 \pm 0.10 $ &  $  61.14 \pm  0.26 $ & $ -23.61 \pm  2.19 $  & $-88.5$  \\ % 1.556  1763.8
0425$+$048  & $ 18.58 \pm 0.06 $ &  $ -2.20 \pm 0.19 $ &  $ 115.05 \pm  0.37 $ & $ -20.66 \pm  1.78 $  & $-100.7$ \\ % 0.517  1277.2
0507$+$179  & $  4.52 \pm 0.05 $ &  $ -1.73 \pm 0.19 $ &  $  24.74 \pm  0.27 $ & $ -13.41 \pm  1.47 $  & $-100.3$ \\ % 0.416  1128.8
0610$+$260  & $  6.34 \pm 0.06 $ &  $ -3.69 \pm 0.20 $ &  $  41.62 \pm  0.39 $ & $ -38.28 \pm  2.07 $  & $-94.5$  \\ % 0.580  1354.2
0839$+$187  & $ 12.21 \pm 0.09 $ &  $ -3.67 \pm 0.31 $ &  $ 102.76 \pm  0.76 $ & $ -70.18 \pm  5.93 $  & $15.2$   \\ % 1.272  1736.0
0952$+$179  & $ 14.46 \pm 0.06 $ &  $ -1.60 \pm 0.19 $ &  $ 123.39 \pm  0.51 $ & $ -33.83 \pm  4.02 $  & $-3.5$   \\ % 1.478  1760.1
1004$+$141  & $ 11.25 \pm 0.04 $ &  $ -2.53 \pm 0.13 $ &  $  90.53 \pm  0.32 $ & $ -75.47 \pm  3.88 $  & $132.2$  \\ % 2.707  1659.8
1011$+$250  & $  6.89 \pm 0.05 $ &  $ -2.10 \pm 0.15 $ &  $  58.96 \pm  0.43 $ & $ -47.37 \pm  3.38 $  & $-107.3$ \\ % 1.636  1765.1
1049$+$215  & $  8.33 \pm 0.02 $ &  $ -1.82 \pm 0.09 $ &  $  70.30 \pm  0.17 $ & $ -35.33 \pm  1.75 $  & $108.0$  \\ % 1.300  1740.7
1219$+$285  & $  6.95 \pm 0.04 $ &  $ -2.38 \pm 0.16 $ &  $  19.06 \pm  0.11 $ & $  -7.58 \pm  0.51 $  & $109.8$  \\ % 0.161   565.8
1406$-$076  & $  7.25 \pm 0.01 $ &  $ -1.25 \pm 0.05 $ &  $  61.90 \pm  0.09 $ & $ -26.61 \pm  1.06 $  & $-103.5$ \\ % 1.493  1761.0
1458$+$718  & $ 24.57 \pm 0.05 $ &  $ -2.38 \pm 0.18 $ &  $ 192.21 \pm  0.39 $ & $ -35.45 \pm  2.68 $  & $163.7$  \\ % 0.904  1613.6
1642$+$690  & $ 10.62 \pm 0.05 $ &  $ -1.68 \pm 0.19 $ &  $  78.01 \pm  0.37 $ & $ -21.61 \pm  2.44 $  & $-167.0$ \\ % 0.751  1515.2
1655$+$077  & $  8.50 \pm 0.02 $ &  $ -1.37 \pm 0.08 $ &  $  57.64 \pm  0.14 $ & $ -15.06 \pm  0.88 $  & $-42.9$  \\ % 0.621  1398.7
1803$+$784  & $  6.83 \pm 0.04 $ &  $ -1.04 \pm 0.14 $ &  $  48.21 \pm  0.28 $ & $ -12.33 \pm  1.66 $  & $-96.0$  \\ % 0.680  1455.9
1830$+$285  & $  6.40 \pm 0.06 $ &  $ -3.15 \pm 0.24 $ &  $  42.50 \pm  0.40 $ & $ -33.34 \pm  2.54 $  & $-38.5$  \\ % 0.594  1369.8
1845$+$797  & $  7.23 \pm 0.09 $ &  $ -2.29 \pm 0.35 $ &  $   7.76 \pm  0.10 $ & $  -2.59 \pm  0.40 $  & $-38.0$  \\ % 0.056   221.3
2201$+$315  & $  6.78 \pm 0.03 $ &  $ -3.22 \pm 0.15 $ &  $  29.83 \pm  0.13 $ & $ -18.39 \pm  0.86 $  & $-135.3$ \\ % 0.298   907.6
2320$+$506  & $  8.52 \pm 0.03 $ &  $ -1.15 \pm 0.09 $ &  $  71.76 \pm  0.25 $ & $ -22.07 \pm  1.73 $  & $-135.8$ \\ % 1.279  1737.2
\hline
\end{tabular}
\end{center}

{\bf Column designation:}
Col.~1~-- IAU source name (B1950),
Col.~2~and~3~-- coefficients of the best-fit curve $r = a + b/\nu$ with
their uncertainties. Col.~3~and~4~-- same coefficients converted to the
projected linear scale. Col.~5~-- median (across all frequencies) position
angle of the reference jet component with respect to the core; 
it marks the direction along which the core shift was measured.

\end{table*}

The values of $k$ presented in Fig.~\ref{fig:k} are similar to those obtained for 1038$+$528\,A by
\cite{1984ApJ...276...56M,1985A&A...142...71M}, 3C\,345 by \cite{1998A&A...330...79L},
3C\,309.1 by \cite{2002astro.ph.11200R}, 0850$+$581 by \cite{2008MmSAI..79.1153K}, 1418$+$546, 2007$+$777, 
BL~Lac by \cite{2009MNRAS.400...26O}, and Mrk~501 by \cite{2010MNRAS.402..259C}. 
These values also agree with the results of
\cite{2008arXiv0811.2926Y}, who estimates $k$ from a statistical
analysis of parsec-scale core sizes of $\sim 3000$ sources. The core size 
is expected to be directly related to the observed position of the core 
along the conical jet \citep{1994ApJ...432..103U}. 
The value of $k<1$ was obtained from analysis of the core size
in 3C\,345 measured between 5 and 22 GHz by \cite{1994ApJ...432..103U}.
Values of $k$ slightly more than unity were estimated for BL~Lac by
analyzing time lags between radio lightcurves obtained at different frequencies
\citep{2006A&A...456..105B}.
We also note that \cite{1998A&A...330...79L} has found an indication that $k$
changes with
frequency up to values greater than $1$ for the quasar 3C\,309.1.
\cite{2004A&A...426..481K} found much higher values of $k$, indicating free-free absorption in the radio galaxy NGC~1052.
The key difference between NGC~1052 and the sources observed in our sample
is that this radio galaxy shows a two-sided parsec-scale jet. The inner part
of the receding jet of NGC~1052 is likely to be obscured by a circumnuclear torus
\citep{2004A&A...426..481K}.

\subsection{Can the same function describe shifts at all frequencies between $1.4$ and $15.4$~GHz?}

To search for a possible change of $k$ with frequency, following
\cite{1998A&A...330...79L}, we introduce a measure of core
position offset
$$
\Omega_{r\nu} \propto \frac{\Delta r_\mathrm{mas}}{\nu_1^{-1/k}-\nu_2^{-1/k}}
$$
between two frequencies ($\nu_1$, $\nu_2>\nu_1$), where $\Delta r_\mathrm{mas}$
is the difference in the apparent core positions at $\nu_1$ and $\nu_2$.
$\Omega_{r\nu}$ can be used for assessing the possible variations of $k$.
If two values $\Omega_{r\nu1,2}$ (measured between frequencies $\nu_1$ and $\nu_2$) and
$\Omega_{r\nu2,3}$ (between $\nu_2$ and $\nu_3$) are different, the relation between
the corresponding $k_{1,2}$ and $k_{2,3}$ is
$$
k_{2,3}\simeq k_{1,2}\frac{\log\Omega_{r\nu1,2}}{\log\Omega_{r\nu1,3}}\,.
$$
Assuming $k_{1,2}=1$ for the lowest pair of frequencies ($\nu_1=1.408$~GHz and
$\nu_2=1.662$~GHz for most cases), we derive the values of $k_{2,3}$,
$k_{2,4}$, and so on. If the assumption about $k=1$ at $L$ band is not
correct, that would affect the resulting values of $k$; however, the test
will still be sensitive to relative changes in $k$ between frequencies.
The obtained values of $k$ were consistent with $k=1$ at a $3\sigma$ 
level. Therefore we found no significant changes in $k$ with frequency.

      The applicability of the $r_c \propto \nu^{-1/k}$ relation to many extragalactic
sources has important consequences for the radio--optical reference frame
alignment. In the particular case of $k = 1$, if the radio source is
strongly core-dominated, its position measured using the group delay
technique at any frequency (such as those currently used to define the the International
Celestial Reference Frame, ICRF,
\citealt{1998AJ....116..516M,2004AJ....127.3587F}, and ICRF2, \citealt{ICRF2})
corresponds to the jet base, not to the radio core position at a given
frequency \citep{2009A&A...505L...1P}. This is due to the additional time delay 
introduced by the core shift across the observing band \citep{2009A&A...505L...1P}. 
The optical emission of an AGN jet is expected to originate much closer to
the jet base then the radio core position. Therefore, if the structure
of the radio source is dominated by the core, there is no need to
introduce core shift corrections to the radio source positions
determined from group delays before matching them with optical positions.
This is not true, however, for radio source positions determined from 
phase referencing \citep{2009A&A...505L...1P}. If a radio source has distinct jet
components comparable in brightness to the core (such as many sources
considered here; Fig.~\ref{images}), the group delay would not be a good
estimator of the jet base position. The source structure needs to be
taken into account prior to comparing the radio and optical source
positions in such cases.

\subsection{Fitting results assuming opacity due to synchrotron self-absorption}

As discussed above, our observations are consistent with the assumption of
synchrotron self-absorption dominated opacity across the frequency range
of this VLBI experiment. We repeated the fit of the
function $r_c(\nu) = a + b \nu^{-1/k}$ to the observational data, fixing the
value of $k = 1$, which is our best estimate. 
The results are presented in Table~\ref{tab:fit_results}
and Fig.~\ref{fig:coreshiftplots}. The values of $a$ and $b$ are given in
Table~\ref{tab:fit_results}. The projected core distance in parsecs is
$$r_c(\nu)~\mathrm{[pc]} = \frac{D_\mathrm{A}}{N_\mathrm{rad}}
\left(a~\mathrm{[mas]} +
b~\mathrm{[mas} \cdot \mathrm{GHz]} \cdot
\frac{1+z}{\nu_\mathrm{em}~\mathrm{(GHz)}}\right)\,,$$ 
where $a~\mathrm{[mas]}$ and $b~\mathrm{[mas} \cdot \mathrm{GHz]}$ are the
coefficients obtained from observations, $D_A$ is the angular size distance to
the source in parsecs, $N_\mathrm{rad} \simeq 206264800$ is the number of milliarcseconds in one radian, 
$z$ the source redshift, and $\nu_\mathrm{em}~\mathrm{(GHz)}$ the emission frequency in the source frame.

\begin{figure}[hbt!]
 \centering
 \includegraphics[width=0.5\textwidth,angle=0,trim=1.5cm 0cm 1.5cm 0cm,clip]{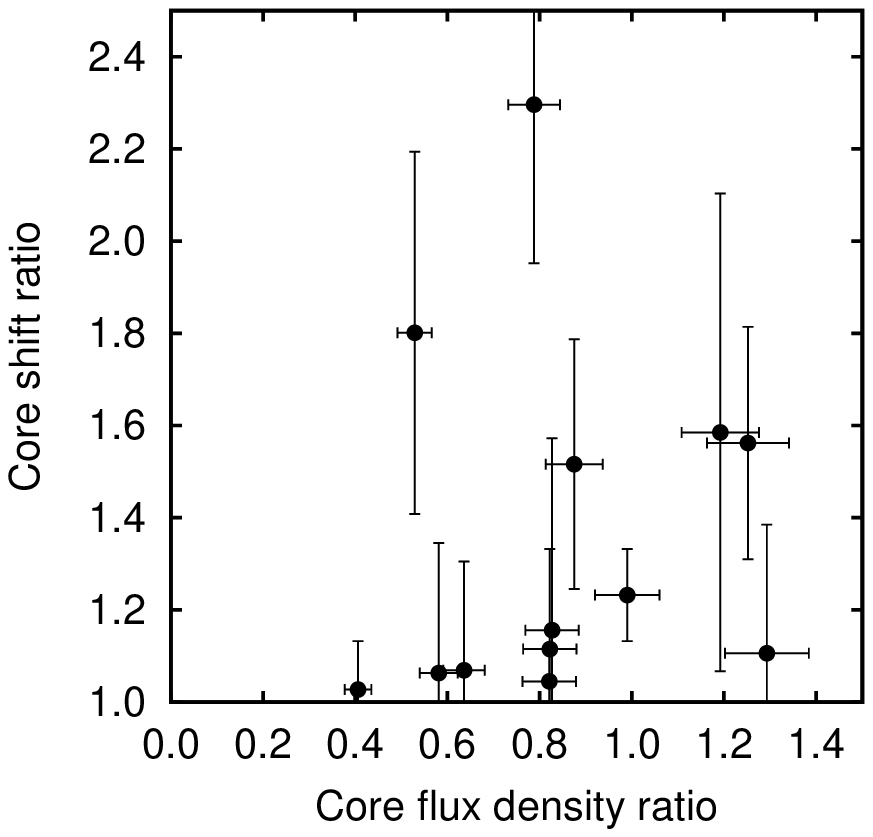} % zoom
 \caption{Ratio of the $X$--$S$ band core position shift measured at 2002 by
\cite{2008A&A...483..759K} to the one measured in 2007 (this paper) as a
function of the $X$ band core flux density ratio at these two epochs.
The core shift ratio is defined to be greater than unity, i.e.,
for each pair of measurements, the higher value of the shift is in the
numerator of the ratio. The one source (not shown in this plot), which
exhibited a large core shift change without a major change in flux 
density is W~Com (1219$+$285); it has a core shift ratio of 
$3.67 \pm 0.36$ and core flux ratio of $1.19 \pm 0.08$.}
 \label{fig:fluxes_shifts}
\end{figure}

We have investigated deviations of the measured distances from the
best-fit curve presented in Fig.~\ref{fig:coreshiftplots} for each
frequency. In all cases the mean difference between the measurement and the
model curve is consistent with zero, and its standard deviation is consistent 
within a factor of two with the adopted typical measurement uncertainties.

It can be seen from Fig.~\ref{fig:coreshiftplots} that, 
while the $r_c(\nu) \propto \nu^{-1}$ law generally provides a good fit to the
data over the whole observed frequency range with no systematic deviations, 
there are indications of possible local deviations.
This may be due to local deviations from equipartition or conical 
shape of the jet.
Another possibility is that the scatter in distance measurements at two
sub-bands used for error estimation does not reflect the total
measurement uncertainty. This may happen if there is a systematic
factor that influences distance measurements at both sub-bands 
in a similar way. However, after a careful review of our analysis
procedures we could not identify any such factor.

\subsection{Search for core shift variability}
\label{s:var}

The new measurements of core shifts between $2.3$ and $8.4$~GHz may be
compared to the previous measurements at similar frequencies obtained using
the RDV data by \cite{2008A&A...483..759K}. As expected for a big enough
sample of variable sources, the mean difference
between the measurements obtained at two epochs (2002 for RDV global VLBI 
data and 2007 for this VLBA dataset) is consistent with zero:
$0.01 \pm 0.08$~mas. The mean absolute value of this difference
is $0.22 \pm 0.05$~mas, 
which is a measure of a typical long-term core-shift variability in
the studied sample within the accuracy of our measurements.

According to formula (1) in \cite{2008A&A...483..759K}, if jet
orientation and velocity remain unchanged, an increasing value of core shift
(due to an increase in particle density and/or jet magnetic field strength)
should be associated with an increase in the flux density of the core.
Figure~\ref{fig:fluxes_shifts} shows the core shift ratio versus $X$ band core flux density ratio 
at two epochs (2002 and 2007). No significant correlation is found
between the core shift and flux ratios at different epochs. This may
reflect that there is no single dominant physical reason for
core flares and subsequent core shift variations, despite the suggestion
by \cite{2008A&A...483..759K}.
A higher accuracy of core-shift ratio measurements is required to probe
this in detail.

%__________________________________________________________________

\section{Summary}
\label{summary}

% What was it?
      Dedicated VLBA observations of twenty AGN jets showing large
frequency-dependent core shifts were conducted at nine frequencies in the $1.4$--$15.4$~GHz range.
      Significant core shifts have been detected and confirmed in all observed sources. The
effect is more pronounced at lower frequencies. The median value of the
core shift in the observed sources is $1.21$~mas if measured between $1.4$ and $15.4$~GHz and
$0.24$~mas between $5.0$ and $15.4$~GHz.
      The core position, $r_c$, shift as a function of frequency, $\nu$, is consistent
with the $r_c(\nu) \propto \nu^{-1}$ pattern expected for a purely synchrotron self-absorbed
conical jet in equipartition \citep{1979ApJ...232...34B,1998A&A...330...79L}.
      The mean value of the coefficient $k$ in the $r_c(\nu) \propto \nu^{-1/k}$
relation was found to be $k = 0.99 \pm 0.14$, while the median is $0.82$.
      These results support the interpretation of the parsec-scale
core as a continuous Blandford-K\"onigl type jet with smooth gradients of
physical properties (including opacity) along it.
      No systematic change with frequency of the power law index in the
$r_c(\nu)$ relation has been convincingly detected. However, some local
changes might be present in a few sources, especially at higher
frequencies.

% Implications
      The general applicability of the $r_c(\nu) \propto \nu^{-1}$ relation to all the
observed sources is a promising indication that, if a radio source is
strongly core-dominated, its positions
obtained using group delays (such as those currently used to define the ICRF2)
may be compared to optical counterpart positions directly, with no need
to apply a correction for the source core shift \citep{2009A&A...505L...1P}.
This may be important in the era of the future space-based optical
astrometry missions aiming for $\mu$as-level accuracy. A correction for the
core shift in the reference source(s) is still required if a
high-precision absolute position of a celestial radio source (or tracked spacecraft) 
is to be determined through a phase-referencing VLBI experiment.

% Future work?
      While in this paper we have tried to concentrate on the observational
results, independent of specific assumptions about jet geometry and 
Lorentz factor, in future work based on the dataset presented
here we plan to estimate geometry, 
magnetic field strength, and total (kinetic plus magnetic field)
power of the flows and to relate the observed shifts to
the properties of the central black hole and broad-line region in its
vicinity. Polarization information will be used to constrain 
the physical interpretation of core shift measurements and to   
investigate the effect of the core shift on Faraday rotation measurements.

It would be important to confirm and investigate possible changes in 
the coefficient $k$ in the $r_c(\nu) \propto \nu^{-1/k}$ relation with
frequency, which may be hinted at in our observations. To
achieve greater accuracy in measuring $k$ one would need to 
obtain more independent core position measurements across the whole
frequency range accessible to VLBI. The use of phase-referenced
observations may also be helpful for minimizing ambiguities inevitably
associated with the model fitting procedure and quantifying 
position-measurement errors more accurately.

Calibrated $uv$-data and models described in this paper are available 
from the first author (KVS) by request.

\begin{acknowledgements}
We thank Richard Porcas and the anonymous referee
for valuable comments that helped to improve
the manuscript.
KVS is supported by the International Max-Planck Research School
(IMPRS) for Astronomy and Astrophysics at the universities of Bonn  
and Cologne. YYK was supported in part by the return fellowship of  
the Alexander von Humboldt foundation and the Russian Foundation for
Basic Research (RFBR) grant 08-02-00545 and 11-02-00368.
This work is based on data
obtained from the National Radio Astronomy Observatory's Very Long 
Baseline Array (VLBA), project BK\,134. The National Radio Astronomy
Observatory is a facility of the National Science Foundation
operated under cooperative agreement by Associated Universities, Inc.
This research has made use of the NASA/IPAC Extragalactic Database (NED),
which is operated by the Jet Propulsion Laboratory, California Institute
of Technology, under contract with the National Aeronautics and Space
Administration. 
This research has made use of NASA's Astrophysics Data System.
\end{acknowledgements}

\end{document}